\documentclass[aps,prb,twocolumn,amsmath,amssymb,superscriptaddress,floatfix,eqsecnum]{revtex4-1}

\usepackage{amsfonts}
\usepackage{cases}
\usepackage{url}
\usepackage{framed}
\usepackage{cancel,soul}
\usepackage{graphicx}
\usepackage{multirow}
\usepackage[pdftex]{color}
\usepackage{bm}
\usepackage{hyperref}
\usepackage[normalem]{ulem}

\newcommand{\bea}{\begin{eqnarray}}
\newcommand{\eea}{\end{eqnarray}}

\newcommand{\opluslimits}{\operatornamewithlimits{\bigoplus}}

\begin{document}

\title{
Non-Abelian topological spin liquids from arrays of quantum wires or spin chains
      }

\author{Po-Hao Huang }
\affiliation{Department of Physics, Boston University, 
Boston, MA, 02215, USA}

\author{Jyong-Hao Chen}
\affiliation{Condensed Matter Theory Group, Paul Scherrer Institute, 
CH-5232 Villigen PSI, Switzerland}

\author{Pedro R. S. Gomes}
\affiliation{Department of Physics, Universidade Estadual de Londrina, 
Caixa Postal 10011, 86057-970, Londrina, PR, Brasil}

\author{Titus Neupert}
\affiliation{Princeton Center for Theoretical Science, 
Princeton University, Princeton, New Jersey 08544, USA}

\author{Claudio Chamon}
\affiliation{Department of Physics, Boston University, 
Boston, MA, 02215, USA}

\author{Christopher Mudry}
\affiliation{Condensed Matter Theory Group, Paul Scherrer Institute, 
CH-5232 Villigen PSI, Switzerland}

\date{\today}

\begin{abstract}
We construct two-dimensional non-Abelian topologically ordered
states by strongly coupling arrays of one-dimensional quantum wires
via interactions. In our scheme, all charge degrees of freedom are
gapped, so the construction can use either quantum wires or 
quantum spin chains as building blocks, with the same end result. The
construction gaps the degrees of freedom in the bulk, while leaving
decoupled states at the edges that are described by 
conformal field theories (CFT) in $(1+1)$-dimensional space and time. 
We consider both the cases where
time-reversal symmetry (TRS) is present or absent. When TRS is
absent, the edge states are chiral and stable. We prescribe, in
particular, how to arrive at all the edge states described by the unitary
CFT minimal models with central charges $c<1$. 
These non-Abelian spin liquid states
have vanishing quantum Hall conductivities, but non-zero thermal
ones. When TRS is present, we describe scenarios where the bulk
state can be a non-Abelian, non-chiral, and gapped quantum spin liquid,
or a gapless one. In the former case, we find that the edge states
are also gapped. The paper provides a brief review of non-Abelian
bosonization and affine current algebras, with the purpose of being
self-contained. To illustrate the methods in a warm-up exercise, we
recover the ten-fold way classification of two-dimensional
non-interacting topological insulators using the Majorana
representation that naturally arises within non-Abelian
bosonization. Within this scheme, the classification reduces to
counting the number of null singular values of a mass matrix, with
gapless edge modes present when left and right null eigenvectors
exist.
\end{abstract}

\maketitle



\section{Introduction}
\label{sec: Introduction}

\subsection{Motivation and strategy}

Topologically ordered states of matter,~\cite{Wen91a} of which the
fractional quantum Hall effect (FQHE) is the quintessential example,
contain rich elementary excitations. A necessary and sufficient
condition for topological order is argued in
Ref.\ \onlinecite{Oshikawa06} to be the existence of point-like
excitations obeying either Abelian~\cite{Leinaas77,Wilczek82} or
non-Abelian~\cite{Froehlich88,Froehlich90a,Froehlich90b,Rehren90,Froehlich91,Moore91,Wen91b,Kitaev06,Nayak08}
anyonic statistics.

The quantum numbers of the
topological anyon excitations are encoded
by a topological quantum field theory (TQFT) in the bulk. 
The type of TQFT in the bulk can imply the existence of 
gapless degrees of freedom on the edge, 
in the form of a conformal field theory (CFT) in $(1+1)$-dimensional 
space and time. While the bulk-boundary correspondence is not one-to-one, 
certain implications can be formulated. 
For example, the fractional part of the central charge~\cite{Kitaev06} 
of the bulk TQFT
has to match that of the chiral central charge of the edge CFT. 
(Changes by integers can always be obtained by gluing an 
integer-quantum-Hall-type phase to the $(2+1)$-dimensional system, 
which does not change the bulk TQFT.)
Thus, the CFT describing the edge excitations is to some extend 
a diagnostic of the bulk topological order. For
instance, the value taken by the central charge of this CFT is sensitive 
to whether it originates from either an Abelian 
or a non-Abelian topological order.
A non-integer chiral central charge of the edge CFT 
implies non-Abelian topological order in the bulk.

The goal of this paper is to establish that a class of models built
out of itinerant electrons, confined to two-dimensional space, display
non-Abelian topological order upon fine-tuning of finite-range
electron-electron interactions. The strategy that we employ is to
couple a one-dimensional array of quantum wires, each of which
supports a finite density of noninteracting electrons, through
electron tunneling and electron-electron interactions.  Prior to
switching on the electron tunneling and electron-electron interactions,
the electrons can only move ballistically along their hosting wire.
There is no electronic motion in the direction transverse to any given
wire.  The one-dimensional array of quantum wires realizes a CFT in
$(1+1)$-dimensional space and time with a central charge
$c^{\,}_{\mathrm{decoupled}}$ twice the number of wires.  After
switching on the electron tunneling and electron-electron interactions, 
a crossover to two-dimensional physics takes place along which the
noninteracting critical theory flows to a CFT with a central charge
$c^{\,}_{\mathrm{coupled}}$ that is either zero or has a nonvanishing
fractional part depending on whether periodic or open boundary
conditions are imposed when coupling the wires. With periodic
boundary conditions along the chain of wires, the ground state is
separated from all excitations by a gap. With open boundary
conditions along the chain of wires, the residual gapless excitations
are necessarily localized along the left and right terminations of the
chain of wires.

In our scheme, the charge degrees of freedom are gapped. For
this reason, instead of using quantum wires as building blocks, we
could equally as well start with a set of coupled quantum spin
chains. This opens the possibility to engineer two-dimensional
non-Abelian quantum spin liquids using coupled spin chains. We
consider both the cases where time-reversal symmetry (TRS) is present
or absent. The fact that we gap the charge degrees of freedom means
that, even when time-reversal symmetry (TRS) is broken, there is no
quantum Hall conductance, but only a quantum thermal Hall conductance;
this is an example of a non-Abelian chiral spin liquid.

\subsection{Summary of main results}

We employ \emph{non-Abelian}
bosonization in order to construct symmetry protected topological (SPT)
phases and topologically ordered phases of matter out of arrays of
interacting quantum wires for, as the name suggests, non-Abelian
bosonization is ideally suited to construct topological orders that
are characterized by a non-Abelian Lie group.

The logic behind our construction is as follows.  An individual
quantum wire with spinful electrons has (in absence of spin-orbit or
Zeeman couplings) an internal symmetry group 
$U^{\,}_{\mathrm{R}}(2)\times U^{\,}_{\mathrm{L}}(2)$, 
where R and L stands for the right-moving and left-moving modes at 
low energies, respectively. A translationally invariant array
of $N$ such wires has the symmetry group 
$U^{\,}_{\mathrm{R}}(2N)\times U^{\,}_{\mathrm{L}}(2N)$. 
The generators of this group and any of its
subgroups can be associated with current operators, which in turn are
products of electron operators.  Consider any subgroup $H$ of
$U(2N)$. The degrees of freedom that are not singlets under the
subgroup can be removed from $U^{\,}_{\mathrm{R}}(2N)$ and
$U^{\,}_{\mathrm{L}}(2N)$ simultaneously via the interaction
\begin{equation}
\lambda^{\,}_{H} 
\sum_{a}
\hat{J}^{a}_{\mathrm{R}}\,
\hat{J}^{a}_{\mathrm{L}},
\qquad
\lambda^{\,}_{H}>0,
\label{eq: schematic H}
\end{equation}
where $a$ runs over all generators of the subgroup $H$, 
while $\hat{J}^{a}_{\mathrm{R}}$
and $\hat{J}^{a}_{\mathrm{L}}$ 
are the associated current operators formed from the
left-moving and right-moving modes respectively. The resulting theory
will have a reduced number of degrees of freedom associated with the 
group quotient (or coset in short)
$\left[U^{\,}_{\mathrm{R}}(2N)/H^{\,}_{\mathrm{R}}\right]
\times
\left[U^{\,}_{\mathrm{L}}(2N)/H^{\,}_{\mathrm{L}}\right]$. 

When choosing possible subgroups $H$, 
the physical constraint of locality has to be observed. 
If the generators of $H$ involve electronic degrees of
freedom from far apart wires, then $H$ is not admissible. Likewise,
while $H^{\,}_{\mathrm{R}}$ and $H^{\,}_{\mathrm{L}}$ are the same mathematical
subgroup, they need not be realized in the same wires. However, they
need to be realized in nearby wires, as interaction \eqref{eq: schematic H} 
would otherwise represent long-range interactions between the wires.

The above procedure is then iterated using the same subgroup $H$
repeatedly, but each time realized on a different set of wires, until
the symmetry group $U^{\,}_{\mathrm{R}}(2N)\times U^{\,}_{\mathrm{L}}(2N)$ is
completely broken in the bulk. Physically this corresponds to gapping
all the low-energy modes in the bulk. In an array of wires with open
boundary conditions, there may remain a protected group coset with
associated currents that are build exclusively from the degrees of
freedom near the edge. For this procedure to be applicable, $H$ must
be chosen such that $U^{\,}_{\mathrm{R}}(2N)/H^{\,}_{\mathrm{R}}$ still
contains $H^{\,}_{\mathrm{R}}$ (shifted by the appropriate number of wires)
as a subgroup, and likewise for L. This is a fundamental compatibility
condition that has to be obeyed by all the current-current
interactions that are used to gap out degrees of freedom. It is
tantamount to the condition that the respective Hamiltonian terms of
the form\ \eqref{eq: schematic H} commute.

Before embarking on this program,
we choose in Sec.\ 
\ref{sec: The ten-fold way via non-Abelian bosonization}
to employ as a warmup the non-Abelian bosonization technique
to construct the noninteracting SPT phases that constitute the tenfold way
for noninteracting topological insulators and superconductors,
(the tenfold way, in short).$\ $%
\cite{Schnyder09,Kitaev09,Schnyder08,Ryu10}
At first sight, this might seem to overcomplicate matters as the
same result has already been obtained with Abelian bosonization.$\ $\,
\cite{Neupert14} 
However, the essential case of the superconducting class D, stabilized by
$\mathbb{Z}^{\,}_{2}$ fermion parity symmetry only, is at odds with the
$U(1)$ group that is fundamentally associated with Abelian
bosonization. One needs to invoke further arguments to obtain the
desired construction.$\ $\,
\cite{Neupert14}  
With non-Abelian bosonization, the construction
follows rather naturally, as we shall see. 

The symmetry group associated with the mean-field description 
of an array of $N$ (spinless) superconducting wires is 
$O^{\,}_{\mathrm{R}}(2N)\times O^{\,}_{\mathrm{L}}(2N)\sim 
U^{\,}_{\mathrm{R}}(N)\times U^{\,}_{\mathrm{L}}(N)$,
where the right-moving and left-moving electronic degrees of freedom
are each decomposed in two Majorana fermions. Via non-Abelian
bosonization, these degrees of freedom are represented by a
$O(2N)$-valued bosonic matrix field $G(t,x)$ that is a function of
time $t$ and the position $x$ along the wire.  A term
\begin{equation}
\lambda^{\,}_{M} \mathrm{tr}\,(G\,M),
\label{eq: schematic H Majorana}
\end{equation}
parametrized by a constant and real-valued $2N\times 2N$ matrix $M$ gaps out all the
modes that are not in the kernel of $M$. More precisely, the remaining
right-moving Majorana modes correspond to the right eigenspace with
eigenvalue $0$ of $M$, while the remaining left-moving Majoranas are
the left eigenspace with eigenvalue 0. For example, if
\begin{equation}
M=
\begin{pmatrix}
0&\cdots &&0\\
1&\ddots& &\\
&\ddots&&\\
0&\cdots&1&0
\end{pmatrix}
\end{equation}
there remains a single left-moving Majorana mode at the left edge and
a single right-moving Majorana mode at the right edge of the wire
array. This realizes the simplest nontrivial example of an SPT state
in class D, equivalent to a chiral $p$-wave superconductor.$\ $%
\cite{Read00,Ivanov01} 
We discuss all nontrivial examples from the tenfold way using this approach.

We then return
in Sec.\ \ref{sec: Non-Abelian topological order out of coupled wires}
to the main part of the paper, namely to intrinsically interacting 
and topologically ordered states of quantum wires.
For this construction, we consider the subgroup $\cdots U(2k)\times
U(2k')\times U(2k)\times U(2k')\cdots$ of the group $U(2N)$ of all
wires by arranging $k$ and $k'$ wires into a bundle in an alternating
fashion.  Then, the low-energy sector of each bundle is reduced to 
the states generated by the nontwisted affine Lie algebra
$\widehat{su}^{\,}(2)^{\,}_{k}$ [and $\widehat{su}(2)^{\,}_{k'}$ respectively].
This is achieved through current-current interactions from the coset
representation
\begin{equation}
\widehat{su}(2)^{\,}_{k}=
\frac{
\widehat{u}(2k)
     }
     {
\widehat{u}(1)\oplus\widehat{su}(k)^{\,}_{2}
     }.
 \label{eq: SU(2)_k relation}
\end{equation}
The identity (\ref{eq: SU(2)_k relation}) is
valid for any integer $k=1,2,\cdots$.
Here, the $U(1)$ subgroup corresponds to
the total charge of the electron modes in the $k$ consecutive wires of a
bundle. To gap only this subgroup without gapping the charge mode of,
e.g., a single wire, a $(2k)$-body interaction is used. In contrast,
all the remaining interactions of the construction are of two-body
nature. For example, the $\widehat{su}(k)^{\,}_{2}$ subalgebra in
Eq.\ \eqref{eq: SU(2)_k relation} corresponds to $k$ flavors within
each bundle and is gapped by the respective current-current
interactions. 
(The same applies to the other $k'$ flavors within each bundle.)  
While these wire flavors in each bundle can be thought of as a pseudo- 
or isospin degree of freedom, the remaining nontwisted affine Lie algebra 
$\cdots
\widehat{su}(2)^{\,}_{k}\oplus\widehat{su}(2)^{\,}_{k'}\oplus
\widehat{su}(2)^{\,}_{k}\oplus\widehat{su}(2)^{\,}_{k'}\cdots$ 
stems from the physical spin of the electrons in the bundles of wires.

\begin{figure}[t]
\begin{center}
\includegraphics[width=0.49\textwidth]{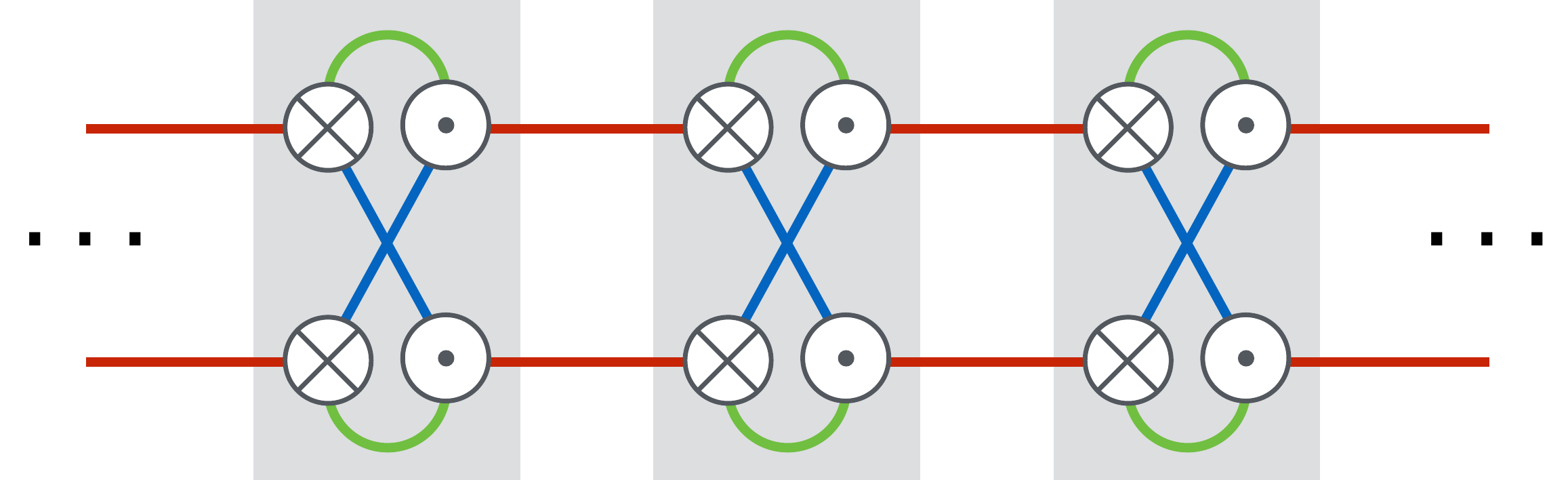}
\caption{(Color online) 
Schematic representation of the Hamiltonian
for a state with non-Abelian topological order arising from
interactions between electronic quantum wires. Each $\otimes$ and
$\odot$ represents the right-moving and left-moving (spinful)
electrons of a quantum wire coming out of the plane of the
page. Each gray area represents an interaction that gaps out charge
fluctuations on the bundle of wires that it encloses. Each line
represents a $SU(2)$ symmetric Heisenberg interaction 
between the spin densities of left-movers and
right-movers that it connects. Lines of the same color represent 
interaction terms of the same strengths.
\label{Fig: SU(2) level 2 example}
         }
\end{center}
\end{figure}

The essential step in our construction consists in coupling these 
coarse-grained $SU(2)$ ``chiral spins'' across the bundles of wires 
in such a way that a pattern of long-range entanglement emerges. 
This is achieved by coupling the right-moving subgroup in one
wire bundle with the left-moving subgroup in the consecutive bundle
with a current-current interaction.
This coupling breaks time-reversal symmetry
and makes our construction chiral.  Our construction thus realizes a
chiral spin liquid, not a fractional quantum Hall state (the Hall
response vanishes). (A one-dimensional array of coupled spin-1/2 chains,
if it is to support such a chiral spin-liquid ground state,
must break time-reversal symmetry either explicitly or spontaneously.)

While gapping all modes in the bulk, 
there remains a right-moving (left-moving) coset-algebra 
$\widehat{su}(2)^{\,}_{k}
\oplus\widehat{su}(2)^{\,}_{k'}/\widehat{su}(2)^{\,}_{k+k'}$ 
on the left (right) edge of the sample. 
It is protected, since it is fully chiral.  
This construction realizes, for different values of $k$ and $k'$, 
edge states associated to different CFTs, with central charges $c^{\,}_{k,k'}$.
In particular, the associated CFTs on the edge include, for $k'=1$, 
all unitary minimal models with central charge
\begin{align}
c^{\,}_{k,1}=&\,
1
-
\frac{6}{(k+2)(k+3)}
\;,
\end{align}
and, for $k'=2$, all superconformal minimal models with central charge
\begin{equation}
c^{\,}_{k,2}=
\frac{3}{2}
-
\frac{12}{(k+2)(k+4)}
\;.
\end{equation}
Figure\ \ref{Fig: SU(2) level 2 example} 
is a schematic illustration of the Hamiltonian 
that leads to the topologically ordered state for $k=k'=1$, 
realizing Ising topological order.  

We emphasize that the large non-Abelian symmetry group 
$U^{\,}_{\mathrm{R}}(2N)\times U^{\,}_{\mathrm{L}}(2N)$
that was invoked prior to coupling the wires should be thought of as a special
limit that allows to use the tools of non-Abelian bosonization. 
It is not the symmetry $U^{\,}_{\mathrm{R}}(2N)\times U^{\,}_{\mathrm{L}}(2N)$
that is protecting the essential topological properties of the phase. 
It is worth noting that our construction
preserves the full $SU(2)$ rotation symmetry of the physical spin.
However, breaking it through the substitutions 
$\lambda^{\,}_{H}\to\lambda^{a}_{H}$ in Eq.\ (\ref{eq: schematic H})
is inconsequential for the stability of the chiral edge states.
Conversely, weakly breaking the full spin-1/2 $SU(2)$ rotation symmetry
prior to coupling the wires is also inconsequential 
for the stability of the chiral edge states. 
(Weakly is defined
relative to the characteristic energy scales involved in our 
$SU(2)$ symmetric construction of a topologically ordered phase.)

In the last parts of 
Sec.\ \ref{sec: Non-Abelian topological order out of coupled wires}, 
we investigate the consequences of imposing a symmetry acting trivially 
in space on such a wire construction. 
We study the case of time-reversal symmetry. One can define a
time-reversal invariant system that is related to the chiral
construction outlined above in one of two ways. 
(I) One adds to the Hamiltonian the time-reversed counterpart 
of each term that is already present. 
(II) One doubles the Hilbert space by invoking an
additional valley degree of freedom that is  exchanged
under reversal of time
and realizes in one valley the chiral construction outlined above and
in the other valley its antichiral partner. Case (I) leads to a
phase transition between two distinct topological ordered states
that cannot be solved using non-Abelian
bosonization. Case (II) is solvable by construction. It realizes a
non-chiral and non-Abelian spin liquid. Since the charge sector is
gapped both in the bulk and on the edges, 
the spin-Hall response vanishes. The edge of this system is
nonchiral, as it hosts the chiral coset CFT in one valley polarization
and the anti-chiral coset CFT in the other valley polarization on one
given edge. It is then imperative to ask to what extend these
edge-modes are stable against local time-reversal symmetric
perturbations at the edge. We find that the edge is not
stable. However, if a certain $U(1)$ symmetry is imposed,
one-body backscattering terms are not sufficient to gap the
edge. We determine the non-Abelian current-current interaction that is
capable of gapping the edge in this case. This result is consistent
with what is known from Abelian wire constructions in the case where a
$U(1)$ subgroup of the $SU(2)$ spin-rotation symmetry is preserved.
Protected edge modes appear only in phases with non-vanishing
spin-Hall conductivity.$\ $%
\cite{Neupert11b}
(See also Ref.\ \onlinecite{Scharfenberger11} 
for a parton construction of non-Abelian
spin liquids that respects time-reversal symmetry.)

\subsection{Comparison with prior works}

Arrays of coupled wires have been applied to many problems 
in statistical and in condensed matter physics.

The multi-channel Kondo effect can be formulated as
an effective array of coupled quantum wires.$\ $%
\cite{Affleck90,Affleck91a,Affleck91b}
We borrow the technology of conformal embedding
from Refs. \onlinecite{Affleck90,Affleck91a,Affleck91b}
in this paper.

Another motivation to study coupled quantum wires stems from the
mystery represented by the pseudogap phase in high-temperature superconductors 
and, more generally, the problem of the breakdown of Fermi liquid theory without
conventional symmetry breaking.~
\cite{Dagotto92,Rice93,Fabrizio93,Dagotto96,Balents96,lin97,Carlson00,%
Essler01,Nersesyan03,Starykh04,Jaefari10}
If each wire is half-filled and decoupled from all other wires, 
the charge sector is gapped while the spin-1/2 degrees of freedom
are gapless. The decoupled array of quantum wires turns into a
decoupled array of quantum spin-1/2 chains. Depending on how these 
spin-1/2 chains are coupled, gapped or gapless magnetic phases emerge 
in two and higher dimensions. Moving away from half-filling allows 
to study the correlated hopping of a small density of electrons or holes
in a strongly correlated background of spins.

The bands of quasi-one-dimensional organic conductors 
such as the Bechgaard salts family are 
characterized by the hierarchy of electronic hopping amplitudes 
$t^{\,}_{a}\gg t^{\,}_{b}\gg t^{\,}_{c}$
along the orthogonal crystalline axis $a$, $b$, and $c$.
This hierarchy justifies modeling the bands by 
weakly coupled quantum wires. 
In the presence of a uniform magnetic field parallel to 
the $c$ crystalline axis, the strongly nested Fermi surface
is unstable to charge- or spin-density wave instabilities triggered by
umklapp instabilities at a commensurate filling fraction. 
The limit, $t^{\,}_{c}/t^{\,}_{b}=0$ realizes
the integer quantum Hall effect (IQHE).$\ $%
\cite{Poilblanc87,Yakovenko91}
The limit, $t^{\,}_{c}/t^{\,}_{b}\ll1$ realizes
a weak topological insulator in the symmetry class A
from the tenfold way.$\ $%
\cite{Halperin87,Yakovenko91,Moore07}

The critical properties of the plateau transitions 
between two consecutive quantized Hall plateaus in the
IQHE are captured by the Chalker-Coddington model 
(see Ref.\ \onlinecite{Chalker88}),
in the limit in which electron-electron interactions are neglected.
It was shown in Ref.\ \onlinecite{Lee94}
how to represent the Chalker-Coddington model as a one-dimensional
array of coupled quantum wires.
More generally, one may assign to any array of quantum wires a transfer matrix 
that maps states that are incoming and outgoing to one end of the wires 
into states that are incoming and outgoing to the other end of the wires.
This is a very useful approach to characterize analytically and numerically
the effects of static disorder on transport along the wires,
ignoring the effects of electron-electron interactions.

Coupling arrays of quantum wires by forward electronic interactions
selects sliding Luttinger Liquid (SLL) phases 
in dimensions larger than one.$\ $%
\cite{Ohern99,Emery00,Vishwanath01,Mukhopadhyay01,Sondhi01}
In two remarkable papers, Refs.\ \onlinecite{Kane02} and \onlinecite{Teo14},
it was shown how to add backward electronic interactions 
in a one-dimensional array of quantum wires so as to gap the SLL phases and
stabilize Abelian and non-Abelian fractional quantum Hall states,
respectively, instead (see also Refs.\  
\onlinecite{Klinovaja14a,Meng14a,Klinovaja14b,Sagi14,Vaezi14,%
Neupert14,Meng14b,Klinovaja15,Santos15,Sagi15}
for one-dimensional arrays of coupled quantum wires and
Ref.\ \onlinecite{Meng15b} 
for a two-dimensional array of coupled quantum wires
stabilizing long-ranged entangled phases of fermionic matter).
Common to all these papers is the fact that only electron-electron
interactions are considered, contrary to the models from 
Refs.\ \onlinecite{Feiguin07,Gils09,Ludwig11,Poilblanc11,%
Mong14,Sahoo15,Hutter15}
in which the fundamental constituents are fractionalized fermions 
(such as Majorana fermions) subject to interactions.

What distinguishes our work from Ref.\ \onlinecite{Teo14}
and ensuing papers is that we do not rely on the charge sector 
of the quantum wires to stabilize a non-Abelian topologically ordered phase.
In Ref.\ \onlinecite{Teo14}, the electrons are spin polarized
by a strong uniform magnetic field, the filling fraction is fine tuned
to the magnitude of the applied magnetic field. Here
and as was done in Ref.\ \onlinecite{Meng15a} when deriving
Abelian and the $SU(2)$ level $k$ Read-Rezayi 
chiral spin liquids from arrays of quantum wires,
we gap the charge sector from the outset 
by breaking translation invariance explicitly
if necessary (i.e., if the filling fraction is not commensurate to the
one-dimensional Fermi wave number), 
leaving only the spin-1/2 degrees of freedom
in the low-energy sector of the theory. 
The non-Abelian topologically ordered 
phase is then selected by fine-tuned spin-spin interactions. 
When these interactions break time-reversal symmetry, 
the non-Abelian topologically ordered 
phase should be compared to the 
Abelian (see Refs.\ \onlinecite{Kalmeyer87,Wen89,Mudry89,Schroeter07,%
Gong14,Bauer14,He15a,Hu15,Gong15,%
Gorohovsky15,Kumar15,Mei14,Mei15,Cincio15,Bieri15,He15b,Hu15})
and non-Abelian (see Refs.\ \onlinecite{Greiter09,Greiter14,Behrmann15})
chiral spin-liquid states that have been
proposed for diverse two-dimensional lattices.

Common to Ref.\ \onlinecite{Teo14} is the belief that
deriving topological ordered states from coupled wires
is useful. First, it provides an intuitive
bridge between the abstract description of topological order
in terms of topological quantum field theories
(see Ref.\ \onlinecite{Wen15}
and references therein)
on the one hand,
and exactly solvable models that are designed from wave functions
or lattice models that can only be studied numerically, 
on the other hand.
Second, it opens the door for engineering materials supporting
topological order.

\section{Review of Non-Abelian bosonization and current algebras}
\label{sec: Review of Non-Abelian bosonization and current algebras}

In order to keep this paper reasonably self-contained, we begin
with a review on non-Abelian bosonization, including non-Abelian
current algebras, which will be of major utility in deriving the
main results of the paper in Sec.\ 
\ref{sec: Non-Abelian topological order out of coupled wires}.  
The reader who is fluent with non-Abelian bosonization is welcome to skip 
this brief summary and may jump to Sec.\ 
\ref{sec: The ten-fold way via non-Abelian bosonization}, 
where, as a warmup exercise,
we rederive the ten-fold classification of topological insulators in
two-dimensional space using the tools here reviewed. 
A particular aspect in this section that is original is how to determine 
the presence of gapless edge modes in systems with boundaries, 
where we introduce a mass matrix whose null singular values signal gapless modes. 
Moreover, the left and right edge modes appear as left and right null
eigenvectors of the mass matrix.

\medskip \medskip

\subsection{Affine Lie algebras}

Non-Abelian bosonization is intimately related to
affine Lie algebras. Affine Lie algebras are generalizations
of Lie algebras.$\ $ %
\cite{DiFrancesco97}
One of the Lie algebras with which physicists
are most familiar is that associated to
the total angular momentum operator $\hat{\bm{J}}$, i.e.,
\begin{equation}
[\hat{J}^{a},\hat{J}^{b}]=
\sum_{c=1}^{3}
\mathrm{i}
\epsilon^{abc}\,
\hat{J}^{c},
\qquad
a,b=1,2,3,
\label{eq: su(N) Lie algebra}
\end{equation}
where we have set the Planck constant $\hbar$ to unity.
The Levi-Civita symbol $\epsilon^{abc}$,
the fully antisymmetric rank three tensor,
is an example of the structure
constants of a Lie algebra. 
The three components of the total angular momentum operator
$\hat{\bm{J}}$
are the generators of the Lie algebra 
(\ref{eq: su(N) Lie algebra}).
This Lie algebra is denoted by $su(2)$,
for the operator 
$\exp(\mathrm{i}\bm{\alpha}\cdot\hat{\bm{J}})$
represents an element 
of the unitary group $SU(2)$
parametrized by the 
vector $\bm{\alpha}\in\mathbb{R}^{3}$.

More generally, a Lie algebra $\mathfrak{g}$ is a vector space
equipped with a binary operation denoted
$[\cdot,\cdot]$ that is called the Lie bracket.
The Lie bracket is a mapping from 
$\mathfrak{g}\times \mathfrak{g}\to \mathfrak{g}$
such that it is 
(i) antisymmetric under interchange of its two entries,
(ii) linear in both entries, 
and (iii) satisfies
the Jacobi identity
\begin{equation}
[X,[Y,Z]]+
[Y,[Z,X]]+
[Z,[X,Y]]=0
\end{equation}
for any $X,Y,Z\in\mathfrak{g}$.

A Lie algebra can be specified by a set of generators
$\hat{J}^{a}$ with $a=1,\cdots,\mathrm{dim}\,\mathfrak{g}$
that are Hermitian operators obeying
the relations
\begin{equation}
[\hat{J}^{a},\hat{J}^{b}]=
\sum_{c=1}^{\mathrm{dim}\,\mathfrak{g}}
\mathrm{i}
f^{ab}_{\hphantom{ab}c}\,\hat{J}^{c}
\end{equation}
for $a,b=1,\cdots,\mathrm{dim}\,\mathfrak{g}$.
The number $\mathrm{dim}\,\mathfrak{g}$ 
of generators is the dimension of the algebra.
The numbers $f^{ab}_{\hphantom{ab}c}$ are real valued
and can be chosen to be antisymmetric under interchange of $a$ and $b$
by virtue of the fact that the Lie bracket is antisymmetric
under exchanging $a$ with $b$. 

A subset $\mathfrak{h}$ of the Lie algebra $\mathfrak{g}$ 
is called a Lie subalgebra
if this subset is closed under the Lie bracket, i.e., if
$[\mathfrak{h},\mathfrak{h}]\subset\mathfrak{h}$. 
A Lie subalgebra $\mathfrak{h}$ of $\mathfrak{g}$ is an ideal
if it satisfies the
stronger constraint that $[\mathfrak{g},\mathfrak{h}]\subset\mathfrak{h}$. 
The null vector and $\mathfrak{g}$ itself
are trivially ideals. A proper ideal of $\mathfrak{g}$ 
is an ideal that is neither the
null vector nor $\mathfrak{g}$ itself. 
A simple Lie algebra has no proper ideal.
A semisimple Lie algebra is
a direct sum of simple Lie algebras. A semisimple Lie algebra generates
a semisimple Lie group, i.e., a direct product of simple Lie groups.

Let $t$ be any real number and
let $\mathbb{C}[t,t^{-1}]$ denote the set of polynomials of the form 
$\sum_{n\in\mathbb{Z}}p^{\,}_{n}\,t^{n}$
with finitely many nonvanishing complex-valued coefficients $p^{\,}_{n}$.
Let $\mathfrak{g}$ denote a Lie algebra. The loop algebra 
\begin{subequations}
\begin{equation}
\tilde{\mathfrak{g}}:=
\mathfrak{g}\otimes\mathbb{C}[t,t^{-1}]
\end{equation}
is a Lie algebra equipped with the Lie bracket
\begin{equation}
[\hat{J}^{a}_{m},\hat{J}^{b}_{n}]=
\sum_{c=1}^{\mathrm{dim}\,\mathfrak{g}}
\mathrm{i}
f^{ab}_{\hphantom{ab}c}\,
\hat{J}^{c}_{m+n},
\end{equation}
where the short-hand notation
\begin{equation}
\hat{J}^{a}_{m}:=
\hat{J}^{a}\otimes t^{m},
\quad
\hat{J}^{b}_{n}:=
\hat{J}^{b}\otimes t^{n},
\quad
\hat{J}^{c}_{m+n}:=
\hat{J}^{c}\otimes t^{m+n},
\end{equation}
\end{subequations}
was introduced 
for any $a,b=1,\cdots,\mathrm{dim}\,\mathfrak{g}$ and for any 
$m,n\in\mathbb{Z}$.

Introduce the one-dimensional vector space
\begin{equation}
\mathbb{C}\hat{k}:=
\{z\,\hat{k}\ |\ z\in\mathbb{C}\}.
\end{equation}
Introduce the operator
\begin{subequations}
\begin{equation}
\hat{L}^{\,}_{0}:=-t\frac{\mathrm{d}}{\mathrm{d}t}
\end{equation}
acting on the vector space of Laurent polynomials $\mathbb{C}[t,t^{-1}]$
through the operation of commutation with the fundamental rule that
\begin{equation}
\hat{L}^{\,}_{0}\,t^{m}-t^{m}\,\hat{L}^{\,}_{0}=-m\,t^{m}
\end{equation}
for any integer $m$ and define the one-dimensional vector space
\begin{equation}
\mathbb{C}\hat{L}^{\,}_{0}:=
\{z\,\hat{L}^{\,}_{0}\ |\ z\in\mathbb{C}\}.
\end{equation}
\end{subequations}
The algebra
\begin{subequations}
\label{eq: def affine lie algebra}
\begin{equation}
\hat{\mathfrak{g}}:=
\tilde{\mathfrak{g}}
\oplus
\mathbb{C}\hat{k}
\oplus
\mathbb{C}\hat{L}^{\,}_{0}
\end{equation}
with the brackets
\begin{equation}
[\hat{J}^{a}_{m},\hat{J}^{b}_{n}]=
\sum_{c=1}^{\mathrm{dim}\,\mathfrak{g}}
\mathrm{i}
f^{ab}_{\hphantom{ab}c}\,\hat{J}^{c}_{m+n}
+
\hat{k}\,n\,\delta^{ab}\,\delta^{\,}_{m+n,0},
\end{equation}
\begin{equation}
[\hat{J}^{a}_{m},\hat{L}^{\,}_{0}]=
m\hat{J}^{a}_{m},
\end{equation}
and
\begin{equation}
[\hat{J}^{a}_{m},\hat{k}]=0
\end{equation}
\end{subequations}
for any $a,b=1,\cdots,\mathrm{dim}\,\mathfrak{g}$ 
and for any $m,n\in\mathbb{Z}$
is called a nontwisted affine Lie algebra. It is an infinite-dimensional
algebra with the generators $\hat{J}^{a}_{m}$, $\hat{k}$, and $\hat{L}^{\,}_{0}$.

The simplest realization of an affine Lie algebra in physics is that
of the normal modes $\hat{a}^{\dag}_{m}$ and $\hat{a}^{\,}_{n}$ 
of the real-valued Klein-Gordon scalar field in $(1+1)$-dimensional 
Minkowski space and time. These obey the canonical Boson algebra
\begin{equation}
[\hat{a}^{\,}_{m},\hat{a}^{\dag}_{n}]=
\delta^{\,}_{m,n},\qquad
[\hat{a}^{\,}_{m},\hat{a}^{\,}_{n}]=[\hat{a}^{\dag}_{m},\hat{a}^{\dag}_{n}]=0.
\end{equation}
The Heisenberg algebra
\begin{subequations}
\label{eq: heisenberg algebra}
\begin{equation}
[\hat{L}^{\,}_{m},\hat{L}^{\,}_{n}]=
[\hat{R}^{\,}_{m},\hat{R}^{\,}_{n}]=
m\,\delta^{\,}_{m+n,0},
\qquad
[\hat{L}^{\,}_{m},\hat{R}^{\,}_{n}]=0,
\label{eq: heisenberg algebra a}
\end{equation}
for the nonvanishing integers $m$ and $n$ 
follows from the definitions
\begin{align}
&
\hat{L}^{\,}_{n}:=
\begin{cases}
-\mathrm{i}\sqrt{+n}\,\hat{a}^{\,}_{+n},&n>0,
\\
+\mathrm{i}\sqrt{-n}\,\hat{a}^{\dag}_{-n},&n<0,
\end{cases}
\label{eq: heisenberg algebra b}
\\
&
\hat{R}^{\,}_{n}:=
\begin{cases}
-\mathrm{i}\sqrt{+n}\,\hat{a}^{\,}_{-n},&n>0,
\\
+\mathrm{i}\sqrt{-n}\,\hat{a}^{\dag}_{+n},&n<0.
\end{cases}
\label{eq: heisenberg algebra c}
\end{align}
\end{subequations}
The Heisenberg algebra is the affine extension of the $\hat{u}(1)$
algebra generated by the zero mode $\hat{a}^{\,}_{0}$. The eigenvalue
of the central operator $\hat{k}$ is not quantized for an Abelian Lie
group as it depends on the multiplicative factor chosen in
the transformations 
(\ref{eq: heisenberg algebra b})
and
(\ref{eq: heisenberg algebra c}).

\subsection{Free fermion realizations of affine Lie algebras}

We define the  partition function 
\begin{subequations}
\label{eq: generic Majorana theory}
\begin{equation}
Z:=
\int
\mathcal{D}[\chi]\,
e^{-S[\chi]}
\end{equation}
over the Grassmann vector field
$\chi^{\mathsf{T}}\equiv
\begin{pmatrix}
\chi^{\mathsf{T}}_{\mathrm{R}}
&
\chi^{\mathsf{T}}_{\mathrm{L}}
\end{pmatrix}$
with the action
\begin{equation}
S[\chi]:=
\frac{\mathrm{i}}{2}
\int\frac{\mathrm{d}\bar{z}\,\mathrm{d}z}{2} 
\left(
\chi^{\mathsf{T}}_{\mathrm{R}}\,
2\partial^{\,}_{\bar{z}}\,
\chi^{\,}_{\mathrm{R}}
+
\chi^{\mathsf{T}}_{\mathrm{L}}\,
2\partial^{\,}_{z}\,
\chi^{\,}_{\mathrm{L}}
\right)
\end{equation}
\end{subequations}
and the complex coordinates
$\bar{z}=x^{\,}_{1}-\mathrm{i}x^{\,}_{2}$
and
$z=x^{\,}_{1}+\mathrm{i}x^{\,}_{2}$
of the complex plane.
(Choosing $x_1\equiv t$ and $x_2\equiv\mathrm{i}x$
relates the complex plane to 
$(1+1)$-dimensional Minkowski space and time.)
The Grassmann vector field 
$\chi^{\mathsf{T}}_{\mathrm{R}}\equiv
(\chi^{\,}_{\mathrm{R},1}
\cdots 
\chi^{\,}_{\mathrm{R},n})$ 
only depends on $z$, 
it is holomorphic.
The Grassmann vector field 
$\chi^{\mathsf{T}}_{\mathrm{L}}\equiv
(\chi^{\,}_{\mathrm{L},1}
\cdots 
\chi^{\,}_{\mathrm{L},n})$ 
only depends on $\bar{z}$, 
it is antiholomorphic. 
Their components obey the Laurent series expansion,
i.e., the operator product expansion (OPE),
\begin{subequations}
\begin{align}
&
\chi^{\,}_{\mathrm{R},\alpha}(z)\,\chi^{\,}_{\mathrm{R},\beta}(0)=
-\frac{\mathrm{i}}{2\pi}\,\frac{\delta^{\,}_{\alpha\beta}}{z}+\cdots,
\\
&
\chi^{\,}_{\mathrm{L},\alpha}(\bar{z})\,\chi^{\,}_{\mathrm{L},\beta}(0)=
-\frac{\mathrm{i}}{2\pi}\,\frac{\delta^{\,}_{\alpha\beta}}{\bar{z}}+\cdots,
\\
&
\chi^{\,}_{\mathrm{R},\alpha}(z)\,\chi^{\,}_{\mathrm{L},\beta}(0)=
0,
\end{align}
\end{subequations}
for any $\alpha,\beta=1,\cdots,n$.

The theory (\ref{eq: generic Majorana theory}) 
is invariant under the local transformation 
\begin{equation}
\chi^{\,}_{\mathrm{R}}(z)\mapsto 
O^{\,}_{\mathrm{R}}(z)\,\chi^{\,}_{\mathrm{R}}(z),
\qquad
\chi^{\,}_{\mathrm{L}}(\bar{z})\mapsto 
O^{\,}_{\mathrm{L}}(\bar{z})\,\chi^{\,}_{\mathrm{L}}(\bar{z}),
\label{eq: generic fermion theory rotation}
\end{equation} 
where $O^{\,}_{\mathrm{R}}(z)$ and $O^{\,}_{\mathrm{L}}(\bar{z})$ 
are matrix fields belonging to $SO(n)$.

Define the corresponding $so(n)$ Noether currents
\begin{subequations}
\label{eq: def so(n) current}
\begin{equation}
J^{a}(z):=
\mathrm{i}\pi\,(\chi^{\mathsf{T}}_{\mathrm{R}}\, T^{a}\,\chi^{\,}_{\mathrm{R}})(z),
\quad
\bar{J}^{a}(\bar{z}):=
\mathrm{i}\pi\,(\chi^{\mathsf{T}}_{\mathrm{L}}\, T^{a}\,\chi^{\,}_{\mathrm{L}})(\bar{z}),
\label{eq: def so(n) current a}
\end{equation}
where the generators $T^{a}\equiv T^{(rs)}$ [with 
the collective label $a$ representing the ordered pair $(r,s)$ with
$1\leq r<s\leq n$] are $n\times n$ Hermitian matrices
with the components
\begin{equation}
T^{(rs)}_{ij}=
\mathrm{i}
\left(
\delta^{\,}_{r,i}\,
\delta^{\,}_{s,j}
-
\delta^{\,}_{r,j}\,
\delta^{\,}_{s,i}
\right).
\label{eq: def so(n) current b}
\end{equation}
\end{subequations}
It then follows that
\begin{subequations}
\label{eq: OPE so(n) currents}  
\begin{align}
&
J^{a}(z)\,
J^{b}(0)=
\sum_{c}
\frac{\mathrm{i} f^{ab}_{\hphantom{ab}c}\,J^{c}(0)}{z}
+
\frac{1}{2}
\frac{\mathrm{tr}\,(T^{a}\,T^{b})}{z^{2}},
\label{eq: OPE so(n) currents a}  
\\
&
\bar{J}^{a}(\bar{z})\,
\bar{J}^{b}(0)=
\sum_{c}
\frac{\mathrm{i} f^{ab}_{\hphantom{ab}c}\,\bar{J}^{c}(0)}{\bar{z}}
+
\frac{1}{2}
\frac{\mathrm{tr}\,(T^{a}\,T^{b})}{\bar{z}^{2}},
\label{eq: OPE so(n) currents b}  
\\
&
J^{a}(z)\,
\bar{J}^{b}(0)=0,
\label{eq: OPE so(n) currents c}  
\end{align}
where
\begin{align}
f^{ab}_{\hphantom{ab}c}\equiv&\,
f^{\,}_{(rs)(pq)(mn)}
\nonumber\\
=&\,
\delta^{\,}_{m,r}
\left(
\delta^{\,}_{n,q}\,
\delta^{\,}_{s,p}
-
\delta^{\,}_{n,p}\,
\delta^{\,}_{s,q}
\right)
\nonumber\\
&
+
\delta^{\,}_{m,s}
\left(
\delta^{\,}_{r,q}\,
\delta^{\,}_{n,p}
-
\delta^{\,}_{n,q}\,
\delta^{\,}_{r,p}
\right)
\label{eq: OPE so(n) currents d}  
\end{align}
are the structure constants of $so(n)$.
Observe that the choice made in Eq.\ (\ref{eq: def so(n) current b})
implies the normalizations
\begin{equation}
\mathrm{tr}\,(T^{a}\,T^{b})=2\,\delta^{ab},
\qquad
\sum_{a,b}
f^{ab}_{\hphantom{ab}c}\,
f^{ab}_{\hphantom{ab}d}=
2(n-2)\,\delta^{\,}_{cd}.
\label{eq: OPE so(n) currents e}  
\end{equation}
\end{subequations}
Insertion of the Laurent expansions
\begin{equation}
J^{a}(z)=:
\sum_{m\in\mathbb{Z}} 
z^{-m-1}\,J^{a}_{m},
\qquad
\bar{J}^{a}(z)=:
\sum_{m\in\mathbb{Z}} 
\bar{z}^{-m-1}\,\bar{J}^{a}_{m},
\end{equation}
into the operator product expansions 
(\ref{eq: OPE so(n) currents a})
and
(\ref{eq: OPE so(n) currents b}),
respectively,
delivers a pair of a holomorphic and an
antiholomorphic affine Lie algebra of the form
(\ref{eq: def affine lie algebra})
with the central term $\hat{k}$ replaced by its eigenvalue,
the level $k=1$.

We close this discussion of free Majorana fermions
with the definition of their central charge.
Without loss of generality, we work in the holomorphic sector
of the theory. The energy-momentum tensor has the light-cone
component
\begin{subequations}
\label{eq: Virasoro OPE}
\begin{align}
T^{\,}_{\mathrm{R}}(z)\equiv&\,
-2\pi\, T^{\,}_{zz}
\nonumber\\
\equiv&\,
-\frac{\pi}{2}\,
T^{\bar{z}\bar{z}}(z)
\nonumber\\
:=&\,
-\mathrm{i}\frac{\pi}{2}
\sum_{\alpha=1}^{n}
2\,
\left(
\frac{\delta S}
{\delta(\partial^{\,}_{\bar{z}}\,\chi^{\,}_{\mathrm{R},\alpha})}
(\partial^{\,}_{z}\chi^{\,}_{\mathrm{R},\alpha})
\right)(z)
\nonumber\\
=&\,
\mathrm{i}\pi
\left(
\chi^{\mathsf{T}}_{\mathrm{R}}\,
\partial^{\,}_{z}\,
\chi^{\,}_{\mathrm{R}}
\right)(z).
\label{eq: Virasoro OPE a}
\end{align}
Its OPE with itself is
\begin{equation}
T^{\,}_{\mathrm{R}}(z)\,T^{\,}_{\mathrm{R}}(0)=
\frac{c/2}{z^{4}}
+
\frac{2\,T^{\,}_{\mathrm{R}}(0)}{z^{2}}
+
\frac{(\partial^{\,}_{z}T^{\,}_{\mathrm{R}})(0)}{z}
+
\cdots,
\label{eq: Virasoro OPE b}
\end{equation}
where the numerator of the term with the fourth-order pole is
\begin{equation}
c=n/2.
\label{eq: Virasoro OPE c}
\end{equation}
\end{subequations}
The number $c$ is called the central charge associated to
the (holomorphic) Virasoro algebra defined by the OPE 
(\ref{eq: Virasoro OPE b}).

\subsection{Bosonic realizations of affine Lie algebras}

Another example of a critical theory is
the Wess-Zumino-Witten (WZW) model defined by the partition function$\ $ %
\cite{Wess71,Witten84}
\begin{subequations}
\label{eq: generic bosonized theory}
\begin{equation}
Z:=
\int\limits\mathcal{D}[G]\,
e^{-S^{\,}_{\mathrm{WZW}}[G]},
\end{equation}
where
$G\in\mathfrak{G}$ denotes a matrix-valued bosonic field,
$\mathfrak{G}$ denotes a compact Lie group, and
$\mathcal{D}[G]$ denotes the Haar measure on $\mathfrak{G}$.
(We shall denote with
$\hat{\mathfrak{g}}^{\,}_{k}$ the affine Lie algebra of integer
level $k$ corresponding to the compact Lie group $\mathfrak{G}$.) 
The WZW action in two-dimensional Euclidean space 
$(x,y)\equiv(x^{\,}_{i})\in\mathbb{R}^{2}$ is
\begin{equation}
S^{\,}_{\mathrm{WZW}}[G]:= 
\frac{k}{16\pi}
\int\mathrm{d}^{2}x\,
\mathrm{tr}
\left(
\partial^{\,}_{i}G\,\partial^{\,}_{i}G^{-1}
\right)
+
k\,
\Gamma[G].
\end{equation}
(The summation convention over the repeated index $i=1,2$ is implied.)
The topological contribution $\Gamma[G]$ is the Wess-Zumino term 
\begin{equation}
\begin{split}
\Gamma[G]:=&\,
-\frac{\mathrm{i}}{24\pi}\int\limits_{B} 
\mathrm{d}^{3}\xi\, 
\epsilon^{ijk}
\\
&\,
\times
\mathrm{tr}
\Bigl[
(\bar{G}^{-1}\partial^{\,}_{i}\bar{G})\,
(\bar{G}^{-1}\partial^{\,}_{j}\bar{G})
(\bar{G}^{-1}\partial^{\,}_{k}\bar{G})
\Bigr].
\end{split}
\label{eq: generic bosonized theory a}
\end{equation}
\end{subequations}
Here, $\bar{G}$ denotes the extension of $G$ to the solid ball 
$B\equiv\{(\xi^{\,}_{1},\xi^{\,}_{2},\xi^{\,}_{3})|\sum_{i=1}^{3}\xi^{2}_{i}\leq1\}$ 
with two-dimensional Euclidean space as its boundary. 
As explained in Refs.\ \onlinecite{Wess71,Witten84},
$k$ must be an integer for the functional
$\exp(-k\,\Gamma[G])$ 
over the compact Lie group $\mathfrak{G}$ to be single valued.

The theory (\ref{eq: generic bosonized theory}) 
is invariant under the local transformation 
\begin{equation}
G(\bar{z},z)\mapsto L(\bar{z})\,G(\bar{z},z)\,R^{\mathsf{T}}(z),
\label{eq: generic bosonized theory rotation}
\end{equation} 
where $R$ and $L$ are matrices belonging to 
$\mathfrak{G}$ and $\bar{z}=x-\mathrm{i}y$ is the complex conjugate
to $z=x+\mathrm{i}y\in\mathbb{C}$.
The OPE of its Noether currents (with proper normalizations)
delivers a pair of a holomorphic and an antiholomorphic affine Lie algebra 
$\hat{\mathfrak{g}}^{\,}_{k}$ of the form 
(\ref{eq: def affine lie algebra})
with the central term $\hat{k}$ replaced by its eigenvalue,
the level $k$.

The central charge $c$ of the bosonic theory 
(\ref{eq: generic bosonized theory}) is  
\begin{equation}
c=
\frac{
k\,\mathrm{dim}(\mathfrak{G})
     }
     {
k+\mathrm{Coxeter}^{\star}(\mathfrak{G})
     },
\end{equation}
where $\mathrm{dim}(\mathfrak{G})$ is the dimension of the 
compact Lie group $\mathfrak{G}$ 
(the dimensionality of its adjoint representation), 
while
$\mathrm{Coxeter}^{\star}(\mathfrak{G})$
is the dual Coxeter (twice the eigenvalue of the Casimir operator
in the adjoint representation when the squared length of the highest root
is normalized to 2).

If 
\begin{subequations}
\begin{equation}
\mathfrak{G}=\mathfrak{G}^{\,}_{1}\times\mathfrak{G}^{\,}_{2},
\end{equation}
it then follows that
\begin{equation}
c=
\sum_{i=1,2}
\frac{
k\,\mathrm{dim}(\mathfrak{G}^{\,}_{i})
     }
     {
k+\mathrm{Coxeter}^{\star}(\mathfrak{G}^{\,}_{i})
     }.
\end{equation}
\end{subequations}
More generally, denote with $\hat{\mathfrak{g}}^{(i)}_{k^{\,}_{i}}$
the WZW theory of level $k^{\,}_{i}$. The WZW theory with the 
semi-simple affine Lie algebra
\begin{subequations}
\begin{equation}
\hat{\mathfrak{g}}:=
\hat{\mathfrak{g}}^{(i)}_{k^{\,}_{1}}\oplus
\cdots\oplus 
\hat{\mathfrak{g}}^{(i)}_{k^{\,}_{i}}\oplus
\cdots
\end{equation}
has the central charge
\begin{equation}
c=
\sum_{i}
\frac{
k^{\,}_{i}\,\mathrm{dim}(\mathfrak{G}^{(i)})
     }
     {
k^{\,}_{i}+\mathrm{Coxeter}^{\star}(\mathfrak{G}^{(i)})
     }.
\end{equation}
\end{subequations}

There are several ways to make contact between
the critical theory (\ref{eq: generic Majorana theory})
and the critical theory (\ref{eq: generic bosonized theory}).
\\
\noindent
\textit{Example 1:} We do the identifications
\begin{subequations}
\begin{equation}
\mathfrak{G}\to
O(n),
\qquad
k\to1,
\end{equation}
for which
\begin{align}
&
\mathrm{dim}(\mathfrak{G})\to
\frac{1}{2}\,n\,(n-1),
\\
&
\mathrm{Coxeter}^{\star}(\mathfrak{G})\to
n-2,
\\
&
c=
\frac{
\mathrm{dim}(\mathfrak{G})
     }
     {
1+\mathrm{Coxeter}^{\star}(\mathfrak{G})
     }
\nonumber\\
&
\hphantom{c}\,
\to
\frac{
(1/2)\,n\,(n-1)
     }
     {
1+n-2     
     }
\nonumber\\
&
\hphantom{c}\,
=
\frac{1}{2}\,n.
\end{align} 
\end{subequations}
\\
\noindent
\textit{Example 2:} We assume that $n=m\,n'$ and do the identifications
\begin{subequations}
\begin{equation}
\hat{\mathfrak{g}}\to
\oplus_{i=1}^{n'}\hat{\mathfrak{g}}^{\,}_{i},
\qquad
\mathfrak{G}^{\,}_{i}=
O(m),
\qquad
k^{\,}_{i}\to1,
\end{equation}
for which
\begin{align}
&
\mathrm{dim}\,\mathfrak{G}^{\,}_{i}\to
\frac{1}{2}\,m\,(m-1),
\\
&
\mathrm{Coxeter}^{\star}(\mathfrak{G}^{\,}_{i})\to
m-2,
\\
&
c=
\sum_{i=1}^{n'}
\frac{
\mathrm{dim}(\mathfrak{G}^{\,}_{i})
     }
     {
k^{\,}_{i}+\mathrm{Coxeter}^{\star}(\mathfrak{G}^{\,}_{i})
     }
\nonumber\\
&
\hphantom{c}
\to
n'
\frac{
(1/2)\,m\,(m-1)
     }
     {
1+m-2
     }
\nonumber\\
&
\hphantom{c}
=
\frac{1}{2}\,m\,n'.
\end{align} 
\end{subequations}
This result for the central charge can be applied to the cases
of $O(1)$ and $O(2)$ even though $O(1)$ is not a continuous Lie group
while $O(2)$ is an Abelian group.

We choose \textit{Example 1}.
The non-Abelian bosonization rule for any local quadratic term 
made from the Majorana fields 
[$\chi^{\,}_{\mathrm{R}}$ ($\chi^{\,}_{\mathrm{L}}$)
denotes the right-moving (left-moving) $n$-component Majorana
vector field]
is 
\begin{equation}
m^{\,}_{\mathrm{uv}}\,G^{\,}_{\alpha\beta}= 
\mathrm{i}\chi^{\,}_{\mathrm{L},\alpha}\,\chi^{\,}_{\mathrm{R},\beta},
\label{eq: non-abelian bosonization formula}
\end{equation}
for $\alpha,\beta=1,\cdots,n$, 
and where $m^{\,}_{\mathrm{uv}}$ 
is the mass parameter that depends on the regularization scheme
(the ultra-violet cutoff), 
and $G^{\,}_{\alpha\beta}$ is a matrix element of $G$.

The central charge of the $O(n)^{\,}_{1}$ WZW model is
\begin{equation}
c=
\frac{n}{2}.
\end{equation}
It coincides with the central charge for $n$ 
Majorana fermions (\ref{eq: generic Majorana theory}), 
as the central charge of 
a single pair of right- and left-moving Majorana channels is $1/2$.

Recall that the central charge counts the effective degrees of freedom 
at criticality, i.e., the effective number of gapless degrees of freedom. 
Thus, if we add some quadratic mass term into our massless fermionic theory 
(\ref{eq: generic Majorana theory})
so as to break a part of the $O(n)$ symmetry, 
the central charge should then be reduced.

For example, if we add the term  
\begin{equation}
\mathrm{i}\chi^{\,}_{\mathrm{L},1}\,\chi^{\,}_{\mathrm{R},2}= 
m^{\,}_{\mathrm{uv}}\,G^{\,}_{12},
\end{equation} 
then the symmetry 
$O^{\,}_{\mathrm{R}}(n)\times O^{\,}_{\mathrm{L}}(n)$ 
breaks down to 
$O^{\,}_{\mathrm{R}}(n-1)\times O^{\,}_{\mathrm{L}}(n-1)$. 
Correspondingly, the central charge reduces to 
\begin{equation}
c=
\frac{n}{2}
- 
\frac{1}{2}.
\end{equation}
A pair of right- and left-moving Majorana modes 
has become massive.

Observe that
\begin{equation}
\mathrm{i}\chi^{\,}_{\mathrm{L},1}\,\chi^{\,}_{\mathrm{R},2} 
+ 
\mathrm{i}\chi^{\,}_{\mathrm{L},1}\,\chi^{\,}_{\mathrm{R},3} = 
m^{\,}_{\mathrm{uv}}\,
(G^{\,}_{12}+G^{\,}_{13})
\label{eq: mass term example}
\end{equation} 
does not reduce the 
$O^{\,}_{\mathrm{R}}(n)\times O^{\,}_{\mathrm{L}}(n)$ symmetry to 
$O^{\,}_{\mathrm{R}}(n-2)\times O^{\,}_{\mathrm{L}}(n-2)$.
To see this, introduce the matrix $M$
\begin{equation}
G^{\,}_{12} + G^{\,}_{13}=: \mathrm{tr}(M\,G).
\end{equation}
A solution is to choose a matrix with the only nonvanishing matrix elements
$M^{\,}_{21}=M^{\,}_{31}=1$ sitting on the same column.
This is to say that $M$ is constructed out of only one
linearly independent column vector out of $n$ column vectors.
Hence, there must exist two orthogonal matrices $R$ and $L$ such that
\begin{equation}
M^{\,}_{\mathrm{d}}=
R^{\mathsf{T}}\,M\,L
\end{equation}
is a diagonal matrix with one and only one nonvanishing diagonal 
matrix element.
We choose this nonvanishing matrix element to be the first diagonal entry, 
$(M^{\,}_{\mathrm{d}})^{\,}_{11}=\Omega^{\,}_{1}\neq0$. 
While the action (\ref{eq: generic bosonized theory}) 
is invariant under the transformation 
(\ref{eq: generic bosonized theory rotation}), the mass term becomes 
\begin{align}
m^{\,}_{\mathrm{uv}}
\mathrm{tr}(M\,G)
\mapsto&\,
m^{\,}_{\mathrm{uv}}\, 
\mathrm{tr}(M\,L\,G^{\,}\,R^{\mathsf{T}})
\nonumber\\
=&\,
m^{\,}_{\mathrm{uv}}\, 
\mathrm{tr}(R^{\mathsf{T}}\,M\,L\,G^{\,})
\nonumber\\
=&\,
m^{\,}_{\mathrm{uv}}\, 
\mathrm{Tr}(M^{\,}_{\mathrm{d}}\,G^{\,})
\nonumber\\
=&\,
m^{\,}_{\mathrm{uv}}\, 
\Omega^{\,}_{1}\,G^{\,}_{11}
\nonumber\\
=&\,
\mathrm{i}\chi^{\,}_{\mathrm{L},1}\,\chi^{\,}_{\mathrm{R},1}
\end{align}
after the transformation (\ref{eq: generic bosonized theory rotation}).
Hence, the mass term (\ref{eq: mass term example})
reduces the symmetry to 
$O^{\,}_{\mathrm{R}}(n-1)\times O^{\,}_{\mathrm{L}}(n-1)$
and not to 
$O^{\,}_{\mathrm{R}}(n-2)\times O^{\,}_{\mathrm{L}}(n-2)$,  
as might have been erroneously deduced by
identifying the ``$2$'' in $n-2$ with two independent
mass terms.

For an arbitrary mass-matrix $M$, 
we can employ the singular-value decomposition 
\begin{equation}\label{eq: svd}
M^{\,}_{\mathrm{d}}=
R^{\mathsf{T}}\,M\,L,
\end{equation}
to get a diagonal matrix of rank $r$, i.e.,
\begin{equation}
M^{\,}_{\mathrm{diag}}= 
\mathrm{diag}
\left(
\underbrace{\Omega^{\,}_{1},\Omega^{\,}_{2},\cdots,\Omega^{\,}_{r}}^{\,}_{r},
\underbrace{0,0,...,0}^{\,}_{n-r}
\right), 
\end{equation}
whereby
$\Omega^{\,}_{1},\Omega^{\,}_{2},\cdots,\Omega^{\,}_{r}\neq0$.
The symmetry is then reduced from
$O^{\,}_{\mathrm{R}}(n)\times O^{\,}_{\mathrm{L}}(n)$
to
$O^{\,}_{\mathrm{R}}(n-r)\times O^{\,}_{\mathrm{L}}(n-r)$  
with the corresponding central charge 
\begin{equation}
c= 
\frac{n}{2} 
- 
\frac{r}{2}.
\end{equation} 

\begin{figure*}[t]
\begin{center}
\includegraphics[scale=0.5]{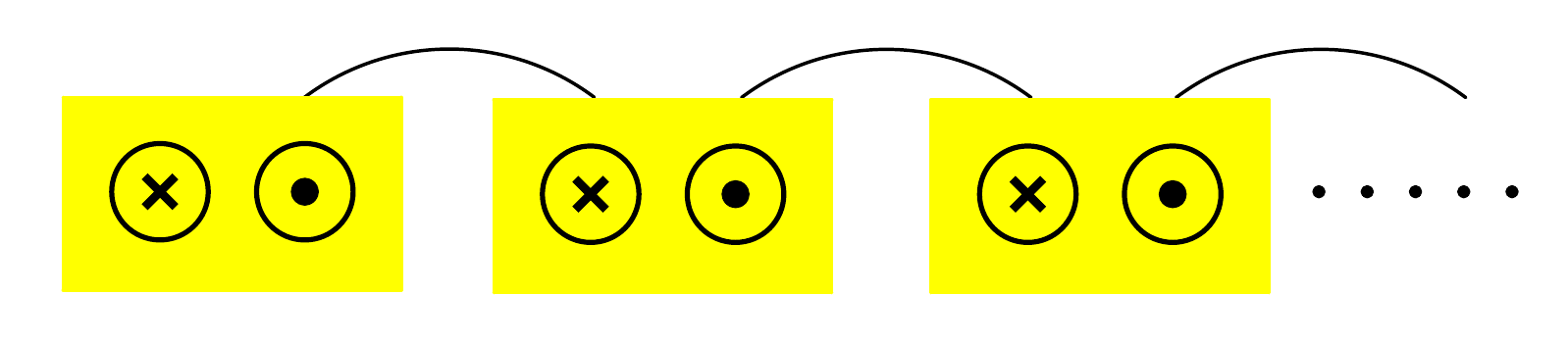}
\caption{(Color online)
Pictorial representation for the selected backscattering 
in the symmetry class D. Each yellow box represents 
a quantum wire composed of two Majorana degrees of freedom. 
The wires are enumerated by $I=1,\cdots,N$ 
in ascending order from  left to right.
For any $I$, the Majorana modes are denoted by
$\chi^{\,}_{\mathrm{R},I}$ and $\chi^{\,}_{\mathrm{L},I}$
reading from left to right, respectively. 
\label{Fig: minimal wire network in D}
         }
\end{center}
\end{figure*}

\subsection{The ten-fold way via non-Abelian bosonization}
\label{sec: The ten-fold way via non-Abelian bosonization}

The goal of this section is to derive the tenfold way in two-dimensional
space by modeling two-dimensional space as an array of wires on which
noninteracting degrees of freedom 
(i) obey the Majorana algebra,
(ii) propagate freely along any wire, 
(iii) while they can hop between consecutive wires.
The novelty in deriving the tenfold way is that we shall use non-Abelian
bosonization techniques, and apply the singular value decomposition on 
the mass matrix, as described above, 
to count the number of gapless edge modes.

We shall consider the symmetry classes D and DIII that, 
together with the symmetry classes C, A, and AII, 
correspond to the topological superconductors and
insulators in two-dimensional space from the tenfold way.$\ $%
\cite{Schnyder09,Kitaev09,Schnyder08,Ryu10}
The symmetry classes C, A, and AII are treated in 
Appendix \ref{appsec: Non-Abelian bosonization for the symmetry classes}.

\subsubsection{The symmetry class D}
\label{subsec: The symmetry class D}

We shall use a path integral representation of the array of quantum wires.
There will be $2MN$ independent Grassmann variables
$\chi^{\,}_{\alpha,\mathrm{f},I}$, where $\alpha=\mathrm{R},\mathrm{L}$ distinguish 
a right- from a left-mover, $\mathrm{f}=1,\cdots,M$ is a flavor index,
and $I=1,\cdots,N$ enumerates the wire.

The simplest model for an array of quantum wires in the symmetry class D
to realize a topological gapped phase assumes
\begin{subequations}
\label{eq: free Majorana class D}
\begin{equation}
M=1,
\qquad
\chi^{\,}_{\alpha,I}(t,x),
\label{eq: free Majorana class D a}
\end{equation}
for $\alpha=\mathrm{R},\mathrm{L}$
and
$I=1,\cdots,N$.
We have thus assigned a pair of Majorana fermions to each wire 
$I=1\,\cdots\,N$. We define the action
\begin{equation}
S^{(\mathrm{D})}_{0}:= 
\int\mathrm{d}t\int\mathrm{d}x\,
\mathcal{L}^{(\mathrm{D})}_{0}
\label{eq: free Majorana class D b}
\end{equation}
with 
\begin{equation}
\mathcal{L}^{(\mathrm{D})}_{0}:=
\frac{\mathrm{i}}{2}
\sum_{I=1}^{N}
\Bigl[
\chi^{\,}_{\mathrm{R},I}
(\partial^{\,}_{t} + \partial^{\,}_{x}) 
\chi^{\,}_{\mathrm{R},I}
+ 
\chi^{\,}_{\mathrm{L},I}
(\partial^{\,}_{t} - \partial^{\,}_{x}) 
\chi^{\,}_{\mathrm{L},I}
\Bigr].
\label{eq: free Majorana class D c}
\end{equation}
We also define the Grassmann partition function
\begin{equation}
Z^{(\mathrm{D})}_{0}:=
\int\mathcal{D}[\chi]\,e^{+\mathrm{i}S^{(\mathrm{D})}_{0}}.
\label{eq: free Majorana class D d}
\end{equation}
\end{subequations}

The theory with the partition function $Z^{(\mathrm{D})}_{0}$
is critical, for there are $2N$ decoupled massless Majorana modes
that are dispersing in $(1+1)$-dimensional Minkowski space and time.
Hence, the central charge $c^{(\mathrm{D})}_{0}$ for the
partition function $Z^{(\mathrm{D})}_{0}$ is 
\begin{subequations}
\begin{equation}
c^{(\mathrm{D})}_{0}=\frac{N}{2}.
\end{equation} 
The partition function $Z^{(\mathrm{D})}_{0}$
is invariant under any local linear transformation
$(O^{(\mathrm{R})},O^{(\mathrm{L})})\in
O^{\,}_{\mathrm{R}}(N)\times O^{\,}_{\mathrm{L}}(N)$
defined by the fundamental rule
\begin{equation}
\begin{split}
&
\chi^{\,}_{\mathrm{R}}(t-x)\mapsto
O^{(\mathrm{R})}(t-x)\,
\chi^{\,}_{\mathrm{R}}(t-x),
\\
&
\chi^{\,}_{\mathrm{L}}(t+x)\mapsto
O^{(\mathrm{L})}(t+x)\,
\chi^{\,}_{\mathrm{L}}(t+x).
\end{split}
\end{equation}
The partition function $Z^{(\mathrm{D})}_{0}$
is also invariant under the antilinear transformation
with the fundamental rule
\begin{equation}
\begin{split}
&
\chi^{\,}_{\mathrm{R}}(t,x)\mapsto
\chi^{\,}_{\mathrm{L}}(-t,x),
\quad
\chi^{\,}_{\mathrm{L}}(t,x)\mapsto
\chi^{\,}_{\mathrm{R}}(-t,x),
\label{eq: critical properties class D minimal c}
\end{split}
\end{equation}
\end{subequations}
that implements reversal of time in such a way
that it squares to the identity
(see Appendix \ref{Appsec: Reversal of time}). 
Even though reversal of time 
(\ref{eq: critical properties class D minimal c})
is a symmetry of the partition function $Z^{(\mathrm{D})}_{0}$,
we shall not impose invariance under
reversal of time (\ref{eq: critical properties class D minimal c})
for a generic representative of the symmetry class D. 

Any partition function $Z^{(\mathrm{D})}$
for the array of quantum wires
is said to belong to the symmetry class D
if $Z^{(\mathrm{D})}$ is invariant under the linear transformation
(fermion parity) with the fundamental rule
\begin{equation}
\chi^{\,}_{\alpha}\mapsto
-\chi^{\,}_{\alpha},
\label{eq: parity conservation D}
\end{equation}
for $\alpha=\mathrm{R},\mathrm{L}$.

We seek a local single-particle perturbation 
$\mathcal{L}^{(\mathrm{D})}_{\mathrm{mass}}$
that satisfies three conditions when added to 
the Lagrangian density (\ref{eq: free Majorana class D c}).

\textbf{Condition D.1}
It must be invariant under the transformation
(\ref{eq: parity conservation D}).

\textbf{Condition D.2}
It must gap completely the theory with the partition function 
$Z^{(\mathrm{D})}_{0}$
if we impose the periodic boundary conditions
\begin{equation}
\chi^{\,}_{\alpha,I}(t,x)=
\chi^{\,}_{\alpha,I+N}(t,x),
\end{equation} 
for 
$\alpha=\mathrm{R},\mathrm{L}$
and
$I=1,\cdots,N$.

\textbf{Condition D.3}
The partition function $Z^{(\mathrm{D})}$
with the Lagrangian density
$\mathcal{L}^{(\mathrm{D})}_{0}+\mathcal{L}^{(\mathrm{D})}_{\mathrm{mass}}$
must be a theory with the central charge 
\begin{equation}
c^{(\mathrm{D})}=\frac{1}{2}
\end{equation}
if open boundary condition are imposed.

\textbf{Conditions D.1},
\textbf{D.2},
and 
\textbf{D.3}
imply that we may assign wire $I=1$ the left-chiral central charge
$1/2$ and wire $I=N$ the right-chiral central charge $1/2$.
For example,
if wire $I=1$ supports a right-moving (i.e., chiral) Majorana edge mode,
then wire $I=N$ supports a left-moving (i.e., chiral) Majorana edge mode.

We make the Ansatz
\begin{equation}
\mathcal{L}^{(\mathrm{D})}_{\mathrm{mass}}:=
\mathrm{i}\lambda\,
\sum_{I=1}^{N-1}
\chi^{\,}_{\mathrm{L},I}\,\chi^{\,}_{\mathrm{R},I+1}
\label{eq: Ansatz for single-particle class D} 
\end{equation}
with $\lambda$ a real-valued coupling.
To establish that the Ansatz
(\ref{eq: Ansatz for single-particle class D})
meets \textbf{Conditions} \textbf{D.2} and \textbf{D.3}, 
we use non-Abelian bosonization.
We choose the non-Abelian bosonization scheme by which
the partition function is given by the path integral 
\begin{subequations}
\begin{equation}
Z^{(\mathrm{D})}=
\int\mathcal{D}[G]\,e^{\mathrm{i}S^{(\mathrm{D})}}.
\end{equation}
The field $G\in O(N)$ is a matrix of bosons.
The measure $\mathcal{D}[G]$ is constructed from the Haar measure on $O(N)$.
The action $S^{(\mathrm{D})}$ is the sum of the actions
$S^{(\mathrm{D})}_{0}$ and $S^{(\mathrm{D})}_{\mathrm{mass}}$.
The action $S^{(\mathrm{D})}_{0}$ is
\begin{equation}
\begin{split}
S^{(\mathrm{D})}_{0}=&\,
\frac{1}{16\pi}
\int\mathrm{d}t 
\int\mathrm{d}x\,
\mathrm{tr}\,
\left(
\partial^{\,}_{\mu}G\,\partial^{\mu}G^{-1}
\right)
\\
&\,
+
\frac{1}{24\pi}\int\limits_{B} 
\mathrm{d}^{3}y\, 
\mathcal{L}^{(\mathrm{D})}_{\mathrm{WZW}},
\end{split} 
\end{equation}
where
\begin{equation}
\mathcal{L}^{(\mathrm{D})}_{\mathrm{WZW}}=
\epsilon^{ijk}\,
\mathrm{tr}\,
\Bigl[
(\bar{G}^{-1}\partial^{\,}_{i}\bar{G})\,
(\bar{G}^{-1}\partial^{\,}_{j}\bar{G})
(\bar{G}^{-1}\partial^{\,}_{k}\bar{G})
\Bigr].
\end{equation}
The action $S^{(\mathrm{D})}_{\mathrm{mass}}$ stems from the Lagrangian density 
\begin{equation}
\mathcal{L}^{(\mathrm{D})}_{\mathrm{mass}}=
\lambda\,
\sum_{I=1}^{N-1}
G^{\,}_{I,I+1} 
\equiv\,
\lambda\,
\mathrm{tr}
\left(M^{(\mathrm{D})}\,G\right).
\label{eq: mass matrix cor D minimal}
\end{equation}
The second equality is established by using the 
non-Abelian bosonization formula 
(\ref{eq: non-abelian bosonization formula})
(we have set the mass parameter $m^{\,}_{\mathrm{uv}}=1$). 
The $N\times N$ matrix $M^{(\mathrm{D})}$ is represented by
\begin{equation}
M^{(\mathrm{D})}:=
\begin{pmatrix}
0 & 0 & 0 & 0 & 0 & 0 &  \cdots \\
1 & 0 & 0 & 0 & 0 & 0 &  \cdots \\
0 & 1 & 0 & 0 & 0 & 0 &  \cdots \\
0 & 0 & 1 & 0 & 0 & 0 &  \cdots \\
0 & 0 & 0 & 1 & 0 & 0 &  \cdots \\
0 & 0 & 0 & 0 & 1 & 0 &  \cdots \\
\vdots & \vdots & \vdots & \vdots & \vdots & \vdots & \ddots\\
\end{pmatrix}.
\label{eq: mass matrix class D}
\end{equation}
\end{subequations}
The singular value decomposition 
of the mass
matrix (\ref{eq: mass matrix class D}) gives
\begin{equation}
M^{(\mathrm{D})}_{\mathrm{diag}}= 
\mathrm{diag}
\left(
\underbrace{
\Omega^{\,}_{1},\Omega^{\,}_{2},\Omega^{\,}_{3},...,\Omega^{\,}_{N-1}
           }^{\,}_{N-1},0
\right).
\label{eq: SVD mass matrix class D}
\end{equation}
The quadratic perturbation (\ref{eq: mass matrix cor D minimal})
thus reduces the central charge $c^{(\mathrm{D})}_{0}=N/2$ by the amount 
$(N-1)/2$, 
i.e., the central charge for the theory with
the partition function $Z^{(\mathrm{D})}$ is
\begin{equation}
c^{(\mathrm{D})} =
\frac{N}{2}-\frac{N-1}{2}=\frac{1}{2}.
\end{equation}
We have constructed a topological superconductor with the
gapless chiral Majorana mode 
$\chi^{\,}_{\mathrm{R},I=1}$
propagating along edge $I=1$ 
(the left eigenstate of the mass matrix)
vand  the gapless chiral Majorana mode of opposite chirality
$\chi^{\,}_{\mathrm{L},I=N}$
propagating along edge $I=N$
(the right eigenstate of the mass matrix).
This construction is summarized by
Fig.\ \ref{Fig: minimal wire network in D}.

The symmetry class D has the $\mathbb{Z}$ topological classification
for the following reason. If one
takes an arbitrary integer number $\nu$ of copies 
of the gapless edge theory, these $\nu$-copies remain gapless. 
The stability of the $\nu$ chiral gapless edge modes within 
either wire 1 or wire $N$ is guaranteed because backscattering 
among these gapless chiral edges modes is not allowed kinematically.

\begin{figure*}[t]
\begin{center}
\includegraphics[scale=0.5]{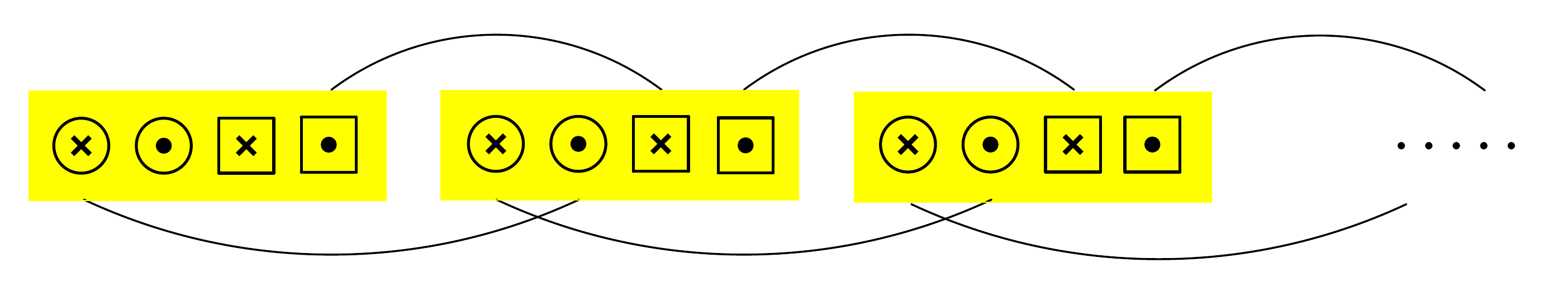}
\caption{(Color online)
Pictorial representation for the selected backscattering 
in the symmetry class DIII. Each yellow box represents 
a quantum wire composed of four-Majorana degrees of freedom.
The wires are enumerated by $I=1,\cdots,N$ 
in ascending order from  left to right.
For any $I$, the Majorana modes are denoted by
$\chi^{\,}_{\mathrm{R},+,I}$,
$\chi^{\,}_{\mathrm{L},+,I}$,
$\chi^{\,}_{\mathrm{R},-,I}$,
and
$\chi^{\,}_{\mathrm{L},-,I}$
reading from left to right,
respectively. 
\label{Fig: minimal wire network in DIII}
         }
\end{center}
\end{figure*}

\subsubsection{The symmetry class DIII}
\label{subsec: The symmetry class DIII}

The simplest model for an array of quantum wires in the symmetry class DIII 
to realize a topological gapped phase assumes
\begin{subequations}
\label{eq: free Majorana class DIII}
\begin{equation}
M=2,
\quad
\chi^{\,}_{\alpha,\mathrm{f},I}(t,x),
\label{eq: free Majorana class DIII a}
\end{equation}
for
$\alpha=\mathrm{R},\mathrm{L}$,
$\mathrm{f}=\pm$,
and
$I=1,\cdots,N$.
We have thus assigned four Majorana fermions to each wire 
$I=1\,\cdots\,N$. We define the action
\begin{equation}
S^{(\mathrm{DIII})}_{0}:= 
\int\mathrm{d}t
\int\mathrm{d}x\,
\mathcal{L}^{(\mathrm{DIII})}_{0}
\label{eq: free Majorana class DIII b}
\end{equation}
with
\begin{equation}
\begin{split}
\mathcal{L}^{(\mathrm{DIII})}_{0}:=&\,
\frac{\mathrm{i}}{2}
\sum_{I=1}^{N}
\sum_{\sigma=\pm}
\Bigl[
\chi^{\,}_{\mathrm{R},\sigma,I}
\left(
\partial^{\,}_{t} + \partial^{\,}_{x}
\right)
\chi^{\,}_{\mathrm{R},\sigma,I}
\\
&\,
+ 
\chi^{\,}_{\mathrm{L},\sigma,I}
\left(
\partial^{\,}_{t} - \partial^{\,}_{x}
\right)
\chi^{\,}_{\mathrm{L},\sigma,I}
\Bigr].
\end{split}
\label{eq: free Majorana class DIII c}
\end{equation}
We also define the Grassmann partition function
\begin{equation}
Z^{(\mathrm{DIII})}_{0}:=
\int\mathcal{D}[\chi]\,e^{\mathrm{i}S^{(\mathrm{DIII})}_{0}}.
\label{eq: free Majorana class DIII d}
\end{equation}
\end{subequations}

The theory with the partition function $Z^{(\mathrm{DIII})}_{0}$ is critical, 
for there are $4N$ decoupled massless Majorana modes
that are dispersing in $(1+1)$-dimensional Minkowski space and time.
Hence, the central charge for the theory with the
partition function $Z^{(\mathrm{DIII})}_{0}$ is 
\begin{subequations}
\label{eq: critical properties class DIII minimal}
\begin{equation}
c^{(\mathrm{DIII})}_{0}=N.
\label{eq: critical properties class DIII minimal a}
\end{equation}
The partition function $Z^{(\mathrm{DIII})}_{0}$
is invariant under any local transformation
$(O^{(\mathrm{R})},O^{(\mathrm{L})})\in
O^{\,}_{\mathrm{R}}(2N)\times O^{\,}_{\mathrm{L}}(2N)$
defined by
\begin{equation}
\begin{split}
&
\chi^{\,}_{\mathrm{R}}(t-x)\mapsto
O^{(\mathrm{R})}(t-x)\,
\chi^{\,}_{\mathrm{R}}(t-x),
\\
&
\chi^{\,}_{\mathrm{L}}(t+x)\mapsto
O^{(\mathrm{L})}(t+x)\,
\chi^{\,}_{\mathrm{L}}(t+x).
\end{split}
\label{eq: critical properties class DIII minimal b}
\end{equation}
It is also invariant under
the antilinear transformation with the fundamental rules
\begin{equation}
\begin{split}
&
\chi^{\,}_{\mathrm{R},+,I}(t,x)\mapsto
+
\chi^{\,}_{\mathrm{L},-,I}(-t,x),
\\
&
\chi^{\,}_{\mathrm{R},-,I}(t,x)\mapsto
-
\chi^{\,}_{\mathrm{L},+,I}(-t,x),
\\
&
\chi^{\,}_{\mathrm{L},+,I}(t,x)\mapsto
+
\chi^{\,}_{\mathrm{R},-,I}(-t,x),
\\
&
\chi^{\,}_{\mathrm{L},-,I}(t,x)\mapsto
-
\chi^{\,}_{\mathrm{R},+,I}(-t,x),
\end{split}
\label{eq: critical properties class DIII minimal c}
\end{equation}
\end{subequations}
that implements reversal of time in such a way that
reversal of time squares to minus the identity
(see Appendix \ref{Appsec: Reversal of time}).

Any partition function $Z^{(\mathrm{DIII})}$
for the array of quantum wires
is said to belong to the symmetry class DIII
if reversal of time is a symmetry represented
by an antilinear and involutive operation that squares to minus the identity,
i.e., Eq.\ (\ref{eq: critical properties class DIII minimal c}),
and if $Z^{(\mathrm{DIII})}$ is invariant under the linear transformation
(fermion parity) with the fundamental rule
\begin{equation}
\chi^{\,}_{\alpha,\mathrm{f},I}\mapsto
-\chi^{\,}_{\alpha,\mathrm{f},I},
\label{eq: parity conservation DIII}
\end{equation}
for $\alpha=\mathrm{R},\mathrm{L}$,
$\mathrm{f}=\pm$,
and
$I=1,\cdots,N$.

We seek a local single-particle perturbation 
$\mathcal{L}^{(\mathrm{DIII})}_{\mathrm{mass}}$
that satisfies three conditions when added to 
the Lagrangian density
(\ref{eq: free Majorana class DIII c}).

\textbf{Condition DIII.1}
It must be invariant under the transformations
(\ref{eq: critical properties class DIII minimal c})
and
(\ref{eq: parity conservation DIII}).

\textbf{Condition DIII.2}
It must gap completely the theory with the partition function 
$Z^{(\mathrm{DIII})}_{0}$
if we impose the periodic boundary conditions
\begin{equation}
\chi^{\,}_{\alpha,\mathrm{f},I}(t,x)=
\chi^{\,}_{\alpha,\mathrm{f},I+N}(t,x)
\end{equation} 
for 
$\alpha=\mathrm{R},\mathrm{L}$,
$\mathrm{f}=\pm$,
and
$I=1,\cdots,N$.

\textbf{Condition DIII.3}
The partition function $Z^{(\mathrm{DIII})}$
with the Lagrangian density
$\mathcal{L}^{(\mathrm{DIII})}_{0}+\mathcal{L}^{(\mathrm{DIII})}_{\mathrm{mass}}$
must be a theory with the central charge 
\begin{equation}
c^{(\mathrm{DIII})}=1
\end{equation}
if open boundary condition are imposed.

\textbf{Conditions DIII.1},
\textbf{DIII.2},
and 
\textbf{DIII.3}
imply that we may assign wire $I=1$ the central charge
$1/2$ and wire $I=N$ the central charge $1/2$,
for wires $I=1$ and $I=N$ both
support a Kramers degenerate pair of right- and left-moving 
Majorana edge modes.

We make the Ansatz 
\begin{equation}
\mathcal{L}^{(\mathrm{DIII})}_{\mathrm{mass}}:=
\sum_{I=1}^{N-1}
\mathrm{i}
\lambda\,
\left(
\chi^{\,}_{\mathrm{L},-,I}\,\chi^{\,}_{\mathrm{R},-,I+1} 
-
\chi^{\,}_{\mathrm{R},+,I}\,\chi^{\,}_{\mathrm{L},+,I+1}
\right)    
\label{eq: Ansatz for single-particle class DIII}
\end{equation}
with $\lambda$ a real-valued coupling. 
\textbf{Condition} \textbf{DIII.1} is met by construction.
To establish that the Ansatz
(\ref{eq: Ansatz for single-particle class DIII})
meets \textbf{Conditions} \textbf{DIII.2} and \textbf{DIII.3}, 
we use non-Abelian bosonization.
We choose the non-Abelian bosonization scheme by which
the partition function is given by the path integral 
\begin{subequations}
\begin{equation}
Z^{(\mathrm{DIII})}=
\int\mathcal{D}[G]\,e^{\mathrm{i}S^{(\mathrm{DIII})}}.
\end{equation}
The field $G\in O(2N)$ is a matrix of bosons.
The measure $\mathcal{D}[G]$ is constructed from the Haar measure on $O(2N)$.
The action $S^{(\mathrm{DIII})}$ is the sum of the actions
$S^{(\mathrm{DIII})}_{0}$ and $S^{(\mathrm{DIII})}_{\mathrm{mass}}$.
The action $S^{(\mathrm{DIII})}_{0}$ is
\begin{equation}
\begin{split}
S^{(\mathrm{DIII})}_{0}=&\,
\frac{1}{16\pi}
\int\mathrm{d}t 
\int\mathrm{d}x\,
\mathrm{tr}\,
\left(
\partial^{\,}_{\mu}G\,\partial^{\mu}G^{-1}
\right)
\\
&\,
+
\frac{1}{24\pi}\int\limits_{B} 
\mathrm{d}^{3}y\, 
\mathcal{L}^{(\mathrm{DIII})}_{\mathrm{WZW}},
\end{split} 
\end{equation}
where
\begin{equation}
\mathcal{L}^{(\mathrm{DIII})}_{\mathrm{WZW}}=
\epsilon^{ijk}\,
\mathrm{tr}\,
\Bigl[
(\bar{G}^{-1}\partial^{\,}_{i}\bar{G})\,
(\bar{G}^{-1}\partial^{\,}_{j}\bar{G})
(\bar{G}^{-1}\partial^{\,}_{k}\bar{G})
\Bigr].
\end{equation}
The action $S^{(\mathrm{DIII})}_{\mathrm{mass}}$ stems from the Lagrangian density
\begin{equation}
\begin{split}
\mathcal{L}^{(\mathrm{DIII})}_{\mathrm{mass}}=&\, 
\sum_{I=1}^{N-1}
\lambda\,
\left(
G^{\,}_{(-,I),(-,I+1)} 
+ 
G^{\,}_{(+,I+1),(+,I)} 
\right) 
\\
\equiv&\,
\lambda\,\mathrm{tr}\,
\left(M^{(\mathrm{DIII})}\,G\right).  
\end{split}
\label{eq: mass matrix cor DIII minimal} 
\end{equation}
The second equality is established by using the 
non-Abelian bosonization formula 
(\ref{eq: non-abelian bosonization formula})
(we have set the mass parameter $m^{\,}_{\mathrm{uv}}=1$). 
The $2N\times2N$ matrix $M^{(\mathrm{DIII})}$ is represented by
\begin{equation}
M^{(\mathrm{DIII})}:= 
\begin{pmatrix}
0 & B & 0 & 0 & 0 & 0 &  \cdots \\
B^{\mathsf{T}}  & 0 & B  & 0 & 0 & 0 & \cdots \\
0 & B^{\mathsf{T}}  & 0 & B & 0 & 0 & \cdots \\
0 & 0 & B^{\mathsf{T}}  & 0 & B  & 0 & \cdots \\
0 & 0 & 0 & B^{\mathsf{T}}  & 0 & B  & \cdots \\
0 & 0 & 0 & 0 & B^{\mathsf{T}}  & 0 &  \cdots \\
\vdots & \vdots & \vdots & \vdots & \vdots & \vdots & \ddots\\
\end{pmatrix}
\label{eq: mass matrix class DIII}
\end{equation}
in the basis for which
$B$ is the $2\times 2$ matrix
\begin{equation}
B:=
\begin{pmatrix}
\overbrace{0}^{(-,+)} & \overbrace{0}^{(-,-)}\\
\underbrace{1}_{(+,+)} & \underbrace{0}_{(+,-)}
\end{pmatrix}.
\end{equation}
\end{subequations} 
For any $N>0$, 
the $2N\times 2N$ matrices $M^{(\mathrm{DIII})}$ defined by 
(\ref{eq: mass matrix class DIII}) has 
two vanishing and $2\times(N-1)$ nonvanishing eigenvalues.

The quadratic perturbation (\ref{eq: mass matrix cor DIII minimal})
thus reduces the central charge $c^{(\mathrm{DIII})}_{0} =2\times N/2$
by the amount $2\times(N-1)/2$,
i.e., the central charge for the theory with the partition function 
$Z^{(\mathrm{DIII})}$ is
\begin{equation}
c^{(\mathrm{DIII})}=
\frac{2\times N}{2}-\frac{2\times(N-1)}{2}=1.
\end{equation}
We have constructed a topological superconductor with 
the gapless pair of helical Majorana modes 
$(\chi^{\,}_{\mathrm{L},+,I},\chi^{\,}_{\mathrm{R},-,I})^{\,}_{I=1}$
propagating along edge $I=1$ and 
the gapless pair of helical Majorana modes 
$(\chi^{\,}_{\mathrm{R},+,I},\chi^{\,}_{\mathrm{L},-,I})^{\,}_{I=N}$
propagating along edge $I=N$. 
This construction is summarized by
Fig.\ \ref{Fig: minimal wire network in DIII}.

The symmetry class DIII has the $\mathbb{Z}^{\,}_{2}$ classification
from the following argument.
We take $\nu$ copies of the gapless edge
theories on the right edge ($I=N$). We drop the index $I=N$
for notational simplicity. The most general backscattering processes are
encoded by
\begin{equation}
\mathcal{L}^{(\mathrm{DIII})}_{N}:=
\sum_{a,b=1}^{\nu}
\mathrm{i}
\chi^{\,}_{\mathrm{L},-,a}\,
\lambda^{\,}_{ab}\,
\chi^{\,}_{\mathrm{R},+,b}.
\label{eq: edge class DIII back scattering} 
\end{equation}
Hermiticity dictates here that 
\begin{equation}
\lambda^{\,}_{ab}=\lambda^{*}_{ab},
\qquad
a,b=1,\cdots,\nu,
\label{eq: lambda ab real for DIII}
\end{equation}
i.e., all matrix elements $\lambda^{\,}_{ab}$ are real valued.
Time-reversal symmetry dictates that
\begin{subequations}
\label{eq: constraints on lambda ab if DIII}
\begin{eqnarray}
\sum_{a,b=1}^{\nu}
\mathrm{i}
\chi^{\,}_{\mathrm{L},-,a}\,\lambda^{\,}_{ab}\,\chi^{\,}_{\mathrm{R},+,b}&\mapsto&
\sum_{a,b=1}^{\nu}
(-\mathrm{i})
(-1)\chi^{\,}_{\mathrm{R},+,a}\,
\lambda^{\,}_{ab}\,
\chi^{\,}_{\mathrm{L},-,b}
\nonumber\\
&=&
\sum_{a,b=1}^{\nu}
\mathrm{i}
\chi^{\,}_{\mathrm{R},+,a}\,
\lambda^{\,}_{ab}\,
\chi^{\,}_{\mathrm{L},-,b} 
\nonumber\\ 
&=&
\sum_{a,b=1}^{\nu}
\mathrm{i}
\chi^{\,}_{\mathrm{R},+,b}\,
\lambda^{\,}_{ba}
\chi^{\,}_{\mathrm{L},-,a}
\nonumber\\ 
&=&
\sum_{a,b=1}^{\nu}
\mathrm{i}\chi^{\,}_{\mathrm{L},-,a}\,(-\lambda^{\,}_{ba})\,\chi^{\,}_{\mathrm{R},+,b},
\nonumber\\
&&
\label{eq: constraints on lambda ab if DIII a}
\end{eqnarray}
i.e., the real-valued matrix elements (\ref{eq: lambda ab real for DIII})
must also be antisymmetric
\begin{equation}
\lambda^{\,}_{ab}=-\lambda^{\,}_{ba},
\qquad
a,b=1,\cdots,\nu.
\label{eq: constraints on lambda ab if DIII b}
\end{equation}
\end{subequations}
Because of the identity
\begin{equation}
\mathrm{det}\,(\lambda^{\,}_{ab})=
\mathrm{det}\,(\lambda^{\,}_{ab})^{\mathsf{T}}=
\mathrm{det}\,(-\lambda^{\,}_{ab})=
(-1)^{\nu}\,
\mathrm{det}\,(\lambda^{\,}_{ab}),
\end{equation}
if follows that the matrix $(\lambda^{\,}_{ab})$ 
has at least one vanishing eigenvalue when $\nu$ is odd.
When $\nu$ is odd, a pair of helical edge modes must remain gapless.
When $\nu$ is even, all pairs of helical edge modes can be gapped.
The topological classification
$\mathbb{Z}^{\,}_{2}$ for the symmetry class DIII in 2D follows.

\section{Non-Abelian topological order out of coupled wires}
\label{sec: Non-Abelian topological order out of coupled wires}

We have shown in 
Sec.~\ref{sec: The ten-fold way via non-Abelian bosonization}
(plus Appendix \ref{appsec: Non-Abelian bosonization for the symmetry classes})
that the tenfold way in two-dimensional space can be derived from
a one-dimensional array of quantum wires, whereby each wire
hosts Majorana fermions (i.e., ``real-valued'' fermions)
that may hop between consecutive wires through one-body
backscattering. This derivation of the tenfold way in two-dimensional space
presumes no more and no less than the existence of 
noninteracting Majorana fermions.

In each of the superconducting symmetry classes D, DIII, and C,
the existence of the numbers 2, 4, and 4 of noninteracting 
Majorana fermions per wire, respectively, was shown to be sufficient
to realize a superconducting ground state with protected edge states.
The numbers 2, 4, and 4 are the same numbers of complex fermions per wire
used in Ref.\ \onlinecite{Neupert14}
to stabilize short-ranged entangled topological superconducting ground states
in the symmetry classes D, DIII, and C for chains of wires 
that were coupled through strictly many-body interactions.
The derivation of the tenfold way in
Sec.\ \ref{sec: The ten-fold way via non-Abelian bosonization}
(plus Appendix \ref{appsec: Non-Abelian bosonization for the symmetry classes})
is thus more economical than that in Ref.\ \onlinecite{Neupert14}. In fact,
the numbers of Majorana fermions per wire that we have postulated
in Sec.\ \ref{sec: The ten-fold way via non-Abelian bosonization}
and in Appendix \ref{appsec: Non-Abelian bosonization for the symmetry classes}
are the minimum numbers of Majorana fermions per wire required
to realize short-ranged entangled gapped phases supporting protected edge 
states in the tenfold way. 

There is a drawback to this derivation, however.
Majorana fermions are not the fundamental fermions
in condensed matter physics. The electron is. 
Majorana fermions only emerge as quasiparticles
out of interactions that electrons undergo with themselves or
with collective modes such as phonons or spin waves.
One plain way to state
the drawback of the derivation in
Sec.~\ref{sec: The ten-fold way via non-Abelian bosonization} is that
it takes as the starting point an already fractionalized electron.

In this section, we are going to modify our strategy as follows.
Each wire in the one-dimensional array of wires supports electrons
instead of Majorana fermions. Second, any interaction that gaps the 
bulk will be built out of one-body or many-body electron-electron interactions
obeying two conditions. First, interactions
explicitly conserve the electron number.
Second, the spatial range of all interactions are bounded from above by one
finite length scale.  In this way, the interaction conserves the 
total electron number and are local. Nevertheless, we shall insist on recovering
Majorana fermions or their generalizations (parafermions) on the boundaries
by properly choosing the many-body electron-electron interactions. 

\begin{figure}[t]
\begin{center}
\includegraphics[scale=0.4]{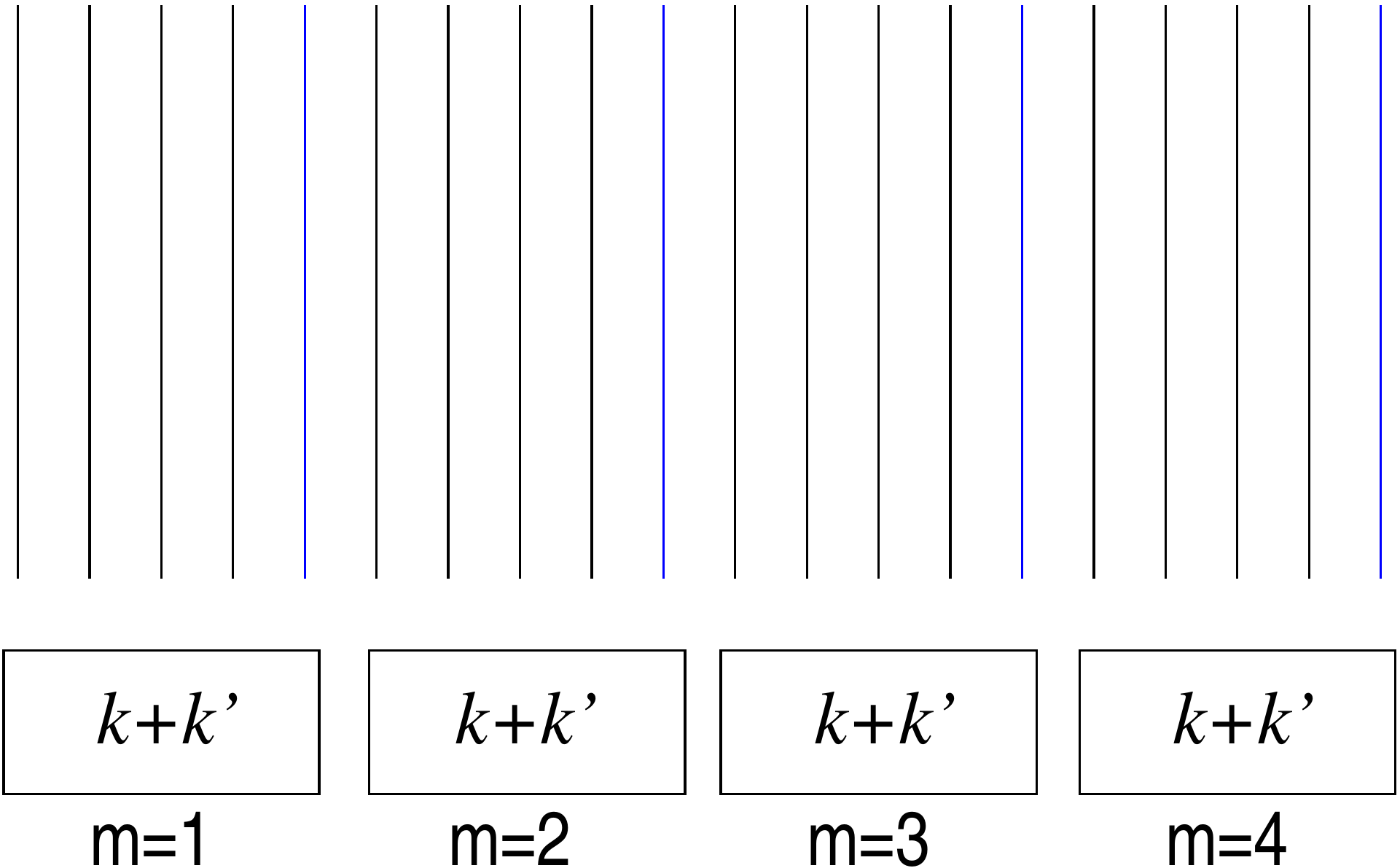}
\caption{(Color online)
Chain of $N$ wires grouped into $n$ bundles of $k+k'$ wires
with $N=20$, $n=4$, $k=4$ and $k'=1$.
The chain of wires is labeled by $I=1,\cdots,N$.
The bundles of $k+k'$ wires are labeled by the teletype
font $\texttt{m}=1,\cdots,n$.
\label{Fig: m bundles of k+k' wires}
         }
\end{center}
\end{figure}

\subsection{One-dimensional arrays of quantum wires with local current-current interactions}
\label{subsec: Arrays of quantum wires with local current-current interactions} 

A chain of $N$ decoupled wires is labeled
with the index $I=1,\cdots,N$. Electrons move freely
along any one of these $N$ wires. Their spin-1/2 projections 
along the quantization axis are $\sigma=\uparrow,\downarrow$.
For simplicity, all wires
are identical. At low energies, we postulate the noninteracting
Lagrangian density$\ $ %
\footnote{\label{eq: footnote on Fermi wave vector}
Any one flavor of a low-energy electron in some given wire is
related to the left- and right-movers from this wire 
by the multiplicative phase factor $e^{\pm\mathrm{i}k^{\,}_{\mathrm{F}}\,x}$,
where the Fermi wave vector $k^{\,}_{\mathrm{F}}$ 
is fixed by the filling fraction for this flavor of electrons in the 
given wire. The product of these multiplicative phase factors 
arising from taking the local product of low-energy electron operators
can always be absorbed by a coupling that is 
modulated with respect to $x$ with the proper periodicity.
With this caveat in mind, we can choose to work with the convention
$k^{\,}_{\mathrm{F}}=0$ without loss of generality.
         }
\begin{subequations}
\label{eq: def N wires with ``one'' electron per wire}
\begin{equation}
\begin{split}
\mathcal{L}^{\,}_{0}:=&\,
\mathrm{i}
\sum_{I=1}^{N}
\sum_{\sigma=\uparrow,\downarrow}
\Big[
\psi^{*}_{\mathrm{R},\sigma,I}\,
(\partial^{\,}_{t}+\partial^{\,}_{x})\,
\psi^{\,}_{\mathrm{R},\sigma,I}
\\
&\,
+
\psi^{*}_{\mathrm{L},\sigma,I}\,
(\partial^{\,}_{t}-\partial^{\,}_{x})\,
\psi^{\,}_{\mathrm{L},\sigma,I}
\Big]
\end{split}
\label{eq: def N wires with ``one'' electron per wire a}
\end{equation}
with the action
\begin{equation}
S^{\,}_{0}:=
\int\mathrm{d}t
\int\mathrm{d}x\,
\mathcal{L}^{\,}_{0}
\label{eq: def N wires with ``one'' electron per wire b}
\end{equation}
and the partition function
\begin{equation}
Z^{\,}_{0}:=
\int
\mathcal{D}[\psi^{*},\psi]\
e^{\mathrm{i}S^{\,}_{0}}.
\end{equation}
\end{subequations}

The partition function $Z^{\,}_{0}$ 
is invariant under any local linear transformation
\begin{subequations}
\label{eq: symmetry group U(2N) U(2N)}
\begin{equation}
(U^{(\mathrm{R})},U^{(\mathrm{L})})\in
U^{\,}_{\mathrm{R}}(2N)\times U^{\,}_{\mathrm{L}}(2N)
\label{eq: symmetry group U(2N) U(2N) a}
\end{equation}
defined by the fundamental rules
\begin{equation}
\begin{split}
&
\psi^{*\mathsf{T}}_{\mathrm{R}}(t-x)\mapsto
\psi^{*\mathsf{T}}_{\mathrm{R}}(t-x)\,
U^{(\mathrm{R})\dag}(t-x),
\\
&
\psi^{*\mathsf{T}}_{\mathrm{L}}(t+x)\mapsto
\psi^{*\mathsf{T}}_{\mathrm{L}}(t+x)\,
U^{(\mathrm{L})\dag}(t+x),
\end{split}
\label{eq: symmetry group U(2N) U(2N) b}
\end{equation}
and
\begin{equation}
\begin{split}
&
\psi^{\,}_{\mathrm{R}}(t-x)\mapsto
U^{(\mathrm{R})}(t-x)\,
\psi^{\,}_{\mathrm{R}}(t-x),
\\
&
\psi^{\,}_{\mathrm{L}}(t+x)\mapsto
U^{(\mathrm{L})}(t+x)\,
\psi^{\,}_{\mathrm{L}}(t+x),
\end{split}
\label{eq: symmetry group U(2N) U(2N) c}
\end{equation}
\end{subequations}
on the Grassmann integration variables.
The corresponding central charge is
\begin{equation}
c^{\,}_{0}=2N.
\label{eq: c0=2N}
\end{equation}

The partition function $Z^{\,}_{0}$ 
is also invariant under reversal of time, whereby this operation
is represented by the antilinear transformation with the fundamental rules
\begin{subequations}
\label{eq: def reversal of time on complex fermions}
\begin{align}
&
\psi^{*}_{\mathrm{R},\uparrow,I}\mapsto
+\psi^{*}_{\mathrm{L},\downarrow,I},
\qquad
\psi^{*}_{\mathrm{R},\downarrow,I}\mapsto
-\psi^{*}_{\mathrm{L},\uparrow,I},
\\
&
\psi^{*}_{\mathrm{L},\uparrow,I}\mapsto
+\psi^{*}_{\mathrm{R},\downarrow,I},
\qquad
\psi^{*}_{\mathrm{L},\downarrow,I}\mapsto
-\psi^{*}_{\mathrm{R},\uparrow,I},
\end{align}
and
\begin{align}
&
\psi^{\,}_{\mathrm{R},\uparrow,I}\mapsto
+\psi^{\,}_{\mathrm{L},\downarrow,I},
\qquad
\psi^{\,}_{\mathrm{R},\downarrow,I}\mapsto
-\psi^{\,}_{\mathrm{L},\uparrow,I},
\\
&
\psi^{\,}_{\mathrm{L},\uparrow,I}\mapsto
+\psi^{\,}_{\mathrm{R},\downarrow,I},
\qquad
\psi^{\,}_{\mathrm{L},\downarrow,I}\mapsto
-\psi^{\,}_{\mathrm{R},\uparrow,I},
\end{align}
\end{subequations}
on the Grassmann integration variables.

The chain-resolved symmetry
$U^{\,}_{\mathrm{R}}(2)\times U^{\,}_{\mathrm{L}}(2)$,
a subgroup of the symmetry group
(\ref{eq: symmetry group U(2N) U(2N) a}),
is broken by coupling consecutive chains 
through one-body tunnelings.

\textit{Example 1.}
The uniform one-body hopping of the electrons between consecutive chains 
\begin{equation}
\label{eq: Example 1 2D gapless phase}
\begin{split}
\mathcal{L}^{\,}_{\mathrm{FS}}:=
-
t
\sum_{I=1}^{N}
\left[
\psi^{*\mathsf{T}}_{\mathrm{R},I}\,
\psi^{\,}_{\mathrm{R},I+1}
+
\psi^{*\mathsf{T}}_{\mathrm{R},I+1}\,
\psi^{\,}_{\mathrm{R},I}
+
(
\mathrm{R}\to
\mathrm{L}
)
\right],
\end{split}
\end{equation}
where $t$ is positive and
periodic boundary conditions by which $I\equiv I+N$ 
are imposed on the Grassmann fields,
turns the one-dimensional critical theory 
(\ref{eq: def N wires with ``one'' electron per wire})
into an anisotropic two-dimensional gas of electrons
in the thermodynamic limit $N\to\infty$.

\textit{Example 2.}
The staggered one-body hopping 
\begin{equation}
\begin{split}
\mathcal{L}^{\,}_{\mathrm{D}}:=&\,
-\mathrm{i}t
\sum_{I=1}^{N}
\Bigl[
\psi^{*\mathsf{T}}_{\mathrm{R},I}\,\psi^{\,}_{\mathrm{L},I+1} 
- 
\psi^{*\mathsf{T}}_{\mathrm{L},I+1}\,\psi^{\,}_{\mathrm{R},I}
\\
&\,
+ 
\psi^{*\mathsf{T}}_{\mathrm{L},I}\,
\psi^{\,}_{\mathrm{R},I+1} 
- 
\psi^{*\mathsf{T}}_{\mathrm{R},I+1}\,
\psi^{\,}_{\mathrm{L},I}  
\Bigr],
\end{split}
\label{eq: Example 2 2D gapless phase}
\end{equation}
where $t$ is positive and
periodic boundary conditions by which $I\equiv I+N$ 
are imposed on the Grassmann fields,
turns the one-dimensional critical theory 
(\ref{eq: def N wires with ``one'' electron per wire})
into an anisotropic two-dimensional Dirac gas of electrons
in the thermodynamic limit $N\to\infty$, i.e., 
a quasi-one-dimensional representation of graphene.

Coupling the chains through many-body tunnelings
that preserve the chain-resolved
$U^{\,}_{\mathrm{R}}(2)\times U^{\,}_{\mathrm{L}}(2)$
subgroup of the symmetry group
(\ref{eq: symmetry group U(2N) U(2N) a})
(i.e., the independent conservation of the right- and left-moving
electronic charge and spin in each wire)
also delivers two-dimensional gapless phases of matter
in the thermodynamic limit $N\to\infty$.$\ $ %
\cite{Ohern99,Emery00,Vishwanath01,Mukhopadhyay01}

\textit{Example 3.}
The four-fermion interactions
\begin{equation}
\mathcal{L}^{\,}_{\mathrm{SLL}}:=
\sum_{I,J=1}^{N}
\left[
\left(
\psi^{*\mathsf{T}}_{\mathrm{R},I}\,
\psi^{\,}_{\mathrm{R},I}
\right)
V^{\,}_{IJ}\,
\left(
\psi^{*\mathsf{T}}_{\mathrm{R},J}\,
\psi^{\,}_{\mathrm{R},J}
\right)
+
\left(
\mathrm{R}\to\mathrm{L}
\right)
\right]
\end{equation}
with $V^{\,}_{IJ}=V^{\,}_{JI}$ a symmetric and real-valued
$N\times N$ matrix and
periodic boundary conditions by which $I\equiv I+N$ 
are imposed on the Grassmann fields,
stabilize a sliding Luttinger liquid (SLL) phase in the thermodynamic limit 
$N\to\infty$,
whose defining properties
is that of algebraic order along the quantum wires
in contrast to exponentially decaying correlation functions
in the direction transverse to that of the quantum wires.$\ $ %
\cite{Ohern99,Emery00,Vishwanath01,Mukhopadhyay01}

We are after two-dimensional phases that are insulating when periodic
boundary conditions hold. This can always be achieved by a suitable
combination of a breaking of translation symmetry, on the one hand,
and of an interaction between left- and right-movers, on the other hand.  
For this reason, we shall ignore couplings of
the quantum wires that deliver gapless two-dimensional phases as in
\textit{Examples 1, 2,} and \textit{3} 
relative to those couplings between left- and
right-movers responsible for an insulating phase when periodic
boundary conditions hold.

This section is organized as follows.
We begin by showing in Sec.\ \ref{subsec: Complete gapping}
how to combine one-body and
current-current interactions that fully gap
the critical theory (\ref{eq: def N wires with ``one'' electron per wire}).
We proceed in Sec.\
\ref{subsec: Partial gapping without time-reversal symmetry}
by coupling the wires through current-current interactions so that
(i) time-reversal symmetry is explicitly broken,
(ii) the critical theory 
(\ref{eq: def N wires with ``one'' electron per wire})
is fully gapped when periodic boundary conditions are imposed 
along the chain of quantum wires,
and (iii) there remains gapless edge states that realize 
chiral conformal field theories with the chiral central charge
\begin{equation}
0<c<3
\label{eq: c<3}
\end{equation}
on any one of the two boundaries close to wire $1$ and $N$, respectively,
when open boundary conditions are imposed 
along the chain of quantum wires.
We close with Sec.\
\ref{subsec: Partial gapping with time-reversal symmetry}
by selecting interactions that are time-reversal symmetric.

\subsection{Complete gapping}
\label{subsec: Complete gapping}

As a warm up, we observe that the symmetry group
(\ref{eq: symmetry group U(2N) U(2N) a})
contains as a subgroup the symmetry group
$\Big(U^{\,}_{\mathrm{R}}(2)\times U^{\,}_{\mathrm{L}}(2)\Big)
\times\cdots\times
\Big(U^{\,}_{\mathrm{R}}(2)\times U^{\,}_{\mathrm{L}}(2)\Big)$.
We are assigning to the unitary group $U(2)$ of $2\times2$ matrices
the label R and L when it acts on the right- and left-moving
electrons, respectively, from a given wire.
The partition function $Z^{\,}_{0}$ is, indeed, invariant under any local
linear transformation
\begin{subequations}
\label{eq: symmetry group U(2) U(2)}
\begin{equation}
\Big(U^{(\mathrm{R})}_{I},U^{(\mathrm{L})}_{I}\Big)\in
U^{\,}_{\mathrm{R}}(2)\times U^{\,}_{\mathrm{L}}(2)
\label{eq: symmetry group U(2) U(2) a}
\end{equation}
defined by the fundamental rules
\begin{equation}
\begin{split}
&
\psi^{*}_{\mathrm{R},\sigma,I}(t-x)\mapsto
\psi^{*}_{\mathrm{R},\sigma',I}(t-x)\,
\left(U^{(\mathrm{R})\dag}_{I}\right)^{\,}_{\sigma'\sigma}(t-x),
\\
&
\psi^{*}_{\mathrm{L},\sigma,I}(t+x)\mapsto
\psi^{*}_{\mathrm{L},\sigma',I}(t+x)\,
\left(U^{(\mathrm{L})\dag}_{I}\right)^{\,}_{\sigma'\sigma}(t+x),
\end{split}
\label{eq: symmetry group U(2) U(2) b}
\end{equation}
and
\begin{equation}
\begin{split}
&
\psi^{\,}_{\mathrm{R},\sigma,I}(t-x)\mapsto
\left(U^{(\mathrm{R})}_{I}\right)^{\,}_{\sigma\sigma'}(t-x)\,
\psi^{\,}_{\mathrm{R},\sigma',I}(t-x),
\\
&
\psi^{\,}_{\mathrm{L},\sigma,I}(t+x)\mapsto
\left(U^{(\mathrm{L})}_{I}\right)^{\,}_{\sigma\sigma'}(t+x)\,
\psi^{\,}_{\mathrm{L},\sigma',I}(t+x),
\end{split}
\label{eq: symmetry group U(2) U(2) c}
\end{equation}
\end{subequations}
for any $\sigma=\uparrow,\downarrow$ and any $I=1,\cdots,N$
on the Grassmann integration variables. These symmetries imply
for the light-cone components
[we choose the multiplicative normalization 
from Ref.\ \onlinecite{Affleck90,Affleck91a,Affleck91b}
rather than the one in 
Eq.\ (\ref{eq: Virasoro OPE})] 
\begin{subequations}
\label{eq: def light-cone components Tpm}
\begin{equation}
T^{\,}_{\mathrm{R},I}:=
\frac{\mathrm{i}}{2\pi}
\sum_{\sigma=\uparrow,\downarrow}
\psi^{*}_{\mathrm{R},\sigma,I}\,
\left(\partial^{\,}_{t}-\partial^{\,}_{x}\right)\,
\psi^{\,}_{\mathrm{R},\sigma,I}
\label{eq: def light-cone components Tpm a}
\end{equation}
and
\begin{equation}
T^{\,}_{\mathrm{L},I}:=
\frac{\mathrm{i}}{2\pi}
\sum_{\sigma=\uparrow,\downarrow}
\psi^{*}_{\mathrm{L},\sigma,I}\,
\left(\partial^{\,}_{t}+\partial^{\,}_{x}\right)\,
\psi^{\,}_{\mathrm{L},\sigma,I}
\label{eq: def light-cone components Tpm b}
\end{equation}
\end{subequations}
of the energy-momentum tensor the Sugawara identities$\ $ %
\cite{Sugawara68}
\begin{subequations}
\label{eq: Sugawara construction for smallest symmetry group}
\begin{equation}
T^{\,}_{\mathrm{R},I}=
T^{\,}_{\mathrm{R},I}[\hat{u}(1)]
+
T^{\,}_{\mathrm{R},I}[\widehat{su}(2)^{\,}_{1}]
\label{eq: Sugawara construction for smallest symmetry group a}
\end{equation}
and
\begin{equation}
T^{\,}_{\mathrm{L},I}=
T^{\,}_{\mathrm{L},I}[\hat{u}(1)]
+
T^{\,}_{\mathrm{L},I}[\widehat{su}(2)^{\,}_{1}],
\label{eq: Sugawara construction for smallest symmetry group b}
\end{equation}
where
\begin{align}
&
T^{\,}_{\mathrm{R},I}[\hat{u}(1)]=
\frac{1}{2c^{\,}_{v}}
\bm{j}^{\,}_{\mathrm{R},I}\,
\bm{j}^{\,}_{\mathrm{R},I},
\label{eq: Sugawara construction for smallest symmetry group c}
\\
&
T^{\,}_{\mathrm{R},I}[\widehat{su}(2)^{\,}_{1}]=
\frac{1}{1+c^{\,}_{v}}\,
\bm{J}^{\,}_{\mathrm{R},I}
\cdot
\bm{J}^{\,}_{\mathrm{R},I},
\label{eq: Sugawara construction for smallest symmetry group d}
\end{align}
and
\begin{align}
&
T^{\,}_{\mathrm{L},I}[\hat{u}(1)]=
\frac{1}{2c^{\,}_{v}}
\bm{j}^{\,}_{\mathrm{L},I}\,
\bm{j}^{\,}_{\mathrm{L},I},
\label{eq: Sugawara construction for smallest symmetry group e}
\\
&
T^{\,}_{\mathrm{L},I}[\widehat{su}(2)^{\,}_{1}]=
\frac{1}{1+c^{\,}_{v}}\,
\bm{J}^{\,}_{\mathrm{L},I}
\cdot
\bm{J}^{\,}_{\mathrm{L},I},
\label{eq: Sugawara construction for smallest symmetry group f}
\end{align} 
\end{subequations}
for $I=1,\cdots,N$, respectively.
Here, we have introduced the charge currents
\begin{subequations}
\label{eq: def charge and spin currents} 
\begin{equation}
\bm{j}^{\,}_{\mathrm{R},I}:=
\psi^{*}_{\mathrm{R},I}\,\sigma^{\,}_{0}\,\psi^{\,}_{\mathrm{R},I},
\qquad
\bm{j}^{\,}_{\mathrm{L},I}:=
\psi^{*}_{\mathrm{L},I}\,\sigma^{\,}_{0}\,\psi^{\,}_{\mathrm{L},I},
\label{eq: def charge and spin currents a} 
\end{equation}
and the spin currents 
\begin{equation}
\bm{J}^{\,}_{\mathrm{R},I}:=
\frac{1}{2}
\psi^{*}_{\mathrm{R},I}\,\bm{\sigma}\,\psi^{\,}_{\mathrm{R},I},
\qquad
\bm{J}^{\,}_{\mathrm{L},I}:=
\frac{1}{2}\,
\psi^{*}_{\mathrm{L},I}\,\bm{\sigma}\,\psi^{\,}_{\mathrm{L},I},
\label{eq: def charge and spin currents b} 
\end{equation} 
\end{subequations}
within any wire $I=1,\cdots,N$.
The unit $2\times2$ matrix acting in spin space is denoted
$\sigma^{\,}_{0}$ and
$\bm{\sigma}$ is the vector made of the three Pauli matrices
acting in spin space. Finally, the eigenvalue 
\begin{subequations}
\label{eq:def casimir su(2)}
\begin{equation}
c^{\,}_{v}:=
\sum_{a,b=1}^{3}\epsilon^{\,}_{1ab}\,\epsilon^{\,}_{1ab}=2
\label{eq:def casimir su(2) a}
\end{equation}
of the $SU(2)$ Casimir operator in the adjoint representation
is also the multiplicative normalization factor that enters in 
\begin{equation}
\mathrm{tr}\,(\sigma^{\,}_{\mu}\sigma^{\,}_{\nu})=
c^{\,}_{v}\,\delta^{\,}_{\mu\nu},
\qquad
\mu,\nu=0,1,2,3.
\label{eq:def casimir su(2) b}
\end{equation}
\end{subequations}

The physics of Luttinger liquids has taught us that
we can gap the charge and the spin sector independently
(spin-charge separation) in any given wire
$I=1,\cdots,N$.$\ $ %
\cite{Nagaosa99,Gogolin04}
For example, umklapp scatterings with the proper periodicities 
open Mott gaps in the charge sector,
while preserving the critical behavior in the spin sector.
Conversely, a generic spin current-current interaction of the form
\begin{equation}
\mathcal{L}^{\,}_{\mathrm{int},I}:=
-
\sum_{a=1}^{3}
\lambda^{a}_{I}\,
J^{a}_{\mathrm{R},I}
J^{a}_{\mathrm{L},I}
\label{eq:  SU(2) current-current interaction}
\end{equation}
for any $I=1,\cdots,N$ 
is argued to gap the spin sector in the $I$th wire
if the coupling constants
$\lambda^{a}_{I}>0$ 
without affecting the charge sector, for the
couplings $\lambda^{a}_{I}>0$ obey the one-loop
RG equation (see Appendix \ref{appsec: Warmup})$\ $ %
\cite{Dashen75}
\footnote{
See chapter 17V from Ref.\ \onlinecite{Gogolin04}
for a more modern derivation of this one-loop RG equation.
         }
\begin{equation}
\frac{\mathrm{d}\lambda^{a}_{I}}{\mathrm{d}\ell}=
\pi\,
\sum_{b,c=1}^{3}
(\epsilon^{\,}_{abc})^{2}\,
\lambda^{b}_{I}\,
\lambda^{c}_{I}
\label{eq: one loop RG flow current-current interaction if SU(2)}
\end{equation}
for $a=1,2,3$
under the rescaling 
$\mathfrak{a}\mapsto(1+\mathrm{d}\ell)\,\mathfrak{a}$
of the short-distance characteristic length $\mathfrak{a}$.$\ $%
\footnote{
The nature of the phase corresponding to the strong
coupling fixed point of the one-loop RG-flow
(\ref{eq: one loop RG flow current-current interaction if SU(2)})
can only be established by solving nonperturbatively
the effects of the strong interaction 
(\ref{eq:  SU(2) current-current interaction})
         }
In the special case when the current-current interactions
preserve the spin $SU(2)$ symmetry, i.e., when
\begin{subequations}
\begin{equation}
\lambda^{a}_{I}\equiv\lambda^{\,}_{I}
\end{equation}
for all $I=1,\cdots,N$ and all $a=1,2,3$,
\begin{equation}
\frac{\mathrm{d}\lambda^{\,}_{I}}{\mathrm{d}\ell}=
\pi\,c^{\,}_{v}\,\lambda^{2}_{I}.
\end{equation}
\end{subequations}

\begin{figure*}[t]
\begin{center}
(a)
\includegraphics[scale=0.6]{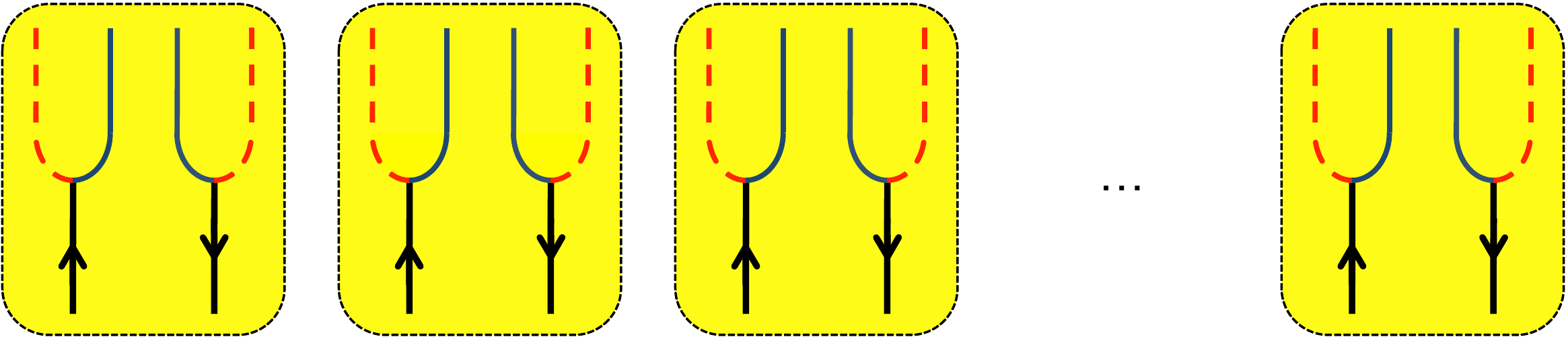}
\\
(b)
\includegraphics[scale=0.6]{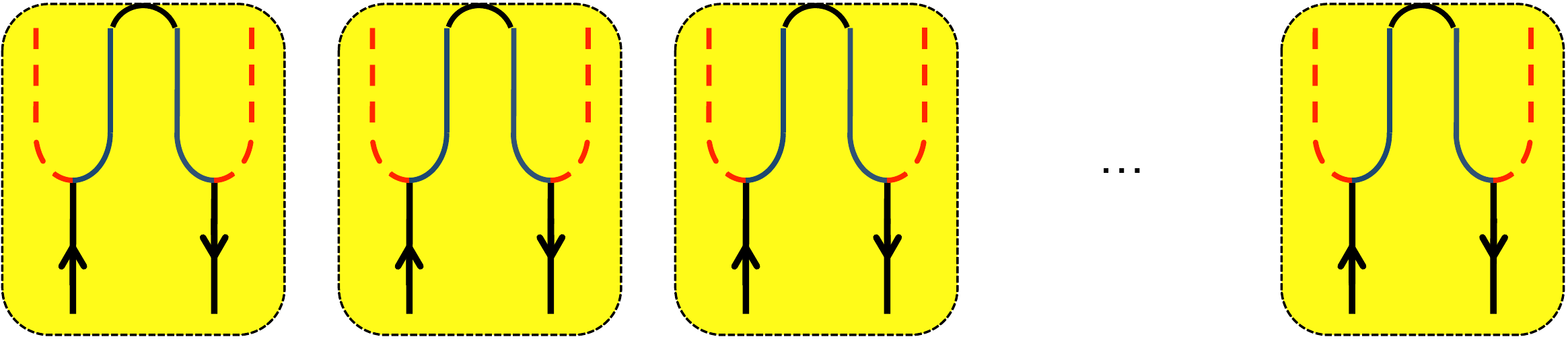}
\\
(c)
\includegraphics[scale=0.6]{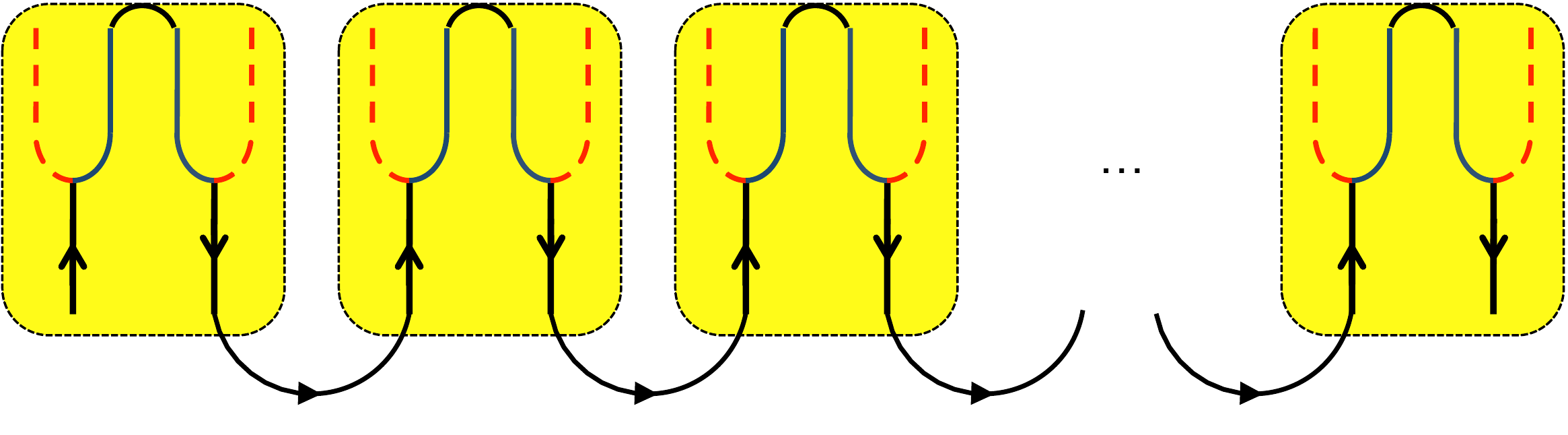}
\\
(d)
\includegraphics[scale=0.6]{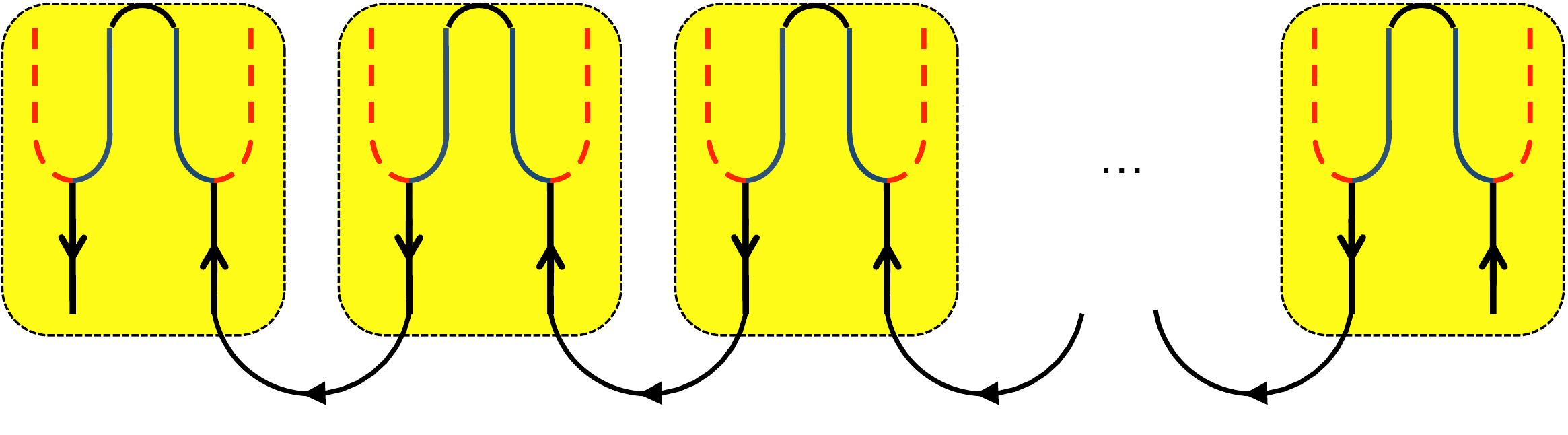}
\\
\caption{(Color online)
(a) A chain of wires is partitioned into bundles.
A bundle is depicted by a domino.
The pattern in the domino corresponds to a right- and left-moving
critical sector with the affine Lie algebra
$\hat{g}^{\,}_{k,k'}:=\widehat{su}(2)^{\,}_{k}\oplus \widehat{su}(2)^{\,}_{k'}$
represented by a thick vertical like supporting an up arrow for right movers
and down arrow for left movers.
Its diagonal subalgebra
$\hat{h}^{\,}_{k,k'}:=\widehat{su}(2)^{\,}_{k+k'}$ 
is represented by the forking into a blue solid line.
The coset
$\hat{g}^{\,}_{k,k'}/\hat{h}^{\,}_{k,k'}$
is represented by the forking into a red dashed line.
(b)
An arc inside each domino depicts a current-current interaction
between the generators of $\hat{h}^{\,}_{k,k'}$.
These arcs gap all the critical modes
generated by $\hat{h}^{\,}_{k,k'}:=\widehat{su}(2)^{\,}_{k+k'}$
within a bundle.
(c)
An arc between two consecutive dominoes depicts 
the current-current interactions
between the generators of $\hat{g}^{\,}_{k,k'}$.
The arrows on these arcs indicate that these
interactions break time-reversal symmetry.
These arcs gap all remaining critical modes
except for the modes generated by the right-moving
$\hat{g}^{\,}_{k,k'}/\hat{h}^{\,}_{k,k'}$ on bundle $\texttt{m}=1$
and the modes  generated by the left-moving
$\hat{g}^{\,}_{k,k'}/\hat{h}^{\,}_{k,k'}$ on bundle $\texttt{m}=n$.
(d)
Reversal of time is represented by reversing all arrows.
\label{Fig: gapping without TRS}
         }
\end{center}
\end{figure*}

\subsection{Partial gapping without time-reversal symmetry}
\label{subsec: Partial gapping without time-reversal symmetry}

We have identified the continuous
symmetry groups 
(\ref{eq: symmetry group U(2N) U(2N)})
and
(\ref{eq: symmetry group U(2) U(2)})
for the free theory 
(\ref{eq: def N wires with ``one'' electron per wire}).
In the latter case, the currents entering the Sugawara construction 
(\ref{eq: Sugawara construction for smallest symmetry group})
corresponding to the symmetry group $U(2)\times\cdots\times U(2)$
obey the semi-simple affine Lie algebra
\begin{equation}
\hat{u}:=
\opluslimits_{I=1}^{N}\hat{u}(2)^{\,}_{1}.
\end{equation}
In the former case, we could also have introduced the 
Sugawara construction with the
affine Lie algebra $\hat{u}(2N)^{\,}_{1}$ of level one
which is associated to the symmetry group $U(2N)$.
In fact, this was done in Appendix 
\ref{subsec: The symmetry class AII}
for the group $U(2N)\sim O(4N)$ when discussing the symmetry
class AII.

We shall now consider a symmetry group 
(and the corresponding Sugawara construction)
that is intermediate between 
$U(2)\times\cdots\times U(2)$
and
$U(2N)$.

The idea is the following. We break
the chain of $N\gg 1$ wires into $n>1$ unit cells (bundles),
each of which is made of $k+k'$ consecutive wires
as is illustrated in Fig.\
\ref{Fig: m bundles of k+k' wires}.
In other words, we assume that
\begin{equation}
N=n\,(k+k')
\label{eq: bundle with TRS breaking}
\end{equation}
with $k$ and $k'$ two nonvanishing positive integers.
The thermodynamic limit $N\to\infty$ is taken holding
$k$ and $k'$ fixed. The spatial range of the
current-current interactions that
we will use to gap partially the spectrum of
the free theory 
(\ref{eq: def N wires with ``one'' electron per wire})
involves at most two consecutive bundles of $k+k'$ wires.
Locality is thus guaranteed.
We assign the teletype font
$\texttt{m}=1,\cdots,n$ when labeling the bundles of $k+k'$
consecutive wires that make up an enlarged unit cell of the chain
of $N$ wires. An important case corresponds to the choice
$k=k'=1$ that amounts to rearranging
the chain of wires into a chain of ladders,
as is depicted in Fig.\ \ref{Fig: SU(2) level 2 example}.

The symmetry that we select when considering any one of
the $n$ bundles of $k+k'$ consecutive wires is the direct product
\begin{subequations}
\label{eq: selcted symmetry when TRS is broken}
\begin{equation}
U:=U(2k)\times U(2k').
\end{equation}
The corresponding semi-simple affine Lie algebra is
\begin{equation}
\hat{u}^{\,}_{1}:=
\hat{u}(2k)^{\,}_{1}
\oplus
\hat{u}(2k')^{\,}_{1}.
\end{equation}
\end{subequations}
By construction, the central charges
$c[\hat{u}^{\,}_{1}]$,
$c[\hat{u}(2k)^{\,}_{1}]$,
and
$c[\hat{u}(2k')^{\,}_{1}]$
are related by
\begin{equation}
c[\hat{u}^{\,}_{1}]=
c[\hat{u}(2k)^{\,}_{1}]
+
c[\hat{u}(2k')^{\,}_{1}].
\end{equation} 
As it should be
\begin{align}
2N=&\,
2\,n\,(k+k')
\nonumber\\
=&\,
n
\left(
c[\hat{u}(2k )^{\,}_{1}]
+
c[\hat{u}(2k')^{\,}_{1}]
\right)
\nonumber\\
=&\,
n\,
c[\hat{u}^{\,}_{1}].
\end{align}
We are in position to take advantage of the
non-Abelian bosonization of a bundle of $k+k'$ consecutive
wires in any of the enlarged unit cell 
labeled by $\texttt{m}=1,\cdots,n$
with the symmetry group $U(2k)\times U(2k')$
making up the chain of $N$ decoupled and identical wires.
To avoid heavy notation, we drop the label $\texttt{m}$
when the bundles are decoupled.

Inspired by the works of Affleck and Ludwig in
connection to the multichannel Kondo effect,$\ $ %
\cite{Affleck90,Affleck91a,Affleck91b}
we use the following generalization of the
Sugawara decomposition
(\ref{eq: Sugawara construction for smallest symmetry group}),
which we only present in the sector with the symmetry group $U(2k)$
without loss of generality. 
The identity
\begin{subequations}
\label{eq: master identity for u(2k) level 1}
\begin{equation}
\hat{u}(2k)^{\,}_{1}=
\hat{u}(1)
\oplus
\widehat{su}(2)^{\,}_{k}
\oplus
\widehat{su}(k)^{\,}_{2}
\end{equation}
between affine Lie algebras is equivalent to stating
that
\begin{align}
&
T^{\,}_{\mathrm{R}}[\hat{u}(2k)^{\,}_{1}]=
T^{\,}_{\mathrm{R}}[\hat{u}(1)]
+
T^{\,}_{\mathrm{R}}[\widehat{su}(2)^{\,}_{k}]
+
T^{\,}_{\mathrm{R}}[\widehat{su}(k)^{\,}_{2}],
\\
&
T^{\,}_{\mathrm{L}}[\hat{u}(2k)^{\,}_{1}]=
T^{\,}_{\mathrm{L}}[\hat{u}(1)]
+
T^{\,}_{\mathrm{L}}[\widehat{su}(2)^{\,}_{k}]
+
T^{\,}_{\mathrm{L}}[\widehat{su}(k)^{\,}_{2}],
\end{align}
where [for simplicity we only present 
this relation in the right-moving sector; we
also choose the multiplicative normalization 
from Ref.\ \onlinecite{Affleck90,Affleck91a,Affleck91b}
rather than the one in 
Eq.\ (\ref{eq: Virasoro OPE})]
\begin{equation}
T^{\,}_{\mathrm{R}}[\hat{u}(2k)^{\,}_{1}]=
\frac{\mathrm{i}}{2\pi}
\sum_{\alpha=1}^{2}
\sum_{A=1}^{k}
\psi^{*}_{\mathrm{R},\alpha,A}\,
\left(\partial^{\,}_{t}-\partial^{\,}_{x}\right)\,
\psi^{\,}_{\mathrm{R},\alpha,A}
\end{equation}
on the one hand, and
\begin{align}
&
T^{\,}_{\mathrm{R}}[\hat{u}(1)]=
\frac{1}{4k}\,
j^{\,}_{\mathrm{R}}\,
j^{\,}_{\mathrm{R}},
\\
&
T^{\,}_{\mathrm{R}}[\widehat{su}(2)^{\,}_{k}]=
\frac{1}{k+2}\,
\sum_{c=1}^{3}
J^{c}_{\mathrm{R}}\,
J^{c}_{\mathrm{R}},
\\
&
T^{\,}_{\mathrm{R}}[\widehat{su}(k)^{\,}_{2}]=
\frac{1}{2+k}\,
\sum_{\mathrm{c}=1}^{k^{2}-1}
\mathrm{J}^{\mathrm{c}}_{\mathrm{R}}\,
\mathrm{J}^{\mathrm{c}}_{\mathrm{R}},
\end{align}
\end{subequations}
on the other hand. The currents are here defined by
\begin{subequations}
\label{eq: def su(2) and su(k) left right currents}
\begin{align}
&
j^{\,}_{\mathrm{R}}:=
\sum_{\alpha=1}^{2}
\sum_{A=1}^{k}
\psi^{*}_{\mathrm{R},\alpha,A}\,
\psi^{\,}_{\mathrm{R},\alpha,A},
\label{eq: master identity for u(2k) level 1 h}
\\
&
J^{c}_{\mathrm{R}}:=
\frac{1}{2}
\sum_{\alpha,\beta=1}^{2}
\sum_{A=1}^{k}
\psi^{*}_{\mathrm{R},\alpha,A}\
\sigma^{c}_{\alpha\beta}\
\psi^{\,}_{\mathrm{R},\beta,A},
\label{eq: master identity for u(2k) level 1 i}
\\
&
\mathrm{J}^{\mathrm{c}}_{\mathrm{R}}:=
\sum_{\alpha=1}^{2}
\sum_{A,B=1}^{k}
\psi^{*}_{\mathrm{R},\alpha,A}\
T^{\mathrm{c}}_{AB}\
\psi^{\,}_{\mathrm{R},\alpha,B},
\label{eq: master identity for u(2k) level 1 j}
\end{align}
for $c=1,2,3$ and $\mathrm{c}=1,\cdots,k^{2}-1$,
respectively. Hereto, we have imposed the normalization
condition
\begin{equation}
\mathrm{tr}\,
\left(
T^{\mathrm{c}}\,
T^{\mathrm{c}'}
\right)=
\frac{1}{2}\,
\delta^{\,}_{\mathrm{c}\mathrm{c}'}
\end{equation}
for $\mathrm{c},\mathrm{c}'=1,\cdots,k^{2}-1$.
This normalization condition
is equivalent to choosing
the structure constants of the unitary Lie algebra $su(k)$ such that
\begin{equation}
\sum_{\mathrm{c}'',\mathrm{c}'''}^{k^{2}-1}
f^{\,}_{\mathrm{c}\mathrm{c}''\mathrm{c}'''}\,
f^{\,}_{\mathrm{c}'\mathrm{c}''\mathrm{c}'''}=
k\,
\delta^{\,}_{\mathrm{c}\mathrm{c}'}
\end{equation}
\end{subequations}
for any $\mathrm{c},\mathrm{c}'=1,\cdots,k^{2}-1$.

The transformation laws of the currents 
(\ref{eq: def su(2) and su(k) left right currents})
under the representation
(\ref{eq: def reversal of time on complex fermions})
of time reversal are
\begin{subequations}
\label{eq: trsf law currents array wires}
\begin{align}
&
j^{\,}_{\mathrm{R}}\mapsto
+
j^{\,}_{\mathrm{L}},
\qquad
j^{\,}_{\mathrm{L}}\mapsto
+
j^{\,}_{\mathrm{R}},
\\
&
J^{c}_{\mathrm{R}}\mapsto
-
J^{c}_{\mathrm{L}},
\qquad
J^{c}_{\mathrm{L}}\mapsto
-
J^{c}_{\mathrm{R}},
\\
&
\mathrm{J}^{\mathrm{c}}_{\mathrm{R}}\mapsto
(-1)^{p(\mathrm{c})}\,
\mathrm{J}^{\mathrm{c}}_{\mathrm{L}},
\qquad
\mathrm{J}^{\mathrm{c}}_{\mathrm{L}}\mapsto
(-1)^{p(\mathrm{c})}\,
\mathrm{J}^{\mathrm{c}}_{\mathrm{R}},
\end{align}
\end{subequations}
for $c=1,2,3$ and $\mathrm{c}=1,\cdots,k^{2}-1$. Here,
$p(\mathrm{c})=0$ if the generator $T^{\mathrm{c}}$ 
is a real-valued matrix
while $p(\mathrm{c})=1$ if the generator $T^{\mathrm{c}}$ 
is an imaginary-valued matrix.

For any given bundle,
the currents
(\ref{eq: master identity for u(2k) level 1 h}),
(\ref{eq: master identity for u(2k) level 1 i}),
(\ref{eq: master identity for u(2k) level 1 j}),
and their counterparts with $k$ replaced by $k'$
are separately conserved, for they all commute pairwise. 
To each of these six pairwise commuting currents,
there corresponds a gapless sector of the free theory
on which these currents act. 
The point-split and normal-ordered Lagrangian density$\ $%
\footnote{
There exists a different ordering of the 
right- and left-moving, $\alpha$, and $A$ labels
than the ordering chosen in Eq.\ (\ref{eq: def umklapp gap mathcal L})
that opens up a superconducting gap in the charge sector.
The ordering chosen in Eq.\ (\ref{eq: def umklapp gap mathcal L})
corresponds to a $k$-th order umklapp process. 
         }
\begin{equation}
\begin{split}
\mathcal{L}^{U(1)}_{\mathrm{int}}:=&\,
-
g^{\,}_{U(1)}\, 
e^{\mathrm{i}\alpha(x)}\,
\left(
\prod_{A=1}^{k}
\prod_{\alpha=1}^{2}
\psi^{*}_{\mathrm{R},\alpha,A}
\right)
\\
&\,
\times
\left(
\prod_{A=k}^{1}
\prod_{\alpha=2}^{1}
\psi^{\,}_{\mathrm{L},\alpha,A}
\right)
\\
&\,
+
[\psi^{*}_{\mathrm{R}}\to\psi^{\,}_{\mathrm{L}},\ 
 \psi^{\,}_{\mathrm{L}}\to\psi^{*}_{\mathrm{R}},\ \alpha(x)\to-\alpha(x)]
\end{split}
\label{eq: def umklapp gap mathcal L}
\end{equation}
gaps the $U(1)$ charge sector for the wires 
$1$ to $k$ from the bundle for $g^{\,}_{U(1)}>0$
sufficiently large.
The $SU(2)$ current-current interaction
\begin{equation}
\mathcal{L}^{SU(2)}_{\mathrm{int}}:=
-
\lambda^{\,}_{SU(2)}
\sum_{c=1}^{3}
J^{c}_{\mathrm{R}}\,
J^{c}_{\mathrm{L}}
\end{equation}
gaps the $SU(2)$ sector for the wires 
$1$ to $k$ from the bundle when $\lambda^{\,}_{SU(2)}>0$.
The $SU(k)$ current-current interaction
\begin{equation}
\mathcal{L}^{SU(k)}_{\mathrm{int}}:=
-
\lambda^{\,}_{SU(k)}
\sum_{\mathrm{c}=1}^{k^{2}-1}
\mathrm{J}^{\mathrm{c}}_{\mathrm{R}}\,
\mathrm{J}^{\mathrm{c}}_{\mathrm{L}}
\end{equation}
gaps the $SU(k)$ sector for the wires 
$1$ to $k$ from the bundle when $\lambda^{\,}_{SU(k)}>0$.
The same reasoning applies in the sector with $U(2k')$ symmetry.

We choose to gap the $U(1)$ and $SU(k)$ sectors
without breaking spontaneously the $SU(k)$ symmetry,
while leaving the sector of the theory associated to the symmetry 
\begin{equation}
G:=SU(2)\times SU(2)
\label{eq: def group G}
\end{equation}
momentarily gapless. 
The low-energy theory is then given by the gapless theory
with an energy-momentum tensor 
of the Sugawara form whereby the currents realize the
semi-simple affine Lie algebra
\begin{equation}
\hat{g}^{(n)}_{k,k'}:=
\opluslimits_{\texttt{m}=1}^{n}
\Bigg(
\widehat{su}(2)^{\,}_{k}
\oplus
\widehat{su}(2)^{\,}_{k'}
\Bigg).
\end{equation}
This gapless theory has the central charge
\begin{align}
c[\hat{g}^{(n)}_{k,k'}]=&\,
\sum_{\texttt{m}=1}^{n}
\Bigg(
c[\widehat{su}(2)^{\,}_{k}]
+
c[\widehat{su}(2)^{\,}_{k'}]
\Bigg)
\nonumber\\
=&\,
3\,n\,
\Bigg(
\frac{k}{k+2}
+
\frac{k'}{k'+2}
\Bigg).
\label{eq: c[su(2)kxsu(2)k'}
\end{align}
As it should be, this central charge is smaller than
the central charge $2\,n\,(k+k')$ from Eq.\ (\ref{eq: c0=2N}).

We consider the diagonal subgroup
\begin{equation}
H:=SU(2)
\label{eq: def subgroup H of G}
\end{equation}
of the group (\ref{eq: def group G}).
The corresponding semi-simple affine Lie algebra,
a semi-simple affine subalgebra of $\hat{g}^{\,}_{k,k'}$, is
\begin{equation}
\hat{h}^{(n)}_{k,k'}:=
\opluslimits_{\texttt{m}=1}^{n}
\widehat{su}(2)^{\,}_{k+k'}.
\label{eq: def subalgebra h of g}
\end{equation}
\medskip

We need to reinstate the label
$\texttt{m}=1,\cdots,n$ for the bundles of $k+k'$ consecutive
wires as well as the left- and right-moving labels
as we are going to couple these sectors.
We denote the generators of $\hat{g}^{(n)}_{k,k'}$ by
$\mathcal{J}^{\mathcal{A}}_{\mathrm{R},\texttt{m}}$
and
$\mathcal{J}^{\mathcal{A}}_{\mathrm{L},\texttt{m}}$,
where $\mathcal{A}=1,\cdots,6$ and
$\texttt{m}=1,\cdots,n$. For example, in the right-moving sector,
we may choose the vector field
\begin{subequations}
\label{eq: def mathcal L hat g / hat h}
\begin{equation}
\bm{\mathcal{J}}^{\,}_{\mathrm{R},\texttt{m}}:=
\frac{1}{2}
\sum_{\alpha,\beta=1}^{2}
\sum_{A=1}^{k}
\psi^{*}_{\mathrm{R},\alpha,A,\texttt{m}}\
\bm{\sigma}^{\,}_{\alpha\beta}\
\psi^{\,}_{\mathrm{R},\beta,A,\texttt{m}},
\label{eq: def mathcal L hat g / hat h a}
\end{equation}
when $\mathcal{A}=1,2,3$ and the vector-field
\begin{equation}
\bm{\mathcal{J}}^{\prime}_{\mathrm{R},\texttt{m}}:=
\frac{1}{2}
\sum_{\alpha,\beta=1}^{2}
\sum_{A'=1}^{k'}
\psi^{*}_{\mathrm{R},\alpha,A',\texttt{m}}\
\bm{\sigma}^{\,}_{\alpha\beta}\
\psi^{\,}_{\mathrm{R},\beta,A',\texttt{m}},
\label{eq: def mathcal L hat g / hat h b}
\end{equation} 
when $\mathcal{A}=4,5,6$.
We denote the generators of $\hat{h}^{(n)}_{k,k'}$ by
$\mathcal{K}^{\mathcal{B}}_{\mathrm{R},\texttt{m}}$
and
$\mathcal{K}^{\mathcal{B}}_{\mathrm{L},\texttt{m}}$,
where $\mathcal{B}=1,\cdots,3$ and
$\texttt{m}=1,\cdots,n$. For example, in the right-moving sector,
we may choose the vector field 
\begin{equation}
\bm{\mathcal{K}}^{\,}_{\mathrm{R},\texttt{m}}:=
\bm{\mathcal{J}}^{\,}_{\mathrm{R},\texttt{m}}
+
\bm{\mathcal{J}}^{\prime}_{\mathrm{R},\texttt{m}}.
\label{eq: def mathcal L hat g / hat h c}
\end{equation} 
We work with open boundary conditions along the chain
of quantum wires and define the interaction
[see Fig.\ \ref{Fig: gapping without TRS}(c)]
\begin{equation}
\begin{split}
\mathcal{L}^{\mathrm{L}\to\mathrm{R}}_{\mathrm{int}}:=&\,
-
\sum_{\texttt{m}=1}^{n-1}
\sum_{\mathcal{A}=1}^{6}
\lambda^{\mathcal{A}}_{\texttt{m}}\,
\mathcal{J}^{\mathcal{A}}_{\mathrm{L},\texttt{m}}\,
\mathcal{J}^{\mathcal{A}}_{\mathrm{R},\texttt{m}+1}
\\
&\,
-
\sum_{\texttt{m}=1}^{n}
\sum_{\mathcal{B}=1}^{3}
\upsilon^{\mathcal{B}}_{\texttt{m}}\,
\mathcal{K}^{\mathcal{B}}_{\mathrm{L},\texttt{m}}\,
\mathcal{K}^{\mathcal{B}}_{\mathrm{R},\texttt{m}},
\end{split}
\label{eq: def mathcal L hat g / hat h d}
\end{equation}
\end{subequations}
where the couplings 
$\lambda^{\mathcal{A}}_{\texttt{m}}$
and
$\upsilon^{\mathcal{B}}_{\texttt{m}}$
are real-valued.
Had we imposed periodic boundary conditions in the direction
of the chain of wires on the Grassmann fields, 
it would be legitimate to extend the sum
over the bundles so as to include the term with $\texttt{m}=n$.

It is shown in Appendix 
\ref{appsubsec: Derivation of the one-loop RG flows if no TRS}
that 
(i) 
all couplings 
in Eq.\ (\ref{eq: def mathcal L hat g / hat h d})
flow to strong coupling when initially nonvanishing and positive,
(ii) 
no new terms involving the right-moving generators from
$\hat{g}^{(n)}_{k,k'}/\hat{h}^{(n)}_{k,k'}$
in the bundle $\texttt{m}=1$ appear to one loop,
and
(iii)
no new terms involving the left-moving generators from
$\hat{g}^{(n)}_{k,k'}/\hat{h}^{(n)}_{k,k'}$
in the bundle $\texttt{m}=n$ appear to one loop.

We make the following conjecture
regarding the strong coupling fixed point 
depending on the initial values of the couplings
in Eq.\ (\ref{eq: def mathcal L hat g / hat h d}).

With open boundary conditions and when all the couplings
in Eq.\ (\ref{eq: def mathcal L hat g / hat h d})
are positive and of the same order, 
the resulting theory remains critical.
As the resulting theory would be fully gapped 
had we opted for periodic boundary conditions,
the critical sectors of the theory with open boundary conditions
must be confined to the boundaries, namely the first bundle
$\texttt{m}=1$ and the last bundle $\texttt{m}=n$.
The first bundle of $k+k'$ wires
hosts the critical theory described
by the right sector of the coset theory
\begin{subequations}
\begin{equation}
\hat{g}^{\,}_{k,k'}/\hat{h}^{\,}_{k,k'}:=
\widehat{su}^{\,}_{}(2)^{\,}_{k}
\oplus
\widehat{su}^{\,}_{}(2)^{\,}_{k'}/
\widehat{su}^{\,}_{}(2)^{\,}_{k+k'}
\end{equation}
with the chiral central charge
\begin{align}
c[(\hat{g}^{\,}_{k,k'}/\hat{h}^{\,}_{k,k'})^{\,}_{\mathrm{R}}]=&\,
3\,
\Bigg(
\frac{k}{k+2}
+
\frac{k'}{k'+2}
\Bigg)
-
3\,\frac{k+k'}{k+k'+2}
\nonumber\\
=&\,
1
-
\frac{6k'}{(k+2)(k+k'+2)}
+
\frac{2(k'-1)}{k'+2}.
\label{eq: choice two for interaction c}
\end{align}
The last bundle of $k+k'$ wires
hosts the critical theory described
by the left sector of the coset theory
\begin{equation}
\hat{g}^{\,}_{k,k'}/\hat{h}^{\,}_{k,k'}:=
\widehat{su}^{\,}_{}(2)^{\,}_{k}
\oplus
\widehat{su}^{\,}_{}(2)^{\,}_{k'}/
\widehat{su}^{\,}_{}(2)^{\,}_{k+k'}
\end{equation}
with the chiral central charge
\begin{align}
c[(\hat{g}^{\,}_{k,k'}/\hat{h}^{\,}_{k,k'})^{\,}_{\mathrm{L}}]=&\,
c[(\hat{g}^{\,}_{k,k'}/\hat{h}^{\,}_{k,k'})^{\,}_{\mathrm{R}}].
\label{eq: choice two for interaction d}
\end{align}
\end{subequations}
The interaction
(\ref{eq: def mathcal L hat g / hat h d})
has broken the time-reversal symmetry,
gapped the bulk, and left in the first and last bundle
two massless coset theories of opposite chiralities.
For the bundle on the left (right) boundary
the critical boundary theory is built from the
holomorphic (antiholomorphic) generators in the quotient
$\widehat{su}(2)^{\,}_{k}\oplus\widehat{su}(2)^{\,}_{k'}
/\widehat{su}(2)^{\,}_{k+k'}$
of affine Lie algebras.

The last term on the right-hand side of
Eqs.\
(\ref{eq: choice two for interaction c})
and
(\ref{eq: choice two for interaction d})
is the central charge 
\begin{equation}
c[\widehat{su}(2)^{\,}_{k'}/\hat{u}(1)]=
\frac{3\,k'}{k'+2}-1=
\frac{2(k'-1)}{k'+2}
\end{equation}  
of the coset $\widehat{su}(2)^{\,}_{k'}/\hat{u}(1)$.
In the local operator content of this theory,
one finds a pair of local parafermionic fields 
$\hat{\psi}^{\dag}_{\mathrm{par}}$
and 
$\hat{\psi}^{\,}_{\mathrm{par}}$
with the scaling dimensions 
$(k'-1)/k'$
and a real-valued bosonic field $\hat{\varphi}$ such that
the generators of the affine Lie algebra
$\widehat{su}(2)^{\,}_{k'}$
are represented by the operators$\ $%
\footnote{
See Chapter 18.5.3 from Ref.\
\onlinecite{DiFrancesco97}.
         }
\begin{subequations}
\begin{align}
&
\hat{J}^{+}(z)=
\sqrt{k'}\
\hat{\psi}^{\,}_{\mathrm{par}}(z)\,
:e^{+\mathrm{i}\sqrt{2/k'}\,\hat{\varphi}(z)}:,
\\
&
\hat{J}^{-}(z)=
\sqrt{k'}\
\hat{\psi}^{\dag}_{\mathrm{par}}(z)\,
:e^{-\mathrm{i}\sqrt{2/k'}\,\hat{\varphi}(z)}:,
\\
&
\hat{J}^{0}(z)=
\mathrm{i}\sqrt{2k'}\,(\partial^{\,}_{z}\hat{\varphi})(z).
\end{align}
\end{subequations}
For $k'=1$, the parafermions reduce to the identity.
For $k'=2$, the parafermions obey the fermion algebra.
For $k'>2$ the parafermions obey a more complicated algebra.
For example, if one writes 
\begin{subequations}
\begin{equation}
\hat{\psi}^{\,}_{\mathrm{par}}\propto
\frac{\mathrm{i}}{2}
\left[
\hat{\chi}^{\,}_{1}
+
\left(\hat{\chi}^{\,}_{1}\right)^{k'-1}
\right]
+
\frac{1}{2}
\left[
\hat{\chi}^{\,}_{2}
+
\left(\hat{\chi}^{\,}_{2}\right)^{k'-1}
\right],
\end{equation}
it then follows that
\begin{align}
&
\left(\hat{\chi}^{\,}_{1}\right)^{k'}=1,
\qquad
\left(\hat{\chi}^{\,}_{2}\right)^{k'}=1,
\\
&
\left(\hat{\chi}^{\,}_{1}\right)^{k'-1}=
\left(\hat{\chi}^{\,}_{1}\right)^{\dag},
\qquad
\left(\hat{\chi}^{\,}_{2}\right)^{k'-1}=
\left(\hat{\chi}^{\,}_{2}\right)^{\dag},
\\
&
\hat{\chi}^{\,}_{1}\,\hat{\chi}^{\,}_{2}=
e^{\mathrm{i}2\pi/k'}\,
\hat{\chi}^{\,}_{2}\,\hat{\chi}^{\,}_{1},
\end{align}
\end{subequations}
holds locally.

It is time to specialize by choosing
\begin{equation}
k'=1.
\end{equation}
With this choice, the chiral central charges
(\ref{eq: choice two for interaction c})
and
(\ref{eq: choice two for interaction d})
are nothing but the central charge
\begin{equation}
c(k)=
1
-
\frac{6}{(k+2)(k+3)}
\label{eq: central charge minimal model}
\end{equation}
for the minimal models of two-dimensional conformal field theories.
This is not a coincidence, for it is known that the coset affine 
Lie algebra
\begin{equation}
\hat{g}^{\,}_{k,1}/\hat{h}^{\,}_{k,1}=
\widehat{su}(2)^{\,}_{k}\oplus\widehat{su}(2)^{\,}_{1}
/
\widehat{su}(2)^{\,}_{k+1}
\end{equation}
realizes the series of minimal models with $k=1,2,\cdots$.$\ $ %
\cite{DiFrancesco97}
The minimal models encode the critical properties of
two-dimensional lattice models at their critical temperature
such as the Ising model ($k=1$), 
the tricritical Ising model ($k=2$),
the three-states Potts model ($k=3$), and so on.
We conclude that we have realized the holomorphic and
antiholomorphic critical sectors of the minimal models
on the opposite boundaries of an open chain of $n$ bundles of wires,
respectively.

\begin{figure}[t!]
\begin{center}
\includegraphics[scale=0.6]{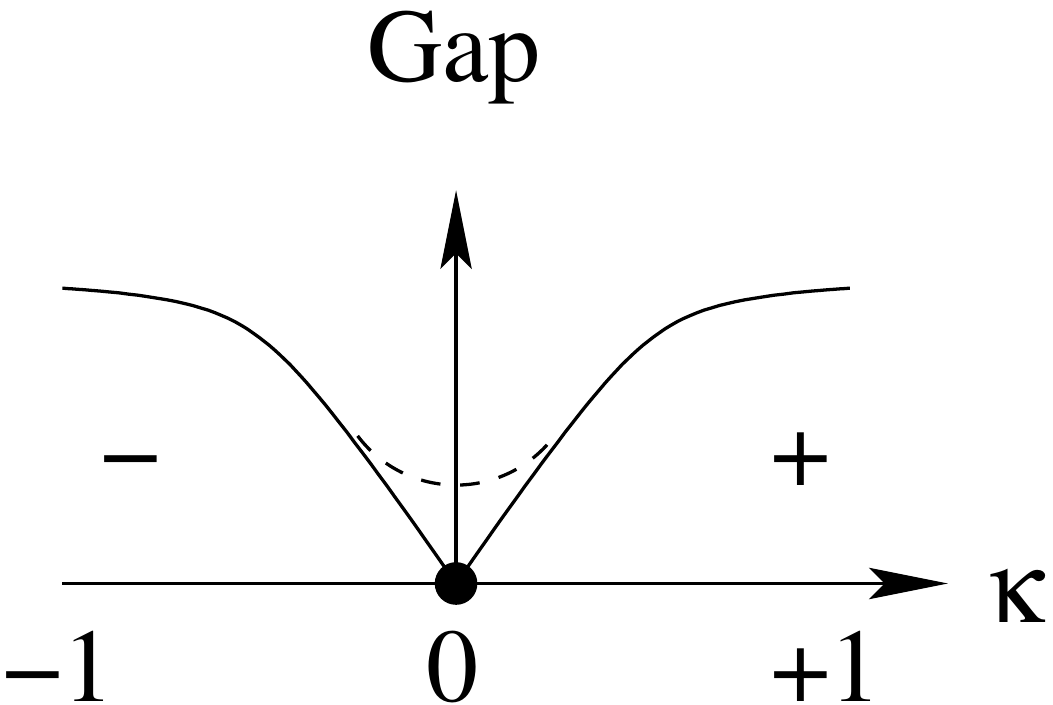}
\\
\caption{
One possible phase diagram with the interaction
(\ref{eq: interpolating L int}).
The time-reversal symmetric point is
parametrized by $\kappa=0$.
The vertical axis is the many-body gap 
between the ground state and all excited states
when periodic boundary conditions are imposed.
The continuous line represents the 
scenario for an exotic spin-liquid quantum critical point. 
The dashed line represents the scenario for 
a first-order quantum phase transition.
The signs $-$ and $+$ distinguish
the two ground states that evolve adiabatically 
as $\kappa\neq0$ changes
and cross precisely at $\kappa=0$.
\label{Fig: phase diagram}
         }
\end{center}
\end{figure}

The choice $k'=2$ turns the chiral central charges
(\ref{eq: choice two for interaction c})
and
(\ref{eq: choice two for interaction d})
into the chiral central charge
\begin{equation}
c(k)=
\frac{3}{2}
\Biggl[1
-
\frac{8}{(k+2)(k+4)}
\Biggr]
\end{equation}
for the minimal models of two-dimensional superconformal field theories.
This is again not a coincidence, for it is known that the coset affine 
Lie algebra
\begin{equation}
\hat{g}^{\,}_{k,2}/\hat{h}^{\,}_{k,2}=
\widehat{su}(2)^{\,}_{k}\oplus\widehat{su}(2)^{\,}_{2}
/
\widehat{su}(2)^{\,}_{k+2}
\end{equation}
realizes the series of superconformal minimal models with
$k=1,2,\cdots$.$\ $ %
\cite{DiFrancesco97}
Notice that, for $k=1$,
$c(k=1) = 7/10$
coincides with the second member ($k=2$) of the minimal model
(\ref{eq: central charge minimal model})
that corresponds to the tricritical Ising model. 
The tricritical Ising model is one example that
realizes supersymmetry in statistical physics.  We conclude that we
have realized the holomorphic and antiholomorphic critical sectors of
the superconformal minimal models on the opposite boundaries of an
open chain of $n$ bundles of wires, respectively.

\subsection{Partial gapping with time-reversal symmetry}
\label{subsec: Partial gapping with time-reversal symmetry}

We shall impose time-reversal symmetry on the array of
quantum wires coupled by current-current interactions
in three different ways.

In Sec.\ \ref{subsubsec: Case I},
we symmetrize the interaction
(\ref{eq: def mathcal L hat g / hat h d})
under reversal of time.

In Sec.\ \ref{subsubsec: Case II},
we double the number of degrees of freedom
in the low-energy sector of the theory
by postulating that this doubling originates from 
degrees of freedom that are exchanged under reversal of time.
We then write down current-current interactions
that preserve time-reversal symmetry, gap the bulk,
but leave gapless boundary states.

In Sec.\ \ref{subsubsec: Case III},
unlike was the case in
Secs.\ 
\ref{subsec: Partial gapping without time-reversal symmetry},
\ref{subsubsec: Case I},
and
\ref{subsubsec: Case II},
we assume that spin-1/2 rotation symmetry is broken prior
to adding current-current interactions. We then explain how
to reproduce the treatment of
Sec.\ \ref{subsubsec: Case II}.

\subsubsection{Case I -- Symmetrized interaction}
\label{subsubsec: Case I}

We assume that the interactions responsible for
gapping the $U(k)\times U(k')$ sector of the theory 
in Sec.\ \ref{subsec: Partial gapping without time-reversal symmetry}
preserve both time-reversal symmetry and spin-1/2 rotation symmetry. 

\begin{figure}[t]
\begin{center}
\begin{center}
(a)$\hphantom{AAAAAAAAAAAAAAAAAAAAAAAAAAAAAAAAAAAA}$ 
\vskip 2 true pt
\includegraphics[scale=0.5]{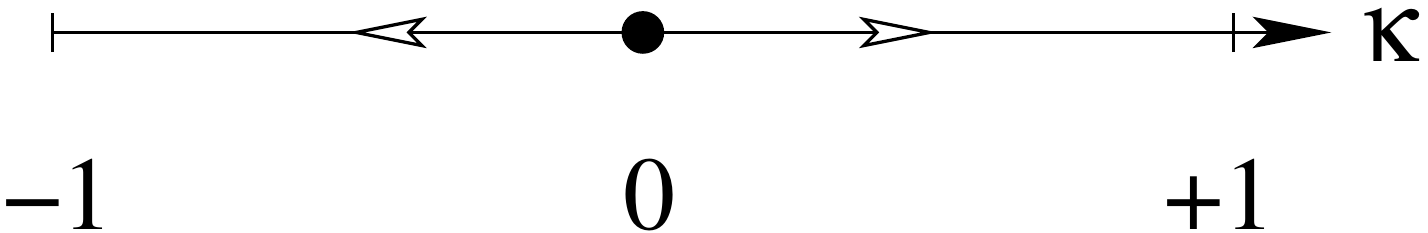}
\end{center}
\vskip 40 true pt $\vphantom{A}$
\begin{center}
(b)$\hphantom{AAAAAAAAAAAAAAAAAAAAAAAAAAAAAAAAAAAA}$ 
\vskip 2 true pt 
\includegraphics[scale=0.5]{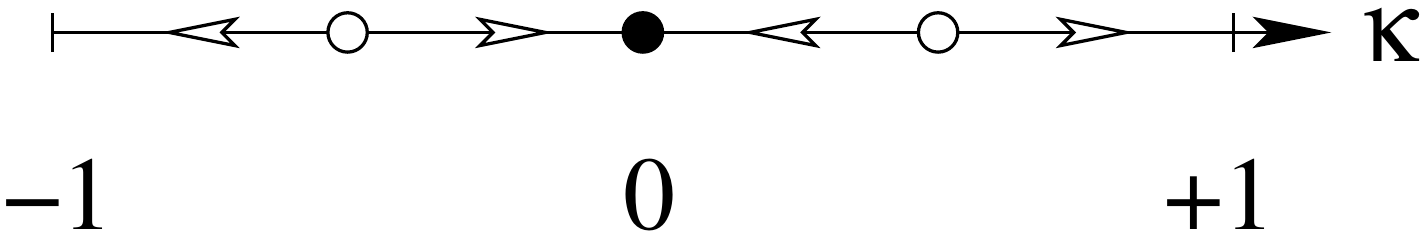}
\end{center}
\caption{
For the case when the phases transitions in the range $-1\leq\kappa\leq+1$ 
at strong values of the couplings 
$\lambda^{\mathcal{A}}_{\texttt{m}}>0$ 
and 
$v^{\mathcal{B}}_{\texttt{m}}>0$
in Eq.\ (\ref{eq: interpolating L int})
are continuous, the critical point at $\kappa=0$
is either unstable or stable depending on
whether the number of critical points for
$-1<\kappa<0$ is even as in (a) or odd
as in (b), respectively. 
\label{Fig: four possible phase diagrams}
         }
\end{center}
\end{figure}

Reversal of time turns the interaction
(\ref{eq: def mathcal L hat g / hat h d})
into the interaction
[see Fig.\ \ref{Fig: gapping without TRS}(d)]
\begin{equation}
\begin{split}
\mathcal{L}^{\mathrm{R}\to\mathrm{L}}_{\mathrm{int}}:=&\,
-
\sum_{\texttt{m}=1}^{n-1}
\sum_{\mathcal{A}=1}^{6}
\lambda^{\mathcal{A}}_{\texttt{m}}\,
\mathcal{J}^{\mathcal{A}}_{\mathrm{R},\texttt{m}}\,
\mathcal{J}^{\mathcal{A}}_{\mathrm{L},\texttt{m}+1}
\\
&\,
-
\sum_{\texttt{m}=1}^{n}
\sum_{\mathcal{B}=1}^{3}
\upsilon^{\mathcal{B}}_{\texttt{m}}\,
\mathcal{K}^{\mathcal{B}}_{\mathrm{R},\texttt{m}}\,
\mathcal{K}^{\mathcal{B}}_{\mathrm{L},\texttt{m}}.
\end{split}
\label{eq: def mathcal L hat g / hat h bis}
\end{equation}
As was the case with the interaction
(\ref{eq: def mathcal L hat g / hat h d}),
we conjecture a gapped bulk with
two massless coset theories of opposite chiralities
on the first and last bundles of wires.
For the left (right) boundary bundle 
the critical boundary theory is built from the
antiholomorphic (holomorphic)
generators in the quotient 
$\widehat{su}(2)^{\,}_{k}\oplus\widehat{su}(2)^{\,}_{k'}
/\widehat{su}(2)^{\,}_{k+k'}$
of affine Lie algebras.

We may then interpolate between the interactions
(\ref{eq: def mathcal L hat g / hat h d})
and
(\ref{eq: def mathcal L hat g / hat h bis})
as a function of the real-valued parameter $\kappa$ by defining
\begin{equation}
\mathcal{L}^{\,}_{\mathrm{int}}(\kappa):=
\frac{1-\kappa}{2}\,
\mathcal{L}^{\mathrm{L}\to\mathrm{R}}_{\mathrm{int}}
+
\frac{1+\kappa}{2}\,
\mathcal{L}^{\mathrm{R}\to\mathrm{L}}_{\mathrm{int}}.
\label{eq: interpolating L int}
\end{equation}
The interactions
(\ref{eq: def mathcal L hat g / hat h d})
and
(\ref{eq: def mathcal L hat g / hat h bis})
compete to impose one of two ways for the breaking of time-reversal symmetry.
When $\kappa\leq-1$, the interaction
$\mathcal{L}^{\mathrm{L}\to\mathrm{R}}_{\mathrm{int}}$
is marginally relevant, while the interaction
$\mathcal{L}^{\mathrm{R}\to\mathrm{L}}_{\mathrm{int}}$
is marginally irrelevant,
as is shown in Appendix \ref{appsubsec: Derivation of the RG flows for Case I}.
It is the fixed point
represented by
Fig.\ \ref{Fig: gapping without TRS}(c)
to which the relevant couplings flow.
When $1\leq\kappa$, it is the fixed point
represented by
Fig.\ \ref{Fig: gapping without TRS}(d)
to which the relevant couplings flow
as is shown in Appendix \ref{appsubsec: Derivation of the RG flows for Case I}.
The analysis of the one-loop RG flows
is more subtle when $\kappa\in[-1,+1]\setminus\{0\}$.
It is shown in Appendix \ref{appsubsec: Derivation of the RG flows for Case I}
that
$\mathcal{L}^{\mathrm{L}\to\mathrm{R}}_{\mathrm{int}}$
and
$\mathcal{L}^{\mathrm{R}\to\mathrm{L}}_{\mathrm{int}}$
are both marginally relevant perturbations.
If one assumes that the point $\kappa=0$ 
at which time-reversal symmetry holds explicitly is singular, 
there are then two logical possibilities pertaining to the nature
of this singularity.

On the one hand, the singularity at $\kappa=0$ 
could signal a continuous quantum phase transition
at which the bulk gap closes and the (thermal) Hall conductivity
switches sign, as occurs with the 
single-particle Dirac Hamiltonian$\ $ %
\cite{Ludwig94}
\begin{equation}
\mathcal{H}^{\,}_{\mathrm{D}}:=
-
\mathrm{i}\sigma^{\,}_{x}\,\hat{p}^{\,}_{x}
-
\mathrm{i}\sigma^{\,}_{y}\,\hat{p}^{\,}_{y}
+
m\, \sigma^{\,}_{z}
\end{equation}
in two-dimensional space
when the mass $m$ changes sign in a continuous fashion.
If so, the gapless bulk phase represents an exotic gapless
spin liquid phase in $(2+1)$-dimensional space and time,
for it emerges from two long-ranged entangled
gapped phases supporting non-Abelian topological order
that are unrelated by a breaking of a local symmetry.

If all phase transitions in the range $-1\leq\kappa\leq+1$
are continuous, the critical point at $\kappa=0$ is either stable
or unstable. The latter case occurs if the number of critical points
in the range $-1<\kappa<0$ is even, as shown in Figs.\
\ref{Fig: four possible phase diagrams}(a).
The former case occurs 
if the number of critical points in the range $-1<\kappa<0$ is odd, 
as shown in Figs.\
\ref{Fig: four possible phase diagrams}(b).
The one-loop RG analysis made in Appendix
\ref{appsubsec: Derivation of the RG flows for Case I}
applies to the vicinity of the noninteracting critical point
when all the couplings
$\lambda^{\mathcal{A}}_{\texttt{m}}$ 
and 
$v^{\mathcal{B}}_{\texttt{m}}$
in Eq.\ (\ref{eq: interpolating L int})
vanish. In the limit
$\lambda^{\mathcal{A}}_{\texttt{m}}\to0$ 
and 
$v^{\mathcal{B}}_{\texttt{m}}\to0$,
the one-loop RG flow for $\kappa$ is the one depicted in
Fig.\ \ref{Fig: four possible phase diagrams}(a).
However, we cannot infer from this weak coupling analysis
whether it is 
Fig.\ \ref{Fig: four possible phase diagrams}(a)
or
Fig.\ \ref{Fig: four possible phase diagrams}(b)
that applies to the relevant limit
$\lambda^{\mathcal{A}}_{\texttt{m}}\to\infty$ 
and 
$v^{\mathcal{B}}_{\texttt{m}}\to\infty$.

On the other hand, the singularity at $\kappa=0$ could signal
a discontinuous transition, as occurs in the Ising model
upon changing the sign of an applied magnetic field.
At $\kappa=0$, the energy eigenvalue of the ground state
for $\kappa<0$ crosses that of the ground state for $\kappa>0$,
while the gap to the excitation spectra for $\kappa<0$ and $\kappa>0$ 
do not close at $\kappa=0$.

\begin{figure*}[t]
\begin{center}
\includegraphics[scale=0.6]{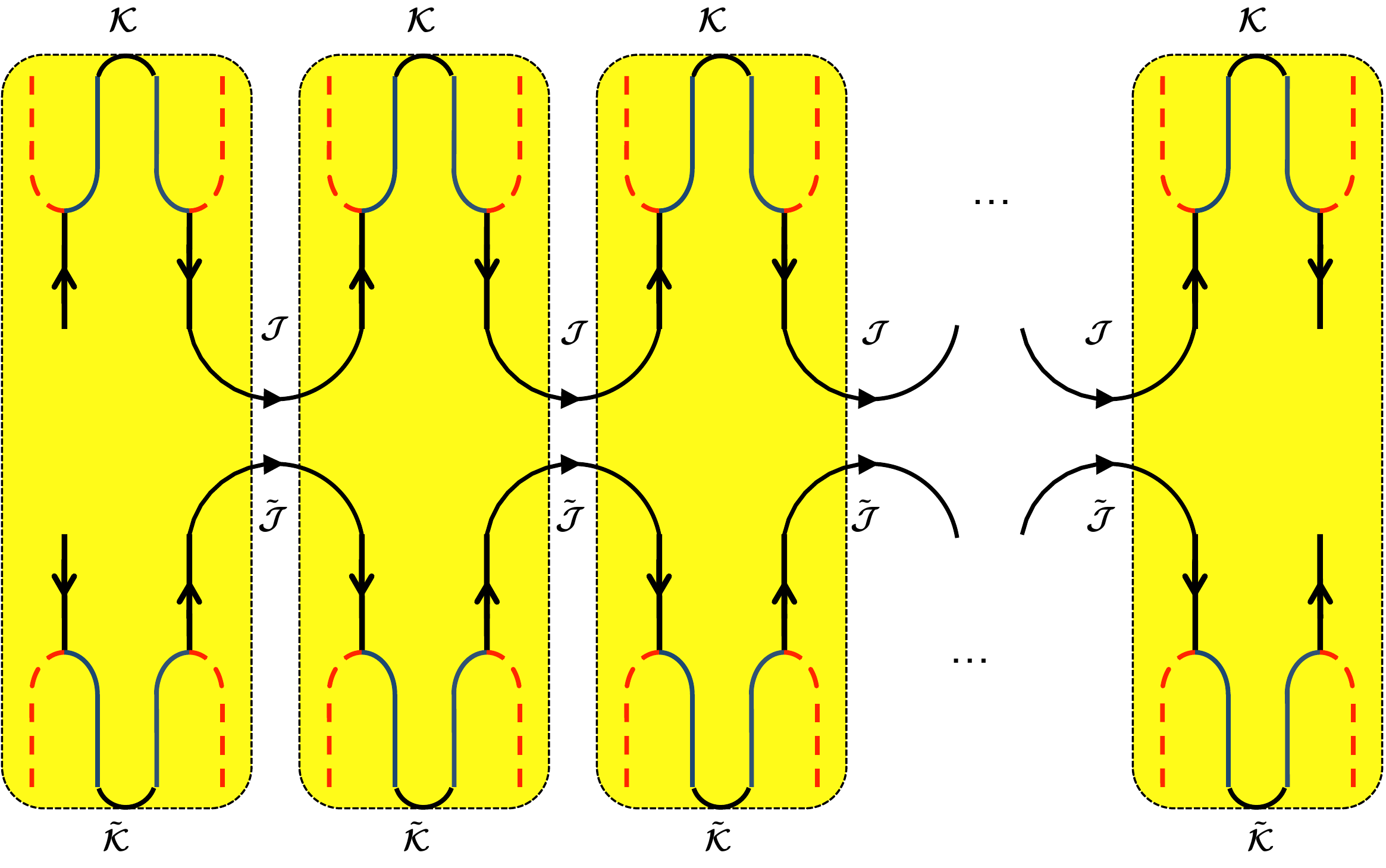}
\hfill
\caption{(Color online)
Chain of $N$ wires, each wire supporting four right- 
and four left-moving flavors,
that are coupled by current-current interactions 
in a way that is explicitly symmetric under reversal of time.
A bundle of $8\times(k+k')$ right- or left-moving
electronic degrees of freedom is represented by a domino.
There are $n=N/(k+k')$ dominoes.
The currents denoted by $\mathcal{J}$ generate the affine Lie algebra
$\widehat{su}(2)^{\,}_{k}\oplus\widehat{su}(2)^{\,}_{k'}$.
The currents denoted by $\mathcal{K}$ generate the affine Lie algebra
$\widehat{su}(2)^{\,}_{k+k'}$.
The currents denoted by $\tilde{\mathcal{J}}$ generate the affine Lie algebra
$\widehat{\widetilde{su}}(2)^{\,}_{k}\oplus\widehat{\widetilde{su}}(2)^{\,}_{k'}$.
The currents denoted by $\tilde{\mathcal{K}}$ generate the affine Lie algebra
$\widehat{\widetilde{su}}(2)^{\,}_{k+k'}$.
\label{Fig: gapping with TRS}
         }
\end{center}
\end{figure*}

\subsubsection{Case II -- Doubled degrees of freedom}
\label{subsubsec: Case II}

We continue assuming that the interactions responsible for
gapping the $U(k)\times U(k')$ sector of the theory 
in Sec.\ \ref{subsec: Partial gapping without time-reversal symmetry}
preserve both time-reversal symmetry and spin-1/2 rotation symmetry.

An alternative implementation of time-reversal symmetry 
consists in (i) doubling the dimensionality of the Fock space
by direct product with a two-dimensional auxiliary Hilbert space
and (ii) demanding that reversal of time is represented by a matrix
that is off-diagonal with respect to this auxiliary 
two-dimensional Hilbert space.
An example of such a two-dimensional auxiliary Hilbert space
is provided by the two valleys of graphene
[recall Example 2 of a quasi-one-dimensional gapless phase 
defined by Eq.\ (\ref{eq: Example 2 2D gapless phase})]. 
According to
Eq.\ (\ref{eq: Example 2 2D gapless phase}),
half of the degrees of freedom encoded in any one of the bundles 
can be interpreted as originating from 
the two-dimensional nonvanishing momenta about 
which the low-energy degrees of freedom are constructed.

Accordingly, we may choose to work with the total of $8\times N$ 
electronic right- or left-moving degrees of freedom, 
which we organize into $n$ bundles, 
each of which supports $8\times(k+k')$ 
electronic right- or left-moving degrees of freedom, 
where $k$ and $k'$ are two nonvanishing positive integers.
In other words, the number $4\times N=4\times n\,(k+k')$
of electronic right- or left-moving degrees of freedom
corresponding to the number of quantum wires 
(\ref{eq: bundle with TRS breaking}) is replaced by
\begin{equation}
8\times N=8n\,(k+k').
\label{eq: bundle with TRS}
\end{equation}
This is to say that we extend
the quadruplet of Grassmann-valued vectors
$\psi^{*}_{\mathrm{R}}$,
$\psi^{*}_{\mathrm{L}}$,
$\psi^{\,}_{\mathrm{R}}$,
and
$\psi^{\,}_{\mathrm{L}}$
with the components
$\psi^{*}_{\mathrm{R},\sigma,I}$,
$\psi^{*}_{\mathrm{L},\sigma,I}$,
$\psi^{\,}_{\mathrm{R},\sigma,I}$,
and
$\psi^{\,}_{\mathrm{L},\sigma,I}$,
respectively, by the quadruplet of Grassmann-valued vectors
$\tilde{\psi}^{*}_{\mathrm{R}}$,
$\tilde{\psi}^{*}_{\mathrm{L}}$,
$\tilde{\psi}^{\,}_{\mathrm{R}}$,
and
$\tilde{\psi}^{\,}_{\mathrm{L}}$
with the components
$\tilde{\psi}^{*}_{\mathrm{R},\sigma,I}$,
$\tilde{\psi}^{*}_{\mathrm{L},\sigma,I}$,
$\tilde{\psi}^{\,}_{\mathrm{R},\sigma,I}$,
and
$\tilde{\psi}^{\,}_{\mathrm{L},\sigma,I}$,
respectively. We then replace the critical theory 
(\ref{eq: def N wires with ``one'' electron per wire})
by the critical theory
\begin{subequations}
\label{eq: def 2N wires with ``one'' electron per wire}
\begin{equation}
\begin{split}
\mathcal{L}^{\,}_{0}:=&\,
\mathrm{i}
\Big[
\psi^{*\mathsf{T}}_{\mathrm{R}}\,
(\partial^{\,}_{t}+\partial^{\,}_{x})\,
\psi^{\,}_{\mathrm{R}}
+
\psi^{*\mathsf{T}}_{\mathrm{L}}\,
(\partial^{\,}_{t}-\partial^{\,}_{x})\,
\psi^{\,}_{\mathrm{L}}
\Big]
\\
&\,
+
\mathrm{i}
\Big[
\tilde{\psi}^{*\mathsf{T}}_{\mathrm{R}}\,
(\partial^{\,}_{t}+\partial^{\,}_{x})\,
\tilde{\psi}^{\,}_{\mathrm{R}}
+
\tilde{\psi}^{*\mathsf{T}}_{\mathrm{L}}\,
(\partial^{\,}_{t}-\partial^{\,}_{x})\,
\tilde{\psi}^{\,}_{\mathrm{L}}
\Big]
\end{split}
\label{eq: def 2N wires with ``one'' electron per wire a}
\end{equation}
with the action
\begin{equation}
S^{\,}_{0}:=
\int\mathrm{d}t
\int\mathrm{d}x\,
\mathcal{L}^{\,}_{0}
\label{eq: def 2N wires with ``one'' electron per wire b}
\end{equation}
and the partition function
\begin{equation}
Z^{\,}_{0}:=
\int\mathcal{D}[\psi^{*},\psi]\
\int\mathcal{D}[\tilde{\psi}^{*},\tilde{\psi}]\
e^{\mathrm{i}S^{\,}_{0}}.
\end{equation}
\end{subequations}

Reversal of time is the antilinear transformation
\textit{defined} by the fundamental rules
\begin{subequations}
\label{eq: def reversal of time for case II}
\begin{align}
&
\psi^{*}_{\mathrm{R},\uparrow,I}\mapsto
+
\tilde{\psi}^{*}_{\mathrm{L},\downarrow,I},
\qquad
\psi^{*}_{\mathrm{R},\downarrow,I}\mapsto
-
\tilde{\psi}^{*}_{\mathrm{L},\uparrow,I},
\\
&
\psi^{*}_{\mathrm{L},\uparrow,I}\mapsto
+
\tilde{\psi}^{*}_{\mathrm{R},\downarrow,I},
\qquad
\psi^{*}_{\mathrm{L},\downarrow,I}\mapsto
-
\tilde{\psi}^{*}_{\mathrm{R},\uparrow,I},
\\
&
\tilde{\psi}^{*}_{\mathrm{R},\uparrow,I}\mapsto
+
\psi^{*}_{\mathrm{L},\downarrow,I},
\qquad
\tilde{\psi}^{*}_{\mathrm{R},\downarrow,I}\mapsto
-
\psi^{*}_{\mathrm{L},\uparrow,I},
\\
&
\tilde{\psi}^{*}_{\mathrm{L},\uparrow,I}\mapsto
+
\psi^{*}_{\mathrm{R},\downarrow,I},
\qquad
\tilde{\psi}^{*}_{\mathrm{L},\downarrow,I}\mapsto
-
\psi^{*}_{\mathrm{R},\uparrow,I},
\end{align}
and
\begin{align}
&
\psi^{\,}_{\mathrm{R},\uparrow,I}\mapsto
+
\tilde{\psi}^{\,}_{\mathrm{L},\downarrow,I},
\qquad
\psi^{\,}_{\mathrm{R},\downarrow,I}\mapsto
-
\tilde{\psi}^{\,}_{\mathrm{L},\uparrow,I},
\\
&
\psi^{\,}_{\mathrm{L},\uparrow,I}\mapsto
+
\tilde{\psi}^{\,}_{\mathrm{R},\downarrow,I},
\qquad
\psi^{\,}_{\mathrm{L},\downarrow,I}\mapsto
-
\tilde{\psi}^{\,}_{\mathrm{R},\uparrow,I},
\\
&
\tilde{\psi}^{\,}_{\mathrm{R},\uparrow,I}\mapsto
+
\psi^{\,}_{\mathrm{L},\downarrow,I},
\qquad
\tilde{\psi}^{\,}_{\mathrm{R},\downarrow,I}\mapsto
-
\psi^{\,}_{\mathrm{L},\uparrow,I},
\\
&
\tilde{\psi}^{\,}_{\mathrm{L},\uparrow,I}\mapsto
+
\psi^{\,}_{\mathrm{R},\downarrow,I},
\qquad
\tilde{\psi}^{\,}_{\mathrm{L},\downarrow,I}\mapsto
-
\psi^{\,}_{\mathrm{R},\uparrow,I}.
\end{align}
\end{subequations}
By this definition, reversal of time squares to minus the identity
and leaves the critical theory
(\ref{eq: def 2N wires with ``one'' electron per wire})
invariant. Moreover, if we define the Grassmann-valued doublets
\begin{subequations}
\label{eq: def  Grassmann-valued doublets Psi}
\begin{equation}
\Psi^{*}_{\mathrm{R}}:=
\begin{pmatrix}
\psi^{*}_{\mathrm{R}}
\\ \\
\tilde{\psi}^{*}_{\mathrm{R}}
\end{pmatrix},
\qquad
\Psi^{*}_{\mathrm{L}}:=
\begin{pmatrix}
\psi^{*}_{\mathrm{L}}
\\ \\
\tilde{\psi}^{*}_{\mathrm{L}}
\end{pmatrix}
\end{equation}
and
\begin{equation}
\Psi^{\,}_{\mathrm{R}}:=
\begin{pmatrix}
\psi^{\,}_{\mathrm{R}}
\\ \\
\tilde{\psi}^{\,}_{\mathrm{R}}
\end{pmatrix},
\qquad
\Psi^{\,}_{\mathrm{L}}:=
\begin{pmatrix}
\psi^{\,}_{\mathrm{L}}
\\ \\
\tilde{\psi}^{\,}_{\mathrm{L}}
\end{pmatrix},
\end{equation}
the representation
\begin{equation}
\mathcal{L}^{\,}_{0}=
\mathrm{i}
\Psi^{*\mathsf{T}}_{\mathrm{R}}
\left(\partial^{\,}_{t}+\partial^{\,}_{x}\right)
\Psi^{\,}_{\mathrm{R}}
+
\mathrm{i}
\Psi^{*\mathsf{T}}_{\mathrm{L}}
\left(\partial^{\,}_{t}-\partial^{\,}_{x}\right)
\Psi^{\,}_{\mathrm{L}}
\end{equation}
\end{subequations}
of the critical theory
(\ref{eq: def 2N wires with ``one'' electron per wire})
makes it explicit that it has the symmetry group
$U^{\,}_{\mathrm{R}}(4N)\times U^{\,}_{\mathrm{L}}(4N)$.

Any one bundle of $8\times(k+k')$ 
electronic right- or left-moving degrees of freedom
is represented by any one domino
from Fig.\ \ref{Fig: gapping with TRS}.
The symmetry that we select when considering any one of
the $n$ bundles of $8\times(k+k')$ 
electronic right- or left-moving degrees of freedom
is the direct product
\begin{subequations} 
\label{eq: selcted symmetry when TRS holds}
\begin{equation}
U:=
\Big(U(2k)\times\widetilde{U}(2k)\Big)
\times 
\Big(U(2k')\times\widetilde{U}(2k')\Big).
\end{equation}
As before, the multiplicative factor of $2$ in $2k$ or $2k'$
stands for the electronic spin-1/2 degrees of freedom. However, 
a second multiplicative  factor of $2$ in $8\times n\,(k+k')$ 
is responsible for the two copies of the
unitary group of $2k$-dimensional matrices
and $2k'$-dimensional matrices, respectively.
The corresponding semi-simple affine Lie algebra is
\begin{equation}
\hat{u}^{\,}_{1}:=
\Big(
\hat{u}(2k)^{\,}_{1}
\oplus
\hat{\tilde{u}}(2k)^{\,}_{1}
\Big)
\oplus
\Big(
\hat{u}(2k')^{\,}_{1}
\oplus
\hat{\tilde{u}}(2k')^{\,}_{1}
\Big).
\end{equation}
\end{subequations}
Equation (\ref{eq: selcted symmetry when TRS holds})
should be compared to 
Eq.\ (\ref{eq: selcted symmetry when TRS is broken}).
As before, we use the conformal embedding
\begin{subequations}
\label{eq: master identity for u(4k) level II}
\begin{align}
&
\hat{u}(2k)^{\,}_{1}=
\hat{u}(1)
\oplus
\widehat{su}(2)^{\,}_{k}
\oplus
\widehat{su}(k)^{\,}_{2},
\label{eq: master identity for u(4k) level II a}
\\
&
\hat{\tilde{u}}(2k)^{\,}_{1}=
\hat{\tilde{u}}(1)
\oplus
\widehat{\widetilde{su}}(2)^{\,}_{k}
\oplus
\widehat{\widetilde{su}}(k)^{\,}_{2},
\label{eq: master identity for u(4k) level II b}
\\
&
\hat{u}(2k')^{\,}_{1}=
\hat{u}(1)
\oplus
\widehat{su}(2)^{\,}_{k'}
\oplus
\widehat{su}(k')^{\,}_{2},
\label{eq: master identity for u(4k) level II c}
\\
&
\hat{\tilde{u}}(2k')^{\,}_{1}=
\hat{\tilde{u}}(1)
\oplus
\widehat{\widetilde{su}}(2)^{\,}_{k'}
\oplus
\widehat{\widetilde{su}}(k')^{\,}_{2},
\label{eq: master identity for u(4k) level II d}
\end{align}
\end{subequations}
between affine Lie algebras. Here, the generators of these
affine Lie algebra are given by Eq.\
(\ref{eq: def su(2) and su(k) left right currents}) 
for the conformal embedding (\ref{eq: master identity for u(4k) level II a}),
by
\begin{subequations} 
\label{eq: def tilde su(2) and tilde su(k) left right currents}
\begin{align}
&
\tilde{j}^{\,}_{\mathrm{R}}:=
\sum_{\alpha=1}^{2}
\sum_{A=1}^{k}
\tilde{\psi}^{*}_{\mathrm{R},\alpha,A}\,
\tilde{\psi}^{\,}_{\mathrm{R},\alpha,A},
\label{eq: master identity for tilde u(2k) level 1 h}
\\
&
\tilde{J}^{c}_{\mathrm{R}}:=
\frac{1}{2}
\sum_{\alpha,\beta=1}^{2}
\sum_{A=1}^{k}
\tilde{\psi}^{*}_{\mathrm{R},\alpha,A}\
\sigma^{c}_{\alpha\beta}\
\tilde{\psi}^{\,}_{\mathrm{R},\beta,A},
\label{eq: master identity for tilde u(2k) level 1 i}
\\
&
\tilde{\mathrm{J}}^{\mathrm{c}}_{\mathrm{R}}:=
\sum_{\alpha=1}^{2}
\sum_{A,B=1}^{k}
\tilde{\psi}^{*}_{\mathrm{R},\alpha,A}\
T^{\mathrm{c}}_{AB}\
\tilde{\psi}^{\,}_{\mathrm{R},\alpha,B},
\label{eq: master identity for tilde u(2k) level 1 j}
\end{align}
\end{subequations}
($c=1,2,3$ and $\mathrm{c}=1,\cdots,k^{2}-1$)
for the conformal embedding (\ref{eq: master identity for u(4k) level II b}),
and similarly for the conformal embeddings
(\ref{eq: master identity for u(4k) level II c})
and 
(\ref{eq: master identity for u(4k) level II d}),
respectively.

We choose to gap the sectors with the symmetries
$U(1)$, $SU(k)$,
$\widetilde{U}(1)$, $\widetilde{SU}(k)$, 
and similarly for $k'$,
while leaving the sector of the theory associated to the symmetry 
\begin{subequations}
\begin{equation}
G:=
\Big(SU(2)\times \widetilde{SU}(2)\Big)
\times 
\Big(SU(2)\times \widetilde{SU}(2)\Big)
\label{eq: def group GII}
\end{equation}
momentarily gapless. 
The semi-simple affine Lie algebra associated to $G$ is
\begin{equation}
\hat{g}^{(n)}_{k,k'}=
\opluslimits_{\texttt{m}=1}^{n}
\Big(
\widehat{su}(2)^{\,}_{k}
\oplus
\widehat{\widetilde{su}}(2)^{\,}_{k}
\Big)
\oplus
\Big(
\widehat{su}(2)^{\,}_{k'}
\oplus
\widehat{\widetilde{su}}(2)^{\,}_{k'}
\Big).
\end{equation}
\end{subequations}
It is now the diagonal subgroup
\begin{subequations}
\begin{equation}
H:=
SU(2) \times \widetilde{SU}(2)
\label{eq: def subgroup H of GII}
\end{equation}
of the group (\ref{eq: def group GII})
that we shall use to construct the gapless theory on the edge.
The corresponding simple affine subalgebra of $\hat{g}^{\,}_{k,k'}$, is
\begin{equation}
\hat{h}^{(n)}_{k,k'}:=
\opluslimits_{\texttt{m}=1}^{n}
\widehat{su}(2)^{\,}_{k+k'}
\oplus
\widehat{\widetilde{su}}(2)^{\,}_{k+k'}.
\label{eq: def subalgebra h of gII}
\end{equation}
\end{subequations}
The currents generating
$\widehat{su}(2)^{\,}_{k}\oplus\widehat{su}(2)^{\,}_{k'}$
are represented by the symbol $J$
in  Fig.\ \ref{Fig: gapping with TRS}.
The currents generating
$\widehat{\widetilde{su}}(2)^{\,}_{k}\oplus\widehat{\widetilde{su}}(2)^{\,}_{k'}$
are represented by the symbol $\tilde{J}$
in Fig.\ \ref{Fig: gapping with TRS}.
The currents generating
$\widehat{su}(2)^{\,}_{k+k'}$
are represented by the symbol $K$
in Fig.\ \ref{Fig: gapping with TRS}.
The currents generating
$\widehat{\widetilde{su}}(2)^{\,}_{k+k'}$
are represented by the symbol $\tilde{K}$
in Fig.\ \ref{Fig: gapping with TRS}.

Current-current interactions are represented in 
Fig.\ \ref{Fig: gapping with TRS}
by arcs that are directed 
when they involve the currents $J$ or $\tilde{J}$,
while they are undirected when they involve the currents $K$ or $\tilde{K}$.
In Fig.\ \ref{Fig: gapping with TRS}, 
the action of reversal of time is twofold. First,
the directions of arrows must be reversed, thereby interchanging right- or 
left-movers. Second, the letters without $\tilde{\hphantom{J}}$ acquire a 
$\tilde{\hphantom{J}}$, 
while letters with $\tilde{\hphantom{J}}$ loose their $\tilde{\hphantom{J}}$.
The corresponding interaction
\begin{widetext}
\begin{subequations}
\label{eq: final interaction if TRS}
\begin{equation}
\begin{split}
\mathcal{L}^{\,}_{\mathrm{int}}:=&\,
-
\sum_{\texttt{m}=1}^{n-1}
\sum_{\mathcal{A}=1}^{6}
\lambda^{\mathcal{A}}_{\texttt{m}}\,
\mathcal{J}^{\mathcal{A}}_{\mathrm{L},\texttt{m}}\,
\mathcal{J}^{\mathcal{A}}_{\mathrm{R},\texttt{m}+1}
-
\sum_{\texttt{m}=1}^{n}
\sum_{\mathcal{B}=1}^{3}
\upsilon^{\mathcal{B}}_{\texttt{m}}\,
\mathcal{K}^{\mathcal{B}}_{\mathrm{L},\texttt{m}}\,
\mathcal{K}^{\mathcal{B}}_{\mathrm{R},\texttt{m}}
-
\sum_{\texttt{m}=1}^{n-1}
\sum_{\mathcal{A}=1}^{6}
\lambda^{\mathcal{A}}_{\texttt{m}}\,
\tilde{\mathcal{J}}^{\mathcal{A}}_{\mathrm{R},\texttt{m}}\,
\tilde{\mathcal{J}}^{\mathcal{A}}_{\mathrm{L},\texttt{m}+1}
-
\sum_{\texttt{m}=1}^{n}
\sum_{\mathcal{B}=1}^{3}
\upsilon^{\mathcal{B}}_{\texttt{m}}\,
\tilde{\mathcal{K}}^{\mathcal{B}}_{\mathrm{R},\texttt{m}}\,
\tilde{\mathcal{K}}^{\mathcal{B}}_{\mathrm{L},\texttt{m}}
\end{split}
\label{eq: final interaction if TRS a}
\end{equation}
with the real-valued couplings
$\lambda^{\mathcal{A}}_{\texttt{m}}$
and
$\upsilon^{\mathcal{B}}_{\texttt{m}}$
is invariant under the rules
(i.e., those for angular momentum)
\begin{equation}
\begin{split}
&
\mathcal{J}^{\mathcal{A}}_{\mathrm{L},\texttt{m}}\mapsto
-
\tilde{\mathcal{J}}^{\mathcal{A}}_{\mathrm{R},\texttt{m}},
\qquad
\mathcal{J}^{\mathcal{A}}_{\mathrm{R},\texttt{m}}\mapsto
-
\tilde{\mathcal{J}}^{\mathcal{A}}_{\mathrm{L},\texttt{m}},
\qquad
\tilde{\mathcal{J}}^{\mathcal{A}}_{\mathrm{R},\texttt{m}}\mapsto
-
\mathcal{J}^{\mathcal{A}}_{\mathrm{L},\texttt{m}},
\qquad
\tilde{\mathcal{J}}^{\mathcal{A}}_{\mathrm{L},\texttt{m}}\mapsto
-
\mathcal{J}^{\mathcal{A}}_{\mathrm{R},\texttt{m}},
\\
&
\mathcal{K}^{\mathcal{B}}_{\mathrm{L},\texttt{m}}\mapsto
-
\tilde{\mathcal{K}}^{\mathcal{B}}_{\mathrm{R},\texttt{m}},
\qquad
\mathcal{K}^{\mathcal{B}}_{\mathrm{R},\texttt{m}}\mapsto
-
\tilde{\mathcal{K}}^{\mathcal{B}}_{\mathrm{L},\texttt{m}},
\qquad
\tilde{\mathcal{K}}^{\mathcal{B}}_{\mathrm{R},\texttt{m}}\mapsto
-
\mathcal{K}^{\mathcal{B}}_{\mathrm{L},\texttt{m}},
\qquad
\tilde{\mathcal{K}}^{\mathcal{B}}_{\mathrm{L},\texttt{m}}\mapsto
-
\mathcal{K}^{\mathcal{B}}_{\mathrm{R},\texttt{m}},
\end{split}
\label{eq: final interaction if TRS b}
\end{equation}
\end{subequations}
\end{widetext}
a consequence of the definition of time reversal made
in Eq.\ (\ref{eq: def reversal of time for case II}).
Observe that iterating the transformation
(\ref{eq: final interaction if TRS b})
twice yields the identity operation.
The time-reversal-symmetric interaction
(\ref{eq: final interaction if TRS})
partially gaps the theory with $8N$ decoupled noninteracting
electronic right- or left-moving degrees of freedom. 
The one-loop RG equations obeyed by the couplings entering
the Lagrangian density (\ref{eq: final interaction if TRS a})
are derived in Sec.\
\ref{appsubsec: Derivation of the RG flows for Case II}
and given in Eqs.\
(\ref{appeq: RGSUN TRS case 2 J})
and
(\ref{appeq: RGSUN TRS case 2 K}). 
They are marginally relevant and flow to strong couplings
if the couplings are initially nonvanishing and positive.

Inclusion of all the spin-rotation symmetric and
time-reversal-symmetric 
interactions responsible for fully gapping 
the $U(1)\times\widetilde{U}(1)$, 
$SU(k)\times\widetilde{SU}(k)$, 
and
$SU(k')\times\widetilde{SU}(k')$
symmetry sectors together with the 
time-reversal-symmetric interaction
(\ref{eq: final interaction if TRS})
results in the critical theory that is built from 
the coset WZW theory
\begin{subequations}
\label{eq: choice two for interaction II}
\begin{equation}
\hat{g}^{\,}_{k,k'}/\hat{h}^{\,}_{k,k'}:= 
\frac{
\Big(\widehat{su}(2)^{\,}_{k}\oplus\widehat{\widetilde{su}}(2)^{\,}_{k}\Big)
\oplus
\Big(\widehat{su}(2)^{\,}_{k'}\oplus\widehat{\widetilde{su}}(2)^{\,}_{k'}\Big)
     }
     {
\widehat{su}(2)^{\,}_{k+k'}\oplus\widehat{\widetilde{su}}(2)^{\,}_{k+k'}
     }
\end{equation}
between affine Lie algebras, with the (nonchiral) central charge
\begin{align}
c[\hat{g}^{\,}_{k,k'}/\hat{h}^{\,}_{k,k'}]=&\,
2
\Biggl[
3\,
\Bigg(
\frac{k}{k+2}
+
\frac{k'}{k'+2}
\Bigg)
-
3\,\frac{k+k'}{k+k'+2}
\Biggr]
\nonumber\\
=&\,
2
\Biggl[
1
-
\frac{6k'}{(k+2)(k+k'+2)}
+
\frac{2(k'-1)}{k'+2}
\Biggr].
\label{eq: choice two for interaction cII}
\end{align}
\end{subequations}
This central charge is twice the value of the
chiral central charge 
(\ref{eq: choice two for interaction d}).
Since imposing periodic boundary conditions
gaps completely the chain of quantum wires,
we infer that both the bundle
$\texttt{m}=1$
and
$\texttt{m}=n$
can be assigned the nonchiral central charge
\begin{equation}
\frac{c[\hat{g}^{\,}_{k,k'}/\hat{h}^{\,}_{k,k'}]}{2}=
\Biggl[
1
-
\frac{6k'}{(k+2)(k+k'+2)}
+
\frac{2(k'-1)}{k'+2}
\Biggr].
\label{eq: nonchiral central charge at one boundary case II}
\end{equation}

The stability analysis of either one of the boundary coset WZW theories 
with the central charge 
(\ref{eq: nonchiral central charge at one boundary case II})
is more subtle than that for
Sec.\ \ref{subsec: Partial gapping without time-reversal symmetry}.
There are relevant primary fields in the
coset WZW theories with the central charge 
(\ref{eq: nonchiral central charge at one boundary case II}).
However, their potential for gapping the critical point 
for the boundaries is not
accounted for in the stability analysis as long as they are not generated
under an RG flow by either one-body or 
many-body electron-electron interactions.

\begin{figure*}[t]
\begin{center}
(a)
\includegraphics[scale=0.5]{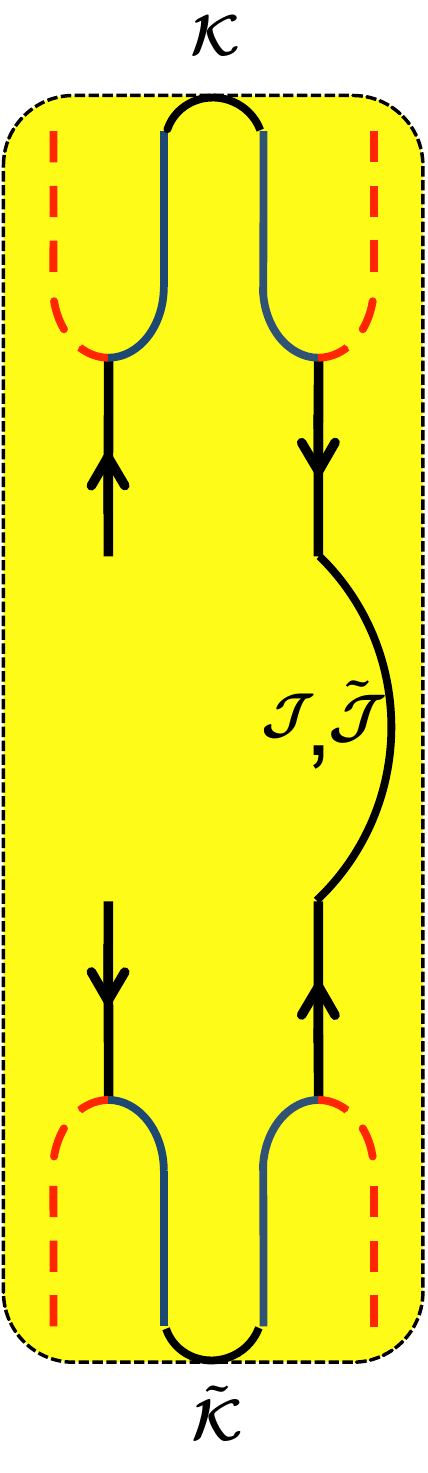}
\hfill
(b)
\includegraphics[scale=0.5]{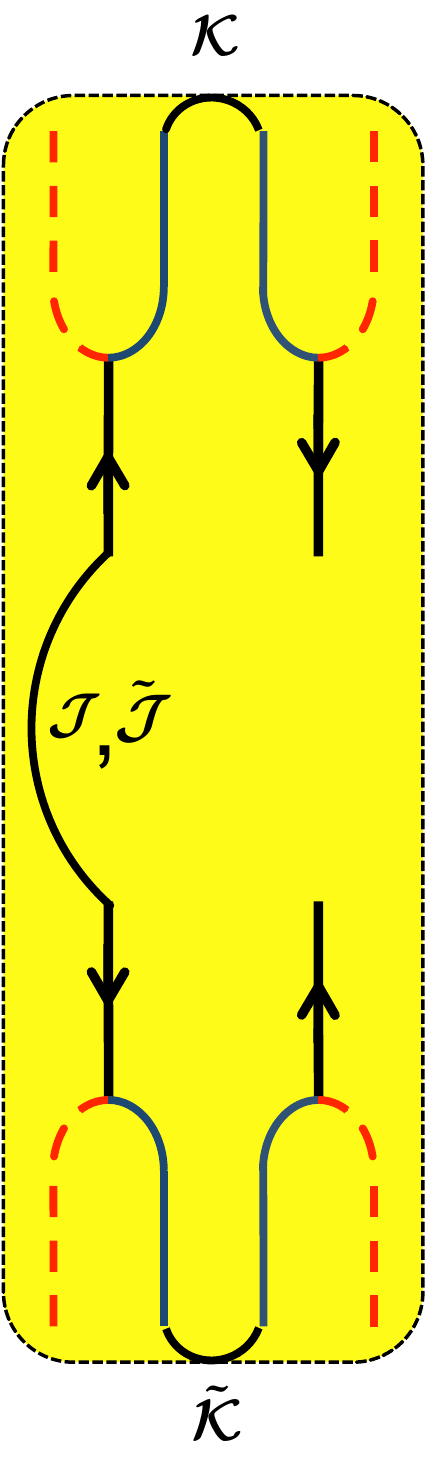}
\hfill
(c)
\includegraphics[scale=0.5]{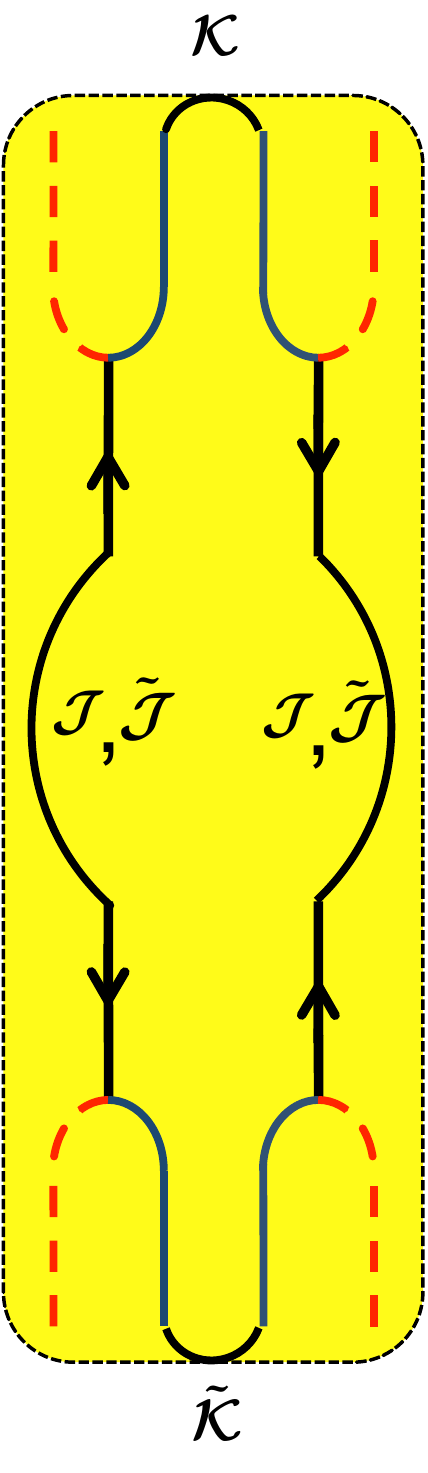}
\caption{(Color online)
The partial gapping of a single domino with local current-current interactions
can be done in two ways (a) and (b).
The complete gapping of a single domino with local current-current interactions
is achieved in (c).
\label{Fig: gapping with TRS of single dominos}
         }
\end{center}
\end{figure*}

There is a crucial difference 
between either one of the boundary coset WZW theories 
with the central charge 
(\ref{eq: nonchiral central charge at one boundary case II})
and the chiral boundary theories from 
Sec.\ \ref{subsec: Partial gapping without time-reversal symmetry}.
Starting from electrons, the latter can only be obtained 
on the one-dimensional boundaries of two-dimensional space. 
Starting from electrons, the former, however, can be obtained 
directly from either one of the strictly one-dimensional models 
represented by the single domino from
Fig.\ \ref{Fig: gapping with TRS of single dominos}(a)
and the single domino from
Fig.\ \ref{Fig: gapping with TRS of single dominos}(b).
For example, Fig.\ \ref{Fig: gapping with TRS of single dominos}(a)
realizes the same critical theory as the 
left boundary critical theory
represented by
Fig.\ \ref{Fig: gapping with TRS}
provided the interaction depicted in
Fig.\ \ref{Fig: gapping with TRS of single dominos}(a)
that is defined by
\begin{equation}
\label{eq: final interaction one domino if TRS}
\begin{split}
\mathcal{L}^{\,}_{\mathrm{int}}:=&\,
-
\sum_{\mathcal{B}=1}^{3}
\upsilon^{\mathcal{B}}_{\mathrm{boundary},1}\,
\mathcal{K}^{\mathcal{B}}_{\mathrm{L},1}\,
\mathcal{K}^{\mathcal{B}}_{\mathrm{R},1}
\\
&\,
-
\sum_{\mathcal{A}=1}^{6}
\lambda^{\mathcal{A}}_{\mathrm{boundary},1}\,
\mathcal{J}^{\mathcal{A}}_{\mathrm{L},1}\,
\widetilde{\mathcal{J}}^{\mathcal{A}}_{\mathrm{R},1}
\\
&
-
\sum_{\mathcal{B}=1}^{3}
\upsilon^{\mathcal{B}}_{\mathrm{boundary},1}\,
\tilde{\mathcal{K}}^{\mathcal{B}}_{\mathrm{R},1}\,
\tilde{\mathcal{K}}^{\mathcal{B}}_{\mathrm{L},1}
\end{split}
\end{equation}
preserves time-reversal symmetry.
This is indeed the case as time reversal is represented by
\begin{equation}
\begin{split}
&
\mathcal{J}^{\mathcal{A}}_{\mathrm{L},1}\mapsto
-
\tilde{\mathcal{J}}^{\mathcal{A}}_{\mathrm{R},1},
\qquad
\mathcal{J}^{\mathcal{A}}_{\mathrm{R},1}\mapsto
-
\tilde{\mathcal{J}}^{\mathcal{A}}_{\mathrm{L},1},
\\
&
\tilde{\mathcal{J}}^{\mathcal{A}}_{\mathrm{R},1}\mapsto
-
\mathcal{J}^{\mathcal{A}}_{\mathrm{L},1},
\qquad
\tilde{\mathcal{J}}^{\mathcal{A}}_{\mathrm{L},1}\mapsto
-
\mathcal{J}^{\mathcal{A}}_{\mathrm{R},1},
\\
&
\mathcal{K}^{\mathcal{A}}_{\mathrm{L},1}\mapsto
-
\tilde{\mathcal{K}}^{\mathcal{A}}_{\mathrm{R},1},
\qquad
\mathcal{K}^{\mathcal{A}}_{\mathrm{R},1}\mapsto
-
\tilde{\mathcal{K}}^{\mathcal{A}}_{\mathrm{L},1},
\\
&
\tilde{\mathcal{K}}^{\mathcal{A}}_{\mathrm{R},1}\mapsto
-
\mathcal{K}^{\mathcal{A}}_{\mathrm{L},1},
\qquad
\tilde{\mathcal{K}}^{\mathcal{A}}_{\mathrm{L},1}\mapsto
-
\mathcal{K}^{\mathcal{A}}_{\mathrm{R},1},
\end{split}
\label{eq: def TR for currents first bundle}
\end{equation}
for $\mathcal{A}=1,\cdots,6$ and $\mathcal{B}=1,\cdots,3$.
We emphasize that the transformation law
(\ref{eq: def TR for currents first bundle})
squares to unity.

On the one hand,
it is shown in Appendix 
\ref{appsec: The stability analysis of the coset WZW theory}
that the time-reversal symmetry alone does not prevent gapping
either one of the boundary coset WZW theories 
with the central charge 
(\ref{eq: nonchiral central charge at one boundary case II})
through one-body mass terms for the electrons.
On the other hand, it is shown in Appendix 
\ref{appsec: The stability analysis of the coset WZW theory}
that the time-reversal symmetry together with the 
$U(1)$ symmetry under the linear transformation
\begin{subequations}
\label{eq: def U(1) symmetry between no tilde and tilde sectors}
\begin{align}
&
\psi^{*}_{\mathrm{R}}\mapsto
\psi^{*}_{\mathrm{R}}\,e^{-\mathrm{i}\theta},
\qquad
&
\tilde{\psi}^{*}_{\mathrm{R}}\mapsto
\tilde{\psi}^{*}_{\mathrm{R}}\,e^{+\mathrm{i}\theta},
\\
&
\psi^{*}_{\mathrm{L}}\mapsto
\psi^{*}_{\mathrm{L}}\,e^{-\mathrm{i}\theta},
\qquad
&
\tilde{\psi}^{*}_{\mathrm{L}}\mapsto
\tilde{\psi}^{*}_{\mathrm{L}}\,e^{+\mathrm{i}\theta},
\\
&
\psi^{\,}_{\mathrm{R}}\mapsto
e^{+\mathrm{i}\theta}\,\psi^{\,}_{\mathrm{R}},
\qquad
&
\tilde{\psi}^{\,}_{\mathrm{R}}\mapsto
e^{-\mathrm{i}\theta}\,\tilde{\psi}^{\,}_{\mathrm{R}},
\\
&
\psi^{\,}_{\mathrm{L}}\mapsto
e^{+\mathrm{i}\theta}\,\psi^{\,}_{\mathrm{L}},
&
\qquad
\tilde{\psi}^{\,}_{\mathrm{L}}\mapsto
e^{-\mathrm{i}\theta}\,\tilde{\psi}^{\,}_{\mathrm{L}},
\end{align}
\end{subequations}
that is parameterized by $0\leq\theta<2\pi$, 
does prevent gapping through one-body mass terms for the electrons.
Observe here that the $U(1)$ symmetry
(\ref{eq: def U(1) symmetry between no tilde and tilde sectors})
of the Lagrangian densities 
(\ref{eq: def 2N wires with ``one'' electron per wire}),
(\ref{eq: final interaction if TRS}),
and (\ref{eq: final interaction one domino if TRS})
is generated from the Ising-like linear transformation 
with the fundamental rules
\begin{subequations}
\label{eq: def Ising generator U(1)}
\begin{align}
&
\psi^{*}_{\mathrm{R}}\mapsto
+\psi^{*}_{\mathrm{R}},
\qquad
&
\tilde{\psi}^{*}_{\mathrm{R}}\mapsto
-\tilde{\psi}^{*}_{\mathrm{R}},
\\
&
\psi^{*}_{\mathrm{L}}\mapsto
+\psi^{*}_{\mathrm{L}},
\qquad
&
\tilde{\psi}^{*}_{\mathrm{L}}\mapsto
-\tilde{\psi}^{*}_{\mathrm{L}},
\\
&
\psi^{\,}_{\mathrm{R}}\mapsto
+\psi^{\,}_{\mathrm{R}},
\qquad
&
\tilde{\psi}^{\,}_{\mathrm{R}}\mapsto
-\tilde{\psi}^{\,}_{\mathrm{R}},
\\
&
\psi^{\,}_{\mathrm{L}}\mapsto
+\psi^{\,}_{\mathrm{L}},
\qquad
&
\tilde{\psi}^{\,}_{\mathrm{L}}\mapsto
-\tilde{\psi}^{\,}_{\mathrm{L}}.
\end{align}
\end{subequations} 
The $U(1)$ symmetry 
(\ref{eq: def U(1) symmetry between no tilde and tilde sectors})
is the analogue to the residual $U(1)$ spin-1/2 symmetry
in the spin quantum Hall effect that insures the quantization
of the spin Hall conductivity.$\ $ %
\cite{Kane05a,Bernevig06}

\begin{figure*}[t]
\begin{center}
\includegraphics[scale=0.5]{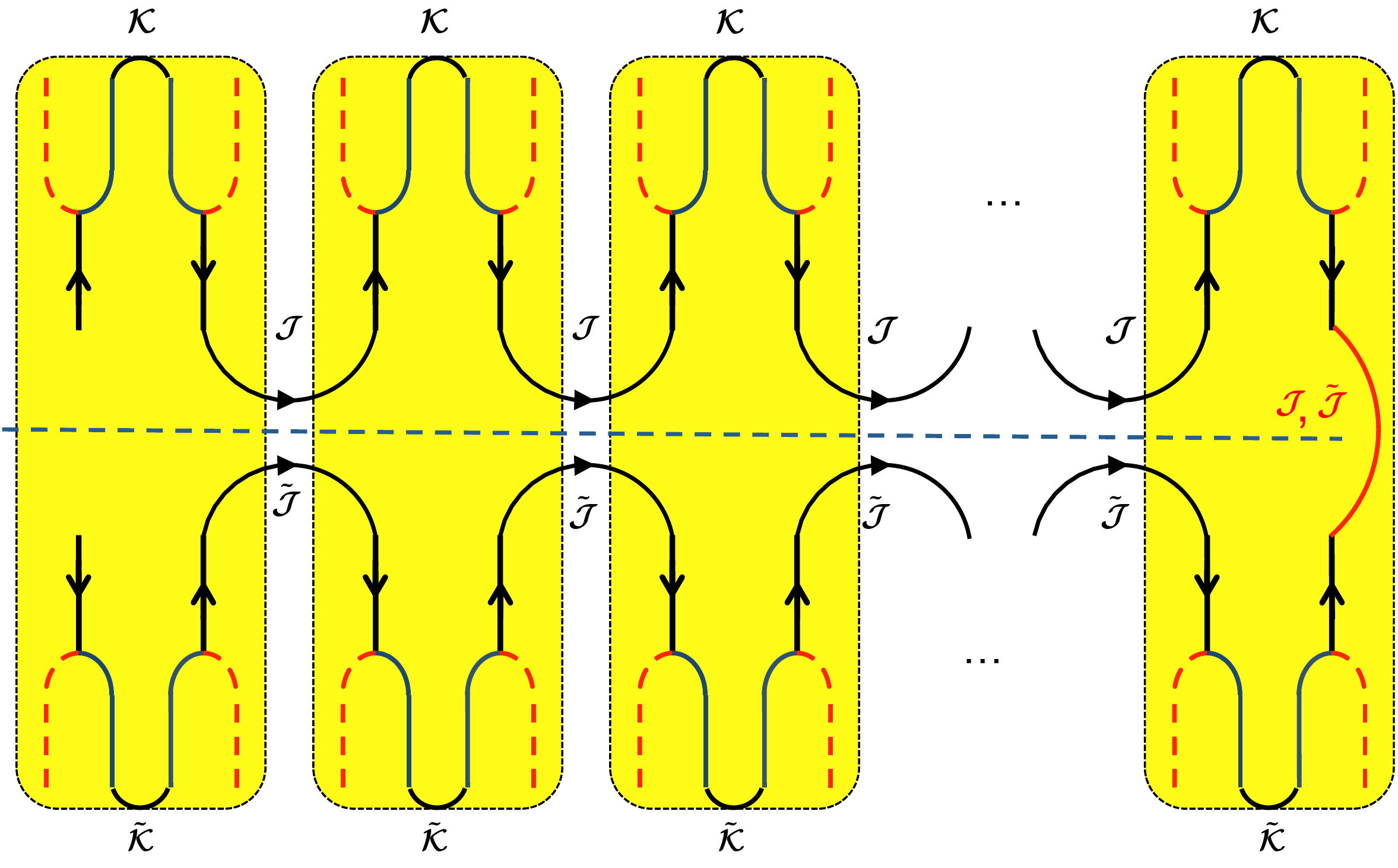}
\caption{
(Color online)
Unfolding a chain of dominoes coupled by time-reversal symmetric
interactions.
\label{Fig: unfolded gapping with TRS}
         }
\end{center}
\end{figure*}

\begin{figure*}[t]
\begin{center}
(a)
\includegraphics[scale=0.5]{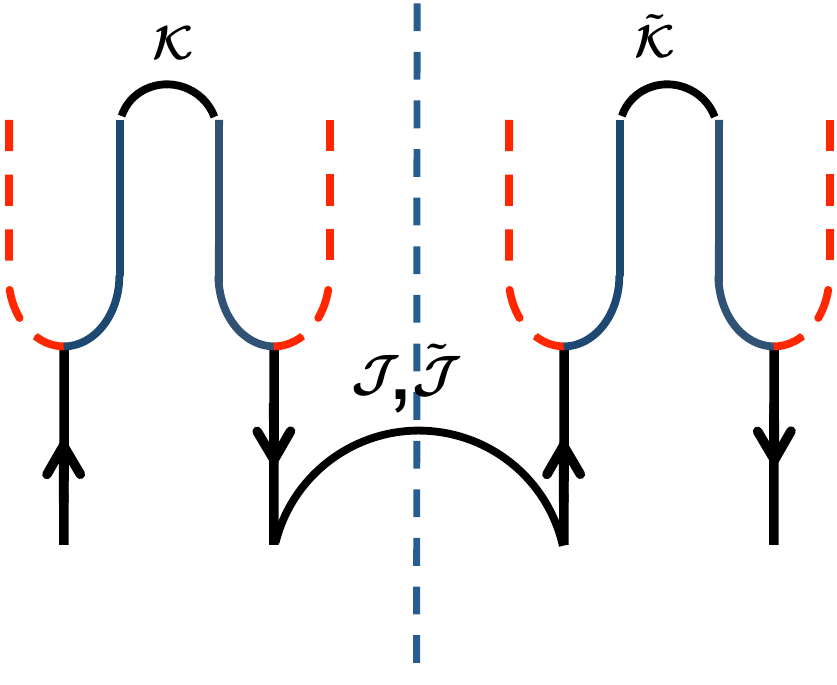}
\hfill
(b)
\includegraphics[scale=0.5]{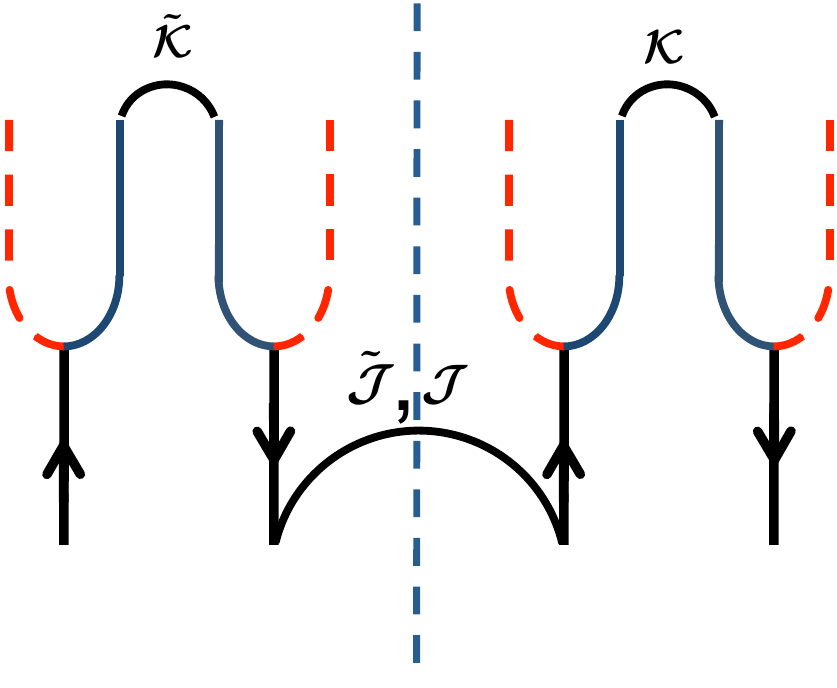}
\hfill
(c)
\includegraphics[scale=0.5]{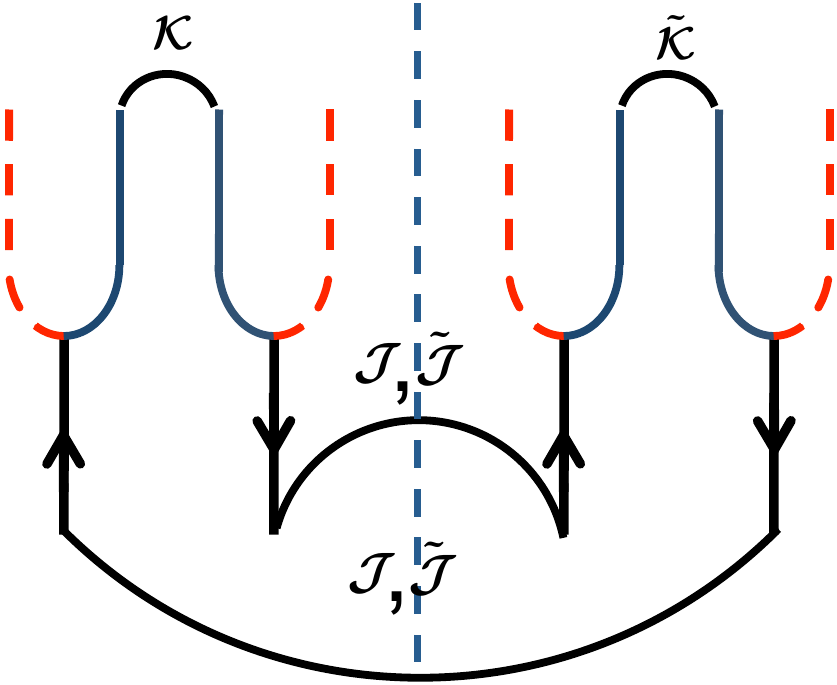}
\caption{
(Color online)
The partial gapping of a bundle made of $2(k+k')$ quantum wires
with local current-current interactions
can be done in two ways (a) and (b).
The complete gapping of a bundle made of $2(k+k')$ quantum wires
with local current-current interactions
is achieved in (c). The dashed vertical line is a mirror axis of symmetry.
\label{Fig: three phases with mirror symmetry of single dominos}
         }
\end{center}
\end{figure*}

However, as is implied by
Fig.\ \ref{Fig: gapping with TRS of single dominos},
it is possible to gap independently the 
coset theory with the central charge
(\ref{eq: nonchiral central charge at one boundary case II})
on any one of the boundary at
$\mathtt{m}=1$ and $\mathtt{m}=n$
by adding either the interaction
\begin{equation}
-
\sum_{\mathcal{A}=1}^{6}
\lambda^{\prime\mathcal{A}}_{\mathrm{boundary},1}\,
\mathcal{J}^{\mathcal{A}}_{\mathrm{R},1}\,
\widetilde{\mathcal{J}}^{\mathcal{A}}_{\mathrm{L},1}
\end{equation}
(with $\lambda^{\prime\mathcal{A}}_{\mathrm{boundary},1}>0$)
or the interaction 
\begin{equation}
-
\sum_{\mathcal{A}=1}^{6}
\lambda^{\prime\mathcal{A}}_{\mathrm{boundary},\texttt{m}}\,
\mathcal{J}^{\mathcal{A}}_{\mathrm{L},n}\,
\widetilde{\mathcal{J}}^{\mathcal{A}}_{\mathrm{R},n}
\end{equation}
(with $\lambda^{\prime\mathcal{A}}_{\mathrm{boundary},\texttt{m}}>0$),
respectively.
The transformation
(\ref{eq: def U(1) symmetry between no tilde and tilde sectors})
acts trivially on the currents
(\ref{eq: def su(2) and su(k) left right currents}),
(\ref{eq: def tilde su(2) and tilde su(k) left right currents}),
etc.
Hence, imposing the symmetry under the transformation 
(\ref{eq: def U(1) symmetry between no tilde and tilde sectors})
is no rescue to prevent the instability of the
helical edge states to local current-current interactions, 
as it was with regard to electronic mass terms.

The instability of the boundary states in Fig.\
\ref{Fig: gapping with TRS}
is not surprising.
The low-energy sector of the theory after gapping
the sectors with the 
$U(k+k')$ and $\tilde{U}(k+k')$
symmetries is of bosonic character, for it is solely expressed in terms
of spin-1/2 currents. Time-reversal in this sector of the conformal embedding 
is represented by an operator that squares to the identity.
If so, time-reversal symmetry is not expected to protect
gapless boundary states. The existence of gapless boundary states
demands fine-tuning of all strong many-body electronic interactions
permitted by time-reversal symmetry. 

This is not to say that the bulk theory in Fig.\
\ref{Fig: gapping with TRS}
is uninteresting. It does support topological order when
two-dimensional space shares the same topology as that of a torus.
When the ground state in Fig.\
\ref{Fig: gapping with TRS}
is the direct product of the ground state corresponding to Fig.\
\ref{Fig: gapping without TRS}(c)
with its time-reversed image, the ground state corresponding to Fig.\
\ref{Fig: gapping without TRS}(d),
the topological degeneracy 
is the square of the topological degeneracy corresponding to Fig.\
\ref{Fig: gapping without TRS}(c).
This counting can be established as follows.
We opt to gap the right-boundary 
in Fig.\
\ref{Fig: gapping with TRS}
as is illustrated
in Fig.\
\ref{Fig: unfolded gapping with TRS}
with the vertical (red) arc.
[We are caping the right boundary with
Fig.\ 
\ref{Fig: gapping with TRS of single dominos}(a).]
We may then unfold the dominoes by cutting them about the dashed blue
line in Fig.\
\ref{Fig: unfolded gapping with TRS}.
The upper and lower parts of all dominoes are now interpreted to be 
distinct (by the presence of absence of the symbol $\tilde{\vphantom{A}}\ $) 
quantum wires. We then recover
Fig.\
\ref{Fig: gapping without TRS}(c)
with $N$ replaced by $2N$.
The operation of time-reversal 
is to be interpreted as a mirror transformation about the dashed line 
(i.e., non-local in space) after unfolding.
We also observe that if we unfold
the dominoes of Figs.\
\ref{Fig: gapping with TRS of single dominos}(a),
\ref{Fig: gapping with TRS of single dominos}(b),
and
\ref{Fig: gapping with TRS of single dominos}(c),
we obtain the bundles made of $2(k+k')$ quantum wires shown in Figs.\
\ref{Fig: three phases with mirror symmetry of single dominos}(a),
\ref{Fig: three phases with mirror symmetry of single dominos}(b),
and
\ref{Fig: three phases with mirror symmetry of single dominos}(c),
respectively. Either of Figs.\
\ref{Fig: three phases with mirror symmetry of single dominos}(a)
and
\ref{Fig: three phases with mirror symmetry of single dominos}(b)
realize a strongly interacting critical point of $2(k+k')$ quantum wires
obtained by fine-tuning of strong many-body electron interactions.
This is reminiscent of the Takhtajan-Babujian critical point in
the spin-1 chain with (competing) bilinear and biquadratic interactions,$\ $%
\cite{Takhtajan82,Babujian82,Haldane83a,Haldane83b,Affleck87,%
Tsvelik90,Chubukov91,Kitazawa99,Ivanov03,Liu12,Lahtinen14,Chen15}
as well as of diverse spin-ladder systems with competing interactions.$\ $%
\cite{Shelton96,White96,Nersesyan98,Allen97,Allen00,Konik10,Tsvelik11,Lecheminant12,Lecheminant15}

\begin{figure}[t]
\begin{center}
\includegraphics[scale=0.35]{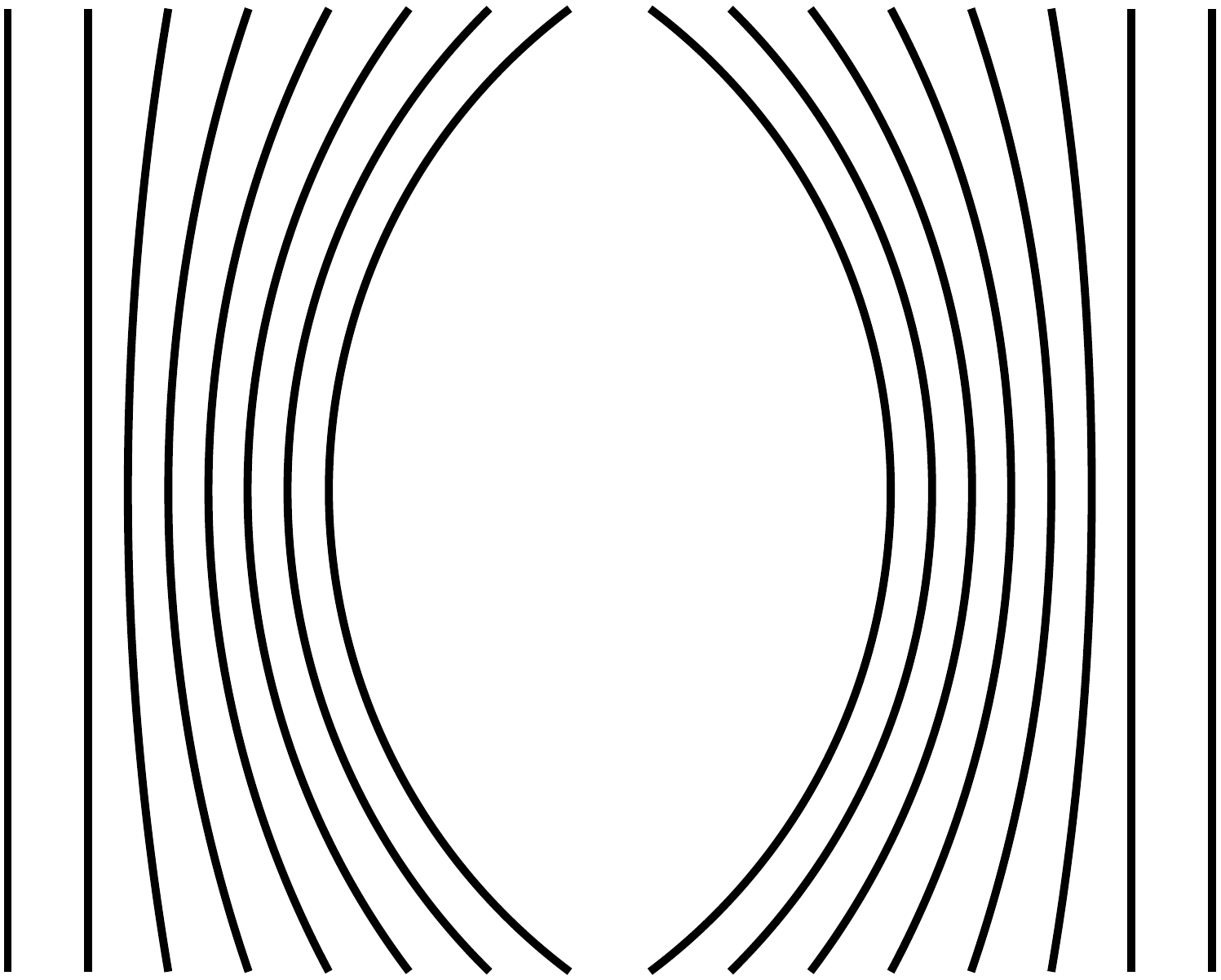}
\caption{
A single ``hole'' in a one-dimensional array of $N$ quantum wires.
Electron tunneling between consecutive wires
is prohibited across the diameter of the hole.
\label{Fig: single hole in chain wires}
         }
\end{center}
\end{figure}

Even if open boundary conditions are imposed on two-dimensional
space (two-dimensional space is the two-dimensional Euclidean plane)
at infinity, ``holes'' in two-dimensional space bring about
a topological degeneracy.$\ $%
\cite{Iadecola14}
A ``Hole'' is here understood to be a path-connected
and large region of two-dimensional space 
in which electrons are precluded 
from entering as is illustrated in Fig.\
\ref{Fig: single hole in chain wires}
within the context of modeling two-dimensional space with a
one-dimensional array of quantum wires. Electrons can neither
tunnel nor interact with electrons across 
a large hole because of locality.

\subsubsection{Case III -- Broken rotation symmetry}
\label{subsubsec: Case III}

The final example with time-reversal symmetry starts from a chain of
decoupled and noninteracting wires that obeys time-reversal symmetry
but with strongly broken spin-1/2 rotation symmetry.  
Instead of the low-energy
theory (\ref{eq: def N wires with ``one'' electron per wire})
with its $SU(2)$ spin-1/2 rotation symmetry, 
we consider the case for which this symmetry is 
strongly broken down to no more than
the $U(1)$ subgroup encoding rotations about a quantization axis. 
In this case, the non-Abelian spin-1/2
current algebra is no longer available to gap the bulk with
current-current interactions. One must rely exclusively on the
$SU(k)\times SU(k')$ sector to gap the bulk with current-current
interactions. 

The construction of 
\ref{subsec: Partial gapping without time-reversal symmetry}
followed by those of
Sec.\ \ref{subsubsec: Case I}
and 
\ref{subsubsec: Case II} 
can then be reproduced if the spin-1/2 $SU(2)$ symmetry group
is replaced by another $SU(2)$. To this end, we choose a bundle with 
$k=2l$ and $k'=2l'$ wires, where $l$ and $l'$ are positive integers. 
We then use the direct-product 
decomposition $U(2k)\times U(2k')$, where any one the two unitary groups
is decomposed according to the rule
$U(2k)=U(4l)=U(1)\times SU(4l)$ and similarly for $k\to k'$. 
We impose on $SU(4l)$ the conformal embedding 
corresponding to  $SU(2)\times SU(2)\times SU(l)$,
where the first $SU(2)$ is generated by the spin-1/2 of the electron.
By assumption, this sector is strongly gapped. Thus, we may ignore it
in the low-energy sector of the theory. We then proceed with the
sectors $U(1)\times SU(2)\times SU(l)$ as we did in Secs.\
\ref{subsec: Partial gapping without time-reversal symmetry}
and
Sec.\ \ref{subsubsec: Case I}.
To duplicate
\ref{subsubsec: Case II}, we furthermore 
introduce the doubling $\widetilde{SU}(2)$ that we may interpret
as assuming that $l=2o$ and $l'=2o'$ with $o$ and $o'$ positive integers. 

Finally, we observe one can also use the $U(1)$ charge sector or the $U(1)$
sector for rotations about the spin-1/2 quantization axis to gap the
bulk while leaving gapless boundaries.$\ $ %
\cite{Bernevig06,Levin09,Neupert11b,Levin12}

\acknowledgements

We warmly thank Alexander Altland for a question that
inspired this work. 
The authors also acknowledge insightful exchanges with
Ian Affleck and Ronny Thomale. 
CMM thanks Akira Furusaki for a constructive criticism.
This work is supported by 
DARPA SPAWARSYSCEN Pacific N66001-11 -1-4110 (T.N.), 
DOE Grant DEF-06ER46316 (P.-H.H. and C.C.), 
FNSNF Grant 2000021\_153648 (J.-H.C. and C.M.), and FAPESP Grant 2012/16082-3 (P.G.).
We acknowledge the Condensed Matter Theory Visitors’ Program 
at Boston University for support.

\appendix

\section{Reversal of time}
\label{Appsec: Reversal of time}

\subsection{Complex fermions}
\label{subsec: Complex fermions}

\subsubsection{Spinless case}
\label{subsubsec: Spinless case}

Denote the fermion annihilation and creation field operators as
$\hat{\psi}^{\dag}_{\mathrm{A}}(t,x)$
and 
$\hat{\psi}^{\,}_{\mathrm{A}}(t,x)$, respectively. 
The index $\mathrm{A}$ belongs to a countable set.
A point in time is denoted by $t$. A point in space is denoted by $x$.
The only nonvanishing equal-time anticommutators are
\begin{subequations}
\label{eq: spinless complex fermions fields}
\begin{equation}
\left\{\hat{\psi}^{\,}_{\mathrm{A}}(t,x),\hat{\psi}^{\dag}_{\mathrm{A}'}(t,x')\right\}=
\delta^{\,}_{\mathrm{A}\mathrm{A}'}\,\delta(x-x').
\end{equation}
For the spinless case,
\begin{equation}
\mathrm{A}=\mathrm{R},\mathrm{L},
\end{equation}
\end{subequations}
i.e., the collective index $\mathrm{A}$ takes the value 
$\mathrm{R}$ and $\mathrm{L}$ with ``$\mathrm{R}$''
standing for a right mover and  ``$\mathrm{L}$'' standing for a left mover.

Reversal of time is the antilinear transformation on the $*$-algebra
generated by the quantum fields 
(\ref{eq: spinless complex fermions fields})
with the fundamental rule
\begin{equation}
\hat{\psi}^{\,}_{\mathrm{R}}(t,x)\mapsto
\hat{\psi}^{\,}_{\mathrm{L}}(-t,x),
\qquad
\hat{\psi}^{\,}_{\mathrm{L}}(t,x)\mapsto
\hat{\psi}^{\,}_{\mathrm{R}}(-t,x).
\end{equation}

Exchange of particle and hole is
the linear transformation on the $*$-algebra
generated by the quantum fields 
(\ref{eq: spinless complex fermions fields})
with the fundamental rule
\begin{equation}
\hat{\psi}^{\,}_{\mathrm{R}}(t,x)\mapsto
\hat{\psi}^{\dag}_{\mathrm{R}}(t,x),
\qquad
\hat{\psi}^{\,}_{\mathrm{L}}(t,x)\mapsto
\hat{\psi}^{\dag}_{\mathrm{L}}(t,x).
\end{equation} 

\subsubsection{Spin-1/2 case}
\label{subsubsec: Spin-1/2 case}

Denote the spin-$1/2$ Dirac field operator as 
$\hat{\psi}^{\,}_{\mathrm{A}}(t,x)$. 
The index $\mathrm{A}$ belongs to a countable set.
A point in time is denoted by $t$. A point in space is denoted by $x$.
The only nonvanishing equal-time anticommutators are 
\begin{subequations}
\label{eq: spin-1/2 complex fermions fields}
\begin{equation}
\left\{
\hat{\psi}^{\,}_{\mathrm{A}}(t,x),
\hat{\psi}^{\dag}_{\mathrm{A}'}(t,x')
\right\}=
\delta^{\,}_{\mathrm{A}\mathrm{A}'}\,\delta(x-x').
\end{equation}
For the spin-$1/2$ case, 
\begin{equation}
\mathrm{A}=(\mathrm{R},+),(\mathrm{R},-),(\mathrm{L},+),(\mathrm{L},-),
\end{equation}
\end{subequations}
i.e., the collective index $\mathrm{A}$ enumerates right and left movers
with an helicity index $\sigma=\pm$ that can be interpreted as
the projection of a spin-$1/2$ quantum number along the Fermi wave vector. 

Reversal of time is the antilinear transformation on the $*$-algebra
generated by the quantum fields 
(\ref{eq: spin-1/2 complex fermions fields})
with the fundamental rule
\begin{equation}
\begin{split}
&
\hat{\psi}^{\,}_{\mathrm{R},+}(t,x)\mapsto
+
\hat{\psi}^{\,}_{\mathrm{L},-}(-t,x),
\\
&
\hat{\psi}^{\,}_{\mathrm{R},-}(t,x)\mapsto
-
\hat{\psi}^{\,}_{\mathrm{L},+}(-t,x),
\\
&
\hat{\psi}^{\,}_{\mathrm{L},+}(t,x)\mapsto
+
\hat{\psi}^{\,}_{\mathrm{R},-}(-t,x),
\\
&
\hat{\psi}^{\,}_{\mathrm{L},-}(t,x)\mapsto
-
\hat{\psi}^{\,}_{\mathrm{R},+}(-t,x).
\end{split}
\end{equation}

Exchange of particle and hole is
the linear transformation on the $*$-algebra
generated by the quantum fields 
(\ref{eq: spin-1/2 complex fermions fields})
with the fundamental rule
\begin{equation}
\hat{\psi}^{\,}_{\mathrm{A}}(t,x)\mapsto
\hat{\psi}^{\dag}_{\mathrm{A}}(t,x)
\end{equation} 
for $\mathrm{A}=(R,+),(R,-),(L,+),(L,-)$.

\subsection{Real Fermions}
\label{subsec: Real Fermions}

\subsubsection{Spinless case} 
\label{subsubsec: Spinless case bis} 

We start from the $*$-algebra defined from 
Eq.~(\ref{eq: spinless complex fermions fields}).
We write
\begin{subequations}
\label{eq: spinless real fermions fields}
\begin{equation}
\hat{\psi}^{\,}_{\mathrm{A}}(t,x)\equiv 
\frac{1}{\sqrt{2}}
\left[
\hat{\chi}^{\,}_{\mathrm{A},1}(t,x) 
+ 
\mathrm{i}
\hat{\chi}^{\,}_{\mathrm{A},2}(t,x)
\right]
\end{equation}
and demand that
\begin{equation}
\hat{\chi}^{\dag}_{\mathrm{A},1}(t,x)=\hat{\chi}^{\,}_{\mathrm{A},1}(t,x),
\qquad
\hat{\chi}^{\dag}_{\mathrm{A},2}(t,x)=\hat{\chi}^{\,}_{\mathrm{A},2}(t,x),
\end{equation}
holds together with the equal-time algebra
\begin{equation}
\left\{\hat{\chi}^{\,}_{\mathrm{A},a}(t,x),\hat{\chi}^{\,}_{\mathrm{A}',a'}(t,x')\right\}=
\delta^{\,}_{\mathrm{A}\mathrm{A}'}\,
\delta^{\,}_{aa'}\,
\delta(x-x'),
\end{equation}
\end{subequations}
for $\mathrm{A},\mathrm{A}'=\mathrm{R},\mathrm{L}$
and $a,a'=1,2$.

Reversal of time is the antilinear transformation on the $*$-algebra
generated by the quantum fields 
(\ref{eq: spinless real fermions fields})
with the fundamental rule 
\begin{equation}
\begin{split}
&
\hat{\chi}^{\,}_{\mathrm{R},a}(t,x)\mapsto
(-1)^{a-1}\,
\hat{\chi}^{\,}_{\mathrm{L},a}(-t,x),
\\
&
\hat{\chi}^{\,}_{\mathrm{L},a}(t,x)\mapsto
(-1)^{a-1}\,
\hat{\chi}^{\,}_{\mathrm{R},a}(-t,x),
\end{split}
\label{eq: spinless real fermions fields time-reversal}
\end{equation} 
for $a=1,2$. The multiplicative negative sign when $a=2$
arises because of the antilinearity. 
Here, reversal of time squares to the identity.

Exchange of particle and hole is
the linear transformation on the $*$-algebra
generated by the quantum fields 
(\ref{eq: spinless real fermions fields})
with the fundamental rule
\begin{equation}
\hat{\chi}^{\,}_{\mathrm{A},1}(t,x)\mapsto
\hat{\chi}^{\,}_{\mathrm{A},1}(t,x),
\qquad
\hat{\chi}^{\,}_{\mathrm{A},2}(t,x)\mapsto
-
\hat{\chi}^{\,}_{\mathrm{A},2}(t,x),
\end{equation} 
for $\mathrm{A}=\mathrm{R},\mathrm{L}$.
Exchange of particle and hole squares to the identity here.

\subsubsection{Spin-1/2 case}
\label{subsubsec: Spin-1/2 case bis} 

We start from $*$-algebra defined from 
Eq.~(\ref{eq: spin-1/2 complex fermions fields}).
We write
\begin{subequations}
\label{eq: spin-1/2 real fermions fields}
\begin{equation}
\hat{\psi}^{\,}_{\mathrm{A}}(t,x)\equiv 
\frac{1}{\sqrt{2}}
\left[
\hat{\chi}^{\,}_{\mathrm{A},1}(t,x) 
+ 
\mathrm{i}
\hat{\chi}^{\,}_{\mathrm{A},2}(t,x)
\right]
\end{equation}
and demand that
\begin{equation}
\hat{\chi}^{\dag}_{\mathrm{A},1}(t,x)=\hat{\chi}^{\,}_{\mathrm{A},1}(t,x),
\qquad
\hat{\chi}^{\dag}_{\mathrm{A},2}(t,x)=\hat{\chi}^{\,}_{\mathrm{A},2}(t,x),
\end{equation}
holds together with the equal-time algebra
\begin{equation}
\left\{
\hat{\chi}^{\,}_{\mathrm{A},a}(t,x),
\hat{\chi}^{\,}_{\mathrm{A}',a'}(t,x')
\right\}=
\delta^{\,}_{\mathrm{A}\mathrm{A}'}\,
\delta^{\,}_{aa'}\,
\delta(x-x'),
\end{equation}
\end{subequations}
for 
$\mathrm{A},\mathrm{A}'=
(\mathrm{R},+),(\mathrm{R},-),(\mathrm{L},+),(\mathrm{L},-)$
and
$a,a'=1,2$.

Reversal of time is the antilinear transformation on the $*$-algebra
generated by the quantum fields 
(\ref{eq: spin-1/2 real fermions fields})
with the fundamental rule
\begin{equation}
\begin{split}
&
\hat{\chi}^{\,}_{\mathrm{R},+,a}(t,x)\mapsto
+
(-1)^{a-1}\,
\hat{\chi}^{\,}_{\mathrm{L},-,a}(-t,x),
\\
&
\hat{\chi}^{\,}_{\mathrm{R},-,a}(t,x)\mapsto
-
(-1)^{a-1}\,
\hat{\chi}^{\,}_{\mathrm{L},+,a}(-t,x),
\\
&
\hat{\chi}^{\,}_{\mathrm{L},+,a}(t,x)\mapsto
+
(-1)^{a-1}\,
\hat{\chi}^{\,}_{\mathrm{R},-,a}(-t,x),
\\
&
\hat{\chi}^{\,}_{\mathrm{L},-,a}(t,x)\mapsto
-
(-1)^{a-1}\,
\hat{\chi}^{\,}_{\mathrm{R},+,a}(-t,x),
\end{split}
\end{equation}
for $a=1,2$.
Here, reversal of time squares to minus the identity.

Exchange of particle and hole can be implemented in two ways.

One may choose the linear transformation on the $*$-algebra
generated by the quantum fields 
(\ref{eq: spin-1/2 real fermions fields})
with the fundamental rule
\begin{equation}
\begin{split}
&
\hat{\chi}^{\,}_{\mathrm{A},a}(t,x)\mapsto
(-1)^{a-1}\,
\hat{\chi}^{\,}_{\mathrm{A},a}(t,x)
\end{split}
\end{equation} 
for
$\mathrm{A}=
(\mathrm{R},+),(\mathrm{R},-),(\mathrm{L},+),(\mathrm{L},-)$
and
$a=1,2$.
This transformation squares to the identity.

One may choose the linear transformation on the $*$-algebra
generated by the quantum fields 
(\ref{eq: spin-1/2 real fermions fields})
with the fundamental rule
\begin{equation}
\begin{split}
&
\hat{\chi}^{\,}_{\alpha,+,a}(t,x)\mapsto
(-1)^{a-1}\,
\hat{\chi}^{\,}_{\alpha,-,a}(t,x),
\\
&
\hat{\chi}^{\,}_{\alpha,-,a}(t,x)\mapsto
-(-1)^{a-1}\,
\hat{\chi}^{\,}_{\alpha,+,a}(t,x),
\end{split}
\end{equation} 
for
$\alpha=\mathrm{R},\mathrm{L}$
and
$a=1,2$.
This transformation squares to minus the identity.

\section{Non-Abelian bosonization for the symmetry classes C, A, and AII}
\label{appsec: Non-Abelian bosonization for the symmetry classes}

\begin{figure*}[t]
\begin{center}
\includegraphics[scale=0.5]{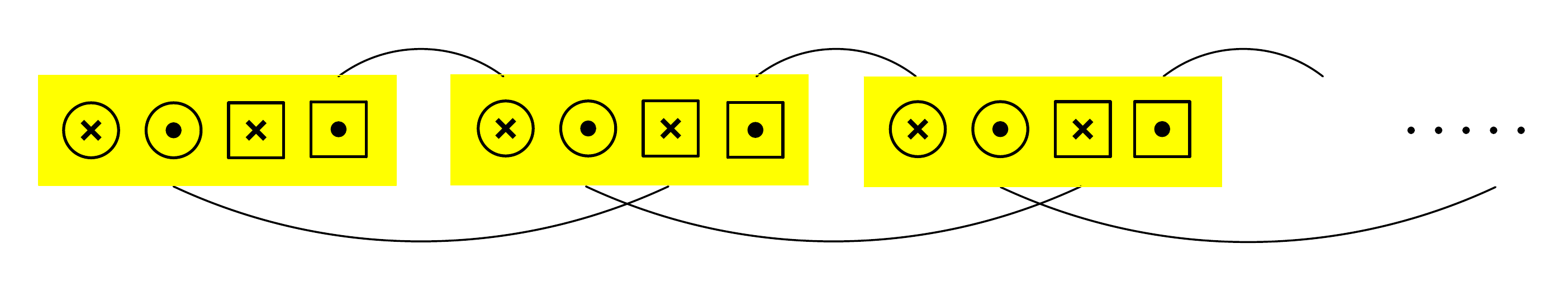}
\caption{(Color online)
Pictorial representation for the selected backscattering 
in the symmetry class C. Each yellow box represents 
a quantum wire composed of four-Majorana degrees of freedom.
The wires are enumerated by $I=1,\cdots,N$ 
in ascending order from  left to right.
For any $I$, the Majorana modes are denoted by
$\chi^{\,}_{\mathrm{R},+,I}$,
$\chi^{\,}_{\mathrm{L},+,I}$,
$\chi^{\,}_{\mathrm{R},-,I}$,
and
$\chi^{\,}_{\mathrm{L},-,I}$
reading from left to right,
respectively. 
\label{Fig: minimal wire network in C}
         }
\end{center}
\end{figure*}

\subsection{The symmetry class C}
\label{subsec: The symmetry class C}
 
The simplest model for an array of quantum wires in the symmetry class C
to realize a topological gapped phase
is defined in two steps.
First, the superscript (DIII) is replaced by (C) in Eq.\
(\ref{eq: free Majorana class DIII}).
Second, we impose the linear transformation 
defined by the fundamental rule 
\begin{equation}
\begin{split}
&
\chi^{\,}_{\mathrm{R},+,I}(t,x)\mapsto
+\,
\chi^{\,}_{\mathrm{R},-,I}(t,x),
\\
&
\chi^{\,}_{\mathrm{R},-,I}(t,x)\mapsto
-\,
\chi^{\,}_{\mathrm{R},+,I}(t,x),
\\
&
\chi^{\,}_{\mathrm{L},+,I}(t,x)\mapsto
+\,
\chi^{\,}_{\mathrm{L},-,I}(t,x),
\\
&
\chi^{\,}_{\mathrm{L},-,I}(t,x)\mapsto
-\,
\chi^{\,}_{\mathrm{L},+,I}(t,x),
\end{split}
\label{eq: critical properties class C minimal c}
\end{equation}
for $I=1,\cdots,N$.$\ $%
\cite{Altland97}
Transformation (\ref{eq: critical properties class C minimal c})
squares to minus the identity. Even though reversal of time
(\ref{eq: critical properties class DIII minimal c})
is a symmetry of the partition function $Z^{(\mathrm{C})}_{0}$,
we shall not impose invariance under
reversal of time 
(\ref{eq: critical properties class DIII minimal c})
for a generic representative of the symmetry class C.

Any partition function $Z^{(\mathrm{C})}$
for the array of quantum wires
is said to belong to the symmetry class C
if $Z^{(\mathrm{C})}$ is invariant under the following transformations.
There is the symmetry (\ref{eq: critical properties class C minimal c}).
There is the symmetry under the linear transformation (fermion parity)
with the fundamental rule
\begin{equation}
\chi^{\,}_{\alpha,\mathrm{f},I}\mapsto 
-\chi^{\,}_{\alpha,\mathrm{f},I}
\label{eq: def symmetry class C parity}
\end{equation}
for any
$\alpha=\mathrm{R},\mathrm{L}$,
$\mathrm{f}=\pm$,
and
$I=1,\cdots,N$. Both symmetries generate the
symmetry under the linear $O(2)$ transformation
\begin{equation}
\chi^{\,}_{\alpha,\mathrm{f},I}\mapsto
O^{\,}_{\mathrm{f}\mathrm{f}'}\,
\chi^{\,}_{\alpha,\mathrm{f}',I}
\label{eq: def symmetry class C O2}
\end{equation}
for any $\alpha=\mathrm{R},\mathrm{L}$,
$\mathrm{f}=\pm$, 
and $I=1,\cdots,N$.
Here, the summation convention over the repeated indices
$\mathrm{f}'=\pm$ is implied and the $2\times2$ matrix
$(O^{\,}_{\mathrm{f}\mathrm{f}'})$
is real-valued and orthogonal.

We seek a local single-particle perturbation 
$\mathcal{L}^{(\mathrm{C})}_{\mathrm{mass}}$
that satisfies three conditions when added to 
the Lagrangian density 
$\mathcal{L}^{(\mathrm{C})}_{0}$.

\textbf{Condition C.1}
It must be invariant under the transformations
(\ref{eq: critical properties class C minimal c})
and
(\ref{eq: def symmetry class C parity}).

\textbf{Condition C.2}
It must gap completely the theory with the partition function 
$Z^{(\mathrm{C})}_{0}$
if we impose the periodic boundary conditions
\begin{equation}
\chi^{\,}_{\alpha,\mathrm{f},I}(t,x)=
\chi^{\,}_{\alpha,\mathrm{f},I+N}(t,x)
\end{equation} 
for 
$\alpha=\mathrm{R},\mathrm{L}$,
$\mathrm{f}=\pm$,
and
$I=1,\cdots,N$.

\textbf{Condition C.3}
The partition function $Z^{(\mathrm{C})}$
with the Lagrangian density
$\mathcal{L}^{(\mathrm{C})}_{0}+\mathcal{L}^{(\mathrm{C})}_{\mathrm{mass}}$
must be a theory with the central charge 
\begin{equation}
c^{(\mathrm{C})}=1
\end{equation}
if open boundary condition are imposed.

\textbf{Conditions C.1},
\textbf{C.2},
and 
\textbf{C.3}
imply that we may assign wire $I=1$ the right-chiral central charge
$1$ and wire $I=N$ the left-chiral central charge $1$,
for wires $I=1$ and $I=N$ both
support a degenerate pair of right- or left-moving 
Majorana edge modes, respectively.

We make the Ansatz 
\begin{equation}
\mathcal{L}^{(\mathrm{C})}_{\mathrm{mass}}:=
\sum_{I=1}^{N-1}
\mathrm{i}
\lambda\,
\left(
\chi^{\,}_{\mathrm{L},-,I}\,\chi^{\,}_{\mathrm{R},+,I+1} 
-
\chi^{\,}_{\mathrm{L},+,I}\,\chi^{\,}_{\mathrm{R},-,I+1}
\right)
\label{eq: Ansatz for single-particle class C}
\end{equation}
with $\lambda$ a real-valued coupling. 
\textbf{Condition} \textbf{C.1} is met by construction.
To establish that the Ansatz
(\ref{eq: Ansatz for single-particle class C})
meets \textbf{Conditions} \textbf{C.2} and \textbf{C.3}, 
we use non-Abelian bosonization.
We choose the non-Abelian bosonization scheme by which
the partition function is given by the path integral 
\begin{subequations}
\begin{equation}
Z^{(\mathrm{C})}=
\int\mathcal{D}[G]\,e^{\mathrm{i}S^{(\mathrm{C})}}.
\end{equation}
The field $G\in O(2N)$ is a matrix of bosons.
The measure $\mathcal{D}[G]$ is constructed from the Haar measure on $O(2N)$.
The action $S^{(\mathrm{C})}$ is the sum of the actions
$S^{(\mathrm{C})}_{0}$ and $S^{(\mathrm{C})}_{\mathrm{mass}}$.
The action $S^{(\mathrm{C})}_{0}$ is
\begin{equation}
\begin{split}
S^{(\mathrm{C})}_{0}=&\,
\frac{1}{16\pi}
\int\mathrm{d}t 
\int\mathrm{d}x\,
\mathrm{tr}\,
\left(
\partial^{\,}_{\mu}G\,\partial^{\mu}G^{-1}
\right)
\\
&\,
+
\frac{1}{24\pi}\int\limits_{B} 
\mathrm{d}^{3}y\, 
\mathcal{L}^{(\mathrm{C})}_{\mathrm{WZW}},
\end{split} 
\end{equation}
where
\begin{equation}
\mathcal{L}^{(\mathrm{C})}_{\mathrm{WZW}}=
\epsilon^{ijk}\,
\mathrm{tr}\,
\Bigl[
(\bar{G}^{-1}\partial^{\,}_{i}\bar{G})\,
(\bar{G}^{-1}\partial^{\,}_{j}\bar{G})
(\bar{G}^{-1}\partial^{\,}_{k}\bar{G})
\Bigr].
\end{equation}
(Recall that $\bar{G}$ denotes the extension of $G$ to the solid 3-ball.)
The action $S^{(\mathrm{C})}_{\mathrm{mass}}$ stems from the Lagrangian density
\begin{equation}
\begin{split}
\mathcal{L}^{(\mathrm{C})}_{\mathrm{mass}}=&\, 
\sum_{I=1}^{N-1}
\lambda\,
\left(
G^{\,}_{(-,I),(-,I+1)} 
+ 
G^{\,}_{(+,I+1),(+,I)} 
\right) 
\\
\equiv&\,
\lambda\,\mathrm{tr}\,
\left(M^{(\mathrm{C})}\,G\right).  
\end{split}
\label{eq: mass matrix cor C minimal} 
\end{equation}
The second equality is established by using the 
non-Abelian bosonization formula 
(\ref{eq: non-abelian bosonization formula})
(we have set the mass parameter $m^{\,}_{\mathrm{uv}}=1$). 
The $2N\times 2N$ matrices $M^{(\mathrm{C})}$
is represented by
\begin{equation}
M^{(\mathrm{C})}:= 
\begin{pmatrix}
0 & 0 & 0 & 0 & 0 & 0 &  \cdots \\
B  & 0 & 0  & 0 & 0 & 0 & \cdots \\
0 & B  & 0 & 0 & 0 & 0 & \cdots \\
0 & 0 & B  & 0 & 0  & 0 & \cdots \\
0 & 0 & 0 & B  & 0 & 0  & \cdots \\
0 & 0 & 0 & 0 & B & 0 &  \cdots \\
\vdots & \vdots & \vdots & \vdots & \vdots & \vdots & \ddots\\
\end{pmatrix}
\label{eq: mass matrix class C}
\end{equation}
in the basis for which
$B$ is the $2\times 2$ matrix
\begin{equation}
B:=
\begin{pmatrix}
\overbrace{-1}^{(-,+)} & \overbrace{0}^{(-,-)}\\
\underbrace{0}_{(+,+)} & \underbrace{1}_{(+,-)}
\end{pmatrix}.
\end{equation}
\end{subequations}
For any $N>2$, 
the $2N\times 2N$ matrices $M^{(\mathrm{C})}$ defined by 
(\ref{eq: mass matrix class C}) has 
two vanishing and $2\times(N-1)$ nonvanishing eigenvalues.

The quadratic perturbation (\ref{eq: mass matrix cor C minimal})
thus reduces the central charge $c^{(\mathrm{C})}_{0} =2\times N/2$
by the amount $2\times(N-1)/2$,
i.e., the central charge for the theory with the partition function 
$Z^{(\mathrm{C})}$ is
\begin{equation}
c^{(\mathrm{C})}=
\frac{2\times N}{2}-\frac{2\times(N-1)}{2}=1.
\end{equation}
We have constructed a topological superconductor with 
the gapless and singlet pair of Majorana modes 
$(\chi^{\,}_{\mathrm{R},+,I},\chi^{\,}_{\mathrm{R},-,I})^{\,}_{I=1}$
propagating along edge $I=1$ and 
the gapless and singlet pair of Majorana modes 
$(\chi^{\,}_{\mathrm{L},+,I},\chi^{\,}_{\mathrm{L},-,I})^{\,}_{I=N}$
propagating along edge $I=N$. 
This construction is summarized by
Fig.\ \ref{Fig: minimal wire network in C}.

The symmetry class C has the $\mathbb{Z}$ topological classification 
for the following reason. If one takes an
arbitrary integer number $\nu$ of copies of the gapless edge theory, 
these $\nu$-copies remain gapless. The stability of the $2\nu$
gapless edge modes within either wire 1 or wire $N$ is guaranteed because
backscattering among gapless chiral edges modes
of the same chirality is not allowed
kinematically within either wire 1 or wire $N$.

\begin{figure*}[t]
\begin{center}
\includegraphics[scale=0.5]{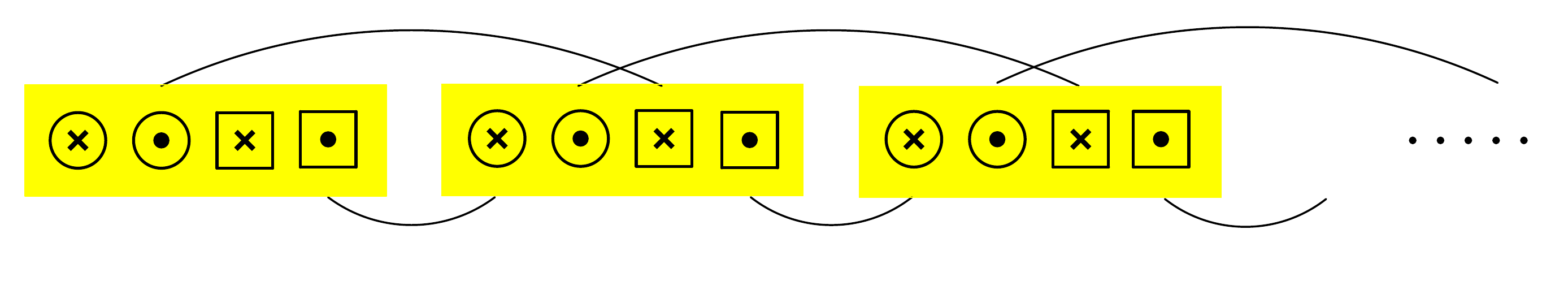}
\caption{(Color online)
Pictorial representation for the selected backscattering 
in the symmetry class A. Each yellow box represents 
a quantum wire composed of four-Majorana degrees of freedom.
The wires are enumerated by $I=1,\cdots,N$ 
in ascending order from  left to right.
For any $I$, the Majorana modes are denoted by
$\chi^{\,}_{\mathrm{R},1,I}$,
$\chi^{\,}_{\mathrm{L},1,I}$,
$\chi^{\,}_{\mathrm{R},2,I}$,
and
$\chi^{\,}_{\mathrm{L},2,I}$
reading from left to right,
respectively.
\label{Fig: minimal wire network in A}
         }
\end{center}
\end{figure*}

\subsection{The symmetry class A}
\label{subsec: The symmetry class A}

The symmetry class A preserves the total number of complex fermions
that can be built out of an even number $M$ of flavors for
the Majorana fermions. The simplest model for an array of quantum wires 
in the symmetry class A to realize a topological gapped phase
assumes
\begin{subequations}
\begin{equation}
M=2,
\qquad
\chi^{\,}_{\alpha,\mathrm{f},I}(t,x),
\end{equation}
for $\alpha=\mathrm{R},\mathrm{L}$,
$\mathrm{f}=1,2$,
and
$I=1,\cdots,N$.
We have thus assigned four Majorana fermions to each wire 
$I=1\,\cdots\,N$. If so, we can interpret
\begin{equation}
\begin{split}
&
\psi^{*}_{\alpha,I}(t,x)\equiv 
\frac{1}{\sqrt{2}}
\left[
{\chi}^{\,}_{\alpha,1,I}(t,x) 
-
\mathrm{i}
{\chi}^{\,}_{\alpha,2,I}(t,x)
\right],
\\
&
\psi^{\,}_{\alpha,I}(t,x)\equiv 
\frac{1}{\sqrt{2}}
\left[
{\chi}^{\,}_{\alpha,1,I}(t,x) 
+ 
\mathrm{i}
{\chi}^{\,}_{\alpha,2,I}(t,x)
\right],
\end{split} 
\label{eq: define dirac in class A}
\end{equation}
\end{subequations}
for $\alpha=\mathrm{R},\mathrm{L}$ and $I=1,\cdots,N$
as the Grassmann representation
of a pair of creation and annihilation fermion operators.

The simplest model for an array of quantum wires in the symmetry class A
to realize a topological gapped phase
is defined in two steps.
First, the superscript (DIII) is replaced by (A) in Eq.\
(\ref{eq: free Majorana class DIII}).
There follows the partition function $Z^{(\mathrm{A})}_{0}$.
Second, we shall represent reversal of time
with the antilinear transformation defined by the fundamental rule
[see Eq.\ (\ref{eq: spinless real fermions fields time-reversal})]
\begin{equation}
\begin{split}
&
\chi^{\,}_{\mathrm{R},\mathrm{f},I}(t,x)\mapsto
(-1)^{\mathrm{f}-1}\,\chi^{\,}_{\mathrm{L},\mathrm{f},I}(-t,x),
\\
&
\chi^{\,}_{\mathrm{L},\mathrm{f},I}(t,x)\mapsto
(-1)^{\mathrm{f}-1}\,\chi^{\,}_{\mathrm{R},\mathrm{f},I}(-t,x),
\end{split}
\label{eq: critical properties class A minimal c}
\end{equation}
for $\mathrm{f}=1,2$ and $I=1,\cdots,N$.
Contrary to the reversal of time defined by Eq.\ 
(\ref{eq: critical properties class DIII minimal c}),
transformation (\ref{eq: critical properties class A minimal c})
squares to the identity. Even though reversal of time 
(\ref{eq: critical properties class A minimal c})
is a symmetry of the partition function $Z^{(\mathrm{A})}_{0}$,
we shall not impose invariance under
reversal of time (\ref{eq: critical properties class A minimal c})
for a generic representative of the symmetry class A.
A symmetry of the partition function $Z^{(\mathrm{A})}_{0}$
that  we shall keep is the $O(2)$ symmetry 
under the transformation 
(\ref{eq: critical properties class A minimal d})
that is parametrized by the angle $0\leq\theta<2\pi$.

The theory with the partition function $Z^{(A)}_{0}$ is critical, 
for there are $4N$ decoupled massless Majorana modes
that are dispersing in $(1+1)$-dimensional Minkowski space and time.
Hence, the central charge for the partition function 
$Z^{(\mathrm{A})}_{0}$ is 
\begin{subequations}
\label{eq: critical properties class A minimal}
\begin{equation}
c^{(\mathrm{A})}_{0}=N.
\label{eq: critical properties class A minimal a}
\end{equation}
The partition function $Z^{(\mathrm{A})}_{0}$ 
is invariant under any local linear transformation from 
$O^{\,}_{\mathrm{R}}(2N)\times O^{\,}_{L}(2N)$
of the form (\ref{eq: critical properties class DIII minimal b}).
For any $0\leq\theta<2\pi$, $Z^{(\mathrm{A})}_{0}$ 
is also invariant under the continuous global linear
transformation with the fundamental rule
\begin{equation}
\begin{split}
&
\chi^{\,}_{\alpha,1,I}(t,x)\mapsto 
\cos\theta\,\chi^{\,}_{\alpha,1,I}(t,x) 
- 
\sin\theta\,\chi^{\,}_{\alpha,2,I}(t,x),
\\
&
\chi^{\,}_{\alpha,2,I}(t,x)\mapsto 
\sin\theta\,\chi^{\,}_{\alpha,1,I}(t,x)
+
\cos\theta\,\chi^{\,}_{\alpha,2,I}(t,x),
\end{split}
\label{eq: critical properties class A minimal d}
\end{equation}
for $\alpha=\mathrm{R},\mathrm{L}$ and $I=1,\cdots,N$.
This transformation implements the $U(1)$ fermion-number conservation law
that follows from the symmetry under the global $U(1)$ transformation
\begin{equation}
\begin{split}
&
\psi^{*}_{\alpha,I}(t,x)\mapsto 
\psi^{*}_{\alpha,I}(t,x)\,
e^{-\mathrm{i}\theta}
\\
&
\psi^{\,}_{\alpha,I}(t,x)\mapsto 
e^{+\mathrm{i}\theta}\,\psi^{\,}_{\alpha,I}(t,x),
\end{split}
\label{eq: critical properties class A minimal d bis}
\end{equation}
\end{subequations}
for any $0\leq\theta<2\pi$, 
$\alpha=\mathrm{R},\mathrm{L}$, 
and $I=1,\cdots,N$.

Any partition function $Z^{(\mathrm{A})}$
for the array of quantum wires
is said to belong to the symmetry class A
if $Z^{(\mathrm{A})}$ is invariant under
the global linear $O(2)$ transformation
(\ref{eq: critical properties class A minimal d}).

Even though the $O(2)$ symmetry in the simplest model
for an array of quantum wires in the symmetry class A
has a very different origin from that in the minimal model
for an array of quantum wires in the symmetry class C,
we may still borrow the analysis of
Sec.\ \ref{subsec: The symmetry class C}
below Eq.\ (\ref{eq: def symmetry class C O2})
verbatim. This construction of a topological phase in the symmetry class A
is summarized by
Fig.\ \ref{Fig: minimal wire network in A}.

The symmetry class A has the $\mathbb{Z}$ topological classification 
for the following reason. If one takes an
arbitrary integer number $\nu$ of copies of the gapless edge theory, 
these $\nu$-copies remain gapless. The stability of the $2\nu$ Majorana
gapless edge modes within either wire 1 or wire $N$ is guaranteed 
for two reasons. First, the $O(2)$ conservation law
allows to group two Majorana fermions into
one complex chiral fermion. Second, 
all Majorana modes within either wire 1 or wire $N$
share the same chirality, 
backscattering among gapless edges modes is not allowed
kinematically within either wire 1 or wire $N$.

\begin{figure*}[t]
\begin{center}
\includegraphics[scale=0.5]{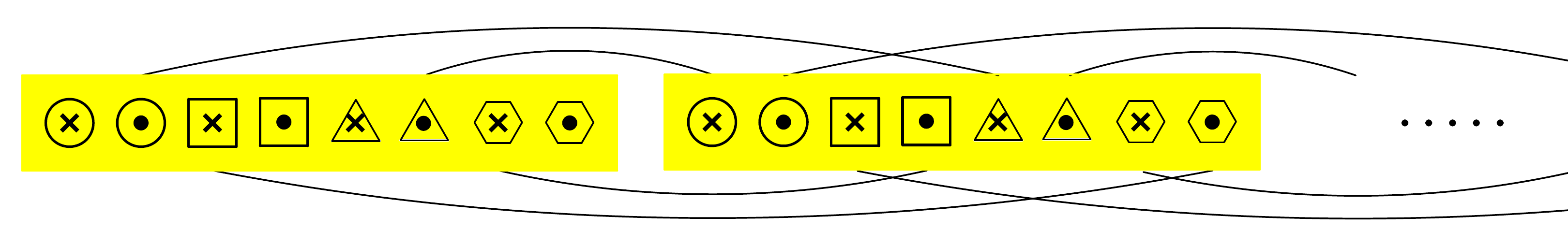}
\caption{(Color online)
Pictorial representation for the selected backscattering 
in the symmetry class AII. Each yellow box represents 
a quantum wire composed of four-Majorana degrees of freedom.
The wires are enumerated by $I=1,\cdots,N$ 
in ascending order from  left to right.
For any $I$, the Majorana modes are denoted by
$\chi^{\,}_{\mathrm{R},+,1,I}$,
$\chi^{\,}_{\mathrm{L},+,1,I}$,
$\chi^{\,}_{\mathrm{R},-,1,I}$,
$\chi^{\,}_{\mathrm{L},-,1,I}$, 
$\chi^{\,}_{\mathrm{R},+,2,I}$,
$\chi^{\,}_{\mathrm{L},+,2,I}$,
$\chi^{\,}_{\mathrm{R},-,2,I}$,
and
$\chi^{\,}_{\mathrm{L},-,2,I}$
reading from left to right,
respectively.
\label{Fig: minimal wire network in AII}
         }
\end{center}
\end{figure*}

\subsection{The symmetry class AII}
\label{subsec: The symmetry class AII}

Fermion-number conservation and a time-reversal symmetry
squaring to minus the identity must hold in the symmetry class AII.
The time-reversal symmetry is that of spin-$1/2$. The minimal model
in the symmetry class AII for an array of quantum wires must thus accommodate
twice as many degrees of freedom as that in
the symmetry class A in order to realize a topological insulating phase.

The simplest model for an array of quantum wires in the symmetry class AII
to realize a topological gapped phase assumes
\begin{subequations}
\label{eq: free Majorana class AII}
\begin{equation}
M=4,
\qquad
\chi^{\,}_{\alpha,\sigma,a,I}(t,x),
\label{eq: free Majorana class AII a}
\end{equation}
with the right- and left-mover labels
$\alpha=\mathrm{R},\mathrm{L}$,
the helicity labels
$\sigma=\pm$,
the complex fermion labels
$
a=1,2,
$
and the wire index
$I=1,\cdots,N$.
This is to say that four complex fermions are represented by
\begin{equation}
{\psi}^{\,}_{A}(t,x)\equiv 
\frac{1}{\sqrt{2}}
\left[
{\chi}^{\,}_{A,1}(t,x) 
+ 
\mathrm{i}
{\chi}^{\,}_{A,2}(t,x)
\right]
\label{eq: free Majorana class AII aa}
\end{equation}
with the collective label
$\mathrm{A}=(\mathrm{R},+),(\mathrm{R},-),(\mathrm{L},+),(\mathrm{L},-)$.
We define the action
\begin{equation}
S^{(\mathrm{AII})}_{0}:= 
\int\mathrm{d}t
\int\mathrm{d}x\,
\mathcal{L}^{(\mathrm{AII})}_{0}
\label{eq: free Majorana class AII b}
\end{equation}
with
\begin{equation}
\begin{split}
\mathcal{L}^{(\mathrm{AII})}_{0}:=&\,
\frac{\mathrm{i}}{2}
\sum_{I=1}^{N}
\sum_{\sigma=\pm}
\sum_{a=1,2}
\Bigl[
\chi^{\,}_{\mathrm{R},\sigma,a,I}
\left(
\partial^{\,}_{t} + \partial^{\,}_{x}
\right)
\chi^{\,}_{\mathrm{R},\sigma,a,I}
\\
&\,
+ 
\chi^{\,}_{\mathrm{L},\sigma,a,I}
\left(
\partial^{\,}_{t} - \partial^{\,}_{x}
\right)
\chi^{\,}_{\mathrm{L},\sigma,a,I}
\Bigr].
\end{split}
\label{eq: free Majorana class AII c}
\end{equation}
We also define the Grassmann partition function
\begin{equation}
Z^{(\mathrm{AII})}_{0}:=
\int\mathcal{D}[\chi]\,e^{+\mathrm{i}S^{(\mathrm{AII})}_{0}}.
\label{eq: free Majorana class AII d}
\end{equation}
\end{subequations}

The theory with the partition function $Z^{(\mathrm{AII})}_{0}$ is critical, 
for there are $8N$ decoupled massless Majorana modes
that are dispersing in $(1+1)$-dimensional Minkowski space and time.
Hence, the central charge for the theory with the partition function 
$Z^{(\mathrm{AII})}_{0}$ is 
\begin{subequations}
\label{eq: critical properties class AII minimal}
\begin{equation}
c^{(\mathrm{AII})}_{0}=2N.
\label{eq: critical properties class AII minimal a}
\end{equation}
The partition function $Z^{(\mathrm{AII})}_{0}$
is invariant under any local transformation
$(O^{(\mathrm{R})},O^{(\mathrm{L})})\in
O^{\,}_{\mathrm{R}}(4N)\times O^{\,}_{\mathrm{L}}(4N)$
defined by
\begin{equation}
\begin{split}
&
\chi^{\,}_{\mathrm{R}}(t-x)\mapsto
O^{(\mathrm{R})}(t-x)\,
\chi^{\,}_{\mathrm{R}}(t-x),
\\
&
\chi^{\,}_{\mathrm{L}}(t+x)\mapsto
O^{(\mathrm{L})}(t+x)\,
\chi^{\,}_{\mathrm{L}}(t+x).
\end{split}
\label{eq: critical properties class AII minimal b}
\end{equation}
It is also invariant under the antilinear transformation 
with the fundamental rules
\begin{equation}
\begin{split}
&
\chi^{\,}_{\mathrm{R},+,a,I}(t,x)\mapsto
+
(-1)^{a-1}\,
\chi^{\,}_{\mathrm{L},-,a,I}(-t,x),
\\
&
\chi^{\,}_{\mathrm{R},-,a,I}(t,x)\mapsto
-
(-1)^{a-1}\,
\chi^{\,}_{\mathrm{L},+,a,I}(-t,x),
\\
&
\chi^{\,}_{\mathrm{L},+,a,I}(t,x)\mapsto
+
(-1)^{a-1}\,
\chi^{\,}_{\mathrm{R},-,a,I}(-t,x),
\\
&
\chi^{\,}_{\mathrm{L},-,a,I}(t,x)\mapsto
-(-1)^{a-1}\,
\chi^{\,}_{\mathrm{R},+,a,I}(-t,x),
\end{split}
\label{eq: critical properties class AII minimal c}
\end{equation}
for $a=1,2$
that implement reversal of time in such a way that
reversal of time squares to minus the identity
(see Appendix \ref{Appsec: Reversal of time}).
Finally, it is invariant under the linear transformation 
with the fundamental rule
\begin{equation}
\begin{split}
&
\chi^{\,}_{\mathrm{A},1,I}(t,x) \mapsto 
\cos\theta\,\chi^{\,}_{\mathrm{A},1,I}(t,x)
-
\sin\theta\,\chi^{\,}_{\mathrm{A},2,I}(t,x),
\\
&
\chi^{\,}_{\mathrm{A},2,I}(t,x) \mapsto 
\sin\theta\,\chi^{\,}_{\mathrm{A},1,I}(t,x)
+
\cos\theta\,\chi^{\,}_{\mathrm{A},2,I}(t,x), 
\label{eq: critical properties class AII minimal d}
\end{split}
\end{equation}
that implements the global $U(1)$ transformation 
\begin{equation}
\begin{split}
&
\psi^{*}_{\mathrm{A},I}(t,x)\mapsto 
\psi^{*}_{\mathrm{A},I}(t,x)\,
e^{-\mathrm{i}\theta},
\\
&
\psi^{\,}_{\mathrm{A},I}(t,x)\mapsto 
e^{+\mathrm{i}\theta}\,\psi^{\,}_{\mathrm{A},I}(t,x),
\end{split}
\label{eq: critical properties class AII minimal dd}
\end{equation}
\end{subequations}
for $\mathrm{A}=(R,+),(R,-),(L,+),(L,-)$ 
and any $0\leq\theta<2\pi$.

Any partition function $Z^{(\mathrm{AII})}$
for the array of quantum wires
is said to belong to the symmetry class AII
if $Z^{(\mathrm{AII})}$ is invariant under the transformations
(\ref{eq: critical properties class AII minimal c})
and
(\ref{eq: critical properties class AII minimal d}).

We seek a local single-particle perturbation 
$\mathcal{L}^{(\mathrm{AII})}_{\mathrm{mass}}$
that satisfies three conditions when added to 
the Lagrangian density (\ref{eq: free Majorana class AII c}).

\textbf{Condition AII.1}
It must be invariant under the transformations
(\ref{eq: critical properties class AII minimal c})
and
(\ref{eq: critical properties class AII minimal d}).

\textbf{Condition AII.2}
It must gap completely the theory with the partition function 
$Z^{(\mathrm{AII})}_{0}$
if we impose the periodic boundary conditions
\begin{equation}
\chi^{\,}_{\alpha,\mathrm{f},a,I}(t,x)=
\chi^{\,}_{\alpha,\mathrm{f},a,I+N}(t,x)
\end{equation} 
for 
$\alpha=\mathrm{R},\mathrm{L}$,
$\mathrm{f}=\pm$,
$a=1,2$,
and
$I=1,\cdots,N$.

\textbf{Condition AII.3}
The partition function $Z^{(\mathrm{AII})}$
with the Lagrangian density
$\mathcal{L}^{(\mathrm{AII})}_{0}+\mathcal{L}^{(\mathrm{AII})}_{\mathrm{mass}}$
must be a theory with the central charge 
\begin{equation}
c^{(\mathrm{AII})}=2
\end{equation}
if open boundary condition are imposed.

\textbf{Conditions AII.1},
\textbf{AII.2},
and 
\textbf{AII.3}
imply that we may assign wire $I=1$ the central charge
$1$ and wire $I=N$ the central charge $1$,
for wires $I=1$ and $I=N$ both
support a single Kramers degenerate pair of edge modes, whereby each mode 
carries the sharp (complex) fermion number of one.

We make the Ansatz 
\begin{equation}
\begin{split}
\mathcal{L}^{(\mathrm{AII})}_{\mathrm{mass}}:=&\, 
\sum_{I=1}^{N-1}
\mathrm{i}
\lambda
\big(
\chi^{\,}_{\mathrm{L},+,1,I}\chi^{\,}_{\mathrm{R},+,2,I+1} 
-
\chi^{\,}_{\mathrm{L},+,2,I}\chi^{\,}_{\mathrm{R},+,1,I+1} 
\\
&\,
+
\chi^{\,}_{\mathrm{R},-,1,I}\,\chi^{\,}_{\mathrm{L},-,2,I+1}
-
\chi^{\,}_{\mathrm{R},-,2,I}\,\chi^{\,}_{\mathrm{L},-,1,I+1} 
\big)
\end{split}
\label{eq: Ansatz for single-particle class AII}
\end{equation}
with $\lambda$ a real-valued coupling. 
\textbf{Condition} \textbf{AII.1} is met by construction.
To establish that the Ansatz
(\ref{eq: Ansatz for single-particle class AII})
meets \textbf{Conditions} \textbf{AII.2} and \textbf{AII.3}, 
we use non-Abelian bosonization.
We choose the non-Abelian bosonization scheme by which
the partition function is given by the path integral 
\begin{subequations}
\begin{equation}
Z^{(\mathrm{AII})}=
\int\mathcal{D}[G]\,e^{+\mathrm{i}S^{(\mathrm{AII})}}.
\end{equation}
The field $G\in O(4N)$ is a matrix of bosons.
The measure $\mathcal{D}[G]$ is constructed from the Haar measure on $O(4N)$.
The action $S^{(\mathrm{AII})}$ is the sum of the actions
$S^{(\mathrm{AII})}_{0}$ and $S^{(\mathrm{AII})}_{\mathrm{mass}}$.
The action $S^{(\mathrm{AII})}_{0}$ is
\begin{equation}
\begin{split}
S^{(\mathrm{AII})}_{0}=&\,
\frac{1}{16\pi}
\int\mathrm{d}t 
\int\mathrm{d}x\,
\mathrm{tr}\,
\left(
\partial^{\,}_{\mu}G\,\partial^{\mu}G^{-1}
\right)
\\
&\,
+
\frac{1}{24\pi}\int\limits_{B} 
\mathrm{d}^{3}y\, 
\mathcal{L}^{(\mathrm{AII})}_{\mathrm{WZW}},
\end{split} 
\end{equation}
where
\begin{equation}
\mathcal{L}^{(\mathrm{AII})}_{\mathrm{WZW}}=
\epsilon^{ijk}\,
\mathrm{tr}\,
\Bigl[
(\bar{G}^{-1}\partial^{\,}_{i}\bar{G})\,
(\bar{G}^{-1}\partial^{\,}_{j}\bar{G})
(\bar{G}^{-1}\partial^{\,}_{k}\bar{G})
\Bigr].
\end{equation}
The action $S^{(\mathrm{AII})}_{\mathrm{mass}}$ stems from the Lagrangian density
\begin{equation}
\begin{split}
\mathcal{L}^{(\mathrm{AII})}_{\mathrm{mass}}=&\, 
\sum_{I=1}^{N-1}
\lambda\,
\left(
G^{\,}_{(-,I),(-,I+1)} 
+ 
G^{\,}_{(+,I+1),(+,I)} 
\right) 
\\
\equiv&\,
\lambda\,\mathrm{tr}\,
\left(M^{(\mathrm{AII})}\,G\right).  
\end{split}
\label{eq: mass matrix cor AII minimal} 
\end{equation}
The second equality is established by using the 
non-Abelian bosonization formula 
(\ref{eq: non-abelian bosonization formula})
(we have set the mass parameter $m^{\,}_{\mathrm{uv}}=1$). 
The $4N\times 4N$ matrices $M^{(\mathrm{AII})}$
is represented by
\begin{equation}
M^{(\mathrm{AII})}:= 
\begin{pmatrix}
0 & B & 0 & 0 & 0 & 0 &  \cdots \\
-B^{\mathsf{T}} & 0 & B & 0 & 0 & 0 & \cdots \\
0 & -B^{\mathsf{T}} & 0 & B & 0 & 0 & \cdots \\
0 & 0 & -B^{\mathsf{T}} & 0 & B & 0 & \cdots \\
0 & 0 & 0 & -B^{\mathsf{T}} & 0 & B & \cdots \\
0 & 0 & 0 & 0 & -B^{\mathsf{T}} & \hphantom{-} 0 \hphantom{-}&  \cdots \\
\vdots & \vdots & \vdots & \vdots & \vdots & \vdots & \ddots\\
\end{pmatrix}
\label{eq: mass matrix class AII}
\end{equation}
in the basis for which
$B$ is the $4\times 4$ matrix
\begin{equation}
B:=
\begin{pmatrix}
\overbrace{0}^{(+,-)} & \overbrace{0}^{(+,+)} 
\\
\underbrace{{-\mathrm{i}\tau_{2}^{\,}} }_{(-,-)}& \underbrace{0}_{(-,+)}
\end{pmatrix}, 
\qquad
-\mathrm{i}\tau^{\,}_{2}= 
\begin{pmatrix}
\overbrace{0}^{(1,1)} & \overbrace{-1}^{(1,2)} \\
\underbrace{1}_{(2,1)} & \underbrace{0}_{(2,2)}
\end{pmatrix}.
\end{equation}
\end{subequations}
For any $N>0$, 
the $4N\times 4N$ matrices $M^{(\mathrm{AII})}$ defined by 
(\ref{eq: mass matrix class AII}) has 
four vanishing and $4\times(N-1)$ nonvanishing eigenvalues.

The quadratic perturbation (\ref{eq: mass matrix cor AII minimal})
thus reduces the central charge $c^{(\mathrm{AII})}_{0} =4\times N/2$
by the amount $4\times(N-1)/2$,
i.e., the central charge for the theory with the partition function 
$Z^{(\mathrm{AII})}$ is
\begin{equation}
c^{(\mathrm{AII})}=
\frac{4\times N}{2}-\frac{4\times(N-1)}{2}=2.
\label{10.15}
\end{equation}

We have constructed a topological insulator with 
the gapless and Kramers degenerate pairs of Majorana modes 
$(\chi^{\,}_{\mathrm{R},+,1,I},
\chi^{\,}_{\mathrm{L},-,1,I})^{\,}_{I=1}$
and
$(\chi^{\,}_{\mathrm{R},+,2,I},
\chi^{\,}_{\mathrm{L},-,2,I})^{\,}_{I=1}$
propagating along edge $I=1$ and 
the gapless and Kramers degenerate pairs of Majorana modes 
$(\chi^{\,}_{\mathrm{L},+,1,I},
\chi^{\,}_{\mathrm{R},-,1,I})^{\,}_{I=N}$
and
$(\chi^{\,}_{\mathrm{L},+,2,I},\chi^{\,}_{\mathrm{R},-,2,I})^{\,}_{I=N}$
propagating along edge $I=N$. 
This construction is summarized by
Fig.\ \ref{Fig: minimal wire network in AII}.

The symmetry class AII
supports a $\mathbb{Z}^{\,}_{2}$ topological classification.
This can be shown along the same lines as was done
for the symmetry class DIII.

\section{One-loop renormalization group flows}
\label{appsec: One-loop renormalization group flow}

In this appendix, we outline how to obtain the one-loop
renormalization group (RG) flows for
the current-current interactions from Sec.\
\ref{sec: Non-Abelian topological order out of coupled wires}.

\subsection{Warmup}
\label{appsec: Warmup}

We start with a $SU(N)$
current-current interaction that breaks the local
$SU^{\,}_{\mathrm{R}}(N)\times SU^{\,}_{\mathrm{L}}(N)$
symmetry at the $SU(N)$ WZW critical point with the Lagrangian density
$\mathcal{L}^{\,}_{0}$. We require that these interactions preserve the
Lorentz symmetry of $S^{\,}_{0}$. 
This assumption insures that the speed
of ``light'', the Fermi velocity, is not renormalized.
Imposing Lorentz symmetry allows to focus solely on
perturbations that open a spectral gap.

We recall that under a Lorentz boost parametrized by 
the rapidity $\beta\in\mathbb{R}$,
the light-cone coordinates transform as
\begin{equation}
t-x\mapsto
e^{-\beta}
(t-x),
\qquad
t+x\mapsto
e^{+\beta}
(t+x),
\end{equation}
while the right- and left-moving $SU(N)$ currents transform as
\begin{equation}
J^{a}_{\mathrm{R}}\to 
e^{+\beta}\,J^{a}_{\mathrm{R}}, 
\qquad
J^{a}_{\mathrm{L}}\to 
e^{-\beta}\,
J^{a}_{\mathrm{L}},
\end{equation}
for $a=1,\cdots, N^{2}-1$.
Hence, the requirement of Lorentz invariance imposes that
the current-current interactions involve products of
left- and right-moving currents with the number of 
left-moving generators equal to the number of right-moving
generators. To quadratic order in the currents,
the most general perturbing Lagrangian density 
that is quadratic in the generators of $SU(N)$ is
\begin{equation}
\mathcal{L}^{\,}_{\mathrm{int}}:=
-
\sum_{a,b=1}^{N^{2}-1}
J^{a}_{\mathrm{R}}\,
\lambda^{ab}\,
J^{b}_{\mathrm{L}},
\label{appeq: def current-current L-R interaction for SU(N)} 
\end{equation}
where $\lambda^{ab}=\lambda^{ba}$ are real-valued.

The partition function is
\begin{subequations}
\begin{equation}
Z:=
\int\limits_{\mathfrak{a}}\mathcal{D}[\psi^{*},\psi]\,
e^{+\mathrm{i}(S^{\,}_{0}+S^{\,}_{\mathrm{int}})},
\end{equation}
where both the measure for the fields and the actions 
\begin{equation}
S^{\,}_{0}=
\int
\frac{\mathrm{d}\bar{z}\mathrm{d}z}{2\pi\mathrm{i}}\,
\mathcal{L}^{\,}_{0},
\qquad
S^{\,}_{\mathrm{int}}=
\int
\frac{\mathrm{d}\bar{z}\mathrm{d}z}{2\pi\mathrm{i}}\,
\mathcal{L}^{\,}_{\mathrm{int}}
\end{equation}
\end{subequations}
depend on a short-distance cutoff $\mathfrak{a}$.
The integral in the partition function is over the fields
chosen to represent the Lagrangian densities.
The integrals in the actions are over 
two-dimensional Minkowski space.

The renormalization of the couplings entering the 
Lagrangian density consists in doing first the expansion
\begin{subequations}
\label{app eq: general RG procedure} 
\begin{equation}
Z=
\int\limits_{\mathfrak{a}}
\mathcal{D}[\psi^{*},\psi]\,
e^{+\mathrm{i}S^{\,}_{0}}
\left(
1
+
\mathrm{i}
S^{\,}_{\mathrm{int}}
+
\frac{\mathrm{i}^{2}}{2}
S^{2}_{\mathrm{int}}
+
\cdots
\right).
\label{app eq: general RG procedure a} 
\end{equation}
The short-distance cutoff $\mathfrak{a}$ is implied by the limits
on the path integral.
Second, high-energy degrees of freedom are integrated over,
\begin{equation}
Z=
\int\limits_{e^{\mathrm{d}\ell}\,\mathfrak{a}}
\mathcal{D}[\psi^{*},\psi]\,
\exp
\left(
+\mathrm{i}
\int
\frac{\mathrm{d}\bar{z}\mathrm{d}z}{2\pi\mathrm{i}}\,
\left(
\mathcal{L}^{\,}_{0}
+
\delta\mathcal{L}
\right)
\right),
\label{app eq: general RG procedure b}  
\end{equation}
\end{subequations}
where $\mathrm{d}\ell$ is a positive infinitesimal and
$\delta\mathcal{L}$ is to  be calculated
to any given order in perturbation theory.
The short-distance cutoff 
$e^{\mathrm{d}\ell}\,\mathfrak{a}$ is implied by the limits
on the path integral.
Third, the RG flows
follow from demanding that 
$\mathcal{L}^{\,}_{0}+\delta\mathcal{L}$
has the same form as
$\mathcal{L}^{\,}_{0}+\mathcal{L}^{\,}_{\mathrm{int}}$.

Now, the first non-vanishing term on the
right-hand side of Eq.\
(\ref{app eq: general RG procedure a})
is
\begin{equation}
\delta Z\equiv
\int\mathcal{D}[\psi^{*},\psi]\,
e^{+\mathrm{i}S^{\,}_{0}}
\frac{\mathrm{i}^{2}}{2}
S^{2}_{\mathrm{int}},
\end{equation}
for the Lorentz symmetry would be broken otherwise.

Without loss of generality, we may perform the one-loop
renormalization of the partition function after performing
a Wick rotation from the two-dimensional Minkowski space
to two-dimensional Euclidean space.

The term (the summation convention over repeated indices is
implied)
\begin{equation}
\begin{split}
\frac{S^{2}_{\mathrm{int}}}{2}=&\,
\frac{1}{2}
\int\frac{\mathrm{d}\bar{z}\mathrm{d}z}{2\pi\mathrm{i}}
\int\frac{\mathrm{d}\bar{w}\mathrm{d}w}{2\pi\mathrm{i}}\,
\lambda^{aa'}\,
\lambda^{bb'}\,
J^{a}_{\mathrm{R}}(w)\,
J^{a'}_{\mathrm{L}}(\bar{w})\,
\\
&\,
\times
J^{b}_{\mathrm{R}}(z)\,
J^{b'}_{\mathrm{L}}(\bar{z}).
\end{split}
\label{appeq: Sint raised power 2 divided two}
\end{equation}
is evaluated in three steps.
First, the $SU(N)$ counterparts 
\begin{subequations}
\begin{align}
&
J^{a}_{\mathrm{R}}(w)\,
J^{b}_{\mathrm{R}}(z)=
\frac{
\mathrm{i}
f^{abc}\,J^{c}_{\mathrm{R}}(z)
     }
     {
w-z
     }
+
\frac{1}{2}
\frac{
\mathrm{tr}\left(T^{a}\,T^{b}\right)
     }
     {
(w-z)^{2}
     },
\\
&
J^{a}_{\mathrm{L}}(\bar{w})\,
J^{b}_{\mathrm{L}}(\bar{z})=
\frac{
\mathrm{i}
f^{abc}\,J^{c}_{\mathrm{L}}(\bar{z})
     }
     {
\bar{w}-\bar{z}
     }
+
\frac{1}{2}
\frac{
\mathrm{tr}\left(T^{a}\,T^{b}\right)
     }
     {
(\bar{w}-\bar{z})^{2}
     },
\\
&
J^{a}_{\mathrm{R}}(w)\,
J^{b}_{\mathrm{L}}(\bar{z})=
J^{a}_{\mathrm{L}}(\bar{w})\,
J^{b}_{\mathrm{R}}(z)=
0,
\end{align}
\end{subequations}
to the OPE (\ref{eq: OPE so(n) currents}) 
are inserted into Eq.\ (\ref{appeq: Sint raised power 2 divided two}).
Because Lorentz invariance of $S^{\,}_{0}$  
is not broken spontaneously,
the leading field-dependent contribution is given by  
\begin{equation}
\begin{split}
\frac{S^{2}_{\mathrm{int}}}{2}\approx&\,
\frac{1}{2}
\int\frac{\mathrm{d}\bar{z}\mathrm{d}z}{2\pi\mathrm{i}}
\int\frac{\mathrm{d}\bar{w}\mathrm{d}w}{2\pi\mathrm{i}}\,
\lambda^{aa'}\,
\lambda^{bb'}
\\
&\,\times
\left[
\frac{\mathrm{i}^{2}\,f^{abc}\,f^{a'b'c'}}{|z-w|^{2}}\,
J^{c}_{\mathrm{R}}(z)\,
J^{c'}_{\mathrm{L}}(\bar{z})
+
\cdots
\right].
\end{split}
\end{equation}
Second, the two-dimensional integration
$\int\frac{\mathrm{d}\bar{w}\,\mathrm{d}w}{2\pi\mathrm{i}}$
is logarithmically divergent unless the second-order pole of the integrand
at $w=z$ is regulated by restricting the integration over it 
to a ring of inner radius
$\mathfrak{a}>0$
and of outer radius 
$e^{\mathrm{d}\ell}\,\mathfrak{a}>\mathfrak{a}$
with $\mathrm{d}\ell$ a positive infinitesimal,
\begin{equation}
\begin{split}
\delta S=&\,
-
\frac{2\pi\,\mathrm{d}\ell}{2}\,
\int\frac{\mathrm{d}\bar{z}\mathrm{d}z}{2\pi\mathrm{i}}\,
\lambda^{aa'}\,
\lambda^{bb'}\,
f^{abc}\,f^{a'b'c'}\,
J^{c}_{\mathrm{R}}(z)\,
J^{c'}_{\mathrm{L}}(\bar{z}).
\end{split}
\end{equation}
Third, the one-loop RG equations
\begin{equation}
\frac{\mathrm{d}\lambda^{cc'}}{\mathrm{d}\mathrm{\ell}}=
\pi\,
f^{abc}\,
f^{a'b'c'}\,
\lambda^{aa'}\,
\lambda^{bb'}
\label{app eq: RGSUN}
\end{equation}
follow for $c,c'=1,\cdots,N^{2}-1$
under the rescaling 
$\mathfrak{a}\mapsto(1+\mathrm{d}\ell)\,\mathfrak{a}$
of the short-distance characteristic length $\mathfrak{a}$.

For $SU(N)$ symmetric interaction, 
$\lambda^{ab}=\lambda\,\delta^{ab}$ with $a,b=1,\cdots,N^{2}-1$, 
the one-loop RG equations (\ref{app eq: RGSUN})
become
\begin{subequations}
\begin{equation}
\frac{\mathrm{d}\lambda}{\mathrm{d}\mathrm{\ell}}=
\pi N\lambda^{2}
\label{app eq: RGSUN sym}
\end{equation}
if the convention
\begin{equation}
\sum_{b,c=1}^{N^{2}-1}
f^{abc}\,
f^{a'bc}
=
N\,
\delta^{aa'}
\end{equation}
\end{subequations}
for $a,a'=1,\cdots,N^{2}-1$
is chosen for the quadratic Casimir eigenvalue.
For $\lambda$ positive, the flow is to strong coupling and is interpreted as
the opening of a gap in the $SU(N)$ sector of the theory
by the left-right non-Abelian current-current interaction. 

We specialize to two Lie groups from now on. 
There is the semi-simple Lie group $SU(2)\times SU(2)$
whose generators 
$\mathcal{J}^{\mathcal{A}}$
with $\mathcal{A}=1,\cdots,6$
are defined in Eqs.\ (\ref{eq: def mathcal L hat g / hat h a})
and (\ref{eq: def mathcal L hat g / hat h b}).
There is the diagonal subgroup $SU(2)$ of $SU(2)\times SU(2)$
whose generators 
$\mathcal{K}^{\mathcal{B}}$
are defined in Eq.\ (\ref{eq: def mathcal L hat g / hat h c}).

\subsection{Derivation of the one-loop RG flows for
Sec.\ \ref{subsec: Partial gapping without time-reversal symmetry}}
\label{appsubsec: Derivation of the one-loop RG flows if no TRS}

To calculate the one-loop RG flows obeyed by the coupling constants entering
the current-current interaction (\ref{eq: def mathcal L hat g / hat h d}),
we start from the square of the action 
(\ref{eq: def mathcal L hat g / hat h d}), 
as we did in (\ref{appeq: Sint raised power 2 divided two}),
\begin{widetext}
\begin{equation}
\begin{split}
\frac{S^{2}_{\mathrm{int}}}{2}=&\,
\frac{1}{2}
\int\frac{\mathrm{d}\bar{z}\mathrm{d}z}{2\pi\mathrm{i}}
\int\frac{\mathrm{d}\bar{w}\mathrm{d}w}{2\pi\mathrm{i}}\,
\Biggl[
\lambda^{\mathcal{A}^{\,}_{1}}_{\texttt{m}^{\,}_{1}}\,
\lambda^{\mathcal{A}^{\,}_{2}}_{\texttt{m}^{\,}_{2}}\,
\mathcal{J}^{\mathcal{A}^{\,}_{1}}_{\mathrm{L},\texttt{m}^{\,}_{1}}(\bar{w})\,
\mathcal{J}^{\mathcal{A}^{\,}_{1}}_{\mathrm{R},\texttt{m}^{\,}_{1}+1}({w})\,
\mathcal{J}^{\mathcal{A}^{\,}_{2}}_{\mathrm{L},\texttt{m}^{\,}_{2}}(\bar{z})\,
\mathcal{J}^{\mathcal{A}^{\,}_{2}}_{\mathrm{R},\texttt{m}^{\,}_{2}+1}({z})\,
\\
&\,
+
\upsilon^{\mathcal{B}^{\,}_{1}}_{\texttt{m}^{\,}_{1}}\,
\upsilon^{\mathcal{B}^{\,}_{2}}_{\texttt{m}^{\,}_{2}}\,
\mathcal{K}^{\mathcal{B}^{\,}_{1}}_{\mathrm{L},\texttt{m}^{\,}_{1}}(\bar{w})\,
\mathcal{K}^{\mathcal{B}^{\,}_{1}}_{\mathrm{R},\texttt{m}^{\,}_{1}}(w)\,
\mathcal{K}^{\mathcal{B}^{\,}_{2}}_{\mathrm{L},\texttt{m}^{\,}_{2}}(\bar{z})\,
\mathcal{K}^{\mathcal{B}^{\,}_{2}}_{\mathrm{R},\texttt{m}^{\,}_{2}}(z)\,
+
2
\lambda^{\mathcal{A}}_{\texttt{m}^{\,}_{1}}\,
\upsilon^{\mathcal{B}}_{\texttt{m}^{\,}_{2}}\,
\mathcal{J}^{\mathcal{A}}_{\mathrm{L},\texttt{m}^{\,}_{1}}(\bar{w})\,
\mathcal{J}^{\mathcal{A}}_{\mathrm{R},\texttt{m}^{\,}_{1}+1}(w)\,
\mathcal{K}^{\mathcal{B}}_{\mathrm{L},\texttt{m}^{\,}_{2}}(\bar{z})\,
\mathcal{K}^{\mathcal{B}}_{\mathrm{R},\texttt{m}^{\,}_{2}}(z)\,
\Biggr]\,.
\end{split}
\label{appeq: Sint no TRS raised power 2 divided two}
\end{equation}
\end{widetext}
Repeated indices will always be summed over, unless stated otherwise.

We are going to use three types of OPE.
First, we need the OPE for the semisimple Lie algebra 
$su(2)\oplus su(2)$. 
They are
\begin{subequations}
\label{appeq: OPE for su(2) oplus su(2) no TRS}
\begin{equation}
\mathcal{J}^{\mathcal{A}^{\,}_{1}}_{\mathrm{R},\texttt{m}^{\,}_{1}}(w)\,
\mathcal{J}^{\mathcal{A}^{\,}_{2}}_{\mathrm{R},\texttt{m}^{\,}_{2}}(z)=
\frac{
\mathrm{i}
f^{\mathcal{A}^{\,}_{1}\mathcal{A}^{\,}_{2}\mathcal{A}^{\,}_{3}}\,
\mathcal{J}^{\mathcal{A}^{\,}_{3}}_{\mathrm{R},\texttt{m}^{\,}_{1}}(z)
     }
     {
w-z
     } \delta^{\,}_{\texttt{m}^{\,}_{1}\texttt{m}^{\,}_{2}}
+ 
\cdots,
\label{appeq: OPE for su(2) oplus su(2) no TRS a}
\end{equation}
where the fact that the structure constants
$f^{\mathcal{A}^{\,}_{1}\mathcal{A}^{\,}_{2}\mathcal{A}^{\,}_{3}}$
are those for the semisimple Lie algebra $su(2)\times su(2)$
is implied by the use of the caligraphic label $\mathcal{A}=1,\cdots,6$.
Second, we need the OPE for the simple Lie algebra $su(2)$. They are
\begin{equation}
\mathcal{K}^{\mathcal{B}^{\,}_{1}}_{\mathrm{R},\texttt{m}^{\,}_{1}}(w)\,
\mathcal{K}^{\mathcal{B}^{\,}_{2}}_{\mathrm{R},\texttt{m}^{\,}_{2}}(z)=
\frac{
\mathrm{i}
f^{\mathcal{B}^{\,}_{1}\mathcal{B}^{\,}_{2}\mathcal{B}^{\,}_{3}}\,
\mathcal{K}^{\mathcal{B}^{\,}_{3}}_{\mathrm{R},\texttt{m}^{\,}_{1}}(z)
     }
     {
w-z
     } \delta^{\,}_{\texttt{m}^{\,}_{1}\texttt{m}^{\,}_{2}}
+ 
\cdots,
\label{appeq: OPE for su(2) oplus su(2) no TRS b}
\end{equation}
where the fact that the structure constants
$f^{\mathcal{B}^{\,}_{1}\mathcal{B}^{\,}_{2}\mathcal{B}^{\,}_{3}}$
reduce to those for the simple Lie algebra $su(2)$
is implied by the use of the caligraphic label $\mathcal{B}=1,2,3$.
Finally, we need the OPE between the generators of
$su(2)\oplus su(2)$
and its diagonal subalgebra $su(2)$.
They are
\begin{equation}
\mathcal{K}^{\mathcal{B}^{\,}_{1}}_{\mathrm{R},\texttt{m}^{\,}_{1}}(w)\,
\mathcal{J}^{\mathcal{A}^{\,}_{2}}_{\mathrm{R},\texttt{m}^{\,}_{2}}(z)=
\frac{
\mathrm{i}
f^{\mathcal{B}^{\,}_{1}\mathcal{A}^{\,}_{2}\mathcal{A}^{\,}_{3}}\,
\mathcal{J}^{\mathcal{A}^{\,}_{3}}_{\mathrm{R},\texttt{m}^{\,}_{1}}(z)
     }
     {
w-z
     } \delta^{\,}_{\texttt{m}^{\,}_{1}\texttt{m}^{\,}_{2}}
+ 
\cdots.
\label{appeq: OPE for su(2) oplus su(2) no TRS c}
\end{equation}
\end{subequations}
Because the diagonal subalgebra $su(2)$ is not an ideal,
the last entry of the structure constant is a calligraphic
$\mathcal{A}^{\,}_{3}=1,\cdots,6$.

\begin{widetext}
Insertion of the OPEs
(\ref{appeq: OPE for su(2) oplus su(2) no TRS})
into the square bracket on the right-hand side of
Eq.\ (\ref{appeq: Sint no TRS raised power 2 divided two})
gives
\begin{equation}
\begin{split}
\Biggl[&\,
\lambda^{\mathcal{A}^{\,}_{1}}_{\texttt{m}^{\,}_{\,}}\,
\lambda^{\mathcal{A}^{\,}_{2}}_{\texttt{m}^{\,}_{\,}}\,
\frac{\mathrm{i}^{2}\,f^{\mathcal{A}^{\,}_{1}\mathcal{A}^{\,}_{2}\mathcal{A}^{\,}_{3}}\,
f^{\mathcal{A}^{\,}_{1}\mathcal{A}^{\,}_{2}\mathcal{A}^{\,}_{4}}}{|z-w|^{2}}\,
\mathcal{J}^{\mathcal{A}^{\,}_{3}}_{\mathrm{L},\texttt{m}}(\bar{z})\,
\mathcal{J}^{\mathcal{A}^{\,}_{4}}_{\mathrm{R},\texttt{m}+1}(z)
+
\upsilon^{\mathcal{B}^{\,}_{1}}_{\texttt{m}}\,
\upsilon^{\mathcal{B}^{\,}_{2}}_{\texttt{m}}\,
\frac{\mathrm{i}^{2}\,f^{\mathcal{B}^{\,}_{1}\mathcal{B}^{\,}_{2}\mathcal{B}^{\,}_{3}}\,
f^{\mathcal{B}^{\,}_{1}\mathcal{B}^{\,}_{2}\mathcal{B}^{\,}_{4}}}{|z-w|^{2}}\,
\mathcal{K}^{\mathcal{B}^{\,}_{3}}_{\mathrm{L},\texttt{m}}(\bar{z})\,
\mathcal{K}^{\mathcal{B}^{\,}_{4}}_{\mathrm{R},\texttt{m}}(z)
\\
&\,
+
2\lambda^{\mathcal{A}}_{\texttt{m}}\,
\upsilon^{\mathcal{B}}_{\texttt{m}}\,
\frac{\mathrm{i}^{\,}\,f^{\mathcal{A}\mathcal{B}\mathcal{A}'}}{\bar{w}-\bar{z}}\,
\mathcal{J}^{\mathcal{A}'}_{\mathrm{L},\texttt{m}}(\bar{z})\,
\mathcal{J}^{\mathcal{A}}_{\mathrm{R},\texttt{m}+1}(w)\,
\mathcal{K}^{\mathcal{B}}_{\mathrm{R},\texttt{m}}(z)
+
2\lambda^{\mathcal{A}}_{\texttt{m}}\,
\upsilon^{\mathcal{B}}_{\texttt{m}+1}\,
\frac{\mathrm{i}^{\,}\,f^{\mathcal{A}\mathcal{B}\mathcal{A}'}}{w-z}\,
\mathcal{J}^{\mathcal{A}}_{\mathrm{L},\texttt{m}}(\bar{w})\,
\mathcal{K}^{\mathcal{B}}_{\mathrm{L},\texttt{m}+1}(\bar{z})\,
\mathcal{J}^{\mathcal{A}'}_{\mathrm{R},\texttt{m}+1}(z) 
+
\cdots
\Biggr].
\end{split}
\label{appeq: Sint no TRS raised power 2 divided two after OPE}
\end{equation}
\end{widetext}
The third and fourth terms in the integrand 
involves products with different numbers of right- and left-moving currents.  
We can ignore such contributions since operators with
nonvanishing conformal spin do not affect the beta functions, given the
Lorentz invariance of the critical point.
Integrating the second-order poles at $w$ and $\bar{w}$
over a ring with the inner radius $\mathfrak{a}$ and the outer radius 
$(1+\mathrm{d}\ell)\,\mathfrak{a}$ gives
\begin{equation}
\begin{split}
\delta S:=&\,
-
\frac{
2\pi\,\mathrm{d}\ell
     }
     {
2
     }\,
\int\frac{\mathrm{d}\bar{z}\mathrm{d}z}{2\pi\mathrm{i}}\,
\\
&\times\,
\Biggl[
\lambda^{\mathcal{A}^{\,}_{1}}_{\texttt{m}}\,
\lambda^{\mathcal{A}^{\,}_{2}}_{\texttt{m}}\,
f^{\mathcal{A}^{\,}_{1}\mathcal{A}^{\,}_{2}\mathcal{A}^{\,}_{3}}\,
f^{\mathcal{A}^{\,}_{1}\mathcal{A}^{\,}_{2}\mathcal{A}^{\,}_{4}}\,
\mathcal{J}^{\mathcal{A}^{\,}_{3}}_{\mathrm{L},\texttt{m}}(\bar{z})\,
\mathcal{J}^{\mathcal{A}^{\,}_{4}}_{\mathrm{R},\texttt{m}+1}(z)\,
\\
&\,
\,+\,
\upsilon^{\mathcal{B}^{\,}_{1}}_{\texttt{m}}\,
\upsilon^{\mathcal{B}^{\,}_{2}}_{\texttt{m}}\,
f^{\mathcal{B}^{\,}_{1}\mathcal{B}^{\,}_{2}\mathcal{B}^{\,}_{3}}\,
f^{\mathcal{B}^{\,}_{1}\mathcal{B}^{\,}_{2}\mathcal{B}^{\,}_{4}}\,
\mathcal{K}^{\mathcal{B}^{\,}_{3}}_{\mathrm{L},\texttt{m}}(\bar{z})\,
\mathcal{K}^{\mathcal{B}^{\,}_{4}}_{\mathrm{R},\texttt{m}}(z)\,
\Biggr]\,.
\end{split}
\label{appeq: Sint no TRS second order}
\end{equation}

For any given $\mathcal{A}=1,\cdots,6$ 
and any given $\texttt{m}=1,\cdots,n-1$,
the one-loop RG equations
\begin{subequations}
\begin{equation}
\begin{split}
\frac{\mathrm{d}\lambda^{\mathcal{A}}_{\texttt{m}}}{\mathrm{d}\mathrm{\ell}}&=
\pi\,
f^{\mathcal{A}\,\mathcal{A}'\,\mathcal{A}''}\,
f^{\mathcal{A}\,\mathcal{A}'\,\mathcal{A}''}\,
\lambda^{\mathcal{A}'}_{\texttt{m}}\,
\lambda^{\mathcal{A}''}_{\texttt{m}}
\end{split}
\label{appeq: RGSUN no TRS A}
\end{equation}
follow for the current-current interactions
with the generators from the semisimple $su(2)\oplus su(2)$ algebra.
For any given $\mathcal{B}=1,\cdots,3$ 
and any given $\texttt{m}=1,\cdots,n$,
the one-loop RG equations
\begin{equation}
\begin{split}
\frac{\mathrm{d}\upsilon^{\mathcal{B}}_{\texttt{m}}}{\mathrm{d}\mathrm{\ell}}&=
\pi\,
f^{\mathcal{B}\,\mathcal{B}'\,\mathcal{B}''}\,
f^{\mathcal{B}\,\mathcal{B}'\,\mathcal{B}''}\,
\upsilon^{\mathcal{B}'}_{\texttt{m}}\,
\upsilon^{\mathcal{B}''}_{\texttt{m}}
\end{split}
\label{appeq: RGSUN no TRS B}
\end{equation} 
\end{subequations}
follow for the current-current interactions
with the generators from the diagonal $su(2)$ subalgebra.

\subsection{Derivation of the one-loop RG flows for Sec.~\ref{subsubsec: Case I}}
\label{appsubsec: Derivation of the RG flows for Case I}

The following manipulations on the explicit form of 
$\mathcal{L}^{\,}_{\mathrm{int}}(\kappa)$ defined by Eq.\
(\ref{eq: interpolating L int}) 
are useful. Indeed 
\begin{widetext}
\begin{align}
\mathcal{L}^{\,}_{\mathrm{int}}(\kappa):=&\,
\frac{1-\kappa}{2}\,
\Bigl[-
\sum_{\texttt{m}=1}^{n-1}
\sum_{\mathcal{A}=1}^{6}
\lambda^{\mathcal{A}}_{\texttt{m}}\,
\mathcal{J}^{\mathcal{A}}_{\mathrm{L},\texttt{m}}\,
\mathcal{J}^{\mathcal{A}}_{\mathrm{R},\texttt{m}+1}
-
\sum_{\texttt{m}=1}^{n}
\sum_{\mathcal{B}=1}^{3}
\upsilon^{\mathcal{B}}_{\texttt{m}}\,
\mathcal{K}^{\mathcal{B}}_{\mathrm{L},\texttt{m}}\,
\mathcal{K}^{\mathcal{B}}_{\mathrm{R},\texttt{m}}
\Bigr]
\nonumber\\
&\,
+
\frac{1+\kappa}{2}\,
\Bigl[-
\sum_{\texttt{m}=1}^{n-1}
\sum_{\mathcal{A}=1}^{6}
\lambda^{\mathcal{A}}_{\texttt{m}}\,
\mathcal{J}^{\mathcal{A}}_{\mathrm{R},\texttt{m}}\,
\mathcal{J}^{\mathcal{A}}_{\mathrm{L},\texttt{m}+1}
-
\sum_{\texttt{m}=1}^{n}
\sum_{\mathcal{B}=1}^{3}
\upsilon^{\mathcal{B}}_{\texttt{m}}\,
\mathcal{K}^{\mathcal{B}}_{\mathrm{R},\texttt{m}}\,
\mathcal{K}^{\mathcal{B}}_{\mathrm{L},\texttt{m}}
\Bigr]
\nonumber\\
=&\,
-
\frac{1-\kappa}{2}\,
\sum_{\texttt{m}=1}^{n-1}
\sum_{\mathcal{A}=1}^{6}
\lambda^{\mathcal{A}}_{\texttt{m}}\,
\mathcal{J}^{\mathcal{A}}_{\mathrm{L},\texttt{m}}\,
\mathcal{J}^{\mathcal{A}}_{\mathrm{R},\texttt{m}+1}
-\frac{1+\kappa}{2}\,
\sum_{\texttt{m}=1}^{n-1}
\sum_{\mathcal{A}=1}^{6}
\lambda^{\mathcal{A}}_{\texttt{m}}\,
\mathcal{J}^{\mathcal{A}}_{\mathrm{R},\texttt{m}}\,
\mathcal{J}^{\mathcal{A}}_{\mathrm{L},\texttt{m}+1}
-
\sum_{\texttt{m}=1}^{n}
\sum_{\mathcal{B}=1}^{3}
\upsilon^{\mathcal{B}}_{\texttt{m}}\,
\mathcal{K}^{\mathcal{B}}_{\mathrm{R},\texttt{m}}\,
\mathcal{K}^{\mathcal{B}}_{\mathrm{L},\texttt{m}}
\label{appeq: Lint case 1}
\end{align}
shows that the couplings of the current-current interactions
with the generators $\mathcal{K}$ from the diagonal subalgebra $su(2)$
are independent of the interpolating real-valued parameter $\kappa$. 
At this stage, it is convenient to introduce the couplings
\begin{subequations}
\begin{equation}
\lambda^{\mathcal{A}}_{\mathrm{LR},\texttt{m}}:=
\frac{1-\kappa}{2}\,\lambda^{\mathcal{A}}_{\texttt{m}},
\qquad
\lambda^{\mathcal{A}}_{\mathrm{RL},\texttt{m}}:=
\frac{1+\kappa}{2}\,\lambda^{\mathcal{A}}_{\texttt{m}},
\qquad
\mathcal{A}=1,\cdots,6,
\qquad
\texttt{m}=1,\cdots,n-1,
\end{equation}
and the interaction
\begin{equation}
\mathcal{L}^{\,}_{\mathrm{int}}:=
-
\sum_{\texttt{m}=1}^{n-1}
\sum_{\mathcal{A}=1}^{6}
\lambda^{\mathcal{A}}_{\mathrm{LR},\texttt{m}}\,
\mathcal{J}^{\mathcal{A}}_{\mathrm{L},\texttt{m}}\,
\mathcal{J}^{\mathcal{A}}_{\mathrm{R},\texttt{m}+1}
-
\sum_{\texttt{m}=1}^{n-1}
\sum_{\mathcal{A}=1}^{6}
\lambda^{\mathcal{A}}_{\mathrm{RL},\texttt{m}}\,
\mathcal{J}^{\mathcal{A}}_{\mathrm{R},\texttt{m}}\,
\mathcal{J}^{\mathcal{A}}_{\mathrm{L},\texttt{m}+1}
-
\sum_{\texttt{m}=1}^{n}
\sum_{\mathcal{B}=1}^{3}
\upsilon^{\mathcal{B}}_{\texttt{m}}\,
\mathcal{K}^{\mathcal{B}}_{\mathrm{R},\texttt{m}}\,
\mathcal{K}^{\mathcal{B}}_{\mathrm{L},\texttt{m}}.
\end{equation}
\end{subequations}

As was the case in Appendix\ 
\ref{appsubsec: Derivation of the one-loop RG flows if no TRS},
the one-loop RG equations obeyed by the couplings
$\upsilon^{\mathcal{B}}_{\texttt{m}}$ 
with $\mathcal{B}=1,2,3$ and $\texttt{m}=1,\cdots,n$
decouple from the one-loop RG equations obeyed by the couplings
$\lambda^{\mathcal{A}}_{\mathrm{LR},\texttt{m}}$ 
and
$\lambda^{\mathcal{A}}_{\mathrm{RL},\texttt{m}}$ 
with $\mathcal{A}=1,\cdots,6$ and $\texttt{m}=1,\cdots,n-1$.
The one-loop RG equations obeyed by the couplings
$\upsilon^{\mathcal{B}}_{\texttt{m}}$ that enter the current-current interaction
(\ref{appeq: Lint case 1}) 
are given by Eq.\ (\ref{appeq: RGSUN no TRS B}).

Now, to calculate the one-loop RG equations for the coupling constants
in the $\mathcal{J}$ sector of the current-current interaction
(\ref{appeq: Lint case 1}), we only need to treat
the $\mathcal{J}$-dependent contribution
\begin{equation}
\begin{split}
\mathcal{L}^{\,}_{\mathrm{int},\mathcal{J}}(\kappa):=&
-
\sum_{\texttt{m}=1}^{n-1}
\sum_{\mathcal{A}=1}^{6}
\lambda^{\mathcal{A}}_{\mathrm{LR},\texttt{m}}\,
\mathcal{J}^{\mathcal{A}}_{\mathrm{L},\texttt{m}}\,
\mathcal{J}^{\mathcal{A}}_{\mathrm{R},\texttt{m}+1}
-
\sum_{\texttt{m}=1}^{n-1}
\sum_{\mathcal{A}=1}^{6}
\lambda^{\mathcal{A}}_{\mathrm{RL},\texttt{m}}\,
\mathcal{J}^{\mathcal{A}}_{\mathrm{R},\texttt{m}}\,
\mathcal{J}^{\mathcal{A}}_{\mathrm{L},\texttt{m}+1}
\end{split}
\label{appeq: Lint case 1 J}
\end{equation}
to the Lagrangian density (\ref{appeq: Lint case 1}).
Expansion of the Boltzmann weight with the action
corresponding to the Lagrangian density 
(\ref{appeq: Lint case 1 J})
gives the second-order contribution 
\begin{equation}
\begin{split}
\frac{S^{2}_{\mathrm{int}}}{2}=&\,
\frac{1}{2}
\int\frac{\mathrm{d}\bar{z}\mathrm{d}z}{2\pi\mathrm{i}}
\int\frac{\mathrm{d}\bar{w}\mathrm{d}w}{2\pi\mathrm{i}}\,
\Biggl[
\lambda^{\mathcal{A}^{\,}_{1}}_{\mathrm{LR},\texttt{m}^{\,}_{1}}\,
\lambda^{\mathcal{A}^{\,}_{2}}_{\mathrm{LR},\texttt{m}^{\,}_{2}}\,
\mathcal{J}^{\mathcal{A}^{\,}_{1}}_{\mathrm{L},\texttt{m}^{\,}_{1}}(\bar{w})\,
\mathcal{J}^{\mathcal{A}^{\,}_{1}}_{\mathrm{R},\texttt{m}^{\,}_{1}+1}(w)\,
\mathcal{J}^{\mathcal{A}^{\,}_{2}}_{\mathrm{L},\texttt{m}^{\,}_{2}}(\bar{z})\,
\mathcal{J}^{\mathcal{A}^{\,}_{2}}_{\mathrm{R},\texttt{m}^{\,}_{2}+1}(z)
\\
&\,
+
\lambda^{\mathcal{A}^{\,}_{1}}_{\mathrm{RL},\texttt{m}^{\,}_{1}}\,
\lambda^{\mathcal{A}^{\,}_{2}}_{\mathrm{RL},\texttt{m}^{\,}_{2}}\,
\mathcal{J}^{\mathcal{A}^{\,}_{1}}_{\mathrm{R},\texttt{m}^{\,}_{1}}(w)\,
\mathcal{J}^{\mathcal{A}^{\,}_{1}}_{\mathrm{L},\texttt{m}^{\,}_{1}+1}(\bar{w})\,
\mathcal{J}^{\mathcal{A}^{\,}_{2}}_{\mathrm{R},\texttt{m}^{\,}_{2}}(z)\,
\mathcal{J}^{\mathcal{A}^{\,}_{2}}_{\mathrm{L},\texttt{m}^{\,}_{2}+1}(\bar{z})
\\
&\,
+
2\,\lambda^{\mathcal{A}^{\,}_{1}}_{\mathrm{LR},\texttt{m}^{\,}_{1}}\,
\lambda^{\mathcal{A}^{\,}_{2}}_{\mathrm{RL},\texttt{m}^{\,}_{2}}\,
\mathcal{J}^{\mathcal{A}^{\,}_{1}}_{\mathrm{L},\texttt{m}^{\,}_{1}}(\bar{w})\,
\mathcal{J}^{\mathcal{A}^{\,}_{1}}_{\mathrm{R},\texttt{m}^{\,}_{1}+1}(w)\,
\mathcal{J}^{\mathcal{A}^{\,}_{2}}_{\mathrm{R},\texttt{m}^{\,}_{2}}(z)\,
\mathcal{J}^{\mathcal{A}^{\,}_{2}}_{\mathrm{L},\texttt{m}^{\,}_{2}+1}(\bar{z})
\Biggr].
\end{split}
\label{appeq: Sint TRS case 1 raised power 2 divided two}
\end{equation}

Insertion of the OPEs
(\ref{appeq: OPE for su(2) oplus su(2) no TRS})
into the square bracket on the right-hand side of
Eq.\ (\ref{appeq: Sint TRS case 1 raised power 2 divided two})
gives
\begin{equation}
\begin{split}
\Biggl[
\lambda^{\mathcal{A}^{\,}_{1}}_{\mathrm{LR},\texttt{m}^{\,}_{\,}}\,
\lambda^{\mathcal{A}^{\,}_{2}}_{\mathrm{LR},\texttt{m}^{\,}_{\,}}\,
\frac{\mathrm{i}^{2}\,f^{\mathcal{A}^{\,}_{1}\mathcal{A}^{\,}_{2}\mathcal{A}^{\,}_{3}}\,
f^{\mathcal{A}^{\,}_{1}\mathcal{A}^{\,}_{2}\mathcal{A}^{\,}_{4}}}{|z-w|^{2}}\,
\mathcal{J}^{\mathcal{A}^{\,}_{3}}_{\mathrm{L},\texttt{m}}(\bar{z})\,
\mathcal{J}^{\mathcal{A}^{\,}_{4}}_{\mathrm{R},\texttt{m}+1}(z)
+
\lambda^{\mathcal{A}^{\,}_{1}}_{\mathrm{RL},\texttt{m}^{\,}_{\,}}\,
\lambda^{\mathcal{A}^{\,}_{2}}_{\mathrm{RL},\texttt{m}^{\,}_{\,}}\,
\frac{\mathrm{i}^{2}\,f^{\mathcal{A}^{\,}_{1}\mathcal{A}^{\,}_{2}\mathcal{A}^{\,}_{3}}\,
f^{\mathcal{A}^{\,}_{1}\mathcal{A}^{\,}_{2}\mathcal{A}^{\,}_{4}}}{|z-w|^{2}}\,
\mathcal{J}^{\mathcal{A}^{\,}_{3}}_{\mathrm{R},\texttt{m}}(z)\,
\mathcal{J}^{\mathcal{A}^{\,}_{4}}_{\mathrm{L},\texttt{m}+1}(\bar{z})
+
\cdots
\Biggr].
\end{split}
\label{appeq: Sint TRS case 1 raised power 2 divided two after OPE}
\end{equation}
We note that the contribution from the OPEs from the third term 
inside the bracket on the right-hand side of Eq.\ 
(\ref{appeq: Sint TRS case 1 raised power 2 divided two}) 
vanishes because $\delta^{\,}_{\texttt{m}^{\,}_{1},\texttt{m}^{\,}_{2}+1}$ and 
$\delta^{\,}_{\texttt{m}^{\,}_{1}+1,\texttt{m}^{\,}_{2}}$ cannot be met simultaneously.
Integrating the second-order poles at $w$ and $\bar{w}$
over a ring with the inner radius $\mathfrak{a}$ and the outer radius 
$(1+\mathrm{d}\ell)\,\mathfrak{a}$ gives
\begin{equation}
\begin{split}
\delta S=&\,
-
\frac{2\pi\,\mathrm{d}\ell}{2}\,
\int\frac{\mathrm{d}\bar{z}\mathrm{d}z}{2\pi\mathrm{i}}\,
\Biggl[
\lambda^{\mathcal{A}^{\,}_{1}}_{\mathrm{L}\mathrm{R},\texttt{m}^{\,}_{\,}}\,
\lambda^{\mathcal{A}^{\,}_{2}}_{\mathrm{L}\mathrm{R},\texttt{m}^{\,}_{\,}}\,
f^{\mathcal{A}^{\,}_{1}\mathcal{A}^{\,}_{2}\mathcal{A}^{\,}_{3}}\,
f^{\mathcal{A}^{\,}_{1}\mathcal{A}^{\,}_{2}\mathcal{A}^{\,}_{4}}\,
\mathcal{J}^{\mathcal{A}^{\,}_{3}}_{\mathrm{L},\texttt{m}}\,
\mathcal{J}^{\mathcal{A}^{\,}_{4}}_{\mathrm{R},\texttt{m}+1}\,
+
\lambda^{\mathcal{A}^{\,}_{1}}_{\mathrm{R}\mathrm{L},\texttt{m}}\,
\lambda^{\mathcal{A}^{\,}_{2}}_{\mathrm{R}\mathrm{L},\texttt{m}}\,
f^{\mathcal{A}^{\,}_{1}\mathcal{A}^{\,}_{2}\mathcal{A}^{\,}_{3}}
f^{\mathcal{A}^{\,}_{1}\mathcal{A}^{\,}_{2}\mathcal{A}^{\,}_{4}}
\mathcal{J}^{\mathcal{A}^{\,}_{3}}_{\mathrm{R},\texttt{m}}\,
\mathcal{J}^{\mathcal{A}^{\,}_{4}}_{\mathrm{L},\texttt{m}+1}
\Biggr].
\end{split}
\label{appeq: Sint TRS case 1 two coupling second order}
\end{equation}
\end{widetext}

For any given $\mathcal{A}\,=1,\cdots,6$ 
and any given $\texttt{m}=1,\cdots,n-1$,
there follows the pair of one-loop RG equations 
\begin{subequations}
\begin{align}
\frac{
\mathrm{d}\lambda^{\mathcal{A}\,}_{\mathrm{L}\mathrm{R},\texttt{m}}
     }
     {
\mathrm{d}\mathrm{\ell}
      }=
\pi\,
f^{\mathcal{A}\,\mathcal{A}'\,\mathcal{A}''}\,
f^{\mathcal{A}\,\mathcal{A}'\,\mathcal{A}''}\,
\lambda^{\mathcal{A}'\,}_{\mathrm{L}\mathrm{R},\texttt{m}^{\,}_{\,}}\,
\lambda^{\mathcal{A}''}_{\mathrm{L}\mathrm{R},\texttt{m}^{\,}_{\,}}
\label{appeq: RGSUN TRS case 1 kappa A}
\end{align}
with the initial conditions
\begin{align}
\lambda^{\mathcal{A}\,}_{\mathrm{L}\mathrm{R},\texttt{m}}(\ell=0)=
\frac{1-\kappa}{2}\,
\lambda^{\mathcal{A}\,}_{\texttt{m}}
\label{appeq: RGSUN TRS case 1 kappa B}
\end{align} 
on the one hand,
and
\begin{align}
&
\frac{
\mathrm{d}\lambda^{\mathcal{A}\,}_{\mathrm{R}\mathrm{L},\texttt{m}}
     }
     {
\mathrm{d}\mathrm{\ell}
     }=
\pi\,
f^{\mathcal{A}\,\mathcal{A}'\,\mathcal{A}''}\,
f^{\mathcal{A}\,\mathcal{A}'\,\mathcal{A}''}\,
\lambda^{\mathcal{A}'\,}_{\mathrm{R}\mathrm{L},\texttt{m}^{\,}_{\,}}\,
\lambda^{\mathcal{A}''}_{\mathrm{R}\mathrm{L},\texttt{m}^{\,}_{\,}}
\label{appeq: RGSUN TRS case 1 kappa C}
\end{align}
with the initial conditions
\begin{align}
\lambda^{\mathcal{A}\,}_{\mathrm{R}\mathrm{L},\texttt{m}}(\ell=0)=
\frac{1+\kappa}{2}\,
\lambda^{\mathcal{A}\,}_{\texttt{m}}
\label{appeq: RGSUN TRS case 1 kappa D}
\end{align}
\end{subequations} 
on the other hand. 
We conclude that 
(i)
Eqs.\
(\ref{appeq: RGSUN TRS case 1 kappa A}) 
and
(\ref{appeq: RGSUN TRS case 1 kappa B}) 
are decoupled from 
Eqs.\ 
(\ref{appeq: RGSUN TRS case 1 kappa C})
and
(\ref{appeq: RGSUN TRS case 1 kappa D})
to one loop,
(ii)
$\mathrm{d}\lambda^{\mathcal{A}\,}_{\mathrm{L}\mathrm{R},\texttt{m}}/
\mathrm{d}\mathrm{\ell}$
($\mathrm{d}\lambda^{\mathcal{A}\,}_{\mathrm{R}\mathrm{L},\texttt{m}}/
\mathrm{d}\mathrm{\ell}$) are positive when
all 
$\lambda^{\mathcal{A}\,}_{\mathrm{L}\mathrm{R},\texttt{m}}$
($\lambda^{\mathcal{A}\,}_{\mathrm{R}\mathrm{L},\texttt{m}}$)
share the same sign, and (iii)
all 
$\lambda^{\mathcal{A}\,}_{\mathrm{L}\mathrm{R},\texttt{m}}$
($\lambda^{\mathcal{A}\,}_{\mathrm{R}\mathrm{L},\texttt{m}}$)
are marginally irrelevant when all
$\lambda^{\mathcal{A}\,}_{\mathrm{L}\mathrm{R},\texttt{m}}$
($\lambda^{\mathcal{A}\,}_{\mathrm{R}\mathrm{L},\texttt{m}}$)
are nonvanishing and negative 
[i.e., $\kappa<-1$ ($\kappa>+1$) and $\lambda^{\mathcal{A}\,}_{\texttt{m}}>0$].

For any $\texttt{m}=1,\cdots,n-1$,
their (formal) solutions for the $SU(2)\times SU(2)$ symmetric initial conditions
\begin{subequations}
\begin{equation}
\lambda^{\mathcal{A}}_{\texttt{m}}\equiv
\lambda^{\,}_{\texttt{m}},
\qquad
\mathcal{A}\,=1,\cdots,6,
\end{equation}
are given by
\begin{equation}
\lambda^{\,}_{\mathrm{L}\mathrm{R},\texttt{m}}(\ell)=
\frac{
\left(\frac{1-\kappa}{2}\right)\,
\lambda^{\,}_{\texttt{m}}
     }
     {
1
-
c\,
\left(\frac{1-\kappa}{2}\right)\,
\lambda^{\,}_{\texttt{m}}\,
\ell
     }
\end{equation}
on the one hand, and
\begin{equation}
\lambda^{\,}_{\mathrm{R}\mathrm{L},\texttt{m}}(\ell)=
\frac{
\left(\frac{1+\kappa}{2}\right)\,
\lambda^{\,}_{\texttt{m}}
     }
     {
1
-
c\,
\left(\frac{1+\kappa}{2}\right)\,
\lambda^{\,}_{\texttt{m}}\,
\ell
     }
\end{equation}
on the other hand.
The positive numerical constant $c$ is
here defined by
\begin{equation}
c:=
\pi\,
\sum_{\mathcal{A}',\mathcal{A}''=1}^{6}
\left(f^{\mathcal{A}\,\mathcal{A}'\,\mathcal{A}''}\right)^{2}.
\end{equation}
\end{subequations}
The standard interpretation of the poles at
\begin{equation}
e^{-\ell}=
e^{
-
\frac{
1
     }
     {
c\left(\frac{1\mp\kappa}{2}\right)\,\lambda^{\,}_{\text{m}}
     }
   }
\end{equation}
is that they signal an instability of the unperturbed ground state
to the interacting channel with the bare coupling constant
\begin{equation}
\left(\frac{1\mp\kappa}{2}\right)\,\lambda^{\,}_{\text{m}}>0,
\qquad
\lambda^{\,}_{\text{m}}>0.
\end{equation}
The dominant instability is defined by 
\begin{equation}
\sup
\left\{\left.
\left(\frac{1-\kappa}{2}\right)\,\lambda^{\,}_{\text{m}},
\left(\frac{1+\kappa}{2}\right)\,\lambda^{\,}_{\text{m}}
\right|
\texttt{m}=1,\cdots,n-1
\right\}.
\end{equation}
Following this line of reasonning, 
the competition between
the interactions 
$\mathcal{L}^{\mathrm{L}\to\mathrm{R}}_{\mathrm{int}}$
and
$\mathcal{L}^{\mathrm{R}\to\mathrm{L}}_{\mathrm{int}}$
in Eq.\ (\ref{eq: interpolating L int})
is won by
$\mathcal{L}^{\mathrm{L}\to\mathrm{R}}_{\mathrm{int}}$
when $-1<\kappa<0$
and
$\mathcal{L}^{\mathrm{R}\to\mathrm{L}}_{\mathrm{int}}$
when $0<\kappa<+1$.
When $\kappa\leq-1$,
$\mathcal{L}^{\mathrm{R}\to\mathrm{L}}_{\mathrm{int}}$
is marginally irrelevant
while
$\mathcal{L}^{\mathrm{L}\to\mathrm{R}}_{\mathrm{int}}$
is marginally relevant.
When $1\leq \kappa$,
$\mathcal{L}^{\mathrm{L}\to\mathrm{R}}_{\mathrm{int}}$
is marginally irrelevant
while
$\mathcal{L}^{\mathrm{R}\to\mathrm{L}}_{\mathrm{int}}$
is marginally relevant. 

\subsection{Derivation of the one-loop RG flows for 
Sec.~\ref{subsubsec: Case II}}
\label{appsubsec: Derivation of the RG flows for Case II}

We proceed by computing the one-loop RG flow equations 
for the couplings
$\lambda^{\mathcal{A}}_{\texttt{m}}$
and
$\upsilon^{\mathcal{B}}_{\texttt{m}}$
in (\ref{eq: final interaction if TRS a}).   
The following manipulation on the explicit form of 
$\mathcal{L}^{\,}_{\mathrm{int}}$ defined by Eq.\
(\ref{eq: final interaction if TRS a}) 
is useful.
\begin{widetext}
\begin{align}
\mathcal{L}^{\,}_{\mathrm{int}}:=&\,
-
\sum_{\texttt{m}=1}^{n-1}
\sum_{\mathcal{A}=1}^{6}
\lambda^{\mathcal{A}}_{\texttt{m}}\,
\left(
\mathcal{J}^{\mathcal{A}}_{\mathrm{L},\texttt{m}}\,
\mathcal{J}^{\mathcal{A}}_{\mathrm{R},\texttt{m}+1}\,
+
\tilde{\mathcal{J}}^{\mathcal{A}}_{\mathrm{R},\texttt{m}}\,
\tilde{\mathcal{J}}^{\mathcal{A}}_{\mathrm{L},\texttt{m}+1}
\right)
-
\sum_{\texttt{m}=1}^{n}
\sum_{\mathcal{B}=1}^{3}
\upsilon^{\mathcal{B}}_{\texttt{m}}\,
\left(
\mathcal{K}^{\mathcal{B}}_{\mathrm{L},\texttt{m}}\,
\mathcal{K}^{\mathcal{B}}_{\mathrm{R},\texttt{m}}\,
+
\tilde{\mathcal{K}}^{\mathcal{B}}_{\mathrm{R},\texttt{m}}\,
\tilde{\mathcal{K}}^{\mathcal{B}}_{\mathrm{L},\texttt{m}}
\right).
\label{appeq: Lint case 2}
\end{align}
As was the case in Appendix\ 
\ref{appsubsec: Derivation of the one-loop RG flows if no TRS},
the one-loop RG equations obeyed by the couplings
$\upsilon^{\mathcal{B}}_{\texttt{m}}$ 
with $\mathcal{B}=1,2,3$ and $\texttt{m}=1,\cdots,n$
decouple from the one-loop RG equations obeyed by the couplings
$\lambda^{\mathcal{A}}_{\texttt{m}}$ 
with $\mathcal{A}=1,\cdots,6$ and $\texttt{m}=1,\cdots,n-1$.
We may thus derive
the one-loop RG equations for the coupling constants
in the $\mathcal{J}$ sector and in the $\mathcal{K}$ sector 
of the current-current interaction
(\ref{appeq: Lint case 2}) separately. 

First, we look at 
the $\mathcal{J}$-dependent contribution
\begin{equation}
\begin{split}
\mathcal{L}^{\,}_{\mathrm{int},\mathcal{J}}:=&
-
\sum_{\texttt{m}=1}^{n-1}
\sum_{\mathcal{A}=1}^{6}
\lambda^{\mathcal{A}}_{\texttt{m}}\,
\left(
\mathcal{J}^{\mathcal{A}}_{\mathrm{L},\texttt{m}}\,
\mathcal{J}^{\mathcal{A}}_{\mathrm{R},\texttt{m}+1}\,
+
\tilde{\mathcal{J}}^{\mathcal{A}}_{\mathrm{R},\texttt{m}}\,
\tilde{\mathcal{J}}^{\mathcal{A}}_{\mathrm{L},\texttt{m}+1}
\right)
\end{split}
\label{appeq: Lint case 2 J}
\end{equation}
to the Lagrangian density (\ref{appeq: Lint case 2}).
Expansion of the Boltzmann weight with the action
corresponding to the Lagrangian density 
(\ref{appeq: Lint case 2 J})
gives the second-order contribution 
\begin{equation}
\begin{split}
\frac{S^{2}_{\mathrm{int}}}{2}=&\,
\frac{1}{2}
\int\frac{\mathrm{d}\bar{z}\mathrm{d}z}{2\pi\mathrm{i}}
\int\frac{\mathrm{d}\bar{w}\mathrm{d}w}{2\pi\mathrm{i}}\,
\Biggl[
\lambda^{\mathcal{A}^{\,}_{1}}_{\texttt{m}^{\,}_{1}}\,
\lambda^{\mathcal{A}^{\,}_{2}}_{\texttt{m}^{\,}_{2}}\,
\mathcal{J}^{\mathcal{A}^{\,}_{1}}_{\mathrm{L},\texttt{m}^{\,}_{1}}(\bar{w})\,
\mathcal{J}^{\mathcal{A}^{\,}_{1}}_{\mathrm{R},\texttt{m}^{\,}_{1}+1}(w)\,
\mathcal{J}^{\mathcal{A}^{\,}_{2}}_{\mathrm{L},\texttt{m}^{\,}_{2}}(\bar{z})\,
\mathcal{J}^{\mathcal{A}^{\,}_{2}}_{\mathrm{R},\texttt{m}^{\,}_{2}+1}(z)
\\
&\,
+
\lambda^{\mathcal{A}^{\,}_{1}}_{\texttt{m}^{\,}_{1}}\,
\lambda^{\mathcal{A}^{\,}_{2}}_{\texttt{m}^{\,}_{2}}\,
\tilde{\mathcal{J}}^{\mathcal{A}^{\,}_{1}}_{\mathrm{R},\texttt{m}^{\,}_{1}}(w)\,
\tilde{\mathcal{J}}^{\mathcal{A}^{\,}_{1}}_{\mathrm{L},\texttt{m}^{\,}_{1}+1}(\bar{w})\,
\tilde{\mathcal{J}}^{\mathcal{A}^{\,}_{2}}_{\mathrm{R},\texttt{m}^{\,}_{2}}(z)\,
\tilde{\mathcal{J}}^{\mathcal{A}^{\,}_{2}}_{\mathrm{L},\texttt{m}^{\,}_{2}+1}(\bar{z})
\\
&\,
+
2\,\lambda^{\mathcal{A}^{\,}_{1}}_{\texttt{m}^{\,}_{1}}\,
\lambda^{\mathcal{A}^{\,}_{2}}_{\texttt{m}^{\,}_{2}}\,
\mathcal{J}^{\mathcal{A}^{\,}_{1}}_{\mathrm{L},\texttt{m}^{\,}_{1}}(\bar{w})\,
\mathcal{J}^{\mathcal{A}^{\,}_{1}}_{\mathrm{R},\texttt{m}^{\,}_{1}+1}(w)\,
\tilde{\mathcal{J}}^{\mathcal{A}^{\,}_{2}}_{\mathrm{R},\texttt{m}^{\,}_{2}}(z)\,
\tilde{\mathcal{J}}^{\mathcal{A}^{\,}_{2}}_{\mathrm{L},\texttt{m}^{\,}_{2}+1}(\bar{z})
\Biggr].
\end{split}
\label{appeq: Sint TRS case 2 J raised power 2 divided two}
\end{equation}

Insertion of the OPEs
(\ref{appeq: OPE for su(2) oplus su(2) no TRS})
into the square bracket on the right-hand side of
Eq.\ (\ref{appeq: Sint TRS case 2 J raised power 2 divided two})
gives
\begin{equation}
\begin{split}
\Biggl[
\lambda^{\mathcal{A}^{\,}_{1}}_{\texttt{m}^{\,}_{\,}}\,
\lambda^{\mathcal{A}^{\,}_{2}}_{\texttt{m}^{\,}_{\,}}\,
\frac{\mathrm{i}^{2}\,f^{\mathcal{A}^{\,}_{1}\mathcal{A}^{\,}_{2}\mathcal{A}^{\,}_{3}}\,
f^{\mathcal{A}^{\,}_{1}\mathcal{A}^{\,}_{2}\mathcal{A}^{\,}_{4}}}{|z-w|^{2}}\,
\mathcal{J}^{\mathcal{A}^{\,}_{3}}_{\mathrm{L},\texttt{m}}(\bar{z})\,
\mathcal{J}^{\mathcal{A}^{\,}_{4}}_{\mathrm{R},\texttt{m}+1}(z)
+
\lambda^{\mathcal{A}^{\,}_{1}}_{\texttt{m}^{\,}_{\,}}\,
\lambda^{\mathcal{A}^{\,}_{2}}_{\texttt{m}^{\,}_{\,}}\,
\frac{\mathrm{i}^{2}\,f^{\mathcal{A}^{\,}_{1}\mathcal{A}^{\,}_{2}\mathcal{A}^{\,}_{3}}\,
f^{\mathcal{A}^{\,}_{1}\mathcal{A}^{\,}_{2}\mathcal{A}^{\,}_{4}}}{|z-w|^{2}}\,
\tilde{\mathcal{J}}^{\mathcal{A}^{\,}_{3}}_{\mathrm{R},\texttt{m}}(z)\,
\tilde{\mathcal{J}}^{\mathcal{A}^{\,}_{4}}_{\mathrm{L},\texttt{m}+1}(\bar{z})
+
\cdots
\Biggr].
\end{split}
\label{appeq: Sint TRS case 2 J raised power 2 divided two after OPE}
\end{equation}
We note that the contribution from the OPEs from the third term 
inside the bracket on the right-hand side of Eq.\ 
(\ref{appeq: Sint TRS case 2 J raised power 2 divided two}) 
vanishes because there are no OPEs between 
$\mathcal{J}$ and $\tilde{\mathcal{J}}$,
for $\mathcal{J}$ and $\tilde{\mathcal{J}}$ 
belong to the pair of commuting algebras
$su(2)\oplus su(2)$ and $\widetilde{su}(2)\oplus\widetilde{su}(2)$, 
respectively. 
Integrating the second-order poles at $w$ and $\bar{w}$
over a ring with the inner radius $\mathfrak{a}$ and the outer radius 
$(1+\mathrm{d}\ell)\,\mathfrak{a}$ gives
\begin{equation}
\begin{split}
\delta S=&\,
-
\frac{2\pi\,\mathrm{d}\ell}{2}\,
\int\frac{\mathrm{d}\bar{z}\mathrm{d}z}{2\pi\mathrm{i}}\,
\lambda^{\mathcal{A}^{\,}_{1}}_{\texttt{m}^{\,}_{\,}}\,
\lambda^{\mathcal{A}^{\,}_{2}}_{\texttt{m}^{\,}_{\,}}\,
f^{\mathcal{A}^{\,}_{1}\mathcal{A}^{\,}_{2}\mathcal{A}^{\,}_{3}}\,
f^{\mathcal{A}^{\,}_{1}\mathcal{A}^{\,}_{2}\mathcal{A}^{\,}_{4}}\,
\left(
\mathcal{J}^{\mathcal{A}^{\,}_{3}}_{\mathrm{L},\texttt{m}}\,
\mathcal{J}^{\mathcal{A}^{\,}_{4}}_{\mathrm{R},\texttt{m}+1}\,
+
\tilde{\mathcal{J}}^{\mathcal{A}^{\,}_{3}}_{\mathrm{R},\texttt{m}}\,
\tilde{\mathcal{J}}^{\mathcal{A}^{\,}_{4}}_{\mathrm{L},\texttt{m}+1}
\right).
\end{split}
\label{appeq: Sint TRS case 2 J two coupling second order}
\end{equation}
For any given $\mathcal{A}\,=1,\cdots,6$ 
and any given $\texttt{m}=1,\cdots,n-1$,
the one-loop RG equations
\begin{align}
\frac{
\mathrm{d}\lambda^{\mathcal{A}\,}_{\texttt{m}}
     }
     {
\mathrm{d}\mathrm{\ell}
      }=
\pi\,
f^{\mathcal{A}\,\mathcal{A}'\,\mathcal{A}''}\,
f^{\mathcal{A}\,\mathcal{A}'\,\mathcal{A}''}\,
\lambda^{\mathcal{A}'\,}_{\texttt{m}^{\,}_{\,}}\,
\lambda^{\mathcal{A}''}_{\texttt{m}^{\,}_{\,}}
\label{appeq: RGSUN TRS case 2 J}
\end{align}
follow for the current-current interactions
with the generators from the semisimple 
$\Bigl(su(2)\oplus su(2)\Bigr)
\oplus
\Bigl(\widetilde{su}(2)\oplus \widetilde{su}(2)\Bigr)$ algebra.

Second, we turn our attention to
the $\mathcal{K}$-dependent contribution
\begin{equation}
\begin{split}
\mathcal{L}^{\,}_{\mathrm{int},\mathcal{K}}:=&
-
\sum_{\texttt{m}=1}^{n}
\sum_{\mathcal{B}=1}^{3}
\upsilon^{\mathcal{B}}_{\texttt{m}}\,
\left(
\mathcal{K}^{\mathcal{B}}_{\mathrm{L},\texttt{m}}\,
\mathcal{K}^{\mathcal{B}}_{\mathrm{R},\texttt{m}+1}\,
+
\tilde{\mathcal{K}}^{\mathcal{B}}_{\mathrm{R},\texttt{m}}\,
\tilde{\mathcal{K}}^{\mathcal{B}}_{\mathrm{L},\texttt{m}+1}
\right)
\end{split}
\label{appeq: Lint case 2 K}
\end{equation}
to the Lagrangian density (\ref{appeq: Lint case 2}).
Expansion of the Boltzmann weight with the action
corresponding to the Lagrangian density 
(\ref{appeq: Lint case 2 K})
gives the second-order contribution 
\begin{equation}
\begin{split}
\frac{S^{2}_{\mathrm{int}}}{2}=&\,
\frac{1}{2}
\int\frac{\mathrm{d}\bar{z}\mathrm{d}z}{2\pi\mathrm{i}}
\int\frac{\mathrm{d}\bar{w}\mathrm{d}w}{2\pi\mathrm{i}}\,
\Biggl[
\upsilon^{\mathcal{B}^{\,}_{1}}_{\texttt{m}^{\,}_{1}}\,
\upsilon^{\mathcal{B}^{\,}_{2}}_{\texttt{m}^{\,}_{2}}\,
\mathcal{K}^{\mathcal{B}^{\,}_{1}}_{\mathrm{L},\texttt{m}^{\,}_{1}}(\bar{w})\,
\mathcal{K}^{\mathcal{B}^{\,}_{1}}_{\mathrm{R},\texttt{m}^{\,}_{1}}(w)\,
\mathcal{K}^{\mathcal{B}^{\,}_{2}}_{\mathrm{L},\texttt{m}^{\,}_{2}}(\bar{z})\,
\mathcal{K}^{\mathcal{B}^{\,}_{2}}_{\mathrm{R},\texttt{m}^{\,}_{2}}(z)
\\
&\,
+
\upsilon^{\mathcal{B}^{\,}_{1}}_{\texttt{m}^{\,}_{1}}\,
\upsilon^{\mathcal{B}^{\,}_{2}}_{\texttt{m}^{\,}_{2}}\,
\tilde{\mathcal{K}}^{\mathcal{B}^{\,}_{1}}_{\mathrm{R},\texttt{m}^{\,}_{1}}(w)\,
\tilde{\mathcal{K}}^{\mathcal{B}^{\,}_{1}}_{\mathrm{L},\texttt{m}^{\,}_{1}}(\bar{w})\,
\tilde{\mathcal{K}}^{\mathcal{B}^{\,}_{2}}_{\mathrm{R},\texttt{m}^{\,}_{2}}(z)\,
\tilde{\mathcal{K}}^{\mathcal{B}^{\,}_{2}}_{\mathrm{L},\texttt{m}^{\,}_{2}}(\bar{z})
\\
&\,
+
2\,\upsilon^{\mathcal{B}^{\,}_{1}}_{\texttt{m}^{\,}_{1}}\,
\upsilon^{\mathcal{B}^{\,}_{2}}_{\texttt{m}^{\,}_{2}}\,
\mathcal{K}^{\mathcal{B}^{\,}_{1}}_{\mathrm{L},\texttt{m}^{\,}_{1}}(\bar{w})\,
\mathcal{K}^{\mathcal{B}^{\,}_{1}}_{\mathrm{R},\texttt{m}^{\,}_{1}}(w)\,
\tilde{\mathcal{K}}^{\mathcal{B}^{\,}_{2}}_{\mathrm{R},\texttt{m}^{\,}_{2}}(z)\,
\tilde{\mathcal{K}}^{\mathcal{B}^{\,}_{2}}_{\mathrm{L},\texttt{m}^{\,}_{2}}(\bar{z})
\Biggr].
\end{split}
\label{appeq: Sint TRS case 2 K raised power 2 divided two}
\end{equation}

Insertion of the OPEs
(\ref{appeq: OPE for su(2) oplus su(2) no TRS})
into the square bracket on the right-hand side of
Eq.\ (\ref{appeq: Sint TRS case 2 K raised power 2 divided two})
gives
\begin{equation}
\begin{split}
\Biggl[
\upsilon^{\mathcal{B}^{\,}_{1}}_{\texttt{m}^{\,}_{\,}}\,
\upsilon^{\mathcal{B}^{\,}_{2}}_{\texttt{m}^{\,}_{\,}}\,
\frac{\mathrm{i}^{2}\,f^{\mathcal{B}^{\,}_{1}\mathcal{B}^{\,}_{2}\mathcal{B}^{\,}_{3}}\,
f^{\mathcal{B}^{\,}_{1}\mathcal{B}^{\,}_{2}\mathcal{B}^{\,}_{4}}}{|z-w|^{2}}\,
\mathcal{K}^{\mathcal{B}^{\,}_{3}}_{\mathrm{L},\texttt{m}}(\bar{z})\,
\mathcal{K}^{\mathcal{B}^{\,}_{4}}_{\mathrm{R},\texttt{m}}(z)
+
\upsilon^{\mathcal{B}^{\,}_{1}}_{\texttt{m}^{\,}_{\,}}\,
\upsilon^{\mathcal{B}^{\,}_{2}}_{\texttt{m}^{\,}_{\,}}\,
\frac{\mathrm{i}^{2}\,f^{\mathcal{B}^{\,}_{1}\mathcal{B}^{\,}_{2}\mathcal{B}^{\,}_{3}}\,
f^{\mathcal{B}^{\,}_{1}\mathcal{B}^{\,}_{2}\mathcal{B}^{\,}_{4}}}{|z-w|^{2}}\,
\tilde{\mathcal{K}}^{\mathcal{B}^{\,}_{3}}_{\mathrm{R},\texttt{m}}(z)\,
\tilde{\mathcal{K}}^{\mathcal{B}^{\,}_{4}}_{\mathrm{L},\texttt{m}}(\bar{z})
+
\cdots
\Biggr].
\end{split}
\label{appeq: Sint TRS case 2 K raised power 2 divided two after OPE}
\end{equation}
We note that the contribution from the OPEs from the third term 
inside the bracket on the right-hand side of Eq.\ 
(\ref{appeq: Sint TRS case 2 K raised power 2 divided two}) 
vanishes because there are no OPEs between 
$\mathcal{K}$ and $\tilde{\mathcal{K}}$,
for $\mathcal{K}$ and $\tilde{\mathcal{K}}$ 
belong to the pair of commuting subalgebras
$su(2)$ and $\widetilde{su}(2)$, 
respectively. 
Integrating the second-order poles at $w$ and $\bar{w}$
over a ring with the inner radius $\mathfrak{a}$ and the outer radius 
$(1+\mathrm{d}\ell)\,\mathfrak{a}$ gives
\begin{equation}
\begin{split}
\delta S=&\,
-
\frac{2\pi\,\mathrm{d}\ell}{2}\,
\int\frac{\mathrm{d}\bar{z}\mathrm{d}z}{2\pi\mathrm{i}}\,
\upsilon^{\mathcal{B}^{\,}_{1}}_{\texttt{m}^{\,}_{\,}}\,
\upsilon^{\mathcal{B}^{\,}_{2}}_{\texttt{m}^{\,}_{\,}}\,
f^{\mathcal{B}^{\,}_{1}\mathcal{B}^{\,}_{2}\mathcal{B}^{\,}_{3}}\,
f^{\mathcal{B}^{\,}_{1}\mathcal{B}^{\,}_{2}\mathcal{B}^{\,}_{4}}\,
\left(
\mathcal{K}^{\mathcal{B}^{\,}_{3}}_{\mathrm{L},\texttt{m}}\,
\mathcal{K}^{\mathcal{B}^{\,}_{4}}_{\mathrm{R},\texttt{m}}\,
+
\tilde{\mathcal{K}}^{\mathcal{B}^{\,}_{3}}_{\mathrm{R},\texttt{m}}\,
\tilde{\mathcal{K}}^{\mathcal{B}^{\,}_{4}}_{\mathrm{L},\texttt{m}}
\right).
\end{split}
\label{appeq: Sint TRS case 2 K two coupling second order}
\end{equation}
\end{widetext}
For any given $\mathcal{B}=1,\cdots,3$ 
and any given $\texttt{m}=1,\cdots,n$,
the one-loop RG equations
\begin{equation}
\begin{split}
\frac{\mathrm{d}\upsilon^{\mathcal{B}}_{\texttt{m}}}{\mathrm{d}\mathrm{\ell}}&=
\pi\,
f^{\mathcal{B}\,\mathcal{B}'\,\mathcal{B}''}\,
f^{\mathcal{B}\,\mathcal{B}'\,\mathcal{B}''}\,
\upsilon^{\mathcal{B}'}_{\texttt{m}}\,
\upsilon^{\mathcal{B}''}_{\texttt{m}}
\end{split}
\label{appeq: RGSUN TRS case 2 K}
\end{equation} 
follow for the current-current interactions
with the generators from the diagonal 
$su(2)\oplus\widetilde{su}(2)$ subalgebra.

For the ​sake of completeness, we should also compute the one-loop RG equations
for the couplings  
$\lambda^{\mathcal{A}}_{\mathrm{boundary},1}$
and
$\upsilon^{\mathcal{B}}_{\mathrm{boundary},1}$
in the case of a single domino, i.e., for the interaction 
[recall Eq.\ (\ref{eq: final interaction one domino if TRS})]   
\begin{widetext}
\begin{equation}
\label{appeq: final interaction one domino if TRS}
\begin{split}
\mathcal{L}^{\,}_{\mathrm{int}}:=&\,
-
\sum_{\mathcal{B}=1}^{3}
\upsilon^{\mathcal{B}}_{\mathrm{boundary},1}\,
\left(
\mathcal{K}^{\mathcal{B}}_{\mathrm{L},1}\,
\mathcal{K}^{\mathcal{B}}_{\mathrm{R},1}\,
+
\tilde{\mathcal{K}}^{\mathcal{B}}_{\mathrm{R},1}\,
\tilde{\mathcal{K}}^{\mathcal{B}}_{\mathrm{L},1}
\right)\,
-
\sum_{\mathcal{A}=1}^{6}
\lambda^{\mathcal{A}}_{\mathrm{boundary},1}\,
\mathcal{J}^{\mathcal{A}}_{\mathrm{L},1}\,
\widetilde{\mathcal{J}}^{\mathcal{A}}_{\mathrm{R},1}.
\end{split}
\end{equation}
As was the case in Appendix\ 
\ref{appsubsec: Derivation of the one-loop RG flows if no TRS},
the one-loop RG equations obeyed by the couplings
$\upsilon^{\mathcal{B}}_{\mathrm{boundary},1}$ 
with $\mathcal{B}=1,2,3$ 
decouple from the one-loop RG equations obeyed by the couplings
$\lambda^{\mathcal{A}}_{\mathrm{boundary},1}$ 
with $\mathcal{A}=1,\cdots,6$.
The one-loop RG equations obeyed by the couplings
$\upsilon^{\mathcal{B}}_{\mathrm{boundary},1}$ that enter the current-current interaction
(\ref{appeq: final interaction one domino if TRS}) 
are given by [recall Eq.\ (\ref{appeq: RGSUN TRS case 2 K})] 
\begin{equation}
\begin{split}
\frac{\mathrm{d}\upsilon^{\mathcal{B}}_{\mathrm{boundary},1}}{\mathrm{d}\mathrm{\ell}}&=
\pi\,
f^{\mathcal{B}\,\mathcal{B}'\,\mathcal{B}''}\,
f^{\mathcal{B}\,\mathcal{B}'\,\mathcal{B}''}\,
\upsilon^{\mathcal{B}}_{\mathrm{boundary},1}\,
\upsilon^{\mathcal{B}}_{\mathrm{boundary},1}\,.
\end{split}
\label{appeq: RGSUN TRS case one domino K}
\end{equation} 

Now, to calculate the one-loop RG equations for the coupling constants
in the $\mathcal{J}$ sector of the current-current interaction
(\ref{appeq: final interaction one domino if TRS}), we only need to treat
the $\mathcal{J}$-dependent contribution
\begin{equation}
\begin{split}
\mathcal{L}^{\,}_{\mathrm{int},\mathcal{J}}:=&
-
\sum_{\mathcal{A}=1}^{6}
\lambda^{\mathcal{A}}_{\mathrm{boundary},1}\,
\mathcal{J}^{\mathcal{A}}_{\mathrm{L},1}\,
\widetilde{\mathcal{J}}^{\mathcal{A}}_{\mathrm{R},1}
\end{split}
\label{Sint TRS one domino J}
\end{equation}
to the Lagrangian density (\ref{appeq: final interaction one domino if TRS}).
Expansion of the Boltzmann weight with the action
corresponding to the Lagrangian density 
(\ref{Sint TRS one domino J})
gives the second-order contribution 
\begin{equation}
\begin{split}
\frac{S^{2}_{\mathrm{int}}}{2}=&\,
\frac{1}{2}
\int\frac{\mathrm{d}\bar{z}\mathrm{d}z}{2\pi\mathrm{i}}
\int\frac{\mathrm{d}\bar{w}\mathrm{d}w}{2\pi\mathrm{i}}\,
\lambda^{\mathcal{A}^{}_{1}}_{\mathrm{boundary},1}\,
\lambda^{\mathcal{A}^{}_{2}}_{\mathrm{boundary},1}\,
\mathcal{J}^{\mathcal{A}^{}_{1}}_{\mathrm{L},1}(\bar{w})\,
\widetilde{\mathcal{J}}^{\mathcal{A}^{}_{1}}_{\mathrm{R},1}({w})\,
\mathcal{J}^{\mathcal{A}^{}_{2}}_{\mathrm{L},1}(\bar{z})\,
\widetilde{\mathcal{J}}^{\mathcal{A}^{}_{2}}_{\mathrm{R},1}(z)\,.
\end{split}
\label{appeq: Sint TRS one domino J raised power 2 divided two}
\end{equation}
Insertion of the OPEs
(\ref{appeq: OPE for su(2) oplus su(2) no TRS})
into the square bracket on the right-hand side of
Eq.\ (\ref{appeq: Sint TRS one domino J raised power 2 divided two})
gives
\begin{equation}
\begin{split}
\Biggl[
\lambda^{\mathcal{A}^{}_{1}}_{\mathrm{boundary},1}\,
\lambda^{\mathcal{A}^{}_{2}}_{\mathrm{boundary},1}\,
\frac{\mathrm{i}^{2}\,f^{\mathcal{A}^{\,}_{1}\mathcal{A}^{\,}_{2}\mathcal{A}^{\,}_{3}}\,
f^{\mathcal{A}^{}_{1}\mathcal{A}^{}_{2}\mathcal{A}^{}_{4}}}{|z-w|^{2}}\,
\mathcal{J}^{\mathcal{A}^{\,}_{3}}_{\mathrm{L},1}(\bar{z})\,
\widetilde{\mathcal{J}}^{\mathcal{A}^{}_{4}}_{\mathrm{R},1}(z)
\cdots
\Biggr].
\end{split}
\label{appeq: Sint TRS one domino J raised power 2 divided two after OPE}
\end{equation}
Integrating the second-order poles at $w$ and $\bar{w}$
over a ring with the inner radius $\mathfrak{a}$ and the outer radius 
$(1+\mathrm{d}\ell)\,\mathfrak{a}$ gives
\begin{equation}
\begin{split}
\delta S=&\,
-
\frac{2\pi\,\mathrm{d}\ell}{2}\,
\int\frac{\mathrm{d}\bar{z}\mathrm{d}z}{2\pi\mathrm{i}}\,
\Biggl[
\lambda^{\mathcal{A}^{}_{1}}_{\mathrm{boundary},1}\,
\lambda^{\mathcal{A}^{}_{2}}_{\mathrm{boundary},1}\,
f^{\mathcal{A}^{\,}_{1}\mathcal{A}^{\,}_{2}\mathcal{A}^{\,}_{3}}\,
f^{\mathcal{A}^{}_{1}\mathcal{A}^{}_{2}\mathcal{A}^{}_{4}}\,
\mathcal{J}^{\mathcal{A}^{\,}_{3}}_{\mathrm{L},1}\,
\widetilde{\mathcal{J}}^{\mathcal{A}^{}_{4}}_{\mathrm{R},1}
\Biggr].
\end{split}
\label{appeq: Sint TRS one domino second order}
\end{equation}
For any given $\mathcal{A}\,=1,\cdots,6$ 
the one-loop RG equations
\begin{align}
\frac{
\mathrm{d}\lambda^{\mathcal{A}\,}_{\mathrm{boundary},1}
     }
     {
\mathrm{d}\mathrm{\ell}
      }=
\pi\,
f^{\mathcal{A}\,\mathcal{A}'\,\mathcal{A}''}\,
f^{\mathcal{A}\,\mathcal{A}'\,\mathcal{A}''}\,
\lambda^{\mathcal{A}'\,}_{\mathrm{boundary},1}\,
\lambda^{\mathcal{A}''}_{\mathrm{boundary},1}
\label{appeq: RGSUN TRS one domino J}
\end{align}
follow for the current-current interactions
with the generators from the semisimple 
$\Bigl(su(2)\oplus su(2)\Bigr)
\oplus
\Bigl(\widetilde{su}(2)\oplus \widetilde{su}(2)\Bigr)$ algebra.
\end{widetext}

\section{Stability analysis of the coset WZW theory 
with the central charge (\ref{eq: choice two for interaction II})}
\label{appsec: The stability analysis of the coset WZW theory}

We consider the first bundle
in Fig.\ \ref{Fig: gapping with TRS}.
With open boundary conditions, 
this bundle supports the strongly interacting
critical theory with the central charge
(\ref{eq: nonchiral central charge at one boundary case II}).
The same strongly interacting critical theory is supported by the bundle
from Fig.\ \ref{Fig: gapping with TRS of single dominos}(a).
For simplicity but without loss of generality, we shall 
ask under what conditions is the strongly interacting critical theory
supported by the bundle
from Fig.\ \ref{Fig: gapping with TRS of single dominos}(a)
stable to one-body interactions.
A detailed answer is given in Sec.\
\ref{appsubsec: Stability to mass terms in the fermion representation}
for one-body mass terms in the fermion representation.
We repeat this exercise in Sec.\
\ref{appsubsec: Stability to mass terms in the bosonized representation}
in the bosonized representation.

We always assume that the (charge) $U(1)$ and $SU(k+k')$ sectors 
in the Fock space corresponding to Fig.\ 
\ref{Fig: gapping with TRS of single dominos}(a)
are gapped in such a way that the
strongly interacting critical theory with the central charge
(\ref{eq: nonchiral central charge at one boundary case II})
is in the (charge) $U(1)$ and $SU(k+k')$ singlet sectors of the Fock space.

\subsection{Stability to mass terms in the fermion representation}
\label{appsubsec: Stability to mass terms in the fermion representation}

Prior to introducing the current-current interactions,
the bundle from Fig.\ \ref{Fig: gapping with TRS of single dominos}(a)
is fully described by the single-particle Hamiltonian
\begin{subequations}
\label{appeq: def single particle mathcal H0}
\begin{equation}
\mathcal{H}^{\,}_{0}:=
-\mathrm{i}\partial^{\,}_{x}\,
X^{\,}_{3000},
\label{appeq: def single particle mathcal H0 a}
\end{equation}
where we are using the notation
\begin{equation}
X^{\,}_{\mu^{\,}_{1}\mu^{\,}_{2}\mu^{\,}_{3}\mathrm{c}}:=
\tau^{\,}_{\mu^{\,}_{1}}
\otimes
\sigma^{\,}_{\mu^{\,}_{2}}
\otimes
\rho^{\,}_{\mu^{\,}_{3}}
\otimes
T^{\,}_{\mathrm{c}}
\label{appeq: def single particle mathcal H0 b}
\end{equation}
\end{subequations}
with $\mu^{\,}_{1},\mu^{\,}_{2},\mu^{\,}_{3}=0,1,2,3$ and
$\mathrm{c}=0,1,\cdots,(k+k')^{2}-1$. 
All matrices with the label $0$
are unit matrices of dimensions two for 
$\mu^{\,}_{1},\mu^{\,}_{2},\mu^{\,}_{3}=0$
and $k+k'$ for $\mathrm{c}=0$.
The triplet of matrices $\bm{\tau}$ are the Pauli matrices
acting on the left- and right-moving indices.
The triplet of matrices $\bm{\sigma}$ are the Pauli matrices
acting on the down and up projections on the quantization axis
of the electronic spin-1/2.
The triplet of matrices $\bm{\rho}$ are the Pauli matrices
acting acting on the doublets defined by Eq.\
(\ref{eq: def  Grassmann-valued doublets Psi}).
The matrices $T^{\,}_{\mathrm{c}}$ with $\mathrm{c}=0,1,\cdots,(k+k')^{2}-1$
generate the unitary group $U(k+k')$. They are chosen to be in the
fundamental representation of $U(k+k')$.

The single-particle Hamiltonian 
(\ref{appeq: def single particle mathcal H0})
obeys two symmetries.
It is invariant under
\begin{subequations}
\begin{equation}
X^{\,}_{1210}\,\mathcal{H}^{*}_{0}\,X^{\,}_{1210}=\mathcal{H}^{\,}_{0}.
\end{equation}
We interpret
\begin{equation}
\mathcal{T}:=
X^{\,}_{1210}\,
\mathsf{K},
\end{equation}
\end{subequations}
where $\mathsf{K}$ denotes complex conjugation, 
as representing reversal of time.
It is also invariant under
\begin{subequations}
\begin{equation}
X^{\,}_{0030}\,\mathcal{H}^{\,}_{0}\,X^{\,}_{0030}=\mathcal{H}^{\,}_{0}.
\end{equation}
We interpret
\begin{equation}
\Upsilon^{\,}_{3}:=
X^{\,}_{0030}
\label{appeq: generator U(1) symmetry}
\end{equation}
\end{subequations}
as the diagonal generator of the $SU(2)$ group 
that mixes the $SU(2)$ and $\widetilde{SU}(2)$ sectors
entering the conformal embedding 
(\ref{eq: master identity for u(4k) level II}).
In other words, the $U(1)$ group with the elements
\begin{equation}
\exp(\mathrm{i}\theta\,\Upsilon^{\,}_{3})
\end{equation} 
for $0\leq\theta<2\pi$
is the counterpart to the $U(1)$ transformation 
(\ref{eq: def U(1) symmetry between no tilde and tilde sectors}).

By assumption, the critical theory with the central charge
(\ref{eq: nonchiral central charge at one boundary case II})
is in the singlet sectors of (charge) $U(1)$ and
$SU(k+k')$. Hence, we seek all the matrices 
\begin{equation}
M^{\,}_{\mu^{\,}_{1}\mu^{\,}_{2}\mu^{\,}_{3}\mathrm{c}}:=
\tau^{\,}_{\mu^{\,}_{1}}
\otimes
\sigma^{\,}_{\mu^{\,}_{2}}
\otimes
\rho^{\,}_{\mu^{\,}_{3}}
\otimes
T^{\,}_{\mathrm{c}}
\end{equation}
with $\mu^{\,}_{1},\mu^{\,}_{2},\mu^{\,}_{3}=0,1,2,3$ and
$\mathrm{c}=0,1,\cdots,(k+k')^{2}-1$ 
that obey the mass condition
\begin{equation}
M^{\,}_{\mu^{\,}_{1}\mu^{\,}_{2}\mu^{\,}_{3}\mathrm{c}}\,
\mathcal{H}^{\,}_{0}\,
M^{\,}_{\mu^{\,}_{1}\mu^{\,}_{2}\mu^{\,}_{3}\mathrm{c}}=
-
\mathcal{H}^{\,}_{0},
\end{equation}
the time-reversal symmetry condition
\begin{equation}
\mathcal{T}\,
M^{\,}_{\mu^{\,}_{1}\mu^{\,}_{2}\mu^{\,}_{3}\mathrm{c}}\,
\mathcal{T}^{-1}=
M^{\,}_{\mu^{\,}_{1}\mu^{\,}_{2}\mu^{\,}_{3}\mathrm{c}}
\end{equation}
for any $\mu^{\,}_{1},\mu^{\,}_{2},\mu^{\,}_{3}=0,1,2,3$ and
$\mathrm{c}=0,1,\cdots,(k+k')^{2}-1$,
and the $SU(k+k')$-singlet condition
\begin{equation}
\mathrm{c}=0.
\end{equation}
One finds the twelve mass matrices 
\begin{widetext}
\begin{equation}
\begin{split}
&
M^{\,}_{1000},
\qquad
M^{\,}_{1010},
\qquad
M^{\,}_{1020},
\qquad
M^{\,}_{1130},
\qquad
M^{\,}_{1230},
\qquad
M^{\,}_{1330},
\\
&
M^{\,}_{2000},
\qquad
M^{\,}_{2010},
\qquad
M^{\,}_{2020},
\qquad
M^{\,}_{2130},
\qquad
M^{\,}_{2230},
\qquad
M^{\,}_{2330}.
\end{split}
\label{appeq: TRS SU(k+k') singlet mass matrices}
\end{equation}
\end{widetext}
Of these, only four are off-diagonal with respect to 
the group $SU(2)$ that mixes the $SU(2)$ and $\widetilde{SU}(2)$ sectors
entering the conformal embedding 
(\ref{eq: master identity for u(4k) level II}). 

Now, if we demand that the mass matrices
(\ref{appeq: TRS SU(k+k') singlet mass matrices})
commute with the diagonal generator 
(\ref{appeq: generator U(1) symmetry}),
we are left with the eight time-reversal-,
$SU(k+k')$-, and $U(1)$-symmetric mass matrices
\begin{equation}
\begin{split}
&
M^{\,}_{1000},
\qquad
M^{\,}_{1130},
\qquad
M^{\,}_{1230},
\qquad
M^{\,}_{1330},
\\
&
M^{\,}_{2000},
\qquad
M^{\,}_{2130},
\qquad
M^{\,}_{2230},
\qquad
M^{\,}_{2330}.
\end{split}
\label{appeq: TRS SU(k+k') singlet U(1) singlet mass matrices}
\end{equation}
These eight masses are all diagonal with respect to the group 
$SU(2)$ that mixes the $SU(2)$ and $\widetilde{SU}(2)$ sectors
entering the conformal embedding 
(\ref{eq: master identity for u(4k) level II}).
It follows that their action on the strongly interacting
critical coset theory with the central charge
(\ref{eq: nonchiral central charge at one boundary case II})
is reducible. Any one of
these eight masses can only mix states 
from the strongly interacting critical chiral coset theory 
with the central charge
(\ref{eq: choice two for interaction c}).
However, such mixing is impossible since
any one of these eigth masses is off-diagonal with respect 
to the right- and left-moving degrees of freedom.
Hence, none of these eight masses can gap 
the strongly interacting critical theory with the central charge
(\ref{eq: nonchiral central charge at one boundary case II}).

\subsection{Stability to mass terms in the bosonized representation}
\label{appsubsec: Stability to mass terms in the bosonized representation}

It is instructive to move from the first-quantized representation
(\ref{appeq: def single particle mathcal H0})
to the second-quantized representation implied by the path integral
(\ref{eq: def 2N wires with ``one'' electron per wire}).
From the Lagrangian density 
(\ref{eq: def 2N wires with ``one'' electron per wire a}),
we deduce the Hamiltonian density represented by
\begin{subequations}
\label{appeq: def 2N wires with ``many'' electron per wire}
\begin{equation}
\begin{split}
\hat{H}^{\,}_{0}:=&
-\mathrm{i}
\sum_{\alpha=1}^{2}
\sum_{A=1}^{k+k'}
\Big(
\hat{\psi}^{\dag}_{\mathrm{R},\alpha,A}\,
\partial^{\,}_{x}\,
\hat{\psi}^{\,}_{\mathrm{R},\alpha,A}
-
\hat{\psi}^{\dag}_{\mathrm{L},\alpha,A}\,
\partial^{\,}_{x}\,
\hat{\psi}^{\,}_{\mathrm{L},\alpha,A}
\Big)
\\
&
-\mathrm{i}
\sum_{\alpha=1}^{2}
\sum_{A=1}^{k+k'}
\Big(
\hat{\tilde{\psi}}^{\dag}_{\mathrm{R},\alpha,A}\,
\partial^{\,}_{x}\,
\hat{\tilde{\psi}}^{\,}_{\mathrm{R},\alpha,A}
-
\hat{\tilde{\psi}}^{\dag}_{\mathrm{L},\alpha,A}\,
\partial^{\,}_{x}\,
\hat{\tilde{\psi}}^{\,}_{\mathrm{L},\alpha,A}
\Big),
\end{split}
\label{appeq: def 2N wires with ``many'' electron per wire a}
\end{equation}
in the operator formalism.
The operator-valued Dirac spinors obey the equal-time anticommutators
\begin{equation}
\begin{split}
&
\left\{
\hat{\psi}^{\,}_{\eta,\alpha,A}(x) 
, 
\hat{\psi}^{\dag}_{\eta',\alpha',A'}(x')
\right\} = 
\delta^{\,}_{\eta,\eta'}\,
\delta^{\,}_{\alpha,\alpha'}\,
\delta^{\,}_{A,A'}\delta(x-x'),
\\
&
\left\{
\hat{\tilde{\psi}}^{\,}_{\eta,\alpha,A}(x) 
, 
\hat{\tilde{\psi}}^{\dag}_{\eta',\alpha',A'}(x')
\right\}= 
\delta^{\,}_{\eta,\eta'}\,
\delta^{\,}_{\alpha,\alpha'}\,
\delta^{\,}_{A,A'}\,
\delta(x-x'),
\end{split}
\label{appeq: def 2N wires with ``many'' electron per wire b}
\end{equation}
\end{subequations}
with $\eta,\eta'=\mathrm{R},\mathrm{L}$,
$\alpha,\alpha'=\uparrow,\downarrow\equiv 1,2$,
and
$A,A'=1,\cdots,k+k'$.

The phenomenon of spin-charge separation is not manifest
in the fermionic representation. It becomes manifest by
the conformal embedding 
(\ref{eq: master identity for u(4k) level II}), 
according to which the decomposition
\begin{subequations}
\begin{equation}
\begin{split}
\hat{H}^{\,}_{0}=&\, 
\hat{H}^{\,}_{0}[\hat{u}(2k)^{\,}_{1}] 
\!+\! 
\hat{H}^{\,}_{0}[\hat{u}(2k')^{\,}_{1}] 
\!+\! 
\hat{H}^{\,}_{0}[\hat{\tilde{u}}(2k)^{\,}_{1}] 
\!+\! 
\hat{H}^{\,}_{0}[\hat{\tilde{u}}(2k')^{\,}_{1}] 
\\
=&\, 
\hat{H}^{\,}_{0}[\hat{u}(1)] 
+ 
\hat{H}^{\,}_{0}[\widehat{su}(2)^{\,}_{k}] 
+ 
\hat{H}^{\,}_{0}[\widehat{su}(k)^{\,}_{2}]
\\
&\,
+ 
\hat{H}^{\,}_{0}[\hat{u}(1)] 
+ 
\hat{H}^{\,}_{0}[\widehat{su}(2)^{\,}_{k'}] 
+ 
\hat{H}^{\,}_{0}[\widehat{su}(k')^{\,}_{2}]
\\
&\,
+ 
\hat{H}^{\,}_{0}[\hat{\tilde{u}}(1)] 
+ 
\hat{H}^{\,}_{0}[\widehat{\widetilde{su}}(2)^{\,}_{k}] 
+ 
\hat{H}^{\,}_{0}[\widehat{\widetilde{su}}(k)^{\,}_{2}]
\\
&\,
+ 
\hat{H}^{\,}_{0}[\hat{\tilde{u}}(1)] 
+ 
\hat{H}^{\,}_{0}[\widehat{\widetilde{su}}(2)^{\,}_{k'}] 
+ 
\hat{H}^{\,}_{0}[\widehat{\widetilde{su}}(k')^{\,}_{2}]
\end{split}
\end{equation}
holds. Here,
[recall Eq.~(\ref{eq: master identity for u(2k) level 1})]
\begin{align}
&
\hat{H}^{\,}_{0}[\hat{u}(1)]:=
\frac{\pi}{2k}\,
\left[
\hat{j}^{\,}_{\mathrm{R}}\,
\hat{j}^{\,}_{\mathrm{R}}\,
+
\hat{j}^{\,}_{\mathrm{L}}\,
\hat{j}^{\,}_{\mathrm{L}}\,
\right],
\\
&
\hat{H}^{\,}_{0}[\widehat{su}(2)^{\,}_{k}]:=
\frac{2\pi}{k+2}\,
\sum_{c=1}^{3}
\left[
\hat{j}^{c}_{\mathrm{R}}\,
\hat{j}^{c}_{\mathrm{R}}\,
+
\hat{j}^{c}_{\mathrm{L}}\,
\hat{j}^{c}_{\mathrm{L}}\,
\right],
\\
&
\hat{H}^{\,}_{0}[\widehat{su}(k)^{\,}_{2}]:=
\frac{2\pi}{2+k}\,
\sum_{\mathrm{c}=1}^{k^{2}-1}
\left[
\hat{\mathrm{J}}^{\mathrm{c}}_{\mathrm{R}}\,
\hat{\mathrm{J}}^{\mathrm{c}}_{\mathrm{R}}\,
+
\hat{\mathrm{J}}^{\mathrm{c}}_{\mathrm{L}}\,
\hat{\mathrm{J}}^{\mathrm{c}}_{\mathrm{L}}\,
\right],
\end{align}
for $\hat{H}^{\,}_{0}[\hat{u}(2k)^{\,}_{1}]$ and
\begin{align}
&
\hat{H}^{\,}_{0}[\hat{\tilde{u}}(1)]:=
\frac{\pi}{2k}\,
\left[
\hat{\tilde{j}}^{\,}_{\mathrm{R}}\,
\hat{\tilde{j}}^{\,}_{\mathrm{R}}\,
+
\hat{\tilde{j}}^{\,}_{\mathrm{L}}\,
\hat{\tilde{j}}^{\,}_{\mathrm{L}}\,
\right],
\\
&
\hat{H}^{\,}_{0}[\widehat{\widetilde{su}}(2)^{\,}_{k}]:=
\frac{2\pi}{k+2}\,
\sum_{c=1}^{3}
\left[
\hat{\tilde{J}}^{c}_{\mathrm{R}}\,
\hat{\tilde{J}}^{c}_{\mathrm{R}}\,
+
\hat{\tilde{J}}^{c}_{\mathrm{L}}\,
\hat{\tilde{J}}^{c}_{\mathrm{L}}\,
\right],
\\
&
\hat{H}^{\,}_{0}[\widehat{\widetilde{su}}(k)^{\,}_{2}]:=
\frac{2\pi}{2+k}\,
\sum_{\mathrm{c}=1}^{k^{2}-1}
\left[
\hat{\tilde{\mathrm{J}}}^{\mathrm{c}}_{\mathrm{R}}\,
\hat{\tilde{\mathrm{J}}}^{\mathrm{c}}_{\mathrm{R}}\,
+
\hat{\tilde{\mathrm{J}}}^{\mathrm{c}}_{\mathrm{L}}\,
\hat{\tilde{\mathrm{J}}}^{\mathrm{c}}_{\mathrm{L}}\,
\right],
\end{align}
\end{subequations}
for $\hat{H}^{\,}_{0}[\hat{\tilde{u}}(2k)^{\,}_{1}]$,
and similarly for 
$\hat{H}^{\,}_{0}[\hat{u}(2k')^{\,}_{1}]$ 
and 
$\hat{H}^{\,}_{0}[\hat{\tilde{u}}(2k')^{\,}_{1}]$.
The currents are defined in 
Eqs.\ (\ref{eq: def su(2) and su(k) left right currents}) 
and (\ref{eq: def tilde su(2) and tilde su(k) left right currents})
and similarly for the $k'$ wires.

We seek the Abelian-bosonized representation of 
the Hamiltonian density
(\ref{appeq: def 2N wires with ``many'' electron per wire}).
To this end, we use the following chiral Abelian bosonization rules.
For any $\alpha=1,2$ and $A=1,\cdots,k$,
\begin{subequations}
\label{appeq: chiral bosonization rule}
\begin{equation}
\begin{split}
&
\hat{\psi}^{\,}_{\mathrm{R},\alpha,A}(x)=
\frac{1}{\sqrt{2\pi\mathfrak{a}}}\,
e^{+\mathrm{i}\sqrt{4\pi}\,\hat{\phi}^{\,}_{\mathrm{R},\alpha,A}(x)},
\\
&
\hat{\psi}^{\,}_{\mathrm{L},\alpha,A}(x)=
\frac{1}{\sqrt{2\pi\mathfrak{a}}}\,
e^{-\mathrm{i}\sqrt{4\pi}\,\hat{\phi}^{\,}_{\mathrm{L},\alpha,A}(x)},
\\
&
\hat{\tilde{\psi}}^{\,}_{\mathrm{R},\alpha,A}(x)=
\frac{1}{\sqrt{2\pi\mathfrak{a}}}\,
e^{+\mathrm{i}\sqrt{4\pi}\,\hat{\tilde{\phi}}^{\,}_{\mathrm{R},\alpha,A}(x)},
\\
&
\hat{\tilde{\psi}}^{\,}_{\mathrm{L},\alpha,A}(x)=
\frac{1}{\sqrt{2\pi\mathfrak{a}}}\,
e^{-\mathrm{i}\sqrt{4\pi}\,\hat{\tilde{\phi}}^{\,}_{\mathrm{L},\alpha,A}(x)}.
\end{split}
\end{equation}
For any $\alpha=1,2$ and $A=k+1,\cdots,k+k'$,
\begin{equation}
\begin{split}
&
\hat{\psi}^{\,}_{\mathrm{R},\alpha,A}(x)= 
\frac{1}{\sqrt{2\pi\mathfrak{a}}}\,
e^{+\mathrm{i}\sqrt{4\pi}\,\hat{\varphi}^{\,}_{\mathrm{R},\alpha,A}(x)},
\\
&
\hat{\psi}^{\,}_{\mathrm{L},\alpha,A}(x)=
\frac{1}{\sqrt{2\pi\mathfrak{a}}}\,
e^{-\mathrm{i}\sqrt{4\pi}\,\hat{\varphi}^{\,}_{\mathrm{L},\alpha,A}(x)},
\\
&
\hat{\tilde{\psi}}^{\,}_{\mathrm{R},\alpha,A}(x)= 
\frac{1}{\sqrt{2\pi\mathfrak{a}}}\,
e^{+\mathrm{i}\sqrt{4\pi}\,\hat{\tilde{\varphi}}^{\,}_{\mathrm{R},\alpha,A}(x)},
\\
&
\hat{\tilde{\psi}}^{\,}_{\mathrm{L},\alpha,A}(x)= 
\frac{1}{\sqrt{2\pi\mathfrak{a}}}\,
e^{-\mathrm{i}\sqrt{4\pi}\,\hat{\tilde{\varphi}}^{\,}_{\mathrm{L},\alpha,A}(x)}.
\end{split}
\end{equation}
\end{subequations}
The length scale $\mathfrak{a}$ denotes the short-distance cutoff.  
By imposing the equal-time commuation relations
\begin{subequations}
\begin{equation}
\begin{split}
&
\left[
\hat{\phi}^{\,}_{\mathrm{R},\alpha,A}(x) 
, 
\hat{\phi}^{\,}_{\mathrm{R},\alpha',A'}(x')
\right]= 
+\frac{\mathrm{i}}{4}\,
\delta^{\,}_{\alpha,\alpha'}\,
\delta^{\,}_{A,A'}\,
\mathrm{sgn}(x-x'),
\\
&
\left[
\hat{\phi}^{\,}_{\mathrm{L},\alpha,A}(x) 
, 
\hat{\phi}^{\,}_{\mathrm{L},\alpha',A'}(x')
\right]= 
-\frac{\mathrm{i}}{4}\,
\delta^{\,}_{\alpha,\alpha'}\,
\delta^{\,}_{A,A'}\,
\mathrm{sgn}(x-x'),
\\
&
\left[
\hat{\phi}^{\,}_{\mathrm{R},\alpha,A}(x) 
, 
\hat{\phi}^{\,}_{\mathrm{L},\alpha',A'}(x')
\right]= 
+\frac{\mathrm{i}}{4}\,
\delta^{\,}_{\alpha,\alpha'}\,
\delta^{\,}_{A,A'},
\end{split}
\end{equation}
for any $\alpha,\alpha'=1,2$ and $A,A'=1,\cdots,k$,
and
\begin{equation}
\begin{split}
&
\left[
\hat{\varphi}^{\,}_{\mathrm{R},\alpha,A}(x) 
, 
\hat{\varphi}^{\,}_{\mathrm{R},\alpha',A'}(x')
\right]= 
+\frac{\mathrm{i}}{4}\,
\delta^{\,}_{\alpha,\alpha'}\,
\delta^{\,}_{A,A'}\,
\mathrm{sgn}(x-x'),
\\
&
\left[
\hat{\varphi}^{\,}_{\mathrm{L},\alpha,A}(x) 
, 
\hat{\varphi}^{\,}_{\mathrm{L},\alpha',A'}(x')
\right]= 
-\frac{\mathrm{i}}{4}\,
\delta^{\,}_{\alpha,\alpha'}\,
\delta^{\,}_{A,A'}\,
\mathrm{sgn}(x-x'),
\\
&
\left[
\hat{\varphi}^{\,}_{\mathrm{R},\alpha,A}(x) 
, 
\hat{\varphi}^{\,}_{\mathrm{L},\alpha',A'}(x')
\right]= 
+\frac{\mathrm{i}}{4}\,
\delta^{\,}_{\alpha,\alpha'}\,
\delta^{\,}_{A,A'},
\end{split}
\end{equation}
\end{subequations}
for any $\alpha,\alpha'=1,2$ and $A,A'=k+1,\cdots,k+k'$,
it follows that the equal-time fermionic algebra 
(\ref{appeq: def 2N wires with ``many'' electron per wire b})
is fulfilled.

We first focus on the sector
\begin{equation}
\hat{u}(2k)^{\,}_{1}=
\hat{u}(1) 
\oplus
\widehat{su}(2)^{\,}_{k} 
\oplus 
\widehat{su}(k)^{\,}_{2}.
\end{equation}  
For the special case of $k=1$, we have the conformal embedding
\begin{equation}
\hat{u}(2)^{\,}_{1}=
\hat{u}(1)\oplus\widehat{su}(2)^{\,}_{1}.
\end{equation}
This conformal embedding is nothing but the phenomenon 
of spin-charge separation in one-dimensional space.  
We specialize to the case of $k=1$.  

We can define the ``charge'' and ``spin''
fields 
[recall that $\alpha=1\equiv\,\uparrow,\alpha=2\equiv\,\downarrow$]
\begin{equation}
\begin{split}
&
\hat{\phi}^{\,}_{\mathrm{R},\mathrm{c}}:=
\frac{1}{\sqrt{2}}
\Big(
\hat{\phi}^{\,}_{\mathrm{R},\uparrow} 
+ 
\hat{\phi}^{\,}_{\mathrm{R},\downarrow}  
\Big),
\quad
\hat{\phi}^{\,}_{\mathrm{R},\mathrm{s}}:= 
\frac{1}{\sqrt{2}}
\Big(
\hat{\phi}^{\,}_{\mathrm{R},\uparrow} 
- 
\hat{\phi}^{\,}_{\mathrm{R},\downarrow}  
\Big),
\\
&
\hat{\phi}^{\,}_{\mathrm{L},\mathrm{c}}:=
\frac{1}{\sqrt{2}}
\Big(
\hat{\phi}^{\,}_{\mathrm{L},\uparrow} 
+ 
\hat{\phi}^{\,}_{\mathrm{L},\downarrow}  
\Big),
\quad
\hat{\phi}^{\,}_{\mathrm{L},\mathrm{s}}:= 
\frac{1}{\sqrt{2}}
\Big(
\hat{\phi}^{\,}_{\mathrm{L},\uparrow} 
- 
\hat{\phi}^{\,}_{\mathrm{L},\downarrow}  
\Big).
\end{split}
\label{appeq: def spin charge chiral boson}
\end{equation} 
The $\hat{u}(1)$ current 
(\ref{eq: master identity for u(2k) level 1 h})
obeys the Abelian-bosonized representation
\begin{subequations}
\label{appeq: u(1) bosonized current}
\begin{align}
&
\hat{j}^{\,}_{\mathrm{R}}= 
\frac{1}{\sqrt{\pi}}\,
\partial^{\,}_{x}
\left(
\hat{\phi}^{\,}_{\mathrm{R},\uparrow} 
+ 
\hat{\phi}^{\,}_{\mathrm{R},\downarrow} 
\right)= 
\sqrt{\frac{2}{\pi}}\,
\partial^{\,}_{x}\hat{\phi}^{\,}_{\mathrm{R},\mathrm{c}},
\\
&
\hat{j}^{\,}_{\mathrm{L}}= 
\frac{1}{\sqrt{\pi}}\,
\partial^{\,}_{x}
\left(
\hat{\phi}^{\,}_{\mathrm{L},\uparrow} 
+ 
\hat{\phi}^{\,}_{\mathrm{L},\downarrow} 
\right)= 
\sqrt{\frac{2}{\pi}}\,
\partial^{\,}_{x}\hat{\phi}^{\,}_{\mathrm{L},\mathrm{c}}.
\end{align}
The $\widehat{su}(2)^{\,}_{1}$ currents 
(\ref{eq: master identity for u(2k) level 1 i})
obey the Abelian-bosonized representation
\begin{equation}
\begin{split}
&
\hat{j}^{3}_{\mathrm{R}}=
\frac{1}{\sqrt{2\pi}}\,
\partial^{\,}_{x}\hat{\phi}^{\,}_{\mathrm{R},\mathrm{s}},
\qquad
\hat{j}^{\pm}_{\mathrm{R}}=
\frac{1}{2\pi\mathfrak{a}}\,
e^{\mp\mathrm{i}2\sqrt{2\pi}\hat{\phi}^{\,}_{\mathrm{R},\mathrm{s}}},
\\
&
\hat{j}^{3}_{\mathrm{L}}=
\frac{1}{\sqrt{2\pi}}\,
\partial^{\,}_{x}\hat{\phi}^{\,}_{\mathrm{L},\mathrm{s}},
\qquad
\hat{j}^{\pm}_{\mathrm{L}}=
\frac{1}{2\pi\mathfrak{a}}\,
e^{\pm\mathrm{i}2\sqrt{2\pi}\hat{\phi}^{\,}_{\mathrm{L},\mathrm{s}}},
\end{split}
\end{equation}
where
\begin{equation}
\hat{j}^{\pm}_{\mathrm{R}}:= 
\hat{j}^{1}_{\mathrm{R}} 
\pm
\mathrm{i} \hat{j}^{2}_{\mathrm{R}},
\qquad
\hat{j}^{\pm}_{\mathrm{L}}:=
\hat{j}^{1}_{\mathrm{L}} 
\pm
\mathrm{i} \hat{j}^{2}_{\mathrm{L}}.
\end{equation}
\end{subequations}
Similarly, we can do the replacements 
\begin{subequations}
\begin{align}
&
\hat{\phi}\to\hat{\tilde{\phi}},
\\
&
\hat{\phi}\to\hat{\varphi},
\\
&
\hat{\phi}\to\hat{\tilde{\varphi}}
\end{align}
\end{subequations}
in Eq.\ (\ref{appeq: u(1) bosonized current}) 
to derive the bosonized currents entering the conformal embeddings
\begin{subequations}
\begin{align}
&
\hat{\tilde{u}}(2k)^{\,}_{1}=
\hat{\tilde{u}}(1) 
\oplus
\widehat{\widetilde{su}}(2)^{\,}_{k} 
\oplus
\widehat{\widetilde{su}}(k)^{\,}_{2},
\\
&
\hat{u}(2k')^{\,}_{1}=
\hat{u}(1)
\oplus 
\widehat{su}(2)^{\,}_{k'} 
\oplus 
\widehat{su}(k')^{\,}_{2},
\\
&
\hat{\tilde{u}}(2k')^{\,}_{1}=
\hat{\tilde{u}}(1) 
\oplus
\widehat{\widetilde{su}}(2)^{\,}_{k'} 
\oplus
\widehat{\widetilde{su}}(k')^{\,}_{2},
\end{align}
\end{subequations}
respectively.  

The noninteracting Hamiltonian density
(\ref{appeq: def 2N wires with ``many'' electron per wire})
for $k=k'=1$ 
has the bosonized representation
\begin{equation}
\begin{split}
\hat{H}^{\,}_{0}=&\, 
(\partial^{\,}_{x}\hat{\phi}^{\,}_{\mathrm{R},\mathrm{c}})^{2} 
+ 
(\partial^{\,}_{x}\hat{\phi}^{\,}_{\mathrm{R},\mathrm{s}})^{2} 
+ 
(\mathrm{R}\rightarrow\mathrm{L}) 
\\
&\,
+\,
(\partial^{\,}_{x}\hat{\tilde{\phi}}^{\,}_{\mathrm{R},\mathrm{c}})^{2} 
+\, 
(\partial^{\,}_{x}\hat{\tilde{\phi}}^{\,}_{\mathrm{R},\mathrm{s}})^{2} 
+\, 
(\mathrm{R}\rightarrow\mathrm{L}) 
\\
&\,
+\,
(\partial^{\,}_{x}\hat{\varphi}^{\,}_{\mathrm{R},\mathrm{c}})^{2} 
+\,
(\partial^{\,}_{x}\hat{\varphi}^{\,}_{\mathrm{R},\mathrm{s}})^{2} 
+\, 
(\mathrm{R}\rightarrow\mathrm{L}) 
\\
&\,
+\,
(\partial^{\,}_{x}\hat{\tilde{\varphi}}^{\,}_{\mathrm{R},\mathrm{c}})^{2} 
+\, 
(\partial^{\,}_{x}\hat{\tilde{\varphi}}^{\,}_{\mathrm{R},\mathrm{s}})^{2} 
+\, 
(\mathrm{R}\rightarrow\mathrm{L}).
\end{split}
\label{appeq: free hamiltonian chiral basis}
\end{equation}
This noninteracting Hamiltonian density is nothing but 
four copies of the non-interacting spin-1/2 Tomonaga-Luttinger model. 
The phenomenon of spin-charge separation is manifest in the
bosonized representation of the noninteracting limit.

It is convenient to define the following non-chiral charge and spin fields
from Eq.~(\ref{appeq: def spin charge chiral boson})
\begin{subequations}
\begin{align}
\hat{\phi}^{\,}_{\mathrm{c}}:= 
\hat{\phi}^{\,}_{\mathrm{L},\mathrm{c}} 
+ 
\hat{\phi}^{\,}_{\mathrm{R},\mathrm{c}}, 
\qquad 
\hat{\phi}^{\,}_{\mathrm{s}}:= 
\hat{\phi}^{\,}_{\mathrm{L},\mathrm{s}} 
+ 
\hat{\phi}^{\,}_{\mathrm{R},\mathrm{s}},
\end{align}
together with their ``duals''
\begin{align}
\hat{\theta}^{\,}_{\mathrm{c}}:= 
\hat{\phi}^{\,}_{\mathrm{L},\mathrm{c}}
-
\hat{\phi}^{\,}_{\mathrm{R},\mathrm{c}}, 
\qquad 
\hat{\theta}^{\,}_{\mathrm{s}}:= 
\hat{\phi}^{\,}_{\mathrm{L},\mathrm{s}}
-
 \hat{\phi}^{\,}_{\mathrm{R},\mathrm{s}}.
\end{align}
\end{subequations}
They obey the equal-time commutators
\begin{equation}
\begin{split}
&
\left[ 
\hat{\phi}^{\,}_{\mathrm{c}}(x), 
\partial^{\,}_{x'}\hat{\theta}^{\,}_{\mathrm{c}}(x')
\right]= 
\mathrm{i}\delta(x-x'),
\\
&
\left[ 
\hat{\phi}^{\,}_{\mathrm{s}}(x) 
, 
\partial^{\,}_{x'}\hat{\theta}^{\,}_{\mathrm{s}}(x')
\right]= 
\mathrm{i}\delta(x-x'),
\\
&
\left[
\hat{\phi}^{\,}_{\mathrm{c}}(x), 
\hat{\phi}^{\,}_{\mathrm{s}}(x')
\right]= 
\left[
\hat{\theta}^{\,}_{\mathrm{c}}(x) 
, 
\hat{\theta}^{\,}_{\mathrm{s}}(x')
\right]= 0. 
\end{split}
\end{equation}
The spatial derivative of the dual field is the canonical conjugate
to the field. We may proceed similarly to define the six dual pairs
\begin{align}
&
\hat{\tilde{\phi}}^{\,}_{\mathrm{c}}, 
\qquad
\hat{\tilde{\theta}}^{\,}_{\mathrm{c}},
\\
&
\hat{\tilde{\phi}}^{\,}_{\mathrm{s}}, 
\qquad
\hat{\tilde{\theta}}^{\,}_{\mathrm{s}}, 
\\
&
\hat{\varphi}^{\,}_{\mathrm{c}}, 
\qquad
\hat{\vartheta}^{\,}_{\mathrm{c}},
\\
& 
\hat{\varphi}^{\,}_{\mathrm{s}}, 
\qquad
\hat{\vartheta}^{\,}_{\mathrm{s}}, 
\\
&
\hat{\tilde{\varphi}}^{\,}_{\mathrm{c}}, 
\qquad
\hat{\tilde{\vartheta}}^{\,}_{\mathrm{c}}, 
\\
&
\hat{\tilde{\varphi}}^{\,}_{\mathrm{s}},
\qquad
\hat{\tilde{\vartheta}}^{\,}_{\mathrm{s}}.
\end{align}
In terms of these charge and spin fields, we can rewrite the
noninteracting many-body Hamiltonian 
(\ref{appeq: free hamiltonian chiral basis}) as
\begin{equation}
\begin{split}
\hat{H}^{\,}_{0}=&\, 
\frac{1}{2}
\left[
(\partial^{\,}_{x}\hat{\phi}^{\,}_{\mathrm{c}})^{2} 
+ 
(\partial^{\,}_{x}\hat{\theta}^{\,}_{\mathrm{c}})^{2}  
\right]
+
\frac{1}{2}
\left[
(\partial^{\,}_{x}\hat{\phi}^{\,}_{\mathrm{s}})^{2} 
+ 
(\partial^{\,}_{x}\hat{\theta}^{\,}_{\mathrm{s}})^{2}  
\right]
\\
&
+
\frac{1}{2}
\left[
(\partial^{\,}_{x}\hat{\tilde{\phi}}^{\,}_{\mathrm{c}})^{2} 
+ 
(\partial^{\,}_{x}\hat{\tilde{\theta}}^{\,}_{\mathrm{c}})^{2}  
\right]
+
\frac{1}{2}
\left[
(\partial^{\,}_{x}\hat{\tilde{\phi}}^{\,}_{\mathrm{s}})^{2} 
+ 
(\partial^{\,}_{x}\hat{\tilde{\theta}}^{\,}_{\mathrm{s}})^{2}  
\right]
\\
&
+
\frac{1}{2}
\left[
(\partial^{\,}_{x}\hat{\varphi}^{\,}_{\mathrm{c}})^{2} 
+ 
(\partial^{\,}_{x}\hat{\vartheta}^{\,}_{\mathrm{c}})^{2}  
\right]
+
\frac{1}{2}
\left[
(\partial^{\,}_{x}\hat{\varphi}^{\,}_{\mathrm{s}})^{2} 
+ 
(\partial^{\,}_{x}\hat{\vartheta}^{\,}_{\mathrm{s}})^{2}  
\right]
\\
&
+
\frac{1}{2}
\left[
(\partial^{\,}_{x}\hat{\tilde{\varphi}}^{\,}_{\mathrm{c}})^{2} 
+ 
(\partial^{\,}_{x}\hat{\tilde{\vartheta}}^{\,}_{\mathrm{c}})^{2}  
\right]
+
\frac{1}{2}
\left[
(\partial^{\,}_{x}\hat{\tilde{\varphi}}^{\,}_{\mathrm{s}})^{2} 
+ 
(\partial^{\,}_{x}\hat{\tilde{\vartheta}}^{\,}_{\mathrm{s}})^{2}  
\right].
\end{split}
\end{equation}
This is the canonical representation of four copies
of the noninteracting spin-1/2 Tomonaga-Luttinger liquid. 

The following Abelian-bosonized representation of various electron
one-body operators is useful.$\ $%
\cite{Fradkin13} 
For any $\alpha=1,2,$ and $A=1,\cdots,k+k'$, denote with
\begin{equation}
\hat{\Psi}^{\,}_{\mathrm{R},\mathrm{f=1},\alpha,A}\equiv
\hat{\psi}^{\,}_{\mathrm{R},\alpha,A},
\qquad
\hat{\Psi}^{\,}_{\mathrm{R},\mathrm{f=2},\alpha,A}\equiv
\hat{\tilde{\psi}}^{\,}_{\mathrm{R},\alpha,A},
\end{equation}
and with
\begin{equation}
\hat{\Psi}^{\,}_{\mathrm{L},\mathrm{f=1},\alpha,A}\equiv
\hat{\psi}^{\,}_{\mathrm{L},\alpha,A},
\qquad
\hat{\Psi}^{\,}_{\mathrm{L},\mathrm{f=2},\alpha,A}\equiv
\hat{\tilde{\psi}}^{\,}_{\mathrm{L},\alpha,A},
\end{equation}
having or not having the tilde symbol.
The label $\mathrm{f}=1,2$ is a two-valued flavor
that can also be intrepreted as distinguishing 
the upper part (layer) from  the lower part (layer) in 
any domino from Fig.\
\ref{Fig: gapping with TRS}.

There are the fermionic bilinears 
\begin{subequations}
\label{appeq: bosonized density operators}
\begin{equation}
\begin{split}
\hat{O}^{\mathrm{CDW}}_{\mathrm{RL},\mathrm{f,A}}:=&\,
\sum^{\,}_{\alpha=\uparrow,\downarrow}
\hat{\Psi}^{\dag}_{\mathrm{R},\alpha,\mathrm{f},A}\,
\hat{\Psi}^{\,}_{\mathrm{L},\alpha,\mathrm{f},A}
\label{appeq: bosonized density operators a}
\\
=&\,
\frac{1}{\pi\mathfrak{a}}\,
\cos\left(\sqrt{2\pi}\,\hat{\Phi}^{\,}_{\mathrm{s},\mathrm{f},A}\right)\,
e^{-\mathrm{i}\sqrt{2\pi}\,\hat{\Phi}^{\,}_{\mathrm{c},\mathrm{f},A}}
\end{split}
\end{equation}
that encode a charge-density wave (CDW)
for any $\mathrm{f=1,2}$ and $A=1,\cdots,k+k'$.

There are the fermion bilinears
\begin{equation}
\begin{split}
\hat{O}^{(3),\mathrm{SDW}}_{\mathrm{RL},\mathrm{f,A}}:=&\,
\sum^{\,}_{\alpha,\alpha'=\uparrow,\downarrow}
\hat{\Psi}^{\dag}_{\mathrm{R},\alpha,\mathrm{f},A}\,
\sigma^{3}_{\alpha\alpha'}\,
\hat{\Psi}^{\,}_{\mathrm{L},\alpha',\mathrm{f},A},
\\
=&\,
\frac{-\mathrm{i}}{\pi\mathfrak{a}}\,
\sin\left(\sqrt{2\pi}\,\hat{\Phi}^{\,}_{\mathrm{s},\mathrm{f},A}\right)\,
e^{-\mathrm{i}\sqrt{2\pi}\,\hat{\Phi}^{\,}_{\mathrm{c},\mathrm{f},A}},
\end{split}
\label{appeq: bosonized density operators b}
\end{equation}
and
\begin{equation}
\begin{split}
\hat{O}^{(\pm),\mathrm{SDW}}_{\mathrm{RL},\mathrm{f,A}}:=&\,
\sum^{\,}_{\alpha,\alpha'=\uparrow,\downarrow}
\hat{\Psi}^{\dag}_{\mathrm{R},\alpha,\mathrm{f},A}\,
\sigma^{\pm}_{\alpha\alpha'}\,
\hat{\Psi}^{\,}_{\mathrm{L},\alpha',\mathrm{f},A},
\\
=&\,
\frac{1}{\pi\mathfrak{a}}\,
e^{\pm\mathrm{i}\sqrt{2\pi}\,\hat{\Theta}^{\,}_{\mathrm{s},\mathrm{f},A}}\,
e^{-\mathrm{i}\sqrt{2\pi}\,\hat{\Phi}^{\,}_{\mathrm{c},\mathrm{f},A}},
\end{split}
\label{appeq: bosonized density operators c}
\end{equation}
that encode a spin-density wave (SDW)
for any $\mathrm{f=1,2}$ and $A=1,\cdots,k+k'$.
For any $\mathrm{f}=1,2$ and $A=1,2$ (i.e., $k=k'=1$),
the pair of bosonic fields entering the trigonometric functions 
on any one of the lines
(\ref{appeq: bosonized density operators a}),
(\ref{appeq: bosonized density operators b}),
and
(\ref{appeq: bosonized density operators c})
is unique. This pair is to be chosen from
\begin{equation}
\begin{split}
&
\hat{\Phi}^{\,}_{\mathrm{c},\mathrm{f},A}\in
\{
\hat{\phi}^{\,}_{\mathrm{c}}, 
\hat{\tilde{\phi}}^{\,}_{\mathrm{c}}, 
\hat{\varphi}^{\,}_{\mathrm{c}}, 
\hat{\tilde{\varphi}}^{\,}_{\mathrm{c}}
\}, 
\\
&
\hat{\Phi}^{\,}_{\mathrm{s},\mathrm{f},A}\in
\{
\hat{\phi}^{\,}_{\mathrm{s}}, 
\hat{\tilde{\phi}}^{\,}_{\mathrm{s}}, 
\hat{\varphi}^{\,}_{\mathrm{s}}, 
\hat{\tilde{\varphi}}^{\,}_{\mathrm{s}}
\},
\\
&
\hat{\Theta}^{\,}_{\mathrm{c},\mathrm{f},A}\in 
\{
\hat{\theta}^{\,}_{\mathrm{c}}, 
\hat{\tilde{\theta}}^{\,}_{\mathrm{c}}, 
\hat{\vartheta}^{\,}_{\mathrm{c}}, 
\hat{\tilde{\vartheta}}^{\,}_{\mathrm{c}}
\},
\\
&
\hat{\Theta}^{\,}_{\mathrm{s},\mathrm{f},A}\in 
\{
\hat{\theta}^{\,}_{\mathrm{s}}, 
\hat{\tilde{\theta}}^{\,}_{\mathrm{s}}, 
\hat{\vartheta}^{\,}_{\mathrm{s}}, 
\hat{\tilde{\vartheta}}^{\,}_{\mathrm{s}}
\},
\end{split}
\label{appeq: bosonized density operators d}
\end{equation}
\end{subequations}
with the rule that
the choice 
$\hat{\Phi}^{\,}_{\mathrm{s},\mathrm{f},A}=
\hat{\phi}^{\,}_{\mathrm{s}}$
implies that 
$\hat{\Phi}^{\,}_{\mathrm{c},\mathrm{f},A}=
\hat{\tilde{\phi}}^{\,}_{\mathrm{c}}$
for Eq.\ (\ref{appeq: bosonized density operators a}), 
say.

The Abelian bosonization of the twelve mass matrices 
(\ref{appeq: TRS SU(k+k') singlet mass matrices})
can be organized as follows.

\noindent
(1) There are two CDW mass matrices
\begin{subequations}
\begin{equation}
M^{\,}_{1000},\qquad M^{\,}_{2000}.
\end{equation}

\noindent
(2) There is the triplet of SDW mass matrices
\begin{equation}
M^{\,}_{1330},\qquad M^{\,}_{1130},\qquad M^{\,}_{1230},
\end{equation}
obeying the spin-1/2 algebra.

\noindent
(3) There is the triplet of SDW mass matrices
\begin{equation}
M^{\,}_{2330},\qquad M^{\,}_{2130},\qquad M^{\,}_{2230},
\end{equation}
obeying the spin-1/2 algebra.

\noindent
(4) There is the pair of layer-mixing mass matrices
\begin{equation}
M^{\,}_{1010},\qquad M^{\,}_{1020},
\end{equation}
obeying the raising and lowering $SU(2)$ algebra.

\noindent
(5) There is a second pair of layer-mixing mass matrices
\begin{equation}
M^{\,}_{2010},\qquad M^{\,}_{2020},
\end{equation}
\end{subequations}
obeying the raising and lowering $SU(2)$ algebra.

\vskip 90 true pt

\begin{widetext}
The most general mass Hamiltonian density is
\begin{equation}
\begin{split}
\hat{H}^{\,}_{\mathrm{mass}}:=&\,
\sum_{\mu^{\,}_{1}=1}^{2}
\sum_{\mu^{\,}_{2}=0}^{3}
\sum_{\mu^{\,}_{3}=0}^{3}
\sum_{\mathrm{c}=0}^{(k+k')^{2}-1}
m^{\,}_{\mu^{\,}_{1}\mu^{\,}_{2}\mu^{\,}_{3}\mathrm{c}}\,
\hat{\Psi}^{\dag}_{\eta,\alpha,\mathrm{f},A}\,
(\tau^{\,}_{\mu^{\,}_{1}})^{\,}_{\eta\eta'}\,
(\sigma^{\,}_{\mu^{\,}_{2}})^{\,}_{\alpha\alpha'}\,
(\rho^{\,}_{\mu^{\,}_{3}})^{\,}_{\mathrm{f}\mathrm{f}'}\,
(T^{\,}_{\mathrm{c}})^{\,}_{AA'}\,
\hat{\Psi}^{\,}_{\eta',\alpha',\mathrm{f}',A'},
\end{split}
\end{equation}
where $m^{\,}_{\mu^{\,}_{1}\mu^{\,}_{2}\mu^{\,}_{3}\mathrm{c}}$ is a real number for any
$\mu^{\,}_{1}=1,2$, 
$\mu^{\,}_{2},\mu^{\,}_{3}=0,1,2,3$,
and $\mathrm{c}=0,\cdots,(k+k')^{2}-1$ 
and the summation convention
over the repeated indices 
$\eta,\eta'=1,2$, 
$\alpha,\alpha'=1,2$, 
$\mathrm{f},\mathrm{f}'=1,2$,
and $A,A'=1,\cdots,k+k'$ is implied.
Time-reversal symmetry restrict the nonvanishing masses 
that are $SU(k+k')$ singlet to be
the ones with the real-valued couplings
\begin{subequations}
\begin{equation}
m^{\,}_{1000},
\qquad
m^{\,}_{1330},
\qquad
m^{\,}_{1130},
\qquad
m^{\,}_{1230},
\qquad
m^{\,}_{1010},
\qquad
m^{\,}_{1020},
\end{equation}
and
\begin{equation}
m^{\,}_{2000},
\qquad
m^{\,}_{2230},
\qquad
m^{\,}_{2330},
\qquad
m^{\,}_{2130},
\qquad
m^{\,}_{2010},
\qquad
m^{\,}_{2020}.
\end{equation}
\end{subequations}
The most general mass Hamiltonian density 
that preserves time-reversal symmetry is thus
\begin{subequations}
\label{eq: most general mass Hamiltonian if TRS and k =k'-1}
\begin{equation}
\begin{split}
\hat{H}^{\mathrm{TRS}}_{\mathrm{mass}}=&\,
m^{\,}_{1000}\,
\hat{H}^{\,}_{1000}
+
m^{\,}_{1330}\,
\hat{H}^{\,}_{1330}
+
m^{\,}_{1130}\,
\hat{H}^{\,}_{1130}
+
m^{\,}_{1230}\,
\hat{H}^{\,}_{1230}
+
m^{\,}_{1010}\,
\hat{H}^{\,}_{1010}
+
m^{\,}_{1020}\,
\hat{H}^{\,}_{1020}
\\
&\,
+
m^{\,}_{2000}\,
\hat{H}^{\,}_{2000}
+
m^{\,}_{2330}\,
\hat{H}^{\,}_{2330}
+
m^{\,}_{2130}\,
\hat{H}^{\,}_{2130}
+
m^{\,}_{2230}\,
\hat{H}^{\,}_{2230}
+
m^{\,}_{2010}\,
\hat{H}^{\,}_{2010}
+
m^{\,}_{2020}\,
\hat{H}^{\,}_{2020}.
\label{eq: most general mass Hamiltonian if TRS and k =k'-1 a}
\end{split}
\end{equation}
The eight contributions 
\begin{equation}
\begin{split}
\hat{H}^{\,}_{1000}:=
\frac{+2}{\pi\mathfrak{a}}\,
\Bigl[&\,
\cos(\sqrt{2\pi}\,\hat{\phi}^{\,}_{\mathrm{s}})\,
\cos(\sqrt{2\pi}\,\hat{\phi}^{\,}_{\mathrm{c}})
+
\cos(\sqrt{2\pi}\,\hat{\varphi}^{\,}_{\mathrm{s}})\,
\cos(\sqrt{2\pi}\,\hat{\varphi}^{\,}_{\mathrm{c}})
\\
&\,
+
\cos(\sqrt{2\pi}\,\hat{\tilde{\phi}}^{\,}_{\mathrm{s}})\,
\cos(\sqrt{2\pi}\,\hat{\tilde{\phi}}^{\,}_{\mathrm{c}})
+
\cos(\sqrt{2\pi}\,\hat{\tilde{\varphi}}^{\,}_{\mathrm{s}})\,
\cos(\sqrt{2\pi}\,\hat{\tilde{\varphi}}^{\,}_{\mathrm{c}})
\Bigr],
\end{split}
\label{eq: most general mass Hamiltonian if TRS and k =k'-1 b}
\end{equation}
\begin{equation}
\begin{split}
\hat{H}^{\,}_{2000}:=
\frac{-2}{\pi\mathfrak{a}}\,
\Bigl[&\,
\cos(\sqrt{2\pi}\,\hat{\phi}^{\,}_{\mathrm{s}})\,
\sin(\sqrt{2\pi}\,\hat{\phi}^{\,}_{\mathrm{c}})
+
\cos(\sqrt{2\pi}\,\hat{\varphi}^{\,}_{\mathrm{s}})\,
\sin(\sqrt{2\pi}\,\hat{\varphi}^{\,}_{\mathrm{c}})
\\
&\,
+
\cos(\sqrt{2\pi}\,\hat{\tilde{\phi}}^{\,}_{\mathrm{s}})\,
\sin(\sqrt{2\pi}\,\hat{\tilde{\phi}}^{\,}_{\mathrm{c}})
+
\cos(\sqrt{2\pi}\,\hat{\tilde{\varphi}}^{\,}_{\mathrm{s}})\,
\sin(\sqrt{2\pi}\,\hat{\tilde{\varphi}}^{\,}_{\mathrm{c}})
\Bigr],
\end{split}
\label{eq: most general mass Hamiltonian if TRS and k =k'-1 c}
\end{equation}
\begin{equation}
\begin{split}
\hat{H}^{\,}_{1330}:=
\frac{-2}{\pi\mathfrak{a}}\,
\Bigl[&\,
\sin(\sqrt{2\pi}\,\hat{\phi}^{\,}_{\mathrm{s}})\,
\sin(\sqrt{2\pi}\,\hat{\phi}^{\,}_{\mathrm{c}})
+
\sin(\sqrt{2\pi}\,\hat{\varphi}^{\,}_{\mathrm{s}})\,
\sin(\sqrt{2\pi}\,\hat{\varphi}^{\,}_{\mathrm{c}})
\\
&\,
-
\sin(\sqrt{2\pi}\,\hat{\tilde{\phi}}^{\,}_{\mathrm{s}})\,
\sin(\sqrt{2\pi}\,\hat{\tilde{\phi}}^{\,}_{\mathrm{c}})
-
\sin(\sqrt{2\pi}\,\hat{\tilde{\varphi}}^{\,}_{\mathrm{s}})\,
\sin(\sqrt{2\pi}\,\hat{\tilde{\varphi}}^{\,}_{\mathrm{c}})
\Bigr],
\end{split}
\label{eq: most general mass Hamiltonian if TRS and k =k'-1d}
\end{equation}
\begin{equation}
\begin{split}
\hat{H}^{\,}_{2330}:=
\frac{-2}{\pi\mathfrak{a}}\,
\Bigl[&\,
\sin(\sqrt{2\pi}\,\hat{\phi}^{\,}_{\mathrm{s}})\,
\cos(\sqrt{2\pi}\,\hat{\phi}^{\,}_{\mathrm{c}})
+
\sin(\sqrt{2\pi}\,\hat{\varphi}^{\,}_{\mathrm{s}})\,
\cos(\sqrt{2\pi}\,\hat{\varphi}^{\,}_{\mathrm{c}})
\\
&\,
-
\sin(\sqrt{2\pi}\,\hat{\tilde{\phi}}^{\,}_{\mathrm{s}})\,
\cos(\sqrt{2\pi}\,\hat{\tilde{\phi}}^{\,}_{\mathrm{c}})
-
\sin(\sqrt{2\pi}\,\hat{\tilde{\varphi}}^{\,}_{\mathrm{s}})\,
\cos(\sqrt{2\pi}\,\hat{\tilde{\varphi}}^{\,}_{\mathrm{c}})
\Bigr],
\end{split}
\label{eq: most general mass Hamiltonian if TRS and k =k'-1 e}
\end{equation}
\begin{equation}
\begin{split}
\hat{H}^{\,}_{1130}:=
\frac{+2}{\pi\mathfrak{a}}\,
\Bigl[&\,
\cos(\sqrt{2\pi}\,\hat{\theta}^{\,}_{\mathrm{s}})\,
\cos(\sqrt{2\pi}\,\hat{\phi}^{\,}_{\mathrm{c}})
+
\cos(\sqrt{2\pi}\,\hat{\vartheta}^{\,}_{\mathrm{s}})\,
\cos(\sqrt{2\pi}\,\hat{\varphi}^{\,}_{\mathrm{c}})
\\
&\,
-
\cos(\sqrt{2\pi}\,\hat{\tilde{\theta}}^{\,}_{\mathrm{s}})\,
\cos(\sqrt{2\pi}\,\hat{\tilde{\phi}}^{\,}_{\mathrm{c}})
-
\cos(\sqrt{2\pi}\,\hat{\tilde{\vartheta}}^{\,}_{\mathrm{s}})\,
\cos(\sqrt{2\pi}\,\hat{\tilde{\varphi}}^{\,}_{\mathrm{c}})
\Bigr],
\end{split}
\label{eq: most general mass Hamiltonian if TRS and k =k'-1 f}
\end{equation}
\begin{equation}
\begin{split}
\hat{H}^{\,}_{1230}:=
\frac{+2}{\pi\mathfrak{a}}\,
\Bigl[&\,
\sin(\sqrt{2\pi}\,\hat{\theta}^{\,}_{\mathrm{s}})\,
\cos(\sqrt{2\pi}\,\hat{\phi}^{\,}_{\mathrm{c}})
+
\sin(\sqrt{2\pi}\,\hat{\vartheta}^{\,}_{\mathrm{s}})\,
\cos(\sqrt{2\pi}\,\hat{\varphi}^{\,}_{\mathrm{c}})
\\
&\,
-
\sin(\sqrt{2\pi}\,\hat{\tilde{\theta}}^{\,}_{\mathrm{s}})\,
\cos(\sqrt{2\pi}\,\hat{\tilde{\phi}}^{\,}_{\mathrm{c}})
-
\sin(\sqrt{2\pi}\,\hat{\tilde{\vartheta}}^{\,}_{\mathrm{s}})\,
\cos(\sqrt{2\pi}\,\hat{\tilde{\varphi}}^{\,}_{\mathrm{c}})
\Bigr],
\end{split}
\label{eq: most general mass Hamiltonian if TRS and k =k'-1 g}
\end{equation}
\begin{equation}
\begin{split}
\hat{H}^{\,}_{2130}:=
\frac{-2}{\pi\mathfrak{a}}\,
\Bigl[&\,
\cos(\sqrt{2\pi}\,\hat{\theta}^{\,}_{\mathrm{s}})\,
\sin(\sqrt{2\pi}\,\hat{\phi}^{\,}_{\mathrm{c}})
+
\cos(\sqrt{2\pi}\,\hat{\vartheta}^{\,}_{\mathrm{s}})\,
\sin(\sqrt{2\pi}\,\hat{\varphi}^{\,}_{\mathrm{c}})
\\
&\,
-
\cos(\sqrt{2\pi}\,\hat{\tilde{\theta}}^{\,}_{\mathrm{s}})\,
\sin(\sqrt{2\pi}\,\hat{\tilde{\phi}}^{\,}_{\mathrm{c}})
-
\cos(\sqrt{2\pi}\,\hat{\tilde{\vartheta}}^{\,}_{\mathrm{s}})\,
\sin(\sqrt{2\pi}\,\hat{\tilde{\varphi}}^{\,}_{\mathrm{c}})
\Bigr],
\end{split}
\label{eq: most general mass Hamiltonian if TRS and k =k'-1 h}
\end{equation}
and
\begin{equation}
\begin{split}
\hat{H}^{\,}_{2230}:=
\frac{-2}{\pi\mathfrak{a}}\,
\Bigl[&\,
\sin(\sqrt{2\pi}\,\hat{\theta}^{\,}_{\mathrm{s}})\,
\sin(\sqrt{2\pi}\,\hat{\phi}^{\,}_{\mathrm{c}})
+
\sin(\sqrt{2\pi}\,\hat{\vartheta}^{\,}_{\mathrm{s}})\,
\sin(\sqrt{2\pi}\,\hat{\varphi}^{\,}_{\mathrm{c}})
\\
&\,
-
\sin(\sqrt{2\pi}\,\hat{\tilde{\theta}}^{\,}_{\mathrm{s}})\,
\sin(\sqrt{2\pi}\,\hat{\tilde{\phi}}^{\,}_{\mathrm{c}})
-
\sin(\sqrt{2\pi}\,\hat{\tilde{\vartheta}}^{\,}_{\mathrm{s}})\,
\sin(\sqrt{2\pi}\,\hat{\tilde{\varphi}}^{\,}_{\mathrm{c}})
\Bigr],
\label{eq: most general mass Hamiltonian if TRS and k =k'-1 i}
\end{split}
\end{equation}
are diagonal in the flavor index $\mathrm{f}=1,2$, i.e.,
they do not mix the bosonic fields without and with the symbol tilde.
Any one of these eight contributions couple the charge to the spin sectors.
Each cosine carries the scaling dimension one in the sector of the theory
on which it acts. Consequently, any one of these eight contributions
gap the noninteracting critical theory with the four independent 
critical sectors, each of which carries the central charge two.
The remaining four contributions
\begin{equation}
\begin{split}
\hat{H}^{\,}_{1010}:=&\,
\frac{+2}{\pi\mathfrak{a}}\,
\Bigl[
\cos
\Big(
\sqrt{2\pi}
(\hat{\phi}^{\,}_{\mathrm{R},\mathrm{s}}+\hat{\tilde{\phi}}_{\mathrm{L},\mathrm{s}})
\Big)\,
\cos
\Big(
\sqrt{2\pi}
(\hat{\phi}^{\,}_{\mathrm{R},\mathrm{c}}+\hat{\tilde{\phi}}_{\mathrm{L},\mathrm{c}})
\Big)
+
\cos
\Big(
\sqrt{2\pi}
(\hat{\tilde{\phi}}_{\mathrm{R},\mathrm{s}}
+
\hat{\phi}^{\,}_{\mathrm{L},\mathrm{s}})
\Big)\,
\cos
\Big(
\sqrt{2\pi}
(\hat{\tilde{\phi}}_{\mathrm{R},\mathrm{c}}+\hat{\phi}^{\,}_{\mathrm{L},\mathrm{c}})
\Big)
\\
&
+
\cos
\Big(
\sqrt{2\pi}
(\hat{\varphi}^{\,}_{\mathrm{R},\mathrm{s}}+\hat{\tilde{\varphi}}_{\mathrm{L},\mathrm{s}})
\Big)
\,
\cos
\Big(
\sqrt{2\pi}
(\hat{\varphi}^{\,}_{\mathrm{R},\mathrm{c}}+\hat{\tilde{\varphi}}_{\mathrm{L},\mathrm{c}})
\Big)
+
\cos
\Big(
\sqrt{2\pi}
(\hat{\tilde{\varphi}}_{\mathrm{R},\mathrm{s}}+\hat{\varphi}^{\,}_{\mathrm{L},\mathrm{s}})
\Big)
\,
\cos
\Big(
\sqrt{2\pi}
(\hat{\tilde{\varphi}}_{\mathrm{R},\mathrm{c}}+\hat{\varphi}^{\,}_{\mathrm{L},\mathrm{c}})
\Big)
\Bigr],
\end{split}
\label{eq: most general mass Hamiltonian if TRS and k =k'-1 j}
\end{equation}
\begin{equation}
\begin{split}
\hat{H}^{\,}_{1020}:=&\,
\frac{-2}{\pi\mathfrak{a}}\,
\Bigl[
\cos
\Big(
\sqrt{2\pi}
(\hat{\phi}^{\,}_{\mathrm{R},\mathrm{s}}+\hat{\tilde{\phi}}_{\mathrm{L},\mathrm{s}})
\Big)
\,
\sin
\Big(
\sqrt{2\pi}
(\hat{\phi}^{\,}_{\mathrm{R},\mathrm{c}}+\hat{\tilde{\phi}}_{\mathrm{L},\mathrm{c}})
\Big)
-
\cos
\Big(
\sqrt{2\pi}
(\hat{\tilde{\phi}}_{\mathrm{R},\mathrm{s}}+\hat{\phi}^{\,}_{\mathrm{L},\mathrm{s}})
\Big)
\,
\sin
\Big(
\sqrt{2\pi}
(\hat{\tilde{\phi}}_{\mathrm{R},\mathrm{c}}+\hat{\phi}^{\,}_{\mathrm{L},\mathrm{c}})
\Big)
\\
&
+
\cos
\Big(
\sqrt{2\pi}
(\hat{\varphi}^{\,}_{\mathrm{R},\mathrm{s}}+\hat{\tilde{\varphi}}_{\mathrm{L},\mathrm{s}})
\Big)
\,
\sin
\Big(
\sqrt{2\pi}
(\hat{\varphi}^{\,}_{\mathrm{R},\mathrm{c}}+\hat{\tilde{\varphi}}_{\mathrm{L},\mathrm{c}})
\Big)
-
\cos
\Big(
\sqrt{2\pi}
(\hat{\tilde{\varphi}}_{\mathrm{R},\mathrm{s}}+\hat{\varphi}^{\,}_{\mathrm{L},\mathrm{s}})
\Big)
\,
\sin
\Big(
\sqrt{2\pi}
(\hat{\tilde{\varphi}}_{\mathrm{R},\mathrm{c}}+\hat{\varphi}^{\,}_{\mathrm{L},\mathrm{c}})
\Big)
\Bigr],
\end{split}
\label{eq: most general mass Hamiltonian if TRS and k =k'-1 k}
\end{equation}
\begin{equation}
\begin{split}
\hat{H}^{\,}_{2010}:=&\,
\frac{-2}{\pi\mathfrak{a}}\,
\Bigl[
\cos
\Big(
\sqrt{2\pi}
(\hat{\phi}^{\,}_{\mathrm{R},\mathrm{s}}+\hat{\tilde{\phi}}_{\mathrm{L},\mathrm{s}})
\Big)
\,
\sin
\Big(
\sqrt{2\pi}
(\hat{\phi}^{\,}_{\mathrm{R},\mathrm{c}}+\hat{\tilde{\phi}}_{\mathrm{L},\mathrm{c}})
\Big)
+
\cos
\Big(
\sqrt{2\pi}
(\hat{\tilde{\phi}}_{\mathrm{R},\mathrm{s}}+\hat{\phi}^{\,}_{\mathrm{L},\mathrm{s}})
\Big)
\,
\sin
\Big(
\sqrt{2\pi}
(\hat{\tilde{\phi}}_{\mathrm{R},\mathrm{c}}+\hat{\phi}^{\,}_{\mathrm{L},\mathrm{c}})
\Big)
\\
&
+
\cos
\Big(
\sqrt{2\pi}
(\hat{\varphi}^{\,}_{\mathrm{R},\mathrm{s}}+\hat{\tilde{\varphi}}_{\mathrm{L},\mathrm{s}})
\Big)
\,
\sin
\Big(
\sqrt{2\pi}
(\hat{\varphi}^{\,}_{\mathrm{R},\mathrm{c}}+\hat{\tilde{\varphi}}_{\mathrm{L},\mathrm{c}})
\Big)
+
\cos
\Big(
\sqrt{2\pi}
(\hat{\tilde{\varphi}}_{\mathrm{R},\mathrm{s}}+\hat{\varphi}^{\,}_{\mathrm{L},\mathrm{s}})
\Big)
\,
\sin
\Big(
\sqrt{2\pi}
(\hat{\tilde{\varphi}}_{\mathrm{R},\mathrm{c}}+\hat{\varphi}^{\,}_{\mathrm{L},\mathrm{c}})
\Big)
\Bigr],
\end{split}
\label{eq: most general mass Hamiltonian if TRS and k =k'-1 i l}
\end{equation}
and
\begin{equation}
\begin{split}
\hat{H}^{\,}_{2020}:=&\,
\frac{+2}{\pi\mathfrak{a}}\,
\Bigl[
\cos
\Big(
\sqrt{2\pi}
(\hat{\phi}^{\,}_{\mathrm{R},\mathrm{s}}+\hat{\tilde{\phi}}_{\mathrm{L},\mathrm{s}})
\Big)
\,
\cos
\Big(
\sqrt{2\pi}
(\hat{\phi}^{\,}_{\mathrm{R},\mathrm{c}}+\hat{\tilde{\phi}}_{\mathrm{L},\mathrm{c}})
\Big)
-
\cos
\Big(
\sqrt{2\pi}
(\hat{\tilde{\phi}}_{\mathrm{R},\mathrm{s}}+\hat{\phi}^{\,}_{\mathrm{L},\mathrm{s}})
\Big)
\,
\cos
\Big(
\sqrt{2\pi}
(\hat{\tilde{\phi}}_{\mathrm{R},\mathrm{c}}+\hat{\phi}^{\,}_{\mathrm{L},\mathrm{c}})
\Big)
\\
&
+
\cos
\Big(
\sqrt{2\pi}
(\hat{\varphi}^{\,}_{\mathrm{R},\mathrm{s}}+\hat{\tilde{\varphi}}_{\mathrm{L},\mathrm{s}})
\Big)
\,
\cos
\Big(
\sqrt{2\pi}
(\hat{\varphi}^{\,}_{\mathrm{R},\mathrm{c}}+\hat{\tilde{\varphi}}_{\mathrm{L},\mathrm{c}})
\Big)
-
\cos
\Big(
\sqrt{2\pi}
(\hat{\tilde{\varphi}}_{\mathrm{R},\mathrm{s}}+\hat{\varphi}^{\,}_{\mathrm{L},\mathrm{s}})
\Big)
\,
\cos
\Big(
\sqrt{2\pi}
(\hat{\tilde{\varphi}}_{\mathrm{R},\mathrm{c}}+\hat{\varphi}^{\,}_{\mathrm{L},\mathrm{c}})
\Big)
\Bigr],    
\end{split}
\label{eq: most general mass Hamiltonian if TRS and k =k'-1 m}
\end{equation}
\end{subequations}
couple the spin and charge sectors as well as the sector
$U(2k)\times U(2k')$
with the sector
$\tilde{U}(2k)\times\tilde{U}(2k')$, where we recall that $k=k'=1$.

The strongly interacting
critical theory with the central charge
(\ref{eq: nonchiral central charge at one boundary case II})
results from adding non-Abelian current-current interaction between
the sector $SU(2k)\times SU(2k')$ and the sector
$\widetilde{SU}(2k)\times\widetilde{SU}(2k')$
to the noninteracting critical theory with central charge $c=8$.
Although we do not know how to represent 
the strongly interacting
critical theory with the central charge
(\ref{eq: nonchiral central charge at one boundary case II})
by a local Hamiltonian density, we may safely infer that
the projection of any one of the eight contributions
(\ref{eq: most general mass Hamiltonian if TRS and k =k'-1 b})%
--(\ref{eq: most general mass Hamiltonian if TRS and k =k'-1 i})
onto the strongly interacting
critical theory with the central charge
(\ref{eq: nonchiral central charge at one boundary case II})
must be vanishing, since movers with opposite chiralities
carry dictinct eigenvalues of the generator $X^{\,}_{0030}$.
This argument fails for any one of the four contributions
(\ref{eq: most general mass Hamiltonian if TRS and k =k'-1 j})%
--(\ref{eq: most general mass Hamiltonian if TRS and k =k'-1 m}), 
since the eigenvalues of the generator $X^{\,}_{0030}$
are not anymore good quantum numbers. Imposing the $U(1)$
symmetry generated by $X^{\,}_{0030}$ forbids
the presence of the four contributions
(\ref{eq: most general mass Hamiltonian if TRS and k =k'-1 j})%
--(\ref{eq: most general mass Hamiltonian if TRS and k =k'-1 m}).
Imposing the $U(1)$ symmetry generated by $X^{\,}_{0030}$
guarantees the stability of the strongly interacting
critical theory with the central charge
(\ref{eq: nonchiral central charge at one boundary case II})
to the eight one-body masses
(\ref{eq: most general mass Hamiltonian if TRS and k =k'-1 b})%
--(\ref{eq: most general mass Hamiltonian if TRS and k =k'-1 i}).

We close this Appendix by representing the transformation law
(\ref{eq: def U(1) symmetry between no tilde and tilde sectors}) 
on the chiral bosonic fields introduced in
Eq.\  (\ref{appeq: chiral bosonization rule}).
They are
\begin{subequations}
\label{eppeq: def U(1) symmetry between bosonic fields}
\begin{equation}
\begin{split}
&
\hat{\phi}^{\,}_{\mathrm{R},\alpha,A} \,\mapsto\, 
\hat{\phi}^{\,}_{\mathrm{R},\alpha,A} + \frac{1}{\sqrt{4\pi}}\,\theta,
\qquad
\hat{\tilde{\phi}}^{\,}_{\mathrm{R},\alpha,A} \,\mapsto\, 
\hat{\tilde{\phi}}^{\,}_{\mathrm{R},\alpha,A} - \frac{1}{\sqrt{4\pi}}\,\theta,
\end{split}
\end{equation}
for the right-moving bosonic fields and 
\begin{equation}
\begin{split}
&
\hat{\phi}^{\,}_{\mathrm{L},\alpha,A}\,
\mapsto\, 
\hat{\phi}^{\,}_{\mathrm{L},\alpha,A} 
- 
\frac{1}{\sqrt{4\pi}}\,\theta,
\qquad
\hat{\tilde{\phi}}^{\,}_{\mathrm{L},\alpha,A}\,
\mapsto\, 
\hat{\tilde{\phi}}^{\,}_{\mathrm{L},\alpha,A} 
+ 
\frac{1}{\sqrt{4\pi}}\,\theta,
\end{split}
\end{equation}
\end{subequations}
for the left moving bosonic fields.
Here, $0\leq\theta<2\pi$.
One verifies the following facts.

First, the eight one-body masses
(\ref{eq: most general mass Hamiltonian if TRS and k =k'-1 b})%
--(\ref{eq: most general mass Hamiltonian if TRS and k =k'-1 i})
are invariant under the transformation law
(\ref{eppeq: def U(1) symmetry between bosonic fields}).

Second, all trigonometric functions depending on the
bosonic fields with the spin label that enter the four
one-body masses
(\ref{eq: most general mass Hamiltonian if TRS and k =k'-1 j})%
--(\ref{eq: most general mass Hamiltonian if TRS and k =k'-1 m})
are invariant under the transformation law
(\ref{eppeq: def U(1) symmetry between bosonic fields}).
For example, the function
\begin{align}
\cos
\Big(
\sqrt{2\pi}
(\hat{\phi}^{\,}_{\mathrm{R},\mathrm{s}}+\hat{\tilde{\phi}}_{\mathrm{L},\mathrm{s}}) 
\Big)= 
\cos
\Big(
\sqrt{\pi}
(
\hat{\phi}^{\,}_{\mathrm{R},\uparrow} 
- 
\hat{\phi}^{\,}_{\mathrm{R},\downarrow} 
+ 
\hat{\tilde{\phi}}^{\,}_{\mathrm{L},\uparrow} 
- 
\hat{\tilde{\phi}}^{\,}_{\mathrm{L},\downarrow}
) 
\Big)
\,\mapsto\,
\cos
\Big(
\sqrt{2\pi}\,
(\hat{\phi}^{\,}_{\mathrm{R},\mathrm{s}}+\hat{\tilde{\phi}}_{\mathrm{L},\mathrm{s}})
\Big) 
\end{align}
from $\hat{H}^{\,}_{1010}$ is unchanged under the transformation law
(\ref{eppeq: def U(1) symmetry between bosonic fields})
for arbitrary $0\leq\theta<2\pi$.

Third, all trigonometric functions depending on the
bosonic fields with the charge label that enter 
the four one-body masses
(\ref{eq: most general mass Hamiltonian if TRS and k =k'-1 j})%
--(\ref{eq: most general mass Hamiltonian if TRS and k =k'-1 m})
are not invariant under the transformation law
(\ref{eppeq: def U(1) symmetry between bosonic fields}).
For example, the function
\begin{align}
\cos
\Big(
\sqrt{2\pi}\,
(\hat{\phi}^{\,}_{\mathrm{R},\mathrm{c}}+\hat{\tilde{\phi}}_{\mathrm{L},\mathrm{c}}) 
\Big)= 
\cos
\Big(
\sqrt{\pi}\,
(
\hat{\phi}^{\,}_{\mathrm{R},\uparrow} 
+ 
\hat{\phi}^{\,}_{\mathrm{R},\downarrow} 
+ 
\hat{\tilde{\phi}}^{\,}_{\mathrm{L},\uparrow} 
+ 
\hat{\tilde{\phi}}^{\,}_{\mathrm{L},\downarrow}
)
\Big) 
\,\mapsto\,
\cos
\Big(
\sqrt{2\pi}\,
(
\hat{\phi}^{\,}_{\mathrm{R},\mathrm{s}}
+
\hat{\tilde{\phi}}_{\mathrm{L},\mathrm{s}})
+
2\theta
\Big)
\end{align}
from $\hat{H}^{\,}_{1010}$ 
is not invariant for arbitrary $0\leq\theta<2\pi$.  Moreover, the
right-hand side does not match any one of the trigonometric functions
entering the four one-body masses 
(\ref{eq: most general mass Hamiltonian if TRS and k =k'-1 j})%
--(\ref{eq: most general mass Hamiltonian if TRS and k =k'-1 m}).  

This is why the $U(1)$ symmetry 
(\ref{eppeq: def U(1) symmetry between bosonic fields})
suffices to prevent the layer-mixing masses
from gapping the strongly interacting
critical theory with the central charge
(\ref{eq: nonchiral central charge at one boundary case II}).

\end{widetext}

\bibliography{bib-non-abelian-wires}


\end{document}